\documentclass[preprint]{aastex62}
\usepackage{multirow}
\usepackage[numbib]{tocbibind}
\usepackage{amsmath,amsthm,amssymb}
\usepackage{longtable}

\newcommand{\adind}{\gamma_{\text{ad}}}

\begin{document}

\title{THOR 2.0: Major Improvements to the Open-Source General Circulation Model}
\author{Russell Deitrick}
\affil{Center for Space and Habitability, University of Bern, Gesellschaftsstrasse 6, CH-3012, Bern, Switzerland}
\email{russell.deitrick@csh.unibe.ch}
\author{Jo\~{a}o M. Mendon\c{c}a}
\affil{Astrophysics and Atmospheric Physics, National Space Institute, Technical University of Denmark, Elektrovej, 2800, Kgs. Lyngby, Denmark}
\author{Urs Schroffenegger}
\affil{Center for Space and Habitability, University of Bern, Gesellschaftsstrasse 6, CH-3012, Bern, Switzerland}
\author{Simon L. Grimm}
\affil{Center for Space and Habitability, University of Bern, Gesellschaftsstrasse 6, CH-3012, Bern, Switzerland}
\author{Shang-Min Tsai}
\affil{Atmospheric, Ocean, and Planetary Physics, Department of Physics, University of Oxford, OX1 3PU, UK}
\author{Kevin Heng}
\affil{Center for Space and Habitability, University of Bern, Gesellschaftsstrasse 6, CH-3012, Bern, Switzerland}
\date{}

\begin{abstract}
\texttt{THOR} is the first open-source general circulation model (GCM) developed from scratch to study the atmospheres and climates of exoplanets, free from Earth- or Solar System-centric tunings. It solves the general non-hydrostatic Euler equations (instead of the primitive equations) on a sphere using the icosahedral grid. In the current study, we report major upgrades to \texttt{THOR}, building upon the work of \cite{Mendonca2016}. First, while the Horizontally Explicit Vertically Implicit (HEVI) integration scheme is the same as that described in \cite{Mendonca2016}, we provide a clearer description of the scheme and improved its implementation in the code. The differences in implementation between the hydrostatic shallow (HSS), quasi-hydrostatic deep (QHD) and non-hydrostatic deep (NHD) treatments are fully detailed. Second, standard physics modules are added: two-stream, double-gray radiative transfer and dry convective adjustment. Third, \texttt{THOR} is tested on additional benchmarks: tidally-locked Earth, deep hot Jupiter, acoustic wave, and gravity wave. Fourth, we report that differences between the hydrostatic and non-hydrostatic simulations are negligible in the Earth case, but pronounced in the hot Jupiter case. Finally, the effects of the so-called ``sponge layer'', a form of drag implemented in most GCMs to provide numerical stability, are examined. Overall, these upgrades have improved the flexibility, user-friendliness, and stability of \texttt{THOR}.
\end{abstract}

\section{Introduction}

\subsection{The atmospheric circulation of exoplanets}
With new technology and data analysis techniques, we are entering an era in which three-dimensional models of exoplanet atmospheres can be tested and validated. As observations improve, it will be important to test a variety of models, all of which make various assumptions in their representations of physical processes, to create the most accurate interpretations of the data. 
 
Numerous theoretical and computational studies have shown that hot Jupiters have large day/night temperature contrasts and equatorial superrotation \citep{Showman2002,Cooper2005,DobbsDixon2008,Showman2009,DobbsDixon2010,Rauscher2010,Heng2011a,Heng2011b,Rauscher2012,DobbsDixon2013,PerezBecker2013,Kataria2015,Amundsen2016}. These features are broadly consistent across a wide range of models and have been validated by observations \citep{Snellen2010,Knutson2012,Louden2015}. The general consensus states that the superrotation is the product of interacting equatorial Rossby and Kelvin waves, which allow angular momentum to be transported toward the equator \citep{Showman2011,Tsai2014,Hammond2018,Mendonca2019}. 

Other works have explored the importance of clouds and hazes \citep{Heng2012cloud, Helling2016, Lee2016, Roman2017, Mendonca2018a}, atmospheric chemistry \citep{Cooper2006, Parmentier2013, Kataria2016, Drummond2018a, Drummond2018b, Mendonca2018b} and shock physics \citep{Goodman2009,Li2010,Heng2014,Fromang2016,Koll2018}. Still others have focused on initial conditions and the details of numerical techniques \citep{Watkins2010,Thrastarson2010,Thrastarson2011,Liu2013,Polichtchouk2014}. These additional features are thought to be important, but the data are less conclusive in this regard \citep{Heng2015}. 

Further studies have focused on the atmospheric dynamics of other types of planets, including cooler Neptune-size planets \citep{Charnay2015,May2016,Mayne2019} and terrestrial planets \citep{Merlis2010,Carone2014,Kaspi2015,Carone2016,Carone2018,Guendelman2018a,Guendelman2018b}, particularly for the purposes of understanding habitability \citep{Williams2002,Williams2003,Abe2011,Leconte2013a,Leconte2013b,Yang2013,Yang2014,Kopparapu2016,Wolf2017,Way2018,Jansen2019}. Atmospheric constraints for smaller and cooler planets, however, remain scarce because of the comparative difficulty of observation. 

Even though constraining the atmospheric processes remains a challenge for extra-solar planets, the wide variety of orbits, masses, and sizes of planets indicates that the atmospheres will be quite diverse. Most studies of exoplanet atmospheres using GCMs (general circulation models) have adapted codes developed for Earth or solar system planets. The virtue of this method is that it relies on models which have been well tested, however, planet-specific tunings are often built into the code and can be difficult to find and generalize. In the development of \texttt{THOR}, we have chosen the opposite path: to develop a code from scratch that would be completely free of planet-specific tunings and would thus provide a flexible tool for the study of a diverse range of atmospheres. The additional benefit of this path is that we develop an intimate understanding of the physical processes at work and how these are represented in the code. The challenge that remains is the workload associated with development and testing new components. 

Presented in \cite{Mendonca2016} and \cite{Mendonca2018c} and utilized in \cite{Mendonca2018a,Mendonca2018b}, \texttt{THOR}\footnote{\texttt{THOR} is available at \url{https://github.com/exoclime/THOR}} is a non-hydrostatic, fully global, 3-D general circulation model developed specifically for the study of exoplanets. As such, it is free from the Earth- and solar system-tunings that often make use of 3-D GCMs a challenge for exoplanets. However, because it is a young model developed from scratch, much development remains in order to make the model applicable to all types of planets. This work represents a step forward along this path.

The goals of this paper are to consolidate descriptions of improvements that have been made to the model since \cite{Mendonca2016}, clarify the model framework, validate the new physics, and compare results from the model using different approximations, with implications for the general circulation of hot Jupiters.

\section{Theory \& Algorithm for the Dynamical Core} \label{sec:theory}
\subsection{Preliminaries} \label{sec:prelimtheory}
The principal equations solved in \texttt{THOR} are the flux forms of the Euler equations:
\begin{align}
    \frac{\partial \rho}{\partial t} + \nabla \cdot (\rho \mathbf{v})& = 0,\label{eqn:continuity}\\
    \frac{\partial \rho \mathbf{v}}{\partial t} + \nabla \cdot (\rho \mathbf{v} \otimes \mathbf{v})& = -\nabla P - \rho g \mathbf{\hat{r}} - 2 \rho \mathbf{\Omega} \times \mathbf{v},\label{eqn:momentum}\\
    \frac{\partial \rho \theta}{\partial t} + \nabla \cdot (\rho \theta \mathbf{v}) &= 0,\label{eqn:entropy}
\end{align}
where $\rho$ is the density, $\mathbf{v}$ is the velocity, $P$ is the pressure, $g$ is the acceleration due to gravity (assumed to be constant), $\mathbf{\Omega}$ is the planet's rotation vector, and $\theta$ is the potential temperature. Potential temperature is defined as 
\begin{equation}
    \theta \equiv T \left( \frac{P}{P_{\text{ref}}} \right)^{-R_d/C_P}. \label{eqn:pottemp}
\end{equation}
The system of equations (Equations \ref{eqn:continuity}-\ref{eqn:entropy}) is closed by the ideal gas law, $P = \rho R_d T$. Additionally, a relationship between the pressure and potential temperature is necessary to update the pressure for use in Equation \ref{eqn:momentum}. From the ideal gas law and the definition of potential temperature, we have
\begin{equation}
    P = P_{\text{ref}} \left( \frac{R_d \rho \theta}{P_{\text{ref}}} \right)^{C_P/C_V}. \label{eqn:pfrompt}
\end{equation}

\texttt{THOR} solves the Euler equations using a finite-volume method on an icosahedral grid \citep{Tomita2001,Tomita2004,Mendonca2016}. The horizontal resolution is controlled by a single parameter, $g_{\text{level}}$, the number of times the icosahedral grid is refined (\emph{i.e.}, the number of times the sides of the icosahedron are subdivided into smaller triangles). The average angular size of the control volumes is given by
\begin{equation}
    \bar{\theta} = \sqrt{\frac{2\pi}{5}} \frac{1}{2^{g_{\text{level}}}}. \label{eqn:angsize}
\end{equation}
The lowest value of $g_{\text{level}}$ used in this work is 4, which results in an angular resolution of $\bar{\theta} \approx 4^{\circ}$. Every increase of $g_{\text{level}}$ by 1 decreases $\bar{\theta}$ by half. The average size (in m) of the control volumes is simply
\begin{equation}
    \bar{d} = r_0 \bar{\theta}, \label{eqn:dbar}
\end{equation}
where $r_0$ is the planet radius. The value of $\bar{d}$ is used to scale the numerical diffusion coefficients (Section \ref{sec:diff}). 

\subsection{Discretizing the equations} \label{sec:discretizing}
A full description of the \texttt{THOR} algorithm was presented in \cite{Mendonca2016}, however there were a number of typographical errors in that paper and some of the details have changed, and so we include a description here. 

As described in \cite{Mendonca2016}, we use a time-splitting algorithm based on \cite{Wicker2002,Tomita2004}, and \cite{Skamarock2008}. In this scheme, the fluid equations are split into fast and slow modes and the fast modes are integrated using a smaller time step than the slow modes. The time stepping loop consists then of an outer loop (the large time step, which has variable length) and an inner loop (small time step, with length designated $\Delta \tau$). 

During the inner loop (at time $\tau$ or $\tau + \Delta \tau$), the deviation of any
quantity from its large time step value (at time $t$) is 
\begin{equation}
    \Phi^{\star[\tau]} = \Phi^{[\tau]} - \Phi^{[t]}, \label{eqn:deviation1}
\end{equation}
or
\begin{equation}
    \Phi^{\star[\tau + \Delta \tau]} = \Phi^{[\tau+\Delta \tau]} - \Phi^{[t]}, \label{eqn:deviationdtau}
\end{equation}
where the $^{\star}$ superscript indicates the deviation and the superscript in square brackets indicates how frequently the values are updated: slow modes ($^{[t]}$) are updated every large time step and fast modes ($^{[\tau]}$ or $^{[\tau+\Delta \tau]}$) are updated every small time step. Broadly, the fast modes are those terms that are associated with acoustic waves and the slow modes are everything else. This time-splitting method allows for a less stringent constraint on the time step and thus a moderate boost in performance as many of the terms do not have to be computed as frequently (the advection terms are particularly costly).
Further, the 3-D operators are split into horizontal and vertical components \citep{Mendonca2016}
\begin{align}
    \nabla \Phi &= \nabla_h \Phi + \frac{\partial \Phi}{\partial r} \mathbf{\hat{r}} \label{eqn:gradop}\\
    \nabla \cdot \boldsymbol{\Phi} &= \nabla_h \cdot \boldsymbol{\Phi} + \frac{1}{r^2}\frac{\partial}{\partial r}(r^2 \Phi_r).\label{eqn:divop}
\end{align}
The vector $\mathbf{\hat{r}}$ represents the vertical direction and $r$ is the radial distance from the center of the planet. The altitude is given by $z = r- r_0$.

Equations \ref{eqn:continuity}-\ref{eqn:entropy} are then discretized as
\begin{align}
    \begin{aligned}
    \frac{\rho^{\star[\tau+\Delta \tau]}-\rho^{\star[\tau]}}{\Delta \tau} + \nabla_h \cdot (\rho \mathbf{v}_h)^{\star[\tau+\Delta \tau]} &+ \frac{1}{r^2}\frac{\partial}{\partial r} r^2 (\rho v_r)^{\star[\tau+\Delta \tau]} = \\
    &-\nabla_h \cdot (\rho \mathbf{v}_h)^{[t]} -\frac{1}{r^2}\frac{\partial}{\partial r} r^2 (\rho v_r)^{[t]} + \mathcal{F}_{\rho}^{[t]}, \label{eqn:disccont}
    \end{aligned}\\
    \begin{aligned}
    \frac{(\rho \mathbf{v}_h)^{\star[\tau+\Delta \tau]}-(\rho\mathbf{v}_h)^{\star[\tau]}}{\Delta \tau} + \nabla_h P^{\star[\tau]} = -\nabla_h P^{[t]} - \boldsymbol{\mathcal{A}}_h^{[t]} - \boldsymbol{\mathcal{C}}_h^{[t]} + \boldsymbol{\mathcal{F}}_{\mathbf{v}_h}^{[t]} + \boldsymbol{\mathcal{G}}_{\mathbf{v}_h}^{[\tau]}, \label{eqn:horizmom}
    \end{aligned}\\
\frac{(\rho v_r)^{\star[\tau+\Delta \tau]} - (\rho v_r)^{\star[\tau]}}{\Delta \tau} + \frac{\partial}{\partial r} P^{\star[\tau+\Delta \tau]}+\rho^{\star[\tau+\Delta \tau]} g = -\frac{\partial}{\partial r} P^{[t]}-\rho^{[t]} g - \mathcal{A}_r^{[t]}- \mathcal{C}_r^{[t]}+\mathcal{F}_{v_r}^{[t]}, \label{eqn:vertmom} \\
    \frac{(\rho\theta)^{[\tau+\Delta \tau]}-(\rho\theta)^{[\tau]}}{\Delta \tau} + \nabla_h \cdot \theta^{[t]} (\rho \mathbf{v}_h)^{[\tau+\Delta \tau]} + \frac{1}{r^2}\frac{\partial}{\partial r} r^2 \theta^{[t]} (\rho v_r)^{[\tau+\Delta \tau]} = 0, \label{eqn:discentropy}
\end{align}
The terms $\boldsymbol{\mathcal{A}}_h$ and $\mathcal{A}_r$ represent the horizontal and vertical components of the advection term, $\nabla \cdot (\rho \mathbf{v} \otimes \mathbf{v})$, and the terms $\boldsymbol{\mathcal{C}}_h$ and $\mathcal{C}_r$ represent the same for the Coriolis, $2\rho\mathbf{\Omega}\times\mathbf{v}$. The terms $\mathcal{F}_{\rho}$, $\boldsymbol{\mathcal{F}}_{\mathbf{v}_h}$, and $\mathcal{F}_{v_r}$ represent fluxes from the ``slow'' drag or numerical dissipation mechanisms, in this case, hyperdiffusion and Rayleigh friction. The additional term, $\boldsymbol{\mathcal{G}}_{\mathbf{v}_h}$ represents the 3D divergence damping, which needs to be evaluated on the small time step. 

An important note regarding the coordinate system used in Equation \ref{eqn:horizmom}: the two-dimension spherical surface represented by this equation is transformed into a three-dimension Cartesian coordinate system centered on the planet's core and rotating with the planet. For example, the horizontal velocity is defined as
\begin{equation}
    \mathbf{v}_h = v_1 \mathbf{\hat{e}_1} + v_2 \mathbf{\hat{e}_2} +  v_3 \mathbf{\hat{e}_3} - \mathbf{v}\cdot \mathbf{\hat{r}}, \label{eqn:vh}
\end{equation}
where $\mathbf{\hat{e}_i}$ represent the axes of this coordinate system and $v_i$ are the total (horizontal and vertical) velocities in the corresponding directions. The radial unit vector is related to the Cartesian coordinate system by 
\begin{equation}
    \mathbf{\hat{r}} = \cos{\phi}\cos{\lambda} ~\mathbf{\hat{e}_1} + \cos{\phi} \sin{\lambda} ~\mathbf{\hat{e}_2} + \sin{\phi}~ \mathbf{\hat{e}_3},
\end{equation}
where $\phi$ is latitude and $\lambda$ is longitude.
In essence, Equation \ref{eqn:vh} defines the horizontal velocity as the total velocity minus the radial component. The advection and Coriolis terms, $\boldsymbol{\mathcal{A}}_h$ and $\boldsymbol{\mathcal{C}}_h$, are defined in the same fashion. The use of a three-dimension Cartesian coordinate system allows the horizontal and vertical components of the advection and Coriolis terms (in Equation \ref{eqn:momentum}) to be cleanly separated, and is also used in the \texttt{NICAM} GCM \citep{Tomita2004}. The separation, in turn, is key to allowing explicit integration in time in the horizontal dimensions and implicit integration in the vertical dimension (see Section \ref{sec:vertmom}).

Our prognostic variables are $\rho$, $P$, $\rho v_r$, and the three Cartesian components of $\rho \mathbf{v}_h$, as well as their deviations. Despite using a thermodynamic equation for potential temperature, pressure is used as a prognostic rather than $\theta$ or $\rho \theta$. As noted in \cite{Mendonca2016}, in \texttt{THOR} we do not calculate the deviation of $\rho \theta$ from the large time step, because such calculation can lead to numerical instability. Instead we calculate the new value, $(\rho\theta)^{[\tau+\Delta \tau]}$, from Equation \ref{eqn:discentropy} and use this to update the pressure deviation, $P^{\star}$. Similarly, the dissipation and heating terms are applied to the pressure deviation, so that
\begin{equation}
    P^{\star[\tau+\Delta \tau]} = P_{\text{ref}} \left( \frac{R_d (\rho\theta)^{[\tau+\Delta\tau]}}{P_{\text{ref}}} \right)^{C_P/C_V} - P^{[t]} + \mathcal{F}_{P}^{[t]} \Delta \tau + \frac{R_d}{C_V} q_{\text{heat}} \Delta \tau, \label{eqn:pressureupdate}
\end{equation}
where $q_{\text{heat}}$ represents all of the additional sources of heating or cooling (currently only radiation or Newtonian cooling). 

Table \ref{tab:progdiagvars} gives an overview of the variables used during integration of the dynamical core, the roles of each, and their properties. All quantities are defined at the horizontal centers of the control volumes, though there is some staggering of the grid in the vertical.

\begin{table*}
\caption{Summary of physical quantities in Equations \ref{eqn:disccont}-\ref{eqn:pressureupdate} and \ref{eqn:helmholtz}-\ref{eqn:helmholtzC0}. The ``Role'' column indicates whether the variable is a primary variable of the integration (prognostic) or a secondary variable (diagnostic). The ``Location'' column indicates whether variables are defined at the center of the vertical layers or the midpoint between two layers (though many quantities are interpolated to the midpoints to solve Equation \ref{eqn:helmholtz}). The ``Update'' column indicates when each quantity is updated (Master step refers to an update outside the dynamical core loop, see Section \ref{sec:integration}).}
\centering
\begin{tabular}{lcccc}
\hline\hline \\ [-1.5ex]
Variable & Role & Location & Update\\ [0.5ex]
\hline \\ [-1.5ex]
$\rho$ & Prog. & Center & Large step\\
$P$ & Prog. & Center & Large step \\
$\rho \mathbf{v}_h$ & Prog. & Center & Large step\\
$\rho v_r$ & Prog. & Midpoint & Large step\\
$\rho^{\star}$ & Prog. & Center & Small step\\
$P^{\star}$ & Prog. & Center & Small step \\
$(\rho \mathbf{v}_h)^{\star}$ & Prog. & Center & Small step\\
$(\rho v_r)^{\star}$ & Prog. & Midpoint & Small step\\
$\theta$ & Diag. & Center & Large step\\
$\rho \theta$ & Diag. & Center & Small step \\
$\boldsymbol{\mathcal{A}}_h$ & Diag. & Center & Large step \\
$\mathcal{A}_r$ & Diag. & Center & Large step \\
$\boldsymbol{\mathcal{C}}_h$ & Diag. & Center & Large step \\
$\mathcal{C}_r$ & Diag. & Center & Large step \\
$h$ & Diag. & Center & Large step \\
$\tilde{g}$ & Diag. & Center & Large step\\
$\mathcal{F}_{\Phi}$ & Diag. & Center & Large step \\
$\boldsymbol{\mathcal{G}}_{\mathbf{v}_h}$ & Diag. & Center & Large step \\
$q_{\text{heat}}$ & Diag. & Center & Master step \\
\hline 
\end{tabular}
\label{tab:progdiagvars}
\end{table*}

\subsection{Solving the vertical momentum equation} \label{sec:vertmom}
Rather than solving Equation \ref{eqn:vertmom} explicitly in time for the vertical momentum, $(\rho v_r)^{\star[\tau+\Delta\tau]}$, we follow \cite{Tomita2004} in combining the continuity equation, vertical momentum equation, and thermodynamic equation to form a 1-D Helmholtz equation that can be solved implicitly. The implicit solution has the advantage of stabilizing the model without having to resolve the time scale associated with vertically propagating acoustic waves. The resulting Helmholtz equation was presented in \cite{Mendonca2016}, however, there were a number of typographical errors 
and some steps of the derivation were not particularly clear.
We take the opportunity to reproduce the full derivation and to correct the prior mistakes here. We must stress that the typos present in \cite{Mendonca2016} did not propagate to the code itself---in other words, the model was coded correctly, despite typos in the manuscript.

\subsubsection{Preparing the thermodynamic equation} \label{sec:prepthermo}
The use of the entropy equation, once discretized (Equation \ref{eqn:discentropy}), does not result in 
the Helmholtz equation presented in \cite{Mendonca2016} (Equation 37 in that paper). 
Rather, it is the energy form of the thermodynamic equation that is used (Equation 3 of \cite{Tomita2004}):
\begin{equation}
    \frac{\partial \rho e}{\partial t} + \nabla \cdot (h \rho \mathbf{v}) = \mathbf{v} \cdot \nabla P + q_{\text{heat}}, \label{eqn:pressure}
\end{equation}
where $\rho$ is the density, $e$ is the specific internal energy, $h$ is the specific
enthalpy, $\mathbf{v}$ is the velocity of the fluid, $P$ is the pressure, and 
$q_{\text{heat}}$ is the diabatic heating rate. It can be shown that this equation is equivalent to Equation \ref{eqn:entropy}, aside from the heating term \citep[see, for example, Section 1.6 of][]{VallisBook}. However, once discretized, the discrete form for the pressure flux cannot be easily derived from the entropy version, so we begin here with the energy form. For reference, the specific internal energy and specific enthalpy are defined as,
\begin{align}
    e = C_V T, \label{eqn:intenergy}\\
    h = C_P T. \label{eqn:enthalpy}
\end{align}

The internal energy can be written $E_{\text{int}} = \rho e$, which is also 
related to the pressure via the adiabatic gas index, $\adind$:
\begin{equation}
    E_{\text{int}} = \frac{P}{\gamma_{\text{ad}}-1}. \label{eqn:Eint}
\end{equation}
This allows us to write our thermodynamic equation as 
\begin{align}
\begin{aligned}
   \frac{\partial P}{\partial t} +& (\adind-1) \nabla \cdot (h \rho \mathbf{v}) =(\adind-1) \mathbf{v} \cdot \nabla P+ (\adind-1)q_{\text{heat}},  \label{eqn:TSadiabat1}
\end{aligned}
\end{align}

or, since $\adind - 1 = R_d/C_V$, 
\begin{equation}
   \frac{\partial P}{\partial t} + \frac{R_d}{C_V} \nabla \cdot (h \rho \mathbf{v}) = \frac{R_d}{C_V} \mathbf{v} \cdot \nabla P+\frac{R_d}{C_V} q_{\text{heat}},  \label{eqn:TSadiabat2}
\end{equation}
where $R_d$ is the specific gas constant and $C_V$ is the heat capacity at constant volume.

Now, we designate ``slow'' and ``fast'' quantities, and discretize as we did for Equations \ref{eqn:disccont}-\ref{eqn:discentropy}. 
Denoting the small time step $\tau$ and the large time step $t$, Equation (\ref{eqn:TSadiabat2}) can be written
\begin{align}
    \begin{aligned}
    \frac{P^{\star[\tau+\Delta \tau]}- P^{\star[\tau]}}{\Delta \tau} +&\frac{R_d}{C_V}
     \Bigl[ \nabla_h \cdot h^{[t]} (\rho \mathbf{v}_h)^{[\tau+\Delta \tau]} + \frac{1}{r^2}\frac{\partial}{\partial r}\left(r^2 h^{[t]} (\rho v_r)^{[\tau+\Delta \tau]}\right) \Bigr] = \\
     &\frac{R_d}{C_V}\biggl[ \frac{(\rho \mathbf{v}_h)^{[\tau+\Delta \tau]}}{\rho^{[t]}} \cdot 
    \nabla_h P^{[t]}+ \frac{(\rho v_r)^{[\tau+\Delta \tau]}}{\rho^{[t]}} \frac{\partial}{\partial r} P^{[t]} + q_{\text{heat}} + \mathcal{F}_P^{[t]}\biggr]. \\ 
    \end{aligned} \label{eqn:TSfinite1}
\end{align}
In this equation we also include the hyper-diffusive pressure flux, $\mathcal{F}_P^{[t]}$.

Using the fact that
$(\rho v_r)^{[\tau + \Delta \tau]}=(\rho v_r)^{\star[\tau + \Delta \tau]}+(\rho v_r)^{[t]}$ (from Equation \ref{eqn:deviationdtau}), Equation (\ref{eqn:TSfinite1}) becomes

\begin{align}
    \begin{aligned}
    &\frac{P^{\star[\tau+\Delta \tau]}-P^{\star[\tau]}}{\Delta \tau} + \frac{R_d}{C_V} \frac{1}{r^2}\frac{\partial}{\partial r} \left(r^2 h^{[t]} (\rho v_r)^{\star[\tau+\Delta \tau]}\right) \\
    &- \frac{R_d}{C_V} \frac{(\rho v_r)^{\star[\tau+\Delta \tau]}}{\rho^{[t]}}\frac{\partial P^{[t]}}{\partial r} =
    \frac{R_d}{C_V} \biggl[ -\nabla_h \cdot \left( h^{[t]} (\rho \mathbf{v}_h)^{[\tau+\Delta \tau]} \right)  \\
    &- \frac{1}{r^2}\frac{\partial}{\partial r} \left( r^2 h^{[t]} (\rho v_r)^{[t]} \right)+ \frac{(\rho \mathbf{v}_h)^{[t]}}{\rho^{[t]}}\cdot \nabla_h P^{[t]} + \frac{(\rho v_r)^{[t]}}{\rho^{[t]}}\frac{\partial P^{[t]}}{\partial r} + q_{\text{heat}} + \mathcal{F}_P^{[t]}\biggr],
    \end{aligned} \label{eqn:TSfinite2}
\end{align}
where we have collected horizontal velocity terms and large time step ($t$
superscripted) terms on the right hand side. Following \cite{Tomita2004}, we evaluate the pressure gradient force (the third term on the RHS) and the buoyancy force (fourth term, RHS) at the large time step. This is not the optimal choice for the conservation of energy \citep{SatohBook}, but the resulting error is small compared to the errors introduced by the diffusion schemes, and it allows the resulting Helmholtz equation (Equation \ref{eqn:helmholtz}) to be solved implicitly. To achieve a more concise form (and stay consistent in notation with 
\cite{Mendonca2016}), we will introduce the effective gravity
\begin{equation}
    \tilde{g}^{[t]} = -\frac{1}{\rho^{[t]}} \frac{\partial P^{[t]}}{\partial r}, \label{eqn:effg}
\end{equation}
and define
\begin{align}
\begin{aligned}
    S_P \equiv \frac{R_d}{C_V} \biggl[ &-\nabla_h \cdot \left( h^{[t]} (\rho \mathbf{v}_h)^{[\tau+\Delta \tau]} \right)  - \frac{1}{r^2}\frac{\partial}{\partial r} \left( r^2 h^{[t]} (\rho v_r)^{[t]} \right)+ \frac{(\rho \mathbf{v}_h)^{[t]}}{\rho^{[t]}}\cdot \nabla_h P^{[t]} \\
    &- \tilde{g}^{[t]} (\rho v_r)^{[t]} + q_{\text{heat}} + \mathcal{F}_{P}^{[t]} \biggr]. \label{eqn:slowpressure}
\end{aligned}
\end{align}
Again, note that there are several typos in Equations (39) and (40) of \cite{Mendonca2016}, which we have corrected in our Equations \ref{eqn:effg} and \ref{eqn:slowpressure}. 
 
Then, Equation \ref{eqn:TSfinite2} becomes
\begin{equation}
    \frac{P^{\star[\tau+\Delta \tau]}-P^{\star[\tau]}}{\Delta \tau} + \frac{R_d}{C_V} \frac{1}{r^2}\frac{\partial}{\partial r} \left(r^2 h^{[t]} (\rho v_r)^{\star[\tau+\Delta \tau]}\right) + \frac{R_d}{C_V} \tilde{g}^{[t]} (\rho v_r)^{\star[\tau+\Delta \tau]}= S_P.
\end{equation}

Next, we must solve for $P^{\star[\tau+ \Delta \tau]}$ and take its derivative in 
$r$ to replace the pressure term in our vertical momentum equation. Noting that

\begin{equation}
    \frac{\partial}{\partial r}\left[ \frac{1}{r^2}\frac{\partial}{\partial r} r^2 \right] = \frac{1}{r^2} \frac{\partial^2}{\partial r^2}r^2 - \frac{2}{r^3}\frac{\partial}{\partial r} r^2, \label{eqn:second_dr}
\end{equation}
 and taking the derivative with respect to $r$, we have
\begin{align}
    &\begin{aligned}
    \frac{\partial}{\partial r}P^{\star[\tau+\Delta \tau]} = & \frac{\partial}{\partial r}P^{\star[\tau]}-\Delta \tau \frac{R_d}{C_V}\left[ \frac{1}{r^2}\frac{\partial^2}{\partial r^2}r^2 h^{[t]} (\rho v_r)^{\star[\tau+\Delta \tau]}-\frac{2}{r^3}\frac{\partial}{\partial r} r^2 h^{[t]} (\rho v_r)^{\star[\tau+ \Delta \tau]} \right]  \\& -\Delta \tau \frac{R_d}{C_V} \frac{\partial}{\partial r} \tilde{g}^{[t]} (\rho v_r)^{\star[\tau + \Delta \tau]} + \Delta \tau\frac{\partial}{\partial r} S_P.
    \end{aligned} \label{eqn:pressurevert}
\end{align} 

The vertical derivative is taken here in order to eliminate the pressure at time $[\tau+\Delta \tau]$ from the vertical momentum equation, as shown below. 

\subsubsection{Preparing the continuity equation} \label{sec:prepcont}
We begin with the discretized form of the continuity equation (\ref{eqn:continuity}), and define
\begin{equation}
    S_{\rho} \equiv -\nabla_h \cdot (\rho \mathbf{v}_h)^{\star[\tau+\Delta \tau]} - \nabla_h \cdot (\rho \mathbf{v}_h)^{[t]} -\frac{1}{r^2}\frac{\partial}{\partial r} r^2 (\rho v_r)^{[t]} + \mathcal{F}_{\rho}^{[t]}, \label{eqn:slowrho}
\end{equation}
which, like $S_P$, incorporates the terms evaluated at the large time step and the horizontal momentum term (which is already evaluated for the current small time step). 

To eliminate the density at time $[\tau+\Delta \tau]$ from the vertical momentum equation, we simply solve for $\rho^{\star[\tau+\Delta \tau]}$:
\begin{equation}
\rho^{\star[\tau+\Delta \tau]} = \rho^{\star[\tau]} -\Delta \tau \frac{1}{r^2}\frac{\partial}{\partial r} r^2 (\rho v_r)^{\star[\tau+\Delta \tau]}  + \Delta \tau S_{\rho}. \label{eqn:densityvert}
\end{equation}

\subsubsection{Solving the vertical momentum equation} \label{sec:vertmom2}
For the vertical momentum equation (\ref{eqn:vertmom}), we again collect the terms evaluated at the large time step into a single term defined as
\begin{equation}
    S_{v_r} \equiv -\frac{\partial}{\partial r} P^{[t]}-\rho^{[t]} g - \mathcal{A}_r^{[t]}- \mathcal{C}_r^{[t]} + \mathcal{F}_{v_r}^{[t]},\label{eqn:slowvr}
\end{equation}
where $\mathcal{C}_r$ is the vertical component of the Coriolis acceleration and $\mathcal{A}_r$ is the vertical component of the advection term $\nabla \cdot (\rho v \otimes v)$. We now substitute Equations (\ref{eqn:pressurevert}) and (\ref{eqn:densityvert} into Equation (\ref{eqn:vertmom}) and arrange on the LHS all terms involving $(\rho v_r)^{\star[\tau +\Delta \tau]}$ (the deviation of the vertical momentum from the large time step $[t]$ at time $[\tau+\Delta \tau]$). To achieve the final desired form, we divide through by $\Delta \tau R_d/C_V$, resulting in:
\begin{align}
    &\begin{aligned}
    -\frac{1}{r^2}\frac{\partial^2}{\partial r^2}r^2 h^{[t]}(\rho v_r)^{\star[\tau+\Delta \tau]} &+\frac{2}{r^3}\frac{\partial}{\partial r}r^2 h^{[t]}(\rho v_r)^{\star[\tau+\Delta \tau]} - \frac{\partial}{\partial r} \tilde{g}^{[t]} (\rho v_r)^{\star[\tau+\Delta \tau]}\\ &-\frac{C_V}{R_d}\frac{g}{r^2}\frac{\partial}{\partial r} r^2 (\rho v_r)^{\star[\tau+\Delta \tau]} +\frac{C_V}{R_d \Delta \tau^2} (\rho v_r)^{\star[\tau+\Delta \tau]} = C_0,
    \end{aligned} \label{eqn:helmholtz}
\end{align}
where 
\begin{equation}
    C_0 = \frac{C_V}{R_d} \left[ \frac{1}{\Delta \tau^2}(\rho v_r)^{\star[\tau]} + \frac{1}{\Delta \tau} \left( S_{v_r} - \frac{\partial}{\partial r} P^{\star[\tau]} - \rho^{\star[\tau]} g \right) - \frac{\partial}{\partial r} S_P - S_{\rho} g \right]. \label{eqn:helmholtzC0}
\end{equation}

Equation (\ref{eqn:helmholtz}) is the final Helmholtz equation for the vertical momentum. Note that the version printed in \cite{Mendonca2016} contains several typos which have been corrected here. 

\subsubsection{The hydrostatic and shallow approximations} \label{sec:hsshallowapprox}
The hydrostatic approximation that is commonly used in atmospheric models is derived by assuming that the vertical pressure gradient and gravitational force are much larger than the other terms in the vertical momentum equation, in other words
\begin{equation}
    \left( \left| \frac{\partial}{\partial r} P \right|, \left| \rho g \right| \right) \gg \left( \left|  \frac{\partial \rho v_r}{\partial t} \right| , \left| \mathcal{A}_r \right|, \left| \mathcal{C}_r \right| \right) .
\end{equation}
Making this assumption results in the well-known equation for hydrostatic equilibrium:
\begin{equation}
  \frac{\partial}{\partial r} P = -\rho g.
\end{equation}
In the work of \cite{Tomita2004}, this assumption was made possible by maintaining a factor $\alpha$ next to all the terms in the vertical momentum equation except the two above (the pressure gradient and gravitational force), which could then be set to unity for non-hydrostatic or zero for hydrostatic, leaving the algorithm otherwise unchanged. This should be used with caution, however, as  \cite{White2005} found that the application of hydrostatic balance solely to the vertical momentum equation produces an inconsistency in the representation of potential vorticity within the model unless the ``shallow'' approximation is also applied. Because of the mixture of coordinate systems in the \cite{Tomita2004} algorithm (Cartesian for the horizontal momentum and spherical for the vertical) and the form of the differential operators on the icosahedral grid, it is not simple to find exact correspondence between equations used in \texttt{THOR} and those studied in \cite{White2005}. However, because the approximations discussed here refer implicitly to a spherical coordinate system, we believe the problem described in that paper is relevant here. Hence, we follow their example in applying the hydrostatic and shallow approximations. 

The first approximation, which we will call ``quasi-hydrostatic'' following the terminology used by \cite{White2005}, involves neglecting the material derivative of the vertical velocity, in other words,
\begin{equation}
    \frac{D v_r}{Dt} = 0.
\end{equation}
In practice, this involves not only the first term (the partial derivative in time) of Equation \ref{eqn:vertmom} but also the advection term, $\mathcal{A}_r$. The method used to calculate $\mathcal{A}$ in Cartesian coordinates from the vector momentum retains the effects of curvature (which result in the well-known ``metric'' terms in a fully spherical coordinate system), thus we cannot merely discard $\mathcal{A}_r$ from Equation \ref{eqn:vertmom}. However, the metric part of $\mathcal{A}_r$ can be simply calculated from the horizontal momentum, as
\begin{equation}
    \mathcal{A}_{r}^{\text{QH}} = \frac{\rho \mathbf{v}_h \cdot \mathbf{v}_h}{r},
\end{equation}
where $\mathbf{v}_h$ is the horizontal momentum vector in Cartesian coordinates. Then, under the quasi-hydrostatic \emph{deep} (QHD) assumption, the vertical momentum equation becomes,
\begin{align}
\frac{\partial}{\partial r} P^{\star[\tau+\Delta \tau]}+\rho^{\star[\tau+\Delta \tau]} g = -\frac{\partial}{\partial r} P^{[t]}-\rho^{[t]} g - (\mathcal{A}_{r}^{\text{QH}})^{[t]}- \mathcal{C}_r^{[t]}. \label{eqn:vertmomqhd}
\end{align}

As before, we use the thermodynamic and continuity equations to substitute the terms on the LHS, which results in the modified Helmholtz equation:
\begin{align}
    &\begin{aligned}
    -\frac{1}{r^2}\frac{\partial^2}{\partial r^2}r^2 h^{[t]}(\rho v_r)^{\star[\tau+\Delta \tau]}+\frac{2}{r^3}\frac{\partial}{\partial r}r^2 h^{[t]}(\rho v_r)^{\star[\tau+\Delta \tau]} - & \frac{\partial}{\partial r} \tilde{g}^{[t]} (\rho v_r)^{\star[\tau+\Delta \tau]}\\ -&\frac{C_V}{R_d}\frac{g}{r^2}\frac{\partial}{\partial r} r^2 (\rho v_r)^{\star[\tau+\Delta \tau]} = C_0^{\text{QH}},
    \end{aligned} \label{eqn:helmholtzqhd}
\end{align}
where 
\begin{equation}
    C_0^{\text{QH}} = \frac{C_V}{R_d} \left[ \frac{1}{\Delta \tau} \left(S^{\text{QH}}_{v_r} - \frac{\partial}{\partial r} P^{\star[\tau]} - \rho^{\star[\tau]} g \right) - \frac{\partial}{\partial r} S_P - S_{\rho} g \right],
\end{equation}
where $S^{\text{QH}}_{v_r}$ is computed using the QH version of the advection term, $\mathcal{A}_{r}^{\text{QH}}$. Note that the hyper-diffusive flux term $\mathcal{F}_{v_r}$ is also zero in this approximation.
We then solve equations in the same fashion as in the non-hydrostatic model. Everywhere else, the equations remain identical to their non-hydrostatic versions. 

The second approximation (the shallow approximation) is more involved. All horizontal differential operators are defined at the reference pressure (the bottom of the model), and these are proportional to $1/r_0$, where $r_0$ is the radius of the planet (the distance from the center to the location of the reference pressure). To account for curvature, the horizontal operators are multiplied by $r_0/r$, where $r = r_0 + z$ and $z$ is the altitude. Simply, this scales the horizontal area of the control volumes with altitude. In the shallow approximation, $r = r_0$, and the scaling is removed from the horizontal operators. The radial 
derivative in the divergence also changes, as below
\begin{equation}
    \frac{1}{r^2}\frac{\partial (r^2 \psi)}{\partial r}  \rightarrow \frac{1}{r_0^2}\frac{\partial (r_0^2 \psi)}{\partial r} = \frac{\partial \psi}{\partial r},
\end{equation}
where $\psi$ is just a representative quantity. In this way, the second derivative (Equation \ref{eqn:second_dr}) becomes,
\begin{equation}
    \frac{\partial}{\partial r}\left[ \frac{1}{r^2}\frac{\partial}{\partial r} r^2 \right] \rightarrow \frac{\partial}{\partial r}\left[ \frac{1}{r_0^2}\frac{\partial}{\partial r} r_0^2 \right] = \frac{\partial^2}{\partial r^2}.
\end{equation}

Using this in the thermodynamic equation and again constructing the Helmholtz equation, we have
\begin{align}
    &\begin{aligned}
    -\frac{\partial^2}{\partial r^2} h^{[t]}(\rho v_r)^{\star[\tau+\Delta \tau]}- \frac{\partial}{\partial r} \tilde{g}^{[t]} (\rho v_r)^{\star[\tau+\Delta \tau]}-\frac{g C_V}{R_d}\frac{\partial}{\partial r} (\rho v_r)^{\star[\tau+\Delta \tau]} +\frac{C_V}{R_d \Delta \tau^2} (\rho v_r)^{\star[\tau+\Delta \tau]} = C^{\text{S}}_0,
    \end{aligned}
\end{align}
where 
\begin{equation}
    C^{\text{S}}_0 = \frac{C_V}{R_d} \left[ \frac{1}{\Delta \tau^2}(\rho v_r)^{\star[\tau]} + \frac{1}{\Delta \tau} \left( S^{\text{S}}_{v_r} - \frac{\partial}{\partial r} P^{\star[\tau]} - \rho^{\star[\tau]} g \right) - \frac{\partial}{\partial r} S^{\text{S}}_P - S^{\text{S}}_{\rho} g \right].
\end{equation}
In this case, $S^{\text{S}}_{v_r}$, $S^{\text{S}}_P$, $S^{\text{S}}_{\rho}$, and all advection terms are calculated using the shallow operators described above. One additional change is made to ensure that the Coriolis acceleration is consistently represented by the horizontal and vertical equations. The Coriolis vector, $\boldsymbol{\mathcal{C}}$, is now
\begin{align}
    \begin{aligned}
    \boldsymbol{\mathcal{C}}^{\text{S}} = 2 \rho \mathbf{\Omega} \times \mathbf{v} = 2 \Omega \rho ~[ &\left( v_3 \sin{\phi} \cos{\phi} \sin{\lambda} - v_2 \sin^2{\phi} \right) \mathbf{\hat{e}_1} \\
                                                                - & \left(v_3 \sin{\phi} \cos{\phi} \cos{\lambda} - v_1 \sin^2{\phi} \right) \mathbf{\hat{e}_2} \\
                                                                + & \left(v_2 \sin{\phi} \cos{\phi} \cos{\lambda} - v_2 \sin{\phi} \cos{\phi} \sin{\lambda} \right) \mathbf{\hat{e}_3} ] , \label{eqn:shallow_cor}
    \end{aligned}
\end{align}
where the unit vectors $\mathbf{\hat{e}_i}$ represent the rotating Cartesian coordinate system fixed at the center of the planet (the spin axis is aligned with $\mathbf{\hat{e}_3}$), and the velocities $v_i$ are the total (horizontal plus vertical) in the corresponding direction. Equation \ref{eqn:shallow_cor} is equivalent to the form of Coriolis that appears in the primitive equations (i.e., the horizontal component of $\mathbf{\Omega}$ is neglected). The radial component, $\mathcal{C}_r$, is now equal to zero. 

\subsubsection{Discretizing and solving the 1-D Helmholtz equation}  \label{sec:helmholtz}
The time-discretized 1-D Helmholtz equation (Eqn. \ref{eqn:helmholtz}) is now spatially-discretized in the following way. The vertical momentum is solved for at the midpoints between layers, in contrast to the horizontal momentum, pressure, and density, which are solved for at the center of the layer. Denoting quantities defined at the center of layers with subscript $c$ and quantities defined at the 
midpoint between layers (the interfaces) with subscript $m$, we write the first term in Eqn. (\ref{eqn:helmholtz}) containing the second derivative in $r$ as
\begin{equation}
    \frac{-1}{r^2}\frac{\partial^2}{\partial r^2} r^2 h_m W_m \approx \frac{-1}{(r_m^i)^2} \left[ \frac{(r_m^{i+1})^2~h_m^{i+1} W_m^{i+1}-(r_m^{i})^2 ~h_m^{i} W_m^{i}}{\Delta r_m^{i+1} \Delta r_c^i} - \frac{(r_m^{i})^2~ h_m^{i} W_m^{i}-(r_m^{i-1})^2~ h_m^{i-1} W_m^{i-1}}{\Delta r_m^i \Delta r_c^i} \right],
\end{equation}
for the $i$th midpoint (between layers), \emph{i.e.} at $r_m^i$. Superscripts indicate the layer/midpoint at which each quantity is defined and are ordered such that the $i$th midpoint is at the bottom of the $i$th layer. Here, $h_m^i = h^{[t]}$ and $W_m^i = (\rho v_r)^{\star[\tau+\Delta \tau]}$ for the $i$th midpoint. For short-hand, the spatial separations are defined as
\begin{equation}
    \begin{aligned}
    \Delta r_c^i = & r_c^{i} - r_c^{i-1},\\
    \Delta r_m^i = & r_m^{i} - r_m^{i-1},\\
    \Delta r_m^{i+1} = & r_m^{i+1} - r_m^{i}.
    \end{aligned}
\end{equation}

First derivatives are calculated by performing a first order finite difference across the layers $i$ and $i-1$ and then interpolating to the midpoint. The resulting proceedure, for terms 2-4 in Equation (\ref{eqn:helmholtz}), is
\begin{equation}
\begin{aligned}
    \frac{\partial}{\partial r} F_m^i \approx & \frac{r_m^i - r_c^{i-1}}{\Delta r_m^{i+1}\Delta r_c^i}F_m^{i+1} + \left[ \frac{r_c^i - r_m^{i}}{\Delta r_m^i \Delta r_c^i} - \frac{r_m^i - r_c^{i-1}}{\Delta r_m^{i+1} \Delta r_c^i} \right] F_m^{i}-\frac{r_c^i - r_m^{i}}{\Delta r_m^i \Delta r_c^i}F_m^{i-1},
\end{aligned}
\end{equation}
where $F_m^i$ represents the various arguments inside the derivatives for the $i$th midpoint.

The resulting spatially-discretized equations for each midpoint $i$ form a system of equations that can be written
\begin{equation}
    a_i W_m^{i+1} + b_i W_m^i + c_i W_m^{i-1} = d_i \text{ for }(1<i<n-1),
\end{equation}
which we then solve using Thomas' algorithm for tri-diagonal matrices and the boundary conditions $(W_m^0, W_m^n) = 0$ ($n$ represents the index of the top layer). After applying the discretization process and rearranging terms, the resulting coefficients are
\begin{equation}
    \begin{aligned}
    a_i = & \frac{1}{\Delta r_m^{i+1} \Delta r_c^i} \left[ -\frac{(r_m^{i+1})^2~h_m^{i+1}}{(r_m^i)^2} + (r_m^i-r_c^{i-1})\left(\frac{2 (r_m^{i+1})^2 ~h_m^{i+1}}{(r_m^i)^3} - \tilde{g}_m^{i+1} - \frac{C_V g (r_m^{i+1})^2}{R_d (r_m^i)^2} \right) \right],\\
    b_i = &  \frac{1}{\Delta r_m^{i+1} \Delta r_c^i} \left[ h_m^{i} - (r_m^i-r_c^{i-1})\left(\frac{2 h_m^{i}}{r_m^i} - \tilde{g}_m^{i} - \frac{C_V g}{R_d} \right)  \right] \\
         &+ \frac{1}{\Delta r_m^{i} \Delta r_c^i} \left[ h_m^{i}- (r_m^i-r_c^{i})\left(\frac{2 h_m^{i}}{r_m^i} - \tilde{g}_m^{i} - \frac{C_V g }{R_d} \right)  \right] + \frac{C_V}{R_d \Delta \tau^2},\\
    c_i = &  \frac{1}{\Delta r_m^{i} \Delta r_c^i} \left[ -\frac{(r_m^{i-1})^2~h_m^{i-1}}{(r_m^i)^2} + (r_m^i-r_c^{i})\left(\frac{2 (r_m^{i-1})^2~ h_m^{i-1}}{(r_m^i)^3} - \tilde{g}_m^{i-1} - \frac{C_V g (r_m^{i-1})^2}{R_d (r_m^i)^2} \right) \right],\\
    d_i = & C_0^i.
    \end{aligned}
\end{equation}

As before, the enthalpy is defined on the large time-step at the $i$th midpoint, \emph{i.e.}, $h_m^i = h^{[t]}$, and likewise for $\tilde{g}_m^i$ and $C_0^i$.

Of course, in the case that the height grid is uniformly spaced, $\Delta r_c^i = \Delta r_m^i = \Delta r_m^{i+1}$ and the above equations can be greatly simplified. The current version of \texttt{THOR} utilizes only a uniform grid, however, the model has been coded according to the above equations so that non-uniform grids can be utilized in the future. A further simplification can be made in the case of the shallow approximation, in which $r_m^{i+1} \approx r_m^i \approx r_m^{i-1}$. This simplification is carried out in the model when the shallow approximation is used. 

\subsection{Numerical dissipation}
\label{sec:diff}

The flux-form hyperdiffusion terms that are applied to Equations \ref{eqn:disccont}, \ref{eqn:horizmom}, \ref{eqn:vertmom}, and \ref{eqn:pressureupdate}, are
\begin{align}
    \mathcal{F}_{\rho} &= -\nabla^2_h K_{\text{hyp}} \nabla^2_h \rho, \label{eqn:hyprho}\\
    \boldsymbol{\mathcal{F}}_{\mathbf{v}_h} &= -\nabla^2_h \rho K_{\text{hyp}} \nabla^2_h \mathbf{v}_h, \label{eqn:hyphoriz}\\ 
    \mathcal{F}_{v_r} &= -\nabla^2_h \rho K_{\text{hyp}} \nabla^2_h v_r, \label{eqn:hypvert}\\
    \mathcal{F}_{P} &= -R_d \nabla^2_h \rho K_{\text{hyp}} \nabla^2_h T.
\end{align}

The divergence damping term in the horizontal momentum equation is
\begin{equation}
    \boldsymbol{\mathcal{G}}_{\mathbf{v}_h} = -K_{\text{div}}\nabla_h \nabla^2_h \left( \nabla_h \cdot (\rho \mathbf{v}_h) + \frac{1}{r^2} \frac{\partial}{\partial r}(\rho v_r r^2) \right). \label{eqn:divdamp}
\end{equation}
Note that the order of the gradient and Laplacian operators was incorrectly reversed in \cite{Mendonca2016}; the model is coded as written in our Equation \ref{eqn:divdamp}. Divergence damping is necessary to eliminate noise produced by the time-splitting integration scheme \citep{Skamarock1992,Mendonca2016}. 

The diffusion coefficients, $K_{\text{hyp}}$ and $K_{\text{div}}$, have the same functional dependence on the grid resolution and time step size, but can be individually adjusted. These are
\begin{align}
    K_{\text{hyp}} &= D_{\text{hyp}} \frac{\bar{d}^4}{\Delta t}\\
    K_{\text{div}} &= D_{\text{div}} \frac{\bar{d}^4}{\Delta t},
\end{align}
where $\bar{d}$ is the average width of the control volumes given by Equation \ref{eqn:dbar}. The hyperdiffusion fluxes are updated on the large time step while the divergence damping fluxes are updated every small time step. 

The boundary conditions for the top and bottom of the model are that the vertical velocity must equal zero; this is the simplest assumption that allows for conservation of energy and axial angular momentum \citep{Staniforth2003}. Unfortunately, this causes the boundary to act as a node for vertically propagating waves, causing them to reflect and potentially amplify.
An additional dissipation mechanism is often needed to eliminate these reflecting waves. In a real atmosphere, vertical propagating waves will break in the upper layers; however, in the model the artificial reflection of these waves becomes an additional source of noise that can trigger numerical instabilities and cause the model to crash. Methods used to damp reflecting waves at the boundaries are often called ``sponge layers''. In \texttt{THOR}, we use the method based on \cite{Skamarock2008} and described in \cite{Mendonca2018b}, which we briefly reiterate here. 

The zonal, meridional, and vertical winds are all damped toward their zonal averages using Rayleigh friction. In principle, damping toward the zonal averages, rather than toward zero, will selectively damp waves while allowing the general flow to persist. In practice, the method is imperfect and the effect of the sponge layer on the flow at the highest layers is discernible as a decrease in the zonal wind speed. For this reason, we minimize the strength and size of the sponge layer in our simulations. 

The damping takes the form 
\begin{equation}
    \frac{d\mathbf{v}}{dt} = -k_{\text{sp}}(\eta) \left(\mathbf{v}-\mathbf{\bar{v}}\right), \label{eqn:sponge}
\end{equation}
where $\mathbf{v}$ represents the vector velocity and $\mathbf{\bar{v}}$ represents the zonal mean of the components, $\eta = z/z_{\text{top}}$ is the fractional altitude, and $k_{\text{sp}}(\eta)$ is the damping strength as a function of the fractional altitude (with units of s$^{-1}$).  

The damping strength is a function of altitude and takes the form
\begin{equation}
    k_{\text{sp}}(\eta) = \begin{cases}
                            0  & \eta < \eta_{\text{sp}},\\
                            k_{\text{sp}}(\eta_{\text{sp}}) \sin^2{\left(\frac{\pi}{2}\frac{\eta-\eta_{\text{sp}}}{1-\eta_{\text{sp}}}\right)} & \eta> \eta_{\text{sp}},
                            \end{cases} \label{eqn:ksponge}
\end{equation}
where $\eta_{\text{sp}}$, the fractional height at which the sponge layer begins, and $k_{\text{sp}}(\eta_{\text{sp}})$ are values set by the user. 

To calculate the zonal mean on our icosahedral grid, we divide the sphere into a user set number of latitude bins, $n_{\text{lat}}$, and compute the average within each bin. This is then considered as the average at the center latitude of each bin. At any given latitude, the zonal-mean $\mathbf{\bar{v}}$ is determined by linearly interpolating from the center of the respective latitude bin. This allows the zonal-mean velocities to vary more smoothly with latitude. 

Equation \ref{eqn:sponge} is calculated for the zonal, meridional, and vertical wind speeds. Since version 2.3 of the code, there are several options for the time integration of the sponge layer friction. In ``implicit mode'', which is used in the simulations here, the calculation of the rate of change in wind speed is done in the ProfX step (see Section \ref{sec:integration}) and the velocities are updated directly, implicitly, during this step. In ``explicit mode'', the rate of change is converted into fluxes that are passed to the dynamical core. For the vertical winds, this flux takes the form $\rho~dv_r/dt$ (the vertical component of Equation \ref{eqn:sponge}), which is added to Equation \ref{eqn:hypvert}. For the horizontal winds, we first compute the zonal and meridional components of $\rho~d\mathbf{v}_h/dt$, convert those to Cartesian coordinates, and then add them to Equation \ref{eqn:hyphoriz}.

\subsection{Time integration}
\label{sec:integration}
\begin{figure*}
\includegraphics[width=\textwidth]{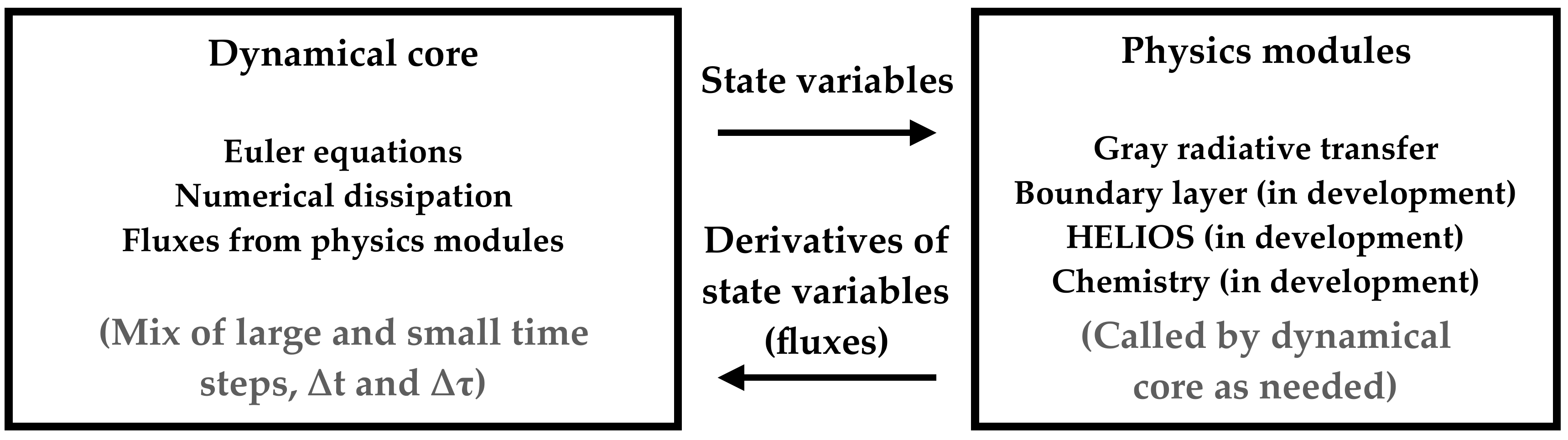}
\caption{Top level code structure of \texttt{THOR}. The dynamical core, which solves the Euler equations, passes state variables to the physics modules, which produces fluxes that are incorporated back into the dynamical core at designated code locations. \label{fig:toplevel}}
\end{figure*}

 For clarity, we outline and summarize the flow of integration. At the top level, each time step contains two components: the dynamical core (THOR) and physics modules (ProfX), see Figure \ref{fig:toplevel}. The ``dynamical core'' refers to the solution of the Euler equations and ``physics modules'' refers to any additional processes. 

The integration of the dynamical core proceeds as follows. For any prognostic quantity, $\Phi$, the value at time $t + \Delta t$ is calculated using three large time steps of variable length (a third-order Runge-Kutta scheme) \citep{Wicker2002, Tomita2004, Klemp2007, Mendonca2016}. Each large time step is broken into smaller time steps of variable length: the first large time step consists of one small time step of length $\Delta \tau = \Delta t/3$; the second large step consists of $n_{\text{max}}/2$ small steps of length $\Delta \tau = \Delta t/n_{\text{max}}$; and the third large step consists of $n_{\text{max}}$ steps of length $\Delta \tau = \Delta t/n_{\text{max}}$. 

\begin{figure*}
\centering
\includegraphics[width=0.5\textwidth]{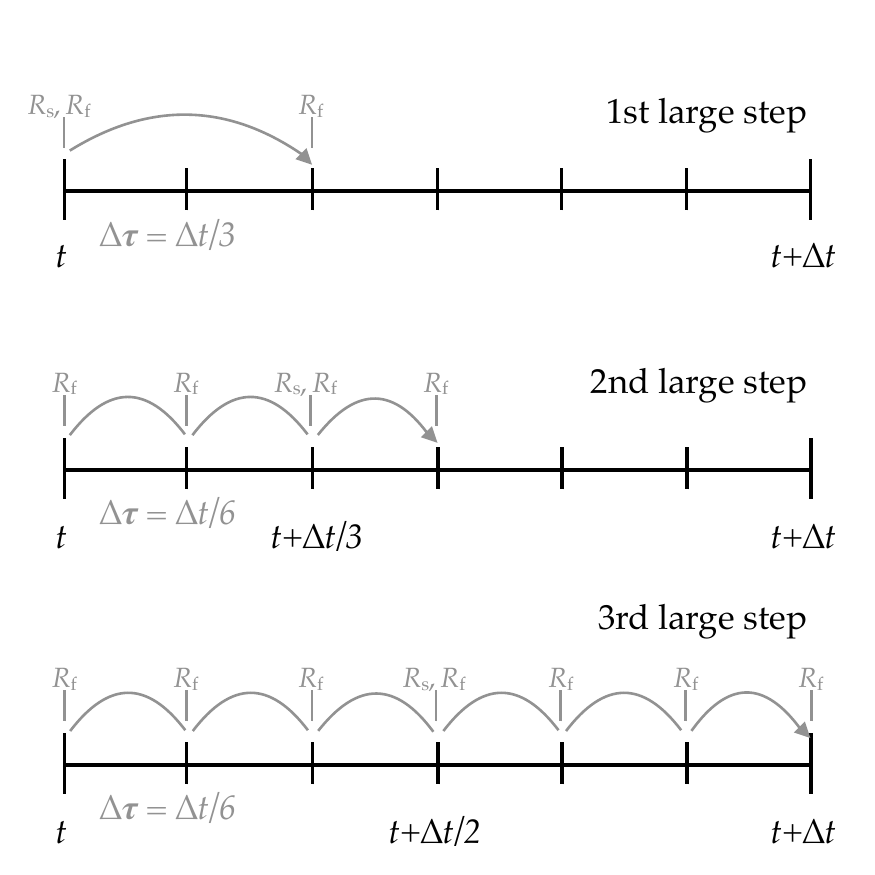}
\caption{Schematic of the three large steps of the Runge-Kutta loop in \texttt{THOR}, for the example with the maximum number of small time steps $n_{\text{max}} = 6$. $R_s$ and $R_f$ represent the slow terms (superscript $^{[t]}$) and fast terms (supescripts $^{[\tau]}$ and $^{[\tau+\Delta \tau]}$), respectively, and indicate when the terms are computed during each large step. \label{fig:timestep}} 
\end{figure*}

A sketch of one time step for variable $\Phi$ follows, for $n_{\text{max}} = 6$ (see Figure \ref{fig:timestep}. The beginning of the time step is time $t$. First, we calculate the slow terms in the derivative $\partial \Phi/\partial t$; these are the terms in Equations \ref{eqn:disccont}-\ref{eqn:discentropy} and \ref{eqn:pressureupdate} designated with $^{[t]}$. Next, we calculate the fast terms (designated with $^{[\tau]}$ or $^{[\tau + \Delta \tau]}$) at times $t$ and $t + \Delta \tau = t + \Delta t/3$. The deviation, $\Phi^{\star}$, is  defined with respect to time $t$, thus $\Phi^{\star}(t) = 0$. The fast and slow terms are used to calculate $\Phi^{\star}(t + \Delta t/3)$, the deviation of $\Phi$ at time $t + \Delta t/3$. Then, we compute $\Phi(t+\Delta t/3) = \Phi(t) + \Phi^{\star}(t+\Delta t/3)$. This completes the first large step.

The second large step begins. First, we recompute the deviations, which are now defined with respect to $t + \Delta t/3$, so that $\Phi^{\star}(t) = \Phi(t) - \Phi(t+\Delta t/3) \neq 0$. The slow terms (superscript $^{[t]}$) are recomputed at time $t + \Delta t/3$. The fast terms are recomputed at times $t$ and $t + \Delta \tau = t + \Delta t/6$ (again, $n_{\text{max}} = 6$ in this example). The fast terms at $(t,t+\Delta t/6)$ and slow terms at $t+\Delta t/3$ are used to advance the deviations by a time step $\Delta t/6$, resulting in the value $\Phi^{\star}(t+\Delta t/6)$. Fast terms are recomputed at $t+\Delta t/6$ and $t + 2 \Delta t/6 = t + \Delta t/3$, and these are used with the slow terms to advance one more step of size $\Delta t/6$, resulting in $\Phi^{\star}(t+\Delta t/3)$. We recompute the fast terms a third time and use them to compute $\Phi^{\star}(t+\Delta t/2)$, and finally $\Phi(t+\Delta t/2) = \Phi(t) + \Phi^{\star}(t+\Delta t/2)$. This completes the second large step.

The third large step begins. Again, the deviations are recalculated, this time with respect to $t+\Delta t/2$. Slow terms are recalculated at $t + \Delta t/2$, fast terms at $t$ and $t + \Delta t/6$. These are used to calculate the deviation $\Phi^{\star}(t+\Delta t/6)$. The fast terms are then recomputed and the deviation updated for a total of $n_{\text{max}} = 6 $ times. We then have $\Phi^{\star}(t + \Delta t)$, from which we calculate $\Phi(t+\Delta t) = \Phi(t) + \Phi^{\star}(t+\Delta t)$. This completes the Runge-Kutta loop.

A more detailed outline of a single time step is as follows:

\begin{enumerate}
    \item ProfX step (additional physics)
        \begin{itemize}
            \item Compute benchmark forcing if applicable. Typically, for benchmark tests, the prognostic variables are updated implicitly or explicitly during this step, rather than computing fluxes that are included in step 2.
            \item Compute radiative transfer fluxes (Section \ref{sec:dgrt}). These are passed to the dynamical core as $q_{\text{heat}}$, rather than updating thermodynamic variables directly during this step.
            \item Update sponge layer quantities (zonal mean winds and resulting drag, Eq. \ref{eqn:sponge}). These can be used to implicitly update the wind speed during this step (implicit mode) or passed as fluxes to the dynamical core and added to the hyper-diffusive terms $\boldsymbol{\mathcal{F}}_{\mathbf{v}_h}$ and $\mathcal{F}_{v_r}$ (explicit mode)
        \end{itemize}
    \item THOR step (dynamical core): solving fluid equations (Eqs. \ref{eqn:disccont}-\ref{eqn:discentropy} and \ref{eqn:pressureupdate})
    \begin{enumerate}
        \item Begin large time step: three steps total, where the first step advances the prognostic variables to $t + \Delta t/3$, the second to $t + \Delta t/2$, and the third to $t + \Delta t$.
        \begin{enumerate}
            \item Compute advection and Coriolis terms $\boldsymbol{\mathcal{A}}_h$, $\mathcal{A}_r$, $\boldsymbol{\mathcal{C}}_h$, $\mathcal{C}_r$ at time $t$, $t+\Delta t/3$, or $t+\Delta t/2$ (for the first, second, and third steps) \label{step:advcor}
            \item Compute enthalpy $h$ (Eq. \ref{eqn:enthalpy}), effective gravity $\tilde{g}$ (Eq. \ref{eqn:effg}), and potential temperature $\theta$ (Eq. \ref{eqn:pottemp}) at time $t$, $t+\Delta t/3$, or $t+\Delta t/2$ \label{step:enthalpygt}
            \item Compute hyper-diffusive and divergence damping fluxes $\mathcal{F}_{\rho}$, $\boldsymbol{\mathcal{F}}_{\mathbf{v}_h}$, $\mathcal{F}_{v_r}$, $\mathcal{F}_{P}$, $\boldsymbol{\mathcal{G}}_{\mathbf{v}_h}$ (Eqs. \ref{eqn:hyprho}-\ref{eqn:divdamp}), at time $t$, $t+\Delta t/3$, or $t+\Delta t/2$. Add sponge layer drag (Eq. \ref{eqn:sponge}), if explicit mode is used for sponge layer. \label{step:diffflux}
            \item Compute slow modes: sums of $[t]$ terms (steps \ref{step:advcor}-\ref{step:diffflux}) in Eqs. \ref{eqn:disccont} - \ref{eqn:pressureupdate}, including $\boldsymbol{\mathcal{G}}_{\mathbf{v}_h}$ and RT fluxes (as $q_{\text{heat}}$) from ProfX step.
            \item Second and third large time steps only: update deviations $\rho^{\star}$, $(\rho \mathbf{v}_h)^{\star}$, $(\rho v_r)^{\star}$, $P^{\star}$ (Eq. \ref{eqn:deviation1}). Deviations are equal to zero on the first step. 
            \item Begin small time step: $n_{\text{step}}$ steps, where $n_{\text{step}} = (1,n_{\text{max}}/2,n_{\text{max}})$ and $\Delta \tau = (\Delta t/3, \Delta t/n_{\text{max}}, \Delta t/n_{\text{max}})$ for the first, second, and third large time step, respectively. For the $n$th iteration, the current time is $\tau = t + n \Delta \tau$.
                \begin{enumerate}
                    \item Update divergence damping $\boldsymbol{\mathcal{G}}_{\mathbf{v}_h}$ (Eq. \ref{eqn:divdamp}) at time $\tau$.
                    \item Compute horizontal momentum deviation $(\rho \mathbf{v}_h)^{\star}$ (Eq. \ref{eqn:horizmom}) at time $\tau + \Delta \tau$.
                    \item Compute $S_{P}$, $S_{\rho}$, $S_{v_r}$ (Eqs. \ref{eqn:slowpressure}, \ref{eqn:slowrho},  and \ref{eqn:slowvr}). These terms encapsulate the slow modes plus the horizontal momentum deviations. 
                    \item Compute vertical momentum deviation $(\rho v_r)^{\star}$ (Eq. \ref{eqn:helmholtz}) using Thomas algorithm described in Section \ref{sec:helmholtz}, at time $\tau + \Delta \tau$.
                    \item Compute density deviations $\rho^{\star}$ (Eq. \ref{eqn:disccont}) at time $\tau + \Delta \tau$.
                    \item Compute potential temperature density $(\rho \theta)$ (Eq. \ref{eqn:discentropy}) at time $\tau + \Delta \tau$
                    \item Compute pressure deviation $P^{\star}$ (Eq. \ref{eqn:pressureupdate}) at time $\tau + \Delta \tau$.
                \end{enumerate}
            \item End small time step
            \item Update prognostic variables $\rho$, $(\rho \mathbf{v}_h)$, $(\rho v_r)$, and $P$ using final deviations from small time step loop. These are now defined at times $t + \Delta t/3$, $t + \Delta t/2$, or $t + \Delta t$, for the first, second, or third large step, respectively.
        \end{enumerate}
        \item End large time step 
    \end{enumerate}
\end{enumerate}


\section{Added Physics}
\label{sec:newphys}
\texttt{THOR}'s double gray radiative transfer module is now publicly available. This module is based on \cite{Lacis1991} and \cite{Frierson2006}. Details of the model are described in \cite{Mendonca2018a} and Section \ref{sec:dgrt}. Note that this version uses a two-stream flux formulation, wherein angle-integrated fluxes are calculated from the Stefan-Boltzmann law and the diffusivity factor is utilized to approximately capture the integral of intensity over angle. In \cite{Mendonca2018a}, the integral of intensity over angle was performed using Gaussian quadrature, however, given the crudeness of the gray approximation, this angle-integration is not strongly motivated, and a good choice of the diffusivity factor provides a solution that is accurate enough for our purposes. The double gray RT code is placed into a modular structure so that it may be replaced by alternative forcing schemes (\emph{e.g}, a more realistic radiative transfer model). Future versions of \texttt{THOR} will utilize the framework to couple to \texttt{HELIOS} \citep{Malik2017}.

Dry convective adjustment \citep{Manabe1965, Hourdin1993} is now included in the public version of \texttt{THOR} and we utilize it here in our simulations with radiative transfer. A mathematical description is given in Section \ref{sec:dryconv}.

Reproductions of the Held-Suarez test \citep{Held1994} and the shallow hot Jupiter benchmark \citep{Menou2009, Heng2011a} using \texttt{THOR} were presented in \cite{Mendonca2016}. Here, we add to the list of benchmark tests the synchronously rotating Earth benchmark \citep{Merlis2010,Heng2011a}, the deep hot Jupiter benchmark \citep{Cooper2005,Cooper2006,Rauscher2010,Heng2011a}, an acoustic wave test \citep{Tomita2004}, and a gravity wave test \citep{Skamarock1994,Tomita2004}. The code for all of the benchmark tests and the configuration files are included in the public repository. 

\subsection{Global diagnostics}
\label{sec:globdiag}
By default, the model now outputs additional diagnostic quantities: total energy, mass, angular momentum, and entropy. The mass, angular momentum, total energy (kinetic + internal + potential), and entropy of the $i$th grid point and $j$th vertical level are, respectively,
\begin{equation}
    \begin{aligned}
    m_{ij} = & \rho_{ij} V_{ij} \\
    \mathbf{l}_{ij} = & \rho_{ij} \mathbf{r}_{ij} \times (\mathbf{v}_h^{ij} + \mathbf{\Omega} \times \mathbf{r}_{ij}) V_{ij} \label{eqn:angmom}\\
    E_{ij}^{\text{tot}} = & \rho_{ij} \left( \frac{1}{2} \mathbf{v}^2_{ij} + C_V T_{ij} + g z_j \right) V_{ij} \\
    s_{ij} = & \rho_{ij} C_P \log{(\theta_{ij})} V_{ij}.
    \end{aligned}
\end{equation}
Above, $\rho$ is the density, $V$ is the volume of the control volume, $\mathbf{r}$ is the Cartesian vector position on the sphere, $\mathbf{\Omega}$ is the vector rotation rate, $\mathbf{v}_h$ is the horizontal wind vector, $\mathbf{v}$ is the \emph{total} wind vector (horizontal + vertical), $C_V$ is the specific heat at constant volume, $T$ is the temperature, $g$ is the gravity (assumed to be constant), $z$ is the altitude, $C_P$ is the specific heat at constant pressure, and $\theta$ is the potential temperature. Vectors are defined in a Cartesian coordinate system centered on the center of the planet and rotating about the $\hat{\mathbf{e}}_3$-axis with rotation rate $\Omega$. The vertical momentum can be ignored in the calculation of $\mathbf{l}_{ij}$ as it is parallel to $\mathbf{r}_{ij}$ by definition. When integrated over the sphere, the non-axial ($\hat{\mathbf{e}}_1$ and $\hat{\mathbf{e}}_2$) components of the angular momentum should vanish; in practice, they are not identically zero because of numerical noise, so these components can provide a useful test of numerical accuracy. 

\begin{figure*}
\includegraphics[width=0.45\textwidth]{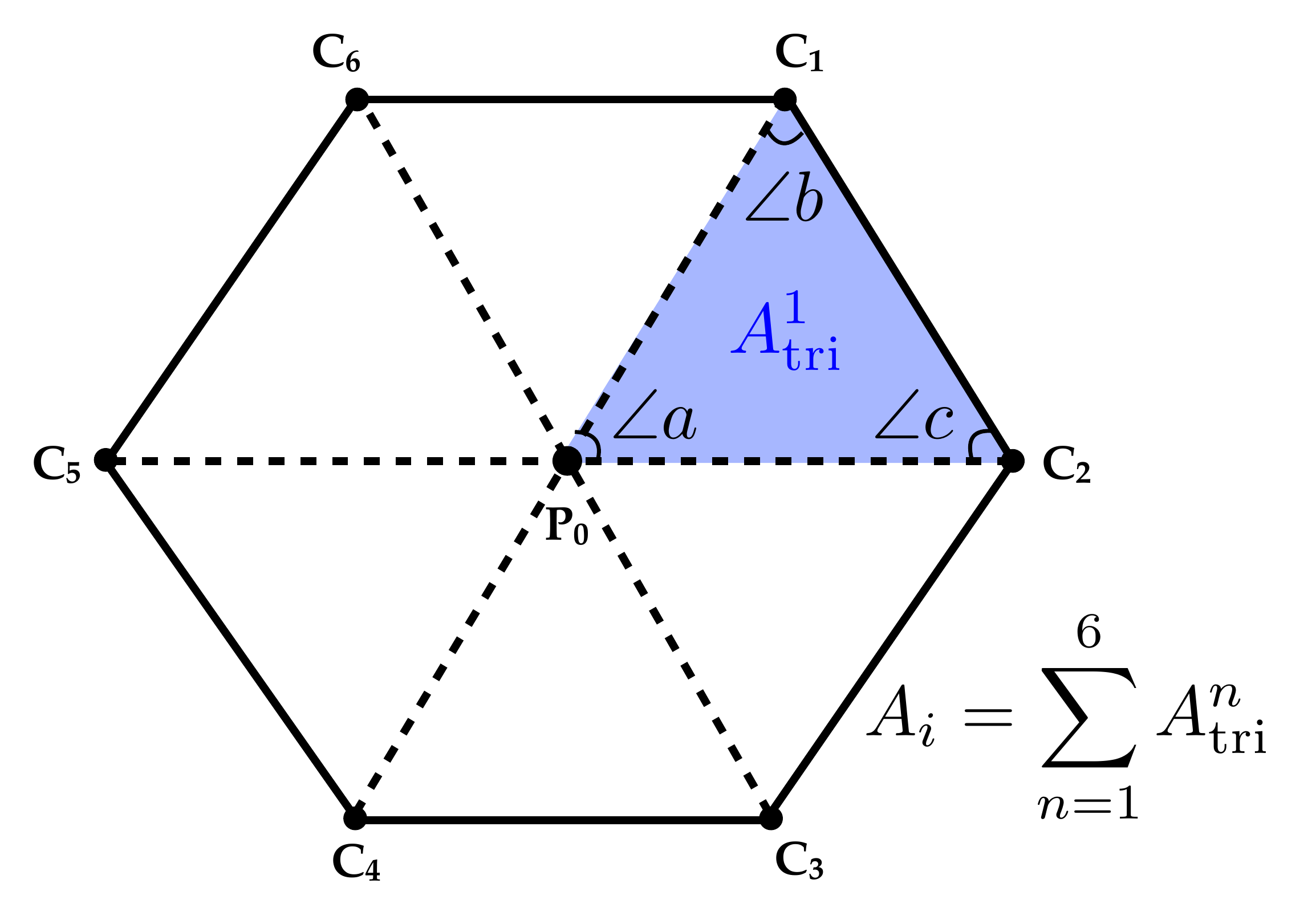}
\includegraphics[width=0.55\textwidth]{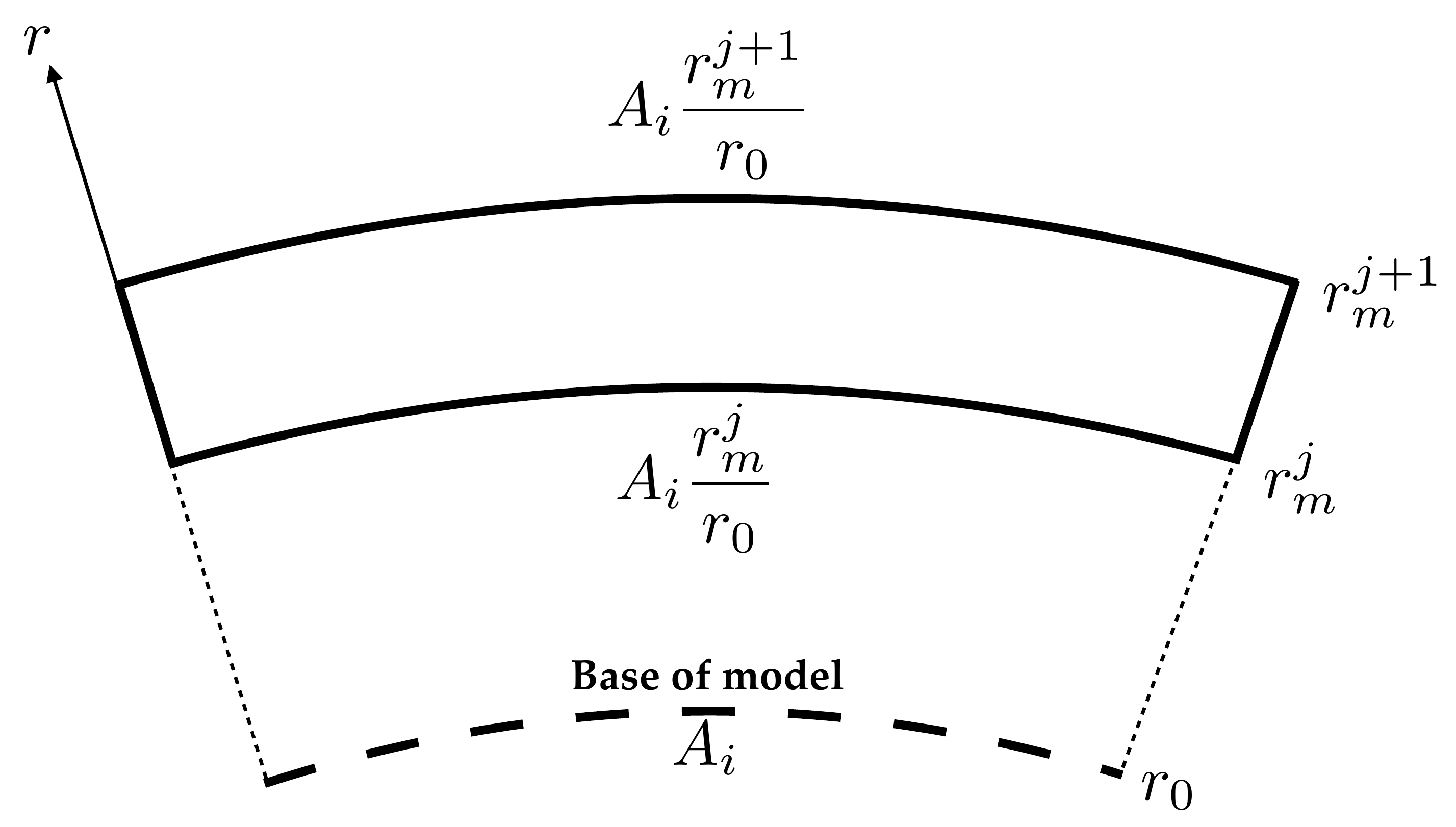}
\caption{Geometry of the control volumes. Left: top-down perspective of a hexagonal control volume (these can also be pentagonal). The total area is calculated by decomposing the volume into triangles, whose area can be found by the formula for spherical triangles and the angles $\angle a$, $\angle b$, and $\angle c$, then summing the areas of the six (or five, for pentagonal cells) triangles. Right: side profile of a control volume some distance $r^j$ above the surface. The total volume is found by integrating the area at the lower boundary, $r_m^j$, to the upper boundary, $r_m^{j+1}$.  \label{fig:areavol}}
\end{figure*}

The global total of each quantity is calculated by summing over $i$ grid points and $j$ vertical levels. In the deep model, the volume of the $i$th grid point and $j$th vertical level is
\begin{equation}
    V_{ij} = \frac{A_i}{3 r_0^2} \left[ (r_m^{j+1})^3 - (r_m^{j})^3 \right],
\end{equation}
where $A_i$ is the area of the $i$th control volume at the lowest boundary of the model, $r_0$ is the planet radius, and $r_m^{j+1}$ and $r_m^{j}$ are the radial coordinates of the top and bottom boundaries of the $j$th layer. Figure \ref{fig:areavol} illustrates the calculation of the size of the control volume, $V_{ij}$. The area, $A_i$, of each control volume is calculated during the grid construction (and adjusted by the spring dynamics process). This is calculated by decomposing the hexagon- or pentagon-shaped control volume into triangles formed by the center and two adjacent vertices \citep[see Figure 2 of][]{Mendonca2016} and summing the areas of the triangles. The areas of each triangle are calculated using the formula for spherical triangles $A_{\text{tri}} = (\angle a + \angle b + \angle c - \pi) r_0^2$, where the angles $\angle a$, $\angle b$, and $\angle c$ are calculated from the vector locations of the center and corresponding vertices. When the shallow approximation is used, the control volumes are treated as flat, and so the volume is
\begin{equation}
    V_{ij} = A_i \left( r_m^{j+1} - r_m^{j} \right).
\end{equation}

\subsection{Dry convective adjustment}
\label{sec:dryconv}

Here we provide a description of the dry convective adjustment scheme, for completeness. Non-hydrostatic models require a parameterization for convection at the coarse resolutions we typically use for exoplanet modeling; typically, scales less than tens of km must be resolved to capture convection with no parameterization in Earth simulations \citep{Jung2004,Arakawa2004,Rio2019}. This scheme is based on \cite{Hourdin1993}. After the dynamical code time step, but prior to the calculation of radiative transfer/Newtonian cooling, each vertical column is searched for unstable layers. Static stability is given by the condition:
\begin{equation}
    \frac{\partial \theta}{\partial r} \geq 0, \label{eqn:convstab}
\end{equation}
where $\theta$ is the potential temperature. When an unstable layer is detected (meaning Equation \ref{eqn:convstab} is violated), we define a ``mixed'' potential temperature, $\theta_{\text{mixed}}$, equal to the average potential temperature in the layer. First, we integrate upward through the unstable layer to find the enthalpy $h$:
\begin{equation}
    h = C_P \int_{P_B}^{P_T} \theta \left( \frac{P}{P_0} \right)^{R_d/C_P} dP,
\end{equation}
where $P_0$, $P_B$, and $P_T$ are the pressure at bottom of the entire column, the pressure at the bottom of the unstable layer, and the pressure at the top of the unstable layer, respectively. Then, the potential temperature across the entire layer is set equal to $\theta_{mixed}$, given by
\begin{equation}
    \theta_{\text{mixed}} = \frac{h}{ C_P \int_{P_B}^{P_T} \left( \frac{P}{P_0} \right)^{R_d/C_P} dP} = \frac{\int_{P_B}^{P_T} \theta \left( \frac{P}{P_0} \right)^{R_d/C_P} dP}{\int_{P_B}^{P_T} \left( \frac{P}{P_0} \right)^{R_d/C_P} dP},
\end{equation}
which is effectively like mixing the entropy across the unstable layer and enforcing an adiabatic profile in that region. After $\theta_{\text{mixed}}$ is calculated, the adjacent layers are tested for static stability again. If the adjacent layers are statically unstable with the new $\theta_{\text{mixed}}$ (\emph{e.g.}, the layer below, altitude-wise, has $\theta > \theta_{\text{mixed}}$), the entire process is repeated, including the additional unstable layers, until the entire column is statically stable. 

\subsection{Double gray radiative transfer}
\label{sec:dgrt}

The algorithm for radiative transfer is based on \cite{Lacis1991} and was described in \cite{Mendonca2018a}. We do not reproduce the entire algorithm in this work, but make several points of clarification. 

We have reverted to using the diffusivity factor, $\mathcal D$, instead of integrating intensities over angle using Gaussian quadrature as was done in \cite{Mendonca2018a}. In the double gray case, this approximation makes very little difference in the calculation of inter-layer fluxes. When using multi-wavelength radiative transfer, as in \cite{Lacis1991}, more accurate integration over angle is probably warranted, but the double gray approximation we use here is likely a much cruder assumption. 

Our double gray scheme is thus a true two-stream approximation, in which the diffusivity factor is used to calculate a characteristic angle for the path of the radiation, and radiation is assumed to be isotropic when the integral over angle is performed. In this approximation, the Planck functions in Equations A1-A6 of \cite{Lacis1991} and Equations 3-8 of \cite{Mendonca2018a} are replaced by the angle integrated flux calculated from the Stefan-Boltzmann law. The angle factor $\mu$ in those equations as well as Equations A9-A10 of \cite{Lacis1991} and Equations 9-10 of \cite{Mendonca2018a} is set to $\mu = 1/\mathcal{D}$, where $\mathcal{D}$ is the diffusivity factor. 

The optical depths are calculated using the form suggested by \cite{Frierson2006, Heng2011b}. For the short-wave, we have a single power law:
\begin{equation}
    \tau_{\text{sw}} = \tau_{\text{sw,}0}\sigma^{n_{\text{sw}}},
\end{equation}
where $\tau_{\text{sw,}0}$ is the optical depth at $P = P_{\text{ref}}$, $\sigma = P/P_{\text{ref}}$, and $n_{\text{sw}}$ is a tuneable factor meant to control the vertical distribution of absorbers. For example, $n_{\text{sw}} = 1$ would represent a uniformly mixed absorber. A value for $n_{\text{sw}}>1$ represents absorbers that are denser in the lower atmosphere. 

The long-wave optical depth is given by 
\begin{equation}
    \tau_{\text{lw}} = \tau_{\text{lw,w}} \sigma + \tau_{\text{lw,s}} \sigma^{n_{\text{lw}}},
\end{equation}
where $\tau_{\text{lw,w}}$ and $\tau_{\text{lw,s}}$ represent the surface optical depths of well-mixed absorbers and vertically segregated absorbers, respectively, and $n_{\text{lw}}$ is again a tuneable factor controlling the vertical distribution of the segregated absorbers. With a factor $f_l$ representing the percent of the surface optical depth attributable to the uniformly mixed absorbers, the optical depths above are given by
\begin{equation}
    \begin{aligned}
    \tau_{\text{lw,w}} &= \tau_{\text{lw,}0} f_l,\\
    \tau_{\text{lw,s}} &= \tau_{\text{lw,}0} (1-f_l).
    \end{aligned}
\end{equation}
The surface optical depth $\tau_{\text{lw,}0}$ can be assumed to be constant (hot Jupiter cases) or given a horizontal distribution. For our Earth-like double gray case, we give $\tau_{\text{lw,}0}$ a latitudinal dependence given by 
\begin{equation}
    \tau_{\text{lw,}0} = \tau_{\text{lw,eq}} + (\tau_{\text{lw,pole}}-\tau_{\text{lw,eq}}) \sin^2{\phi},
\end{equation}
where $\tau_{\text{lw,eq}}$ and $\tau_{\text{lw,pole}}$ are the surface optical depths at the equator and the poles, respectively. This latitudinal dependence approximates the effect of decreased water vapor concentration in the polar regions \citep{Frierson2006}.

The total fluxes passing through each layer are calculated from Equations 1-10 of \cite{Mendonca2018a}. The heating in the $n$th layer, used in Equations \ref{eqn:pressureupdate} and \ref{eqn:slowpressure} is 
\begin{equation}
    q_{\text{heat},n} = \frac{F^{\downarrow \text{sw}}_{n+1} - F^{\downarrow \text{sw}}_{n} - F^{\uparrow \text{lw}}_{n+1} +F^{\uparrow \text{lw}}_{n} + F^{\downarrow \text{lw}}_{n+1}-F^{\downarrow \text{lw}}_{n}}{\Delta z_n},
\end{equation}
where $F^{\downarrow \text{sw}}$ is the downward propagating short wave (stellar) flux, $F^{\uparrow \text{lw}}$ is the upward propagating long wave (thermal) flux, $F^{\downarrow \text{lw}}$ is the downward propagating long wave flux, and $\Delta z_n$ is the vertical thickness of the layer. Here, $q_{\text{heat}}$ has units of W m$^{-3}$, equivalent to kg m$^{-1}$ s$^{-3}$ or Pa s$^{-1}$, in line with Equations \ref{eqn:TSadiabat1} and \ref{eqn:TSadiabat2}.

As in \cite{Heng2011b}, the surface (when used) is treated as a slab with a constant heat capacity, $C_{\text{surf}}$. The temperature is  modeled using the relation
\begin{equation}
    C_{\text{surf}} \frac{\partial T_{\text{surf}}}{\partial t} = F^{\downarrow \text{sw}}_0 -F^{\uparrow \text{lw}}_0 + F^{\downarrow \text{lw}}_0,
\end{equation}
where $F^{\downarrow \text{sw}}_0$, $F^{\downarrow \text{lw}}_0$, and $F^{\uparrow \text{lw}}_0$ are the shortwave and longwave fluxes passing downward from and upward into the lowest atmospheric layer. 

When no surface is included, as in our hot Jupiter simulations, the flux into and out of the lower boundary is zero, unless flux due to internal heating is included. We do not include flux from the interior in any of the simulations in this work, however, in the case where this is desired, it can be included by specifying an internal flux temperature, $T_{\text{int}}$. 

The user can also set the planetary albedo, $A_0$. In this double gray scheme, the albedo represents a top-of-atmosphere albedo, and is thus only applied as a modification of the incoming stellar flux, $Q$.

\section{Further Benchmarks} \label{sec:furtherbench}

\subsection{Synchronously rotating Earth}
\label{sec:syncearth}

\begin{table*}
\caption{Model parameters for Newtonian cooling simulations}
\centering
\begin{tabular}{lp{0.45\linewidth}ccc}
\hline\hline \\ [-1.5ex]
Symbol & Description & Units & Synch. Earth & Deep hot Jupiter\\ [0.5ex]
\hline \\ [-1.5ex]
$r_0$ & Planet radius & m & $6371000$ & 94400000 \\ 
$g$ & Gravity & m s$^{-2}$ & $9.8$ & 9.42 \\
$\Omega$ & Rotation rate & rad s$^{-1}$ & $1.996\times10^{-7}$ & $2.06\times10^{-5}$ \\
$R_d$ & Gas constant & J K$^{-1}$ kg$^{-1}$ & $287$ & $4593$ \\
$C_P$ & Atmospheric heat capacity & J K$^{-1}$ kg$^{-1}$  & $1005$ & $14308.4$ \\
$P_{\text{ref}}$ & Reference pressure (bottom boundary) & bar & 1 & 220 \\
$T_{\text{init}}$ & Initial temperature of atmosphere & K & 300 & 1759 \\
\hline \\ [-1.5ex]
$\Delta t_M$ & Time step & s & $600$ & 300  \\
$z_{\text{top}}$ & Altitude of model top & m & 36000 & $8\times10^6$ \\
$g_{\text{level}}$ & Grid refinement level & - & 4/5 & 4 \\
$v_{\text{level}}$ & Number of vertical levels & - & 32 & 40 \\
$D_{\text{hyp}}$ & Hyperdiffusion coefficient & - & $4.8\times10^{-3}$ & $0.02$ \\
$D_{\text{div}}$ & Divergence damping coefficient & - & $4.8\times10^{-3}$  & $0.03$  \\
\hline \\ [-1.5ex]
$k_{\text{surf}}$ & Friction coefficient of lower boundary & s$^{-1}$ & $1.1574\times10^{-5}$ & - \\ 
$\sigma_b$ & Boundary layer top (fraction of $P_{\text{ref}}$) & - & 0.7 & - \\
\hline 
\end{tabular}
\label{tab:modelparaNC}
\end{table*}

A benchmark test case for a synchronously rotating Earth using prescribed thermal forcing (Newtonian cooling) was suggested by \cite{Heng2011a} for comparison with the model of \cite{Merlis2010}, which used gray radiative transfer. We repeat this test simulation here, with input parameters given in Table \ref{tab:modelparaNC}.

The temperature is forced toward an equilibrium profile given by 
\begin{equation}
    T_{\text{eq}} = \max{(200\text{ K}, T_{\text{HS}})}, \label{eqn:heldsuarezT1}
\end{equation}
where
\begin{equation}
    T_{\text{HS}} = \left[ 315\text{ K} + \Delta T_{\text{EP}} \cos{(\lambda-180^{\circ})} \cos{\phi} - \Delta T_z \ln{\left(\frac{P}{P_{\text{ref}}}\right)} \cos^2{\phi} \right] \left(\frac{P}{P_{\text{ref}}} \right)^{\kappa_{\text{ad}}}, \label{eqn:heldsuarezT2}
\end{equation}
where $\Delta T_{\text{EP}}$ is the temperature difference between the sub-stellar and anti-stellar points (rather than the equator to pole difference), $\Delta T_z$ is a characteristic scale for vertical temperature differences, and $\kappa_{\text{ad}} = R_d/C_P$ is the adiabatic coefficient.
This formulation is identical to the Held-Suarez test except for the second term of $T_{\text{HS}}$, which now depends on longitude, $\lambda$, to emulate the effect of having the sub-stellar point permanently located at $\lambda = 180^{\circ}$ and $\phi = 0^{\circ}$. For the simulation here, $\Delta T_{\text{EP}}= 60$ K and $\Delta T_z = 10$ K, the same as in the Held-Suarez test. The damping time-scale for temperature is also identical to the Held-Suarez test, and is given by
\begin{equation}
    \tau_T = \left[ k_a + (k_s - k_a) \max{\left(0,\frac{\sigma-\sigma_b}{1-\sigma_b}\right)} \cos^4{\phi}\right]^{-1},
\end{equation}
where $k_a = 1/40$ day$^{-1}$, $k_s = 1/4$ day$^{-1}$, $\sigma = P/P_{\text{surf}}$ and $\sigma_b = 0.7$. The surface pressure, $P_{\text{surf}}$, is calculated at the lower boundary of the lowest layer. 

The horizontal winds are damped toward zero with the time scale
\begin{equation}
    \tau_v = \left[ k_{\text{surf}} \max{\left(0,\frac{\sigma-\sigma_b}{1-\sigma_b}\right)} \right]^{-1},
\end{equation}
where $k_{\text{surf}} = 1$ day$^{-1}$.

\begin{figure*}
\includegraphics[width=0.5\textwidth]{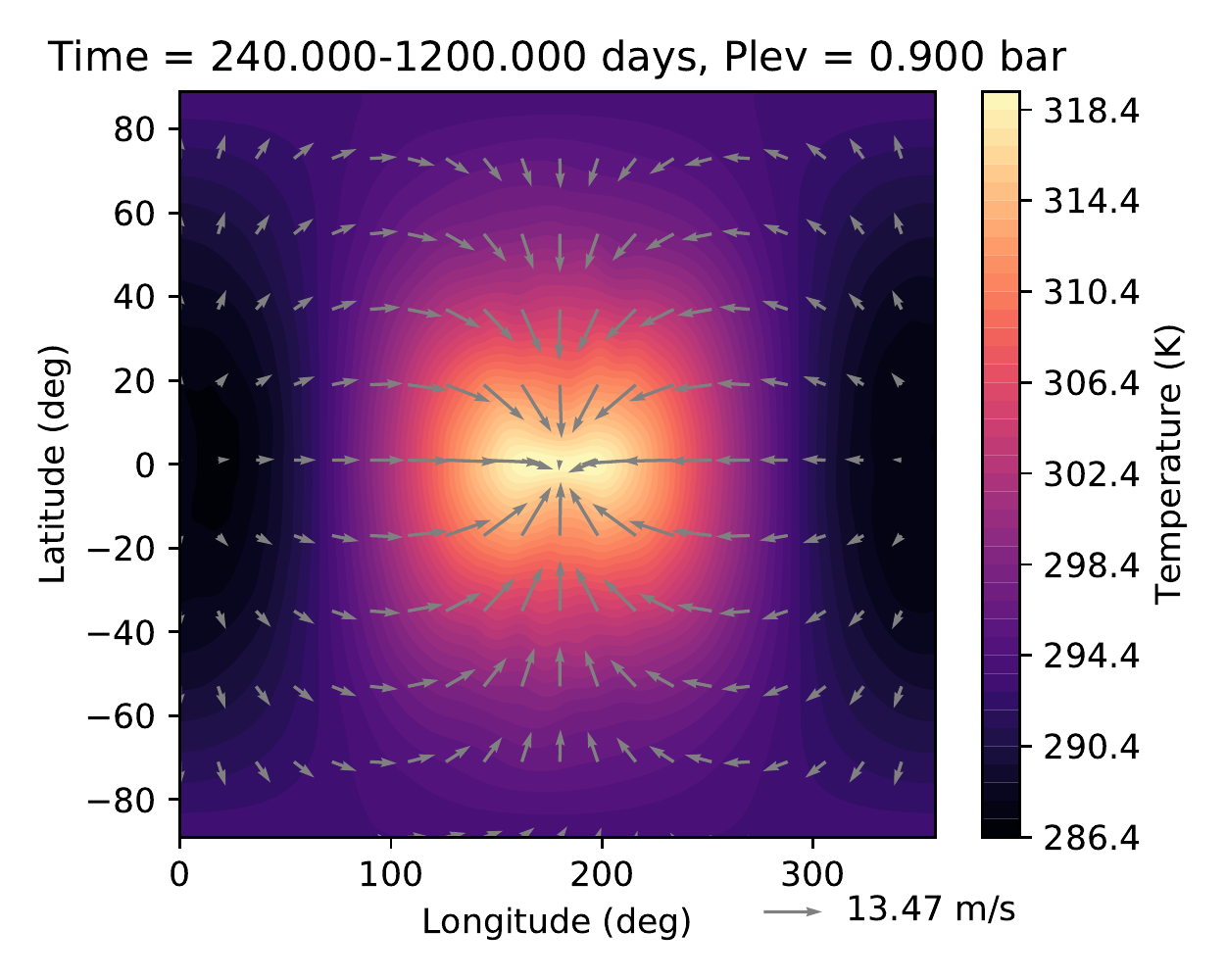}
\includegraphics[width=0.5\textwidth]{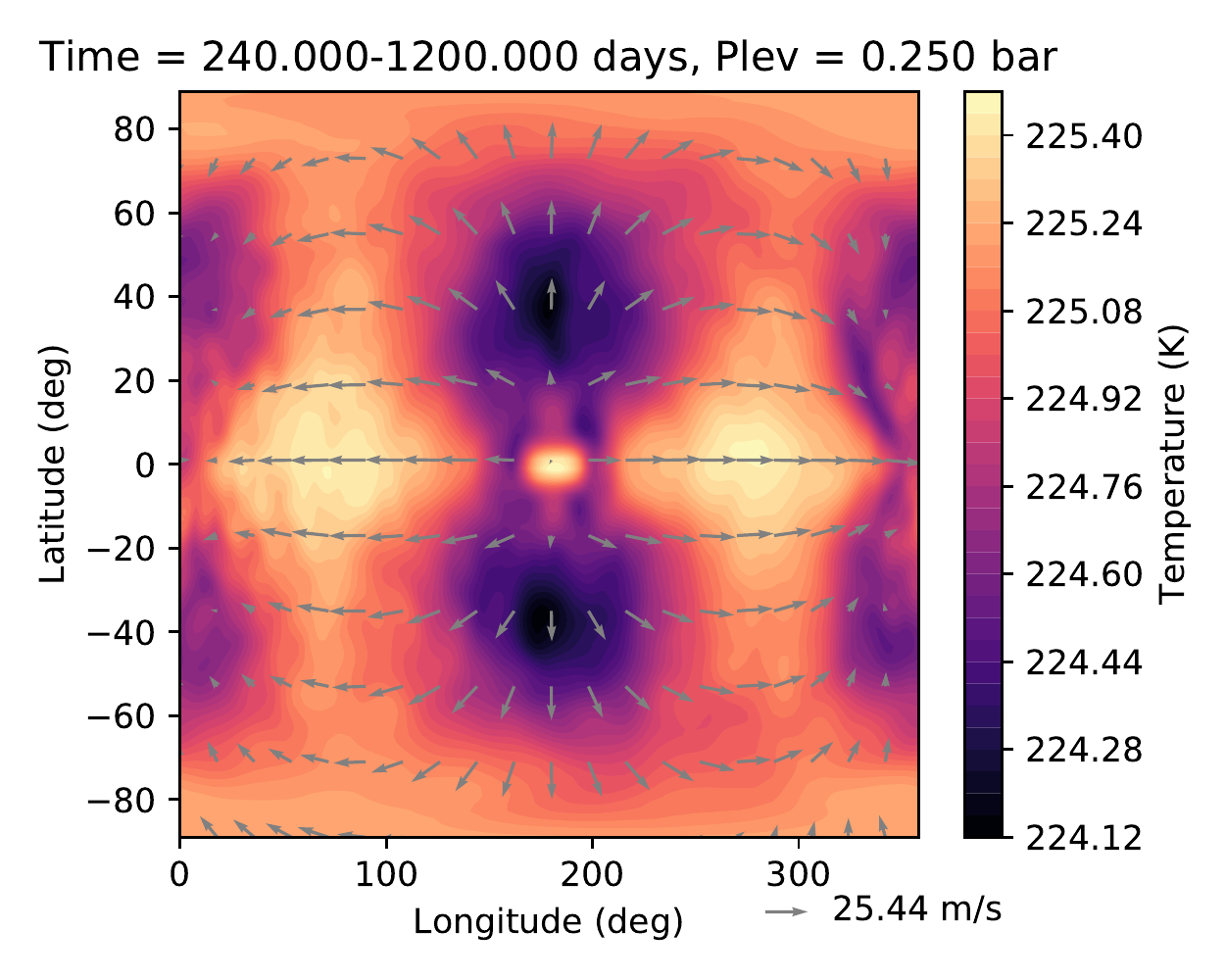}\\
\includegraphics[width=0.5\textwidth]{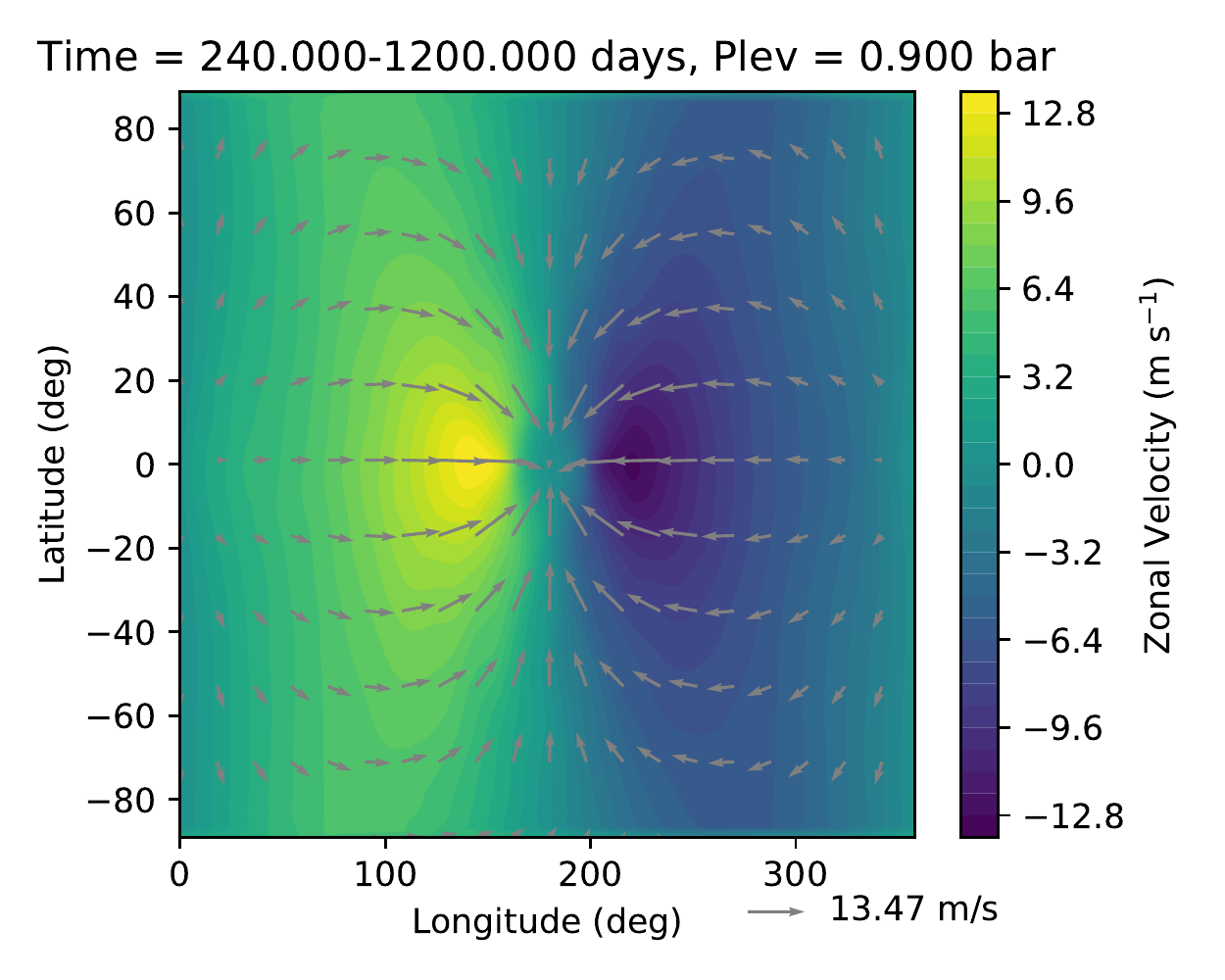}
\includegraphics[width=0.5\textwidth]{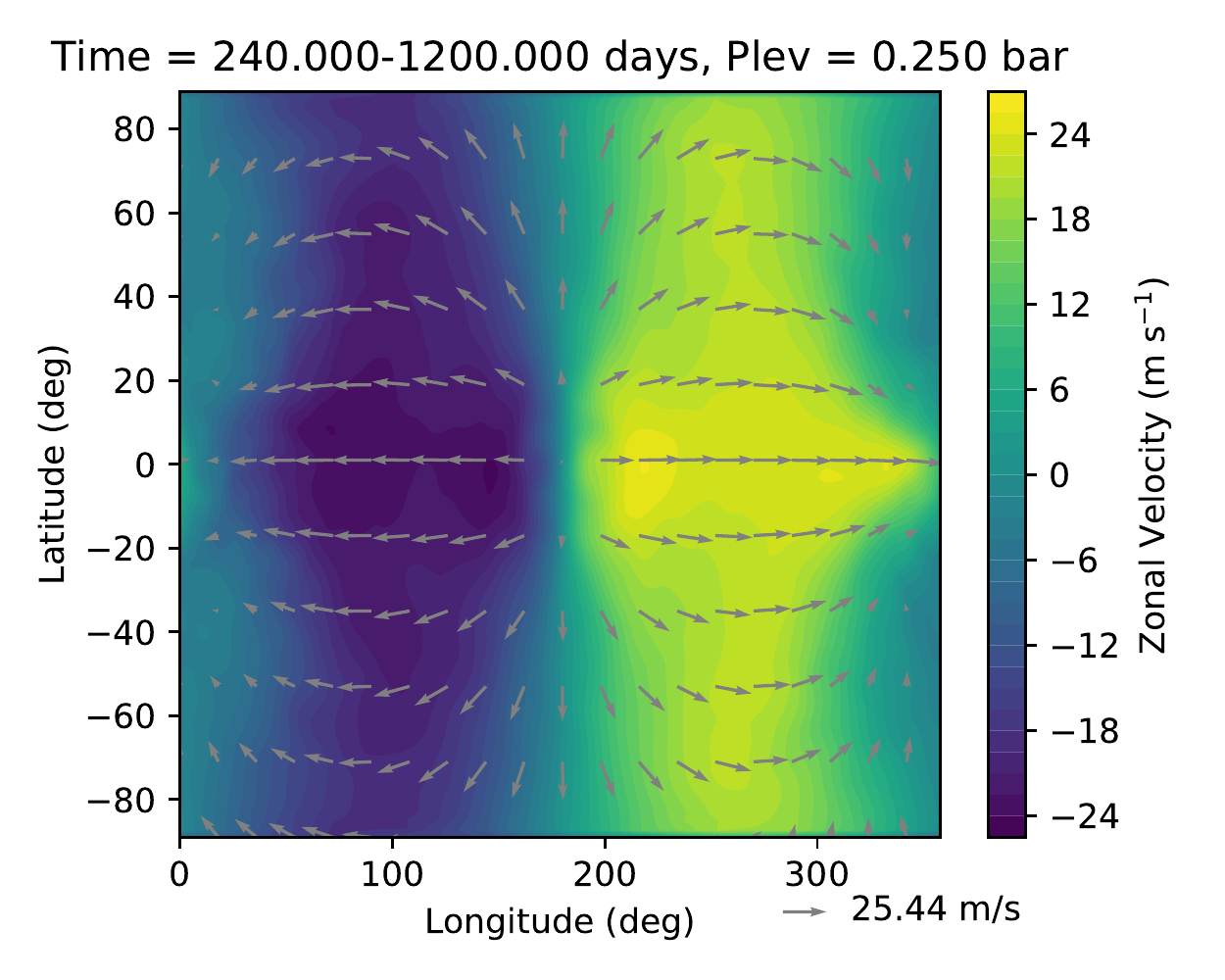}\\
\includegraphics[width=0.5\textwidth]{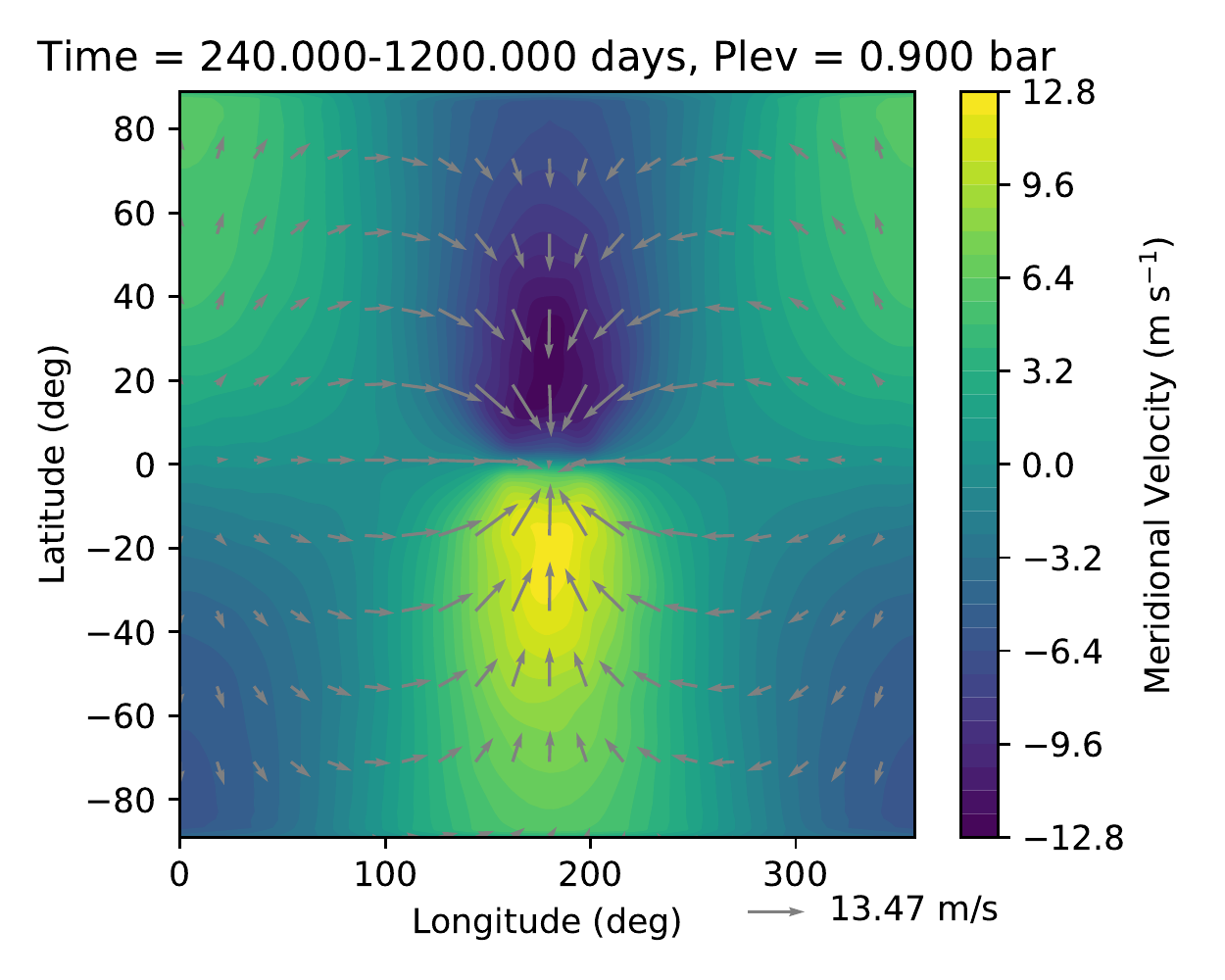}
\includegraphics[width=0.5\textwidth]{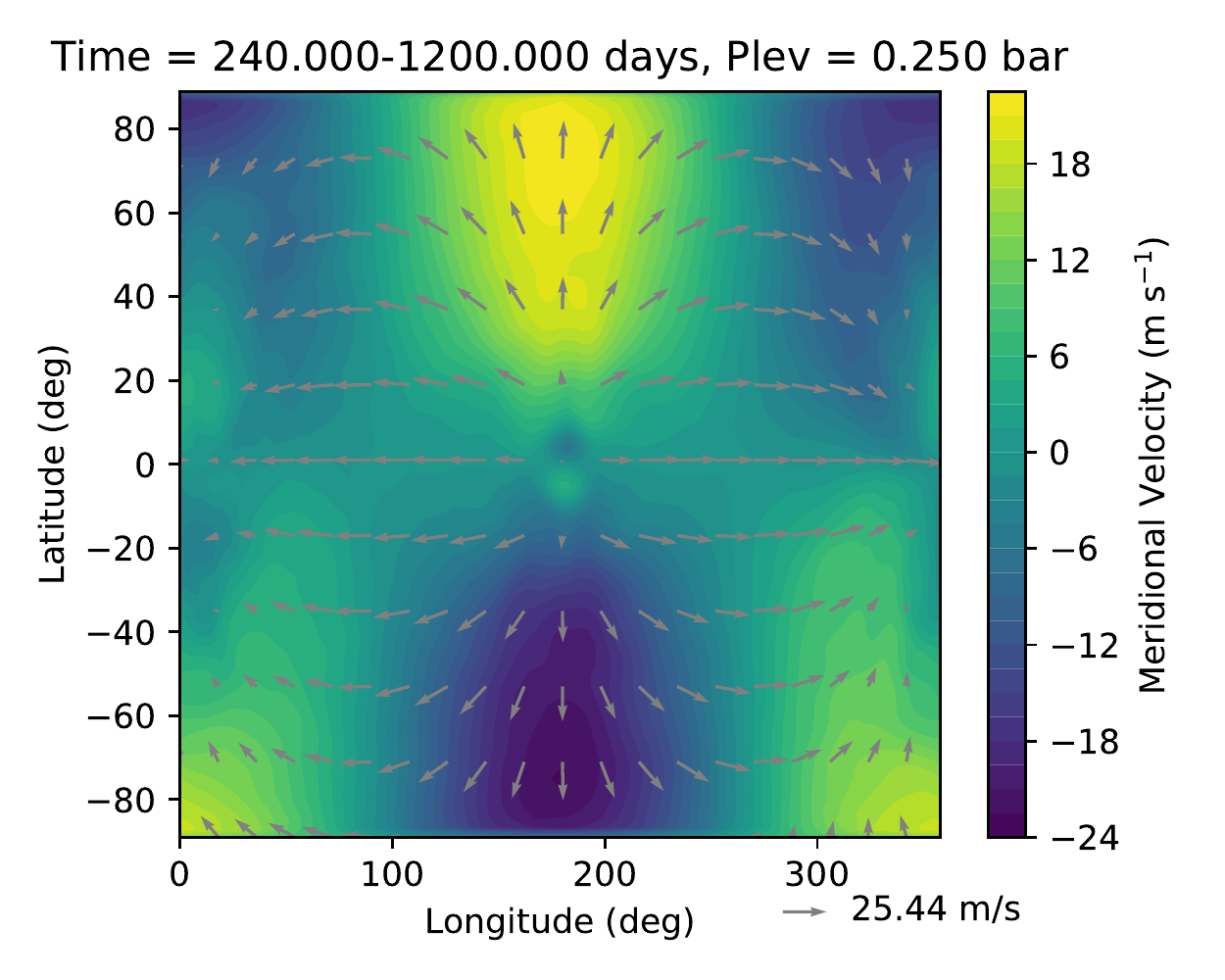}
\caption{Output from the synchronously rotating Earth benchmark, temporally averaged from 240 to 1200 days. In color, the upper panels show the temperature, the middle panels show the zonal wind speed,  and the lower panels show the meridional wind speed. The total horizontal winds are overplotted as arrows. The left column corresponds to a pressure level of 0.9 bar, the right to 0.25 bar.} \label{fig:tidalearth1}
\end{figure*}

Figure \ref{fig:tidalearth1} shows results from the simulation with $g_{\text{level}} = 5$ (a horizontal resolution of $\sim 2^{\circ}$), plotted on two isobaric surfaces. Results are very similar for $g_{\text{level}} = 4$. At $P=0.9$ bar, near the surface, we see convergence toward the sub-stellar point (located at longitude $\lambda = 180^{\circ}$), associated with the rising motion due to the intense heating at this location. At $P=0.25$ bar, the flow diverges from the sub-stellar point, then converges on the night side of the planet. 

\begin{figure*}
\includegraphics[width=0.5\textwidth]{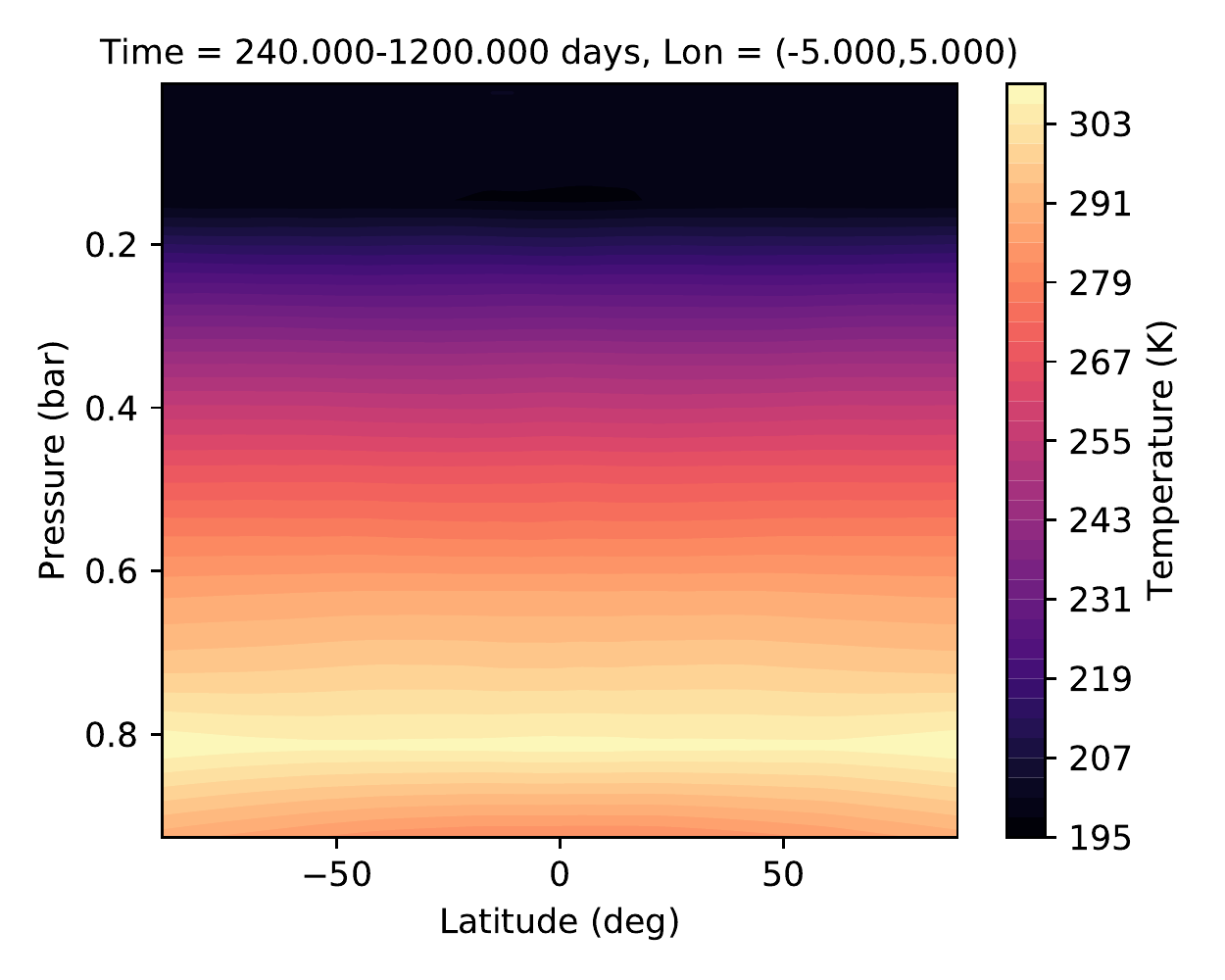}
\includegraphics[width=0.5\textwidth]{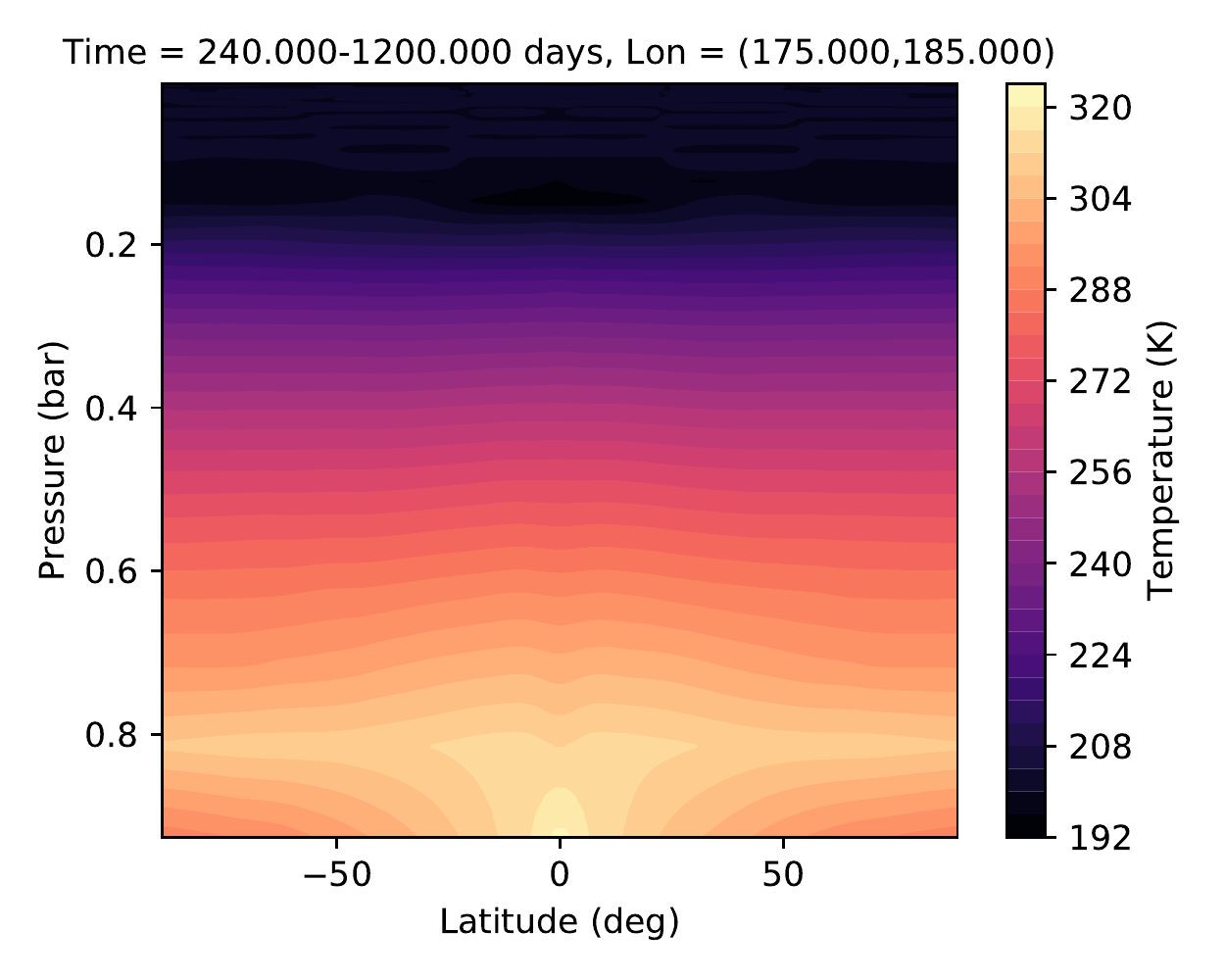}\\
\includegraphics[width=0.5\textwidth]{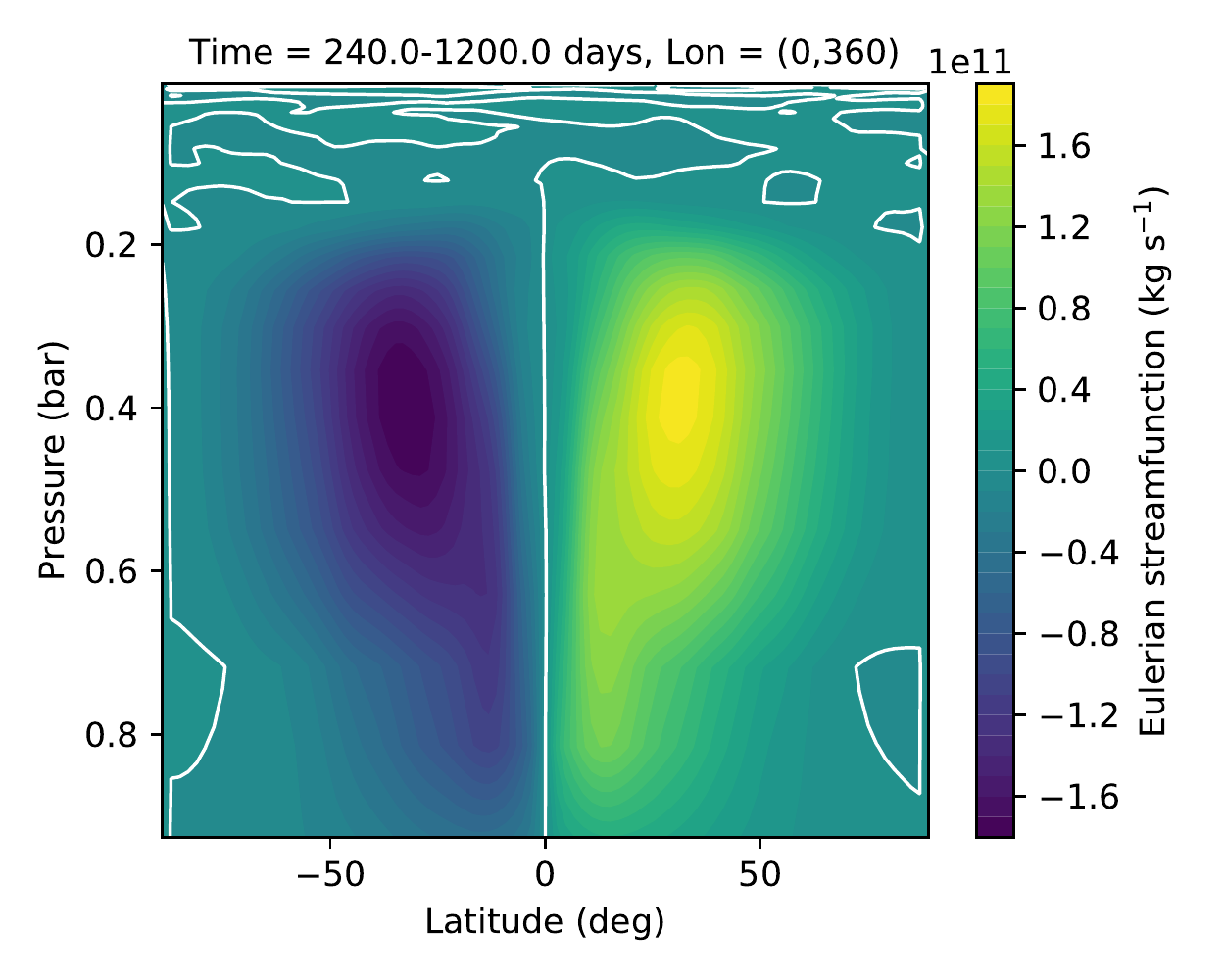}
\includegraphics[width=0.5\textwidth]{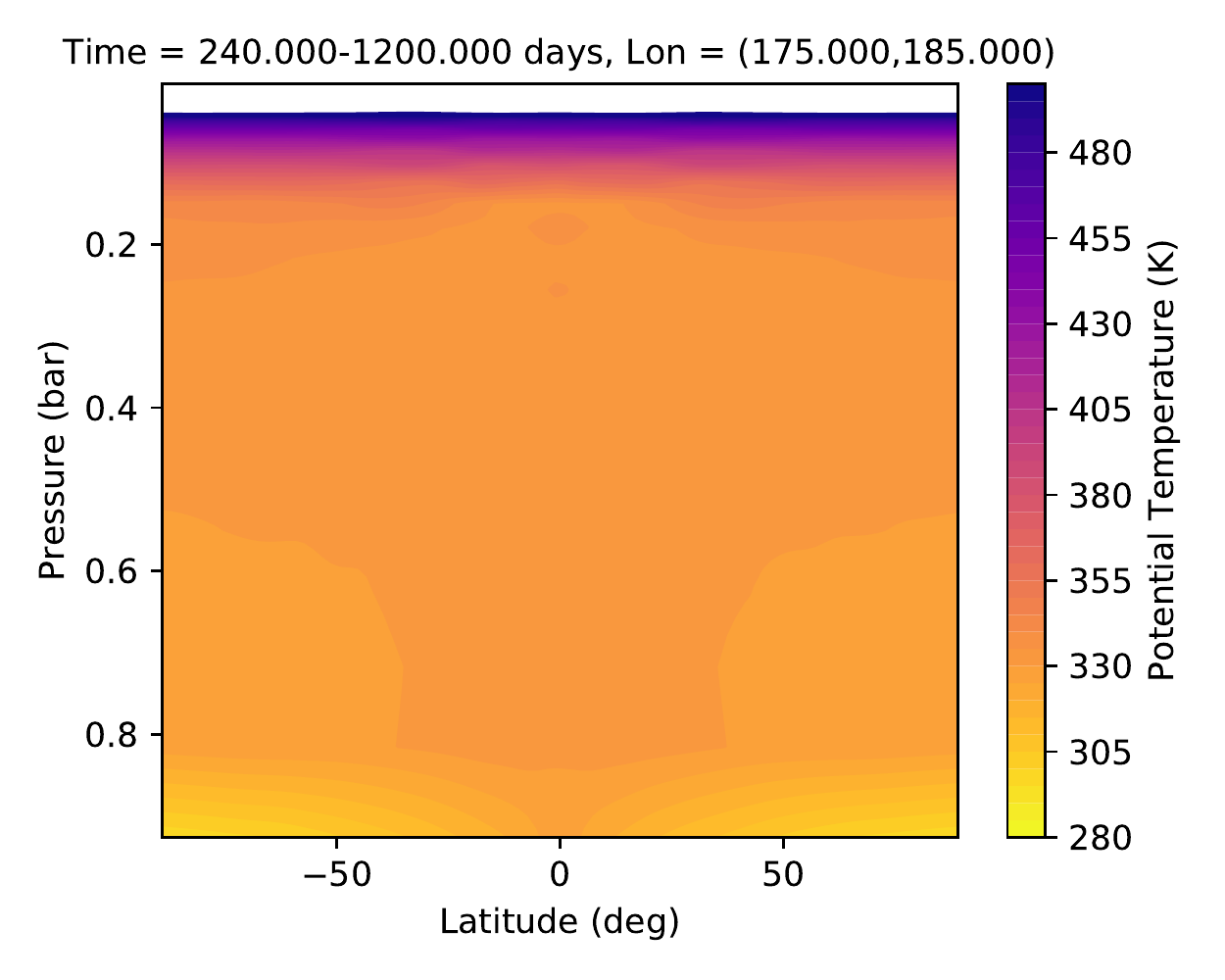}
\caption{Additional quantities from the synchronously rotating Earth benchmark, viewed as a function of latitude and pressure level. The upper left panel is the temperature averaged over a 10$^{\circ}$ slice over the anti-stellar point; the upper right is the temperature averaged over a 10$^{\circ}$ slice over the sub-stellar point; the lower left is the Eulerian mean streamfunction (positive values indicate clockwise motion); the lower right is the potential temperature averaged over 10$^{\circ}$ over the sub-stellar point. In the plot for potential temperature, a narrow region near the top is masked to allow the structure in the lower atmosphere to be discernible---the potential temperature increases sharply up to $\sim 1000$ K in the masked region. As in Figure \ref{fig:tidalearth1}, values are averaged over the interval 240 to 1200 days.} \label{fig:tidalearth2}
\end{figure*}

Figure \ref{fig:tidalearth2} shows several properties as a function of latitude. On the night side, the temperature is highest at $P\sim 0.8$ bar, well above the surface. On the day side, the temperature is highest at the equatorial surface. The transport of heat upward and away from the sub-stellar point is apparent from the temperature distribution centered on $\lambda = 180^{\circ}$ and the streamfunction, which shows a large equator-to-pole Hadley cell in each hemisphere. The potential temperature distribution shows that the atmosphere is marginally unstable near the sub-stellar point. Convective adjustment was not included here.  

For this test, we have not included heating and cooling of the surface. The pressure in the lowest atmospheric layer varies from $\sim 0.93$ bar near the anti-stellar point to $\sim 0.91$ bar near the sub-stellar point. Extrapolating the atmospheric temperature to the reference pressure (roughly the location of the implied surface) for plotting purposes produces poor results because of the steep gradients at the bottom of the model that are not well resolved by our uniform vertical mesh, hence we do not attempt to extend our figures to this region. 

In general, the results compare well with \cite{Heng2011a} and \cite{Merlis2010}. The temperatures and wind speeds are similar to that of \cite{Heng2011a}, with a temperature peak at $\sim 320$ K at the equator, near the surface, and max wind speeds $\sim 13$ m s$^{-1}$ and $\sim 25$ m s$^{-1}$ at $P = 0.9$ and $P = 0.25$ bar, respectively. The fact that the night side temperature in Figure \ref{fig:tidalearth1} is warmer than that  which appears in Figure 3 of \cite{Heng2011a} is due to the difference in pressure level, owing to the poor vertical resolution near the surface in our simulation. The temperatures are $\sim 20 $ K lower across the globe in the simulation of \cite{Merlis2010}, which utilized a radiative transfer scheme; the fact that our temperatures agree with those in \cite{Heng2011a} suggests that this is due to the tuning of the temperature forcing in Equations \ref{eqn:heldsuarezT1} and \ref{eqn:heldsuarezT2} rather than an error in the code. The temperature distribution is otherwise similar to Figure 9 in \cite{Merlis2010}, with a maximum near 0.8 bar on the night side. The Hadley cells are similar in size to theirs, though it is about a factor of 2 weaker in our simulation. 

\subsection{Deep hot Jupiter}
\label{sec:deephj}
Here, we attempt to reproduce the deep hot Jupiter benchmark test from \cite{Heng2011a}. Input parameters are given in Table \ref{tab:modelparaNC}. Like the Held-Suarez test, the benchmark is run with an idealized forcing to the temperature (the winds, in this case, are unforced). As noted in \cite{Mayne2014}, this benchmark has several challenges. First, in a non-hydrostatic model, where the vertical coordinate is altitude rather than pressure, the night side of the planet tends to extend to several orders of magnitude lower pressure than the day side. The exact temperature-pressure profiles suggested by \cite{Heng2011a} tend to lead to runaway cooling on the night side at the start of the simulation, causing the model to crash. \cite{Mayne2014} successfully mitigated this issue in the UK Met Office model by increasing the temperatures at low pressures. We have attempted the same here, with less success. The second issue is the discontinuity in the temperature-pressure profiles at 10 bar, which can lead to numerical instabilities. To avoid the issue, we refit the temperature-pressure profiles used in \cite{Heng2011a} with a new set of polynomials, excluding the pressures near 10 bar. This produces a new profile which varies smoothly in this region. The new polynomial fits are shown in Figure \ref{fig:deephjprof} and presented in Appendix \ref{sec:deephjapp}.

Figure \ref{fig:deephjTu} shows the zonal-mean temperature and zonal wind for the resulting simulation. Unfortunately, our model domain is limited to pressures $\gtrsim 10$ mbar on the day side of the planet. Raising the model top any further causes the night side instability (noted by \cite{Mayne2014} and described above) to occur. At present, it is unclear why the UK Met Office model is able to successfully extend the model domain to lower pressures while \texttt{THOR} is not. However, based on the issues discussed here and in \cite{Mayne2014}, it appears that the deep hot Jupiter benchmark is a challenge to reproduce in altitude-grid models.  

\begin{figure}
\centering
\includegraphics[width=0.5\textwidth]{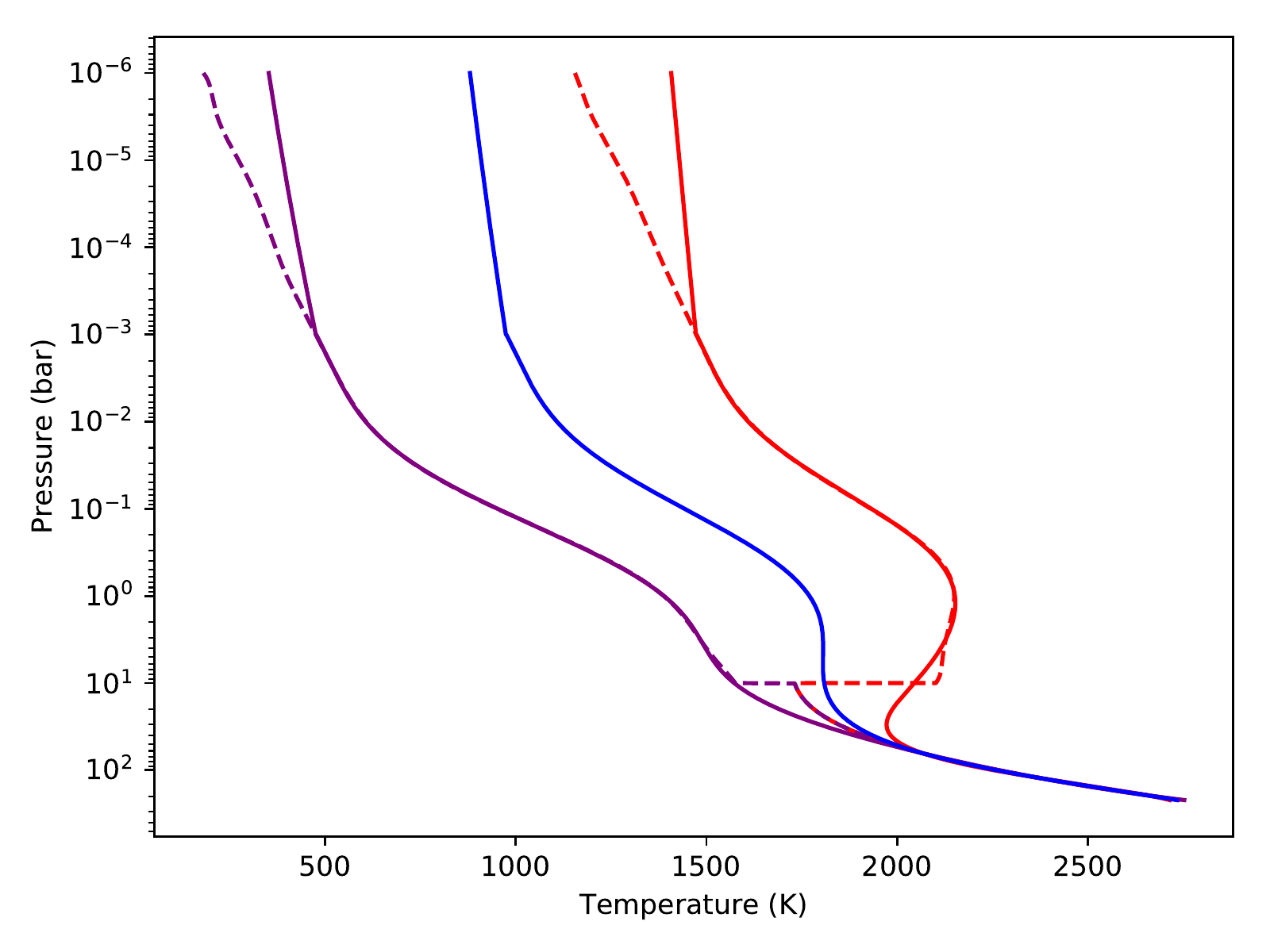}
\caption{Temperature-pressure profiles used in the deep hot Jupiter benchmark test. The equilibrium temperature is equal to $T_{\text{day}}$ (red) at the sub-stellar point, and equal to $T_{\text{night}}$ (purple) at the anti-stellar point. For each location on the planet, the equilibrium temperature is interpolated between $T_{\text{day}}$ and $T_{\text{night}}$ based on the latitude and longitude. Dashed curves are the original profiles from \cite{Heng2011a}, solid are the profiles used in this work. The blue curve represents the average.}  \label{fig:deephjprof}
\end{figure}

We have made several attempts to extend the model domain including: initializing the atmosphere from the average $T_{\text{eq}}$ (blue curve in Figure \ref{fig:deephjprof}) rather than isothermal conditions; including a sponge layer (see Section \ref{sec:diff}) at the top of the model; tuning the strength of divergence damping and hyperdiffusion; and initializing with an zonal wind profile to prompt dayside-nightside mixing. None of the above efforts were successful at preventing the runaway cooling on the nightside that leads to a model instability. 

Even with the model domain limited to $P\gtrsim10 $ mbar on the day side, we do reproduce many of the features of the experiment done in previous works. We see a zonal-mean temperature that is similar in magnitude and structure to that of \cite{Heng2011a} and \cite{Mayne2014}. We see equatorial super-rotation and return flow at the mid-latitudes. The jet speed we find here is weaker ($\sim 3600$ m s$^{-1}$) than in the finite-volume simulation of \cite{Heng2011a} ($\sim 5000$ m s$^{-1}$) and the simulation in \cite{Mayne2014} ($\sim 6000$ m s$^{-1}$). This may be caused by the lower location of the model top (the velocities tend to increase with altitude), though it is also possibly explained by the low horizontal resolution we used for the simulation here (we revisit the problem of resolution in Section \ref{sec:approx}). 

\begin{figure*}
\includegraphics[width=0.5\textwidth]{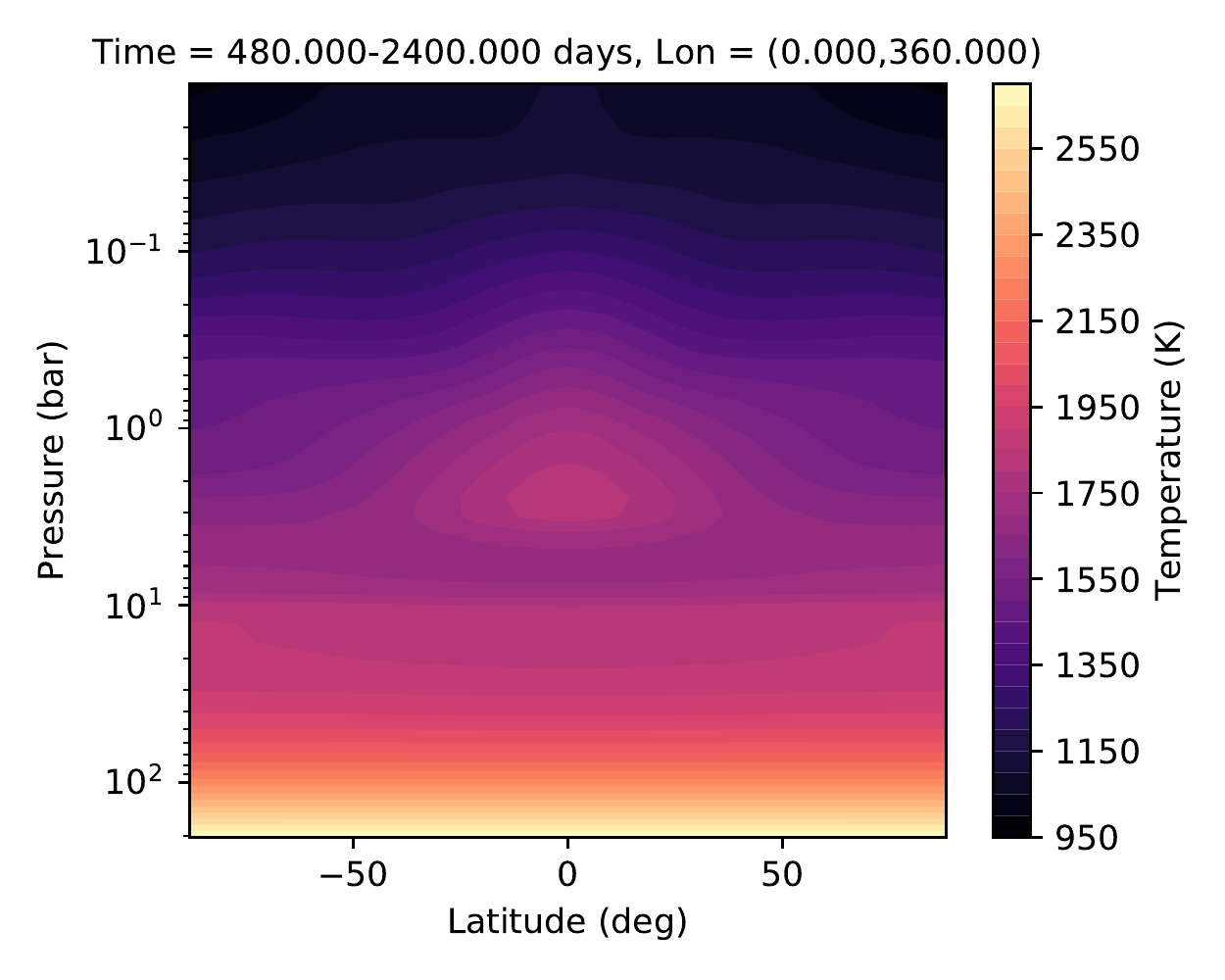}
\includegraphics[width=0.5\textwidth]{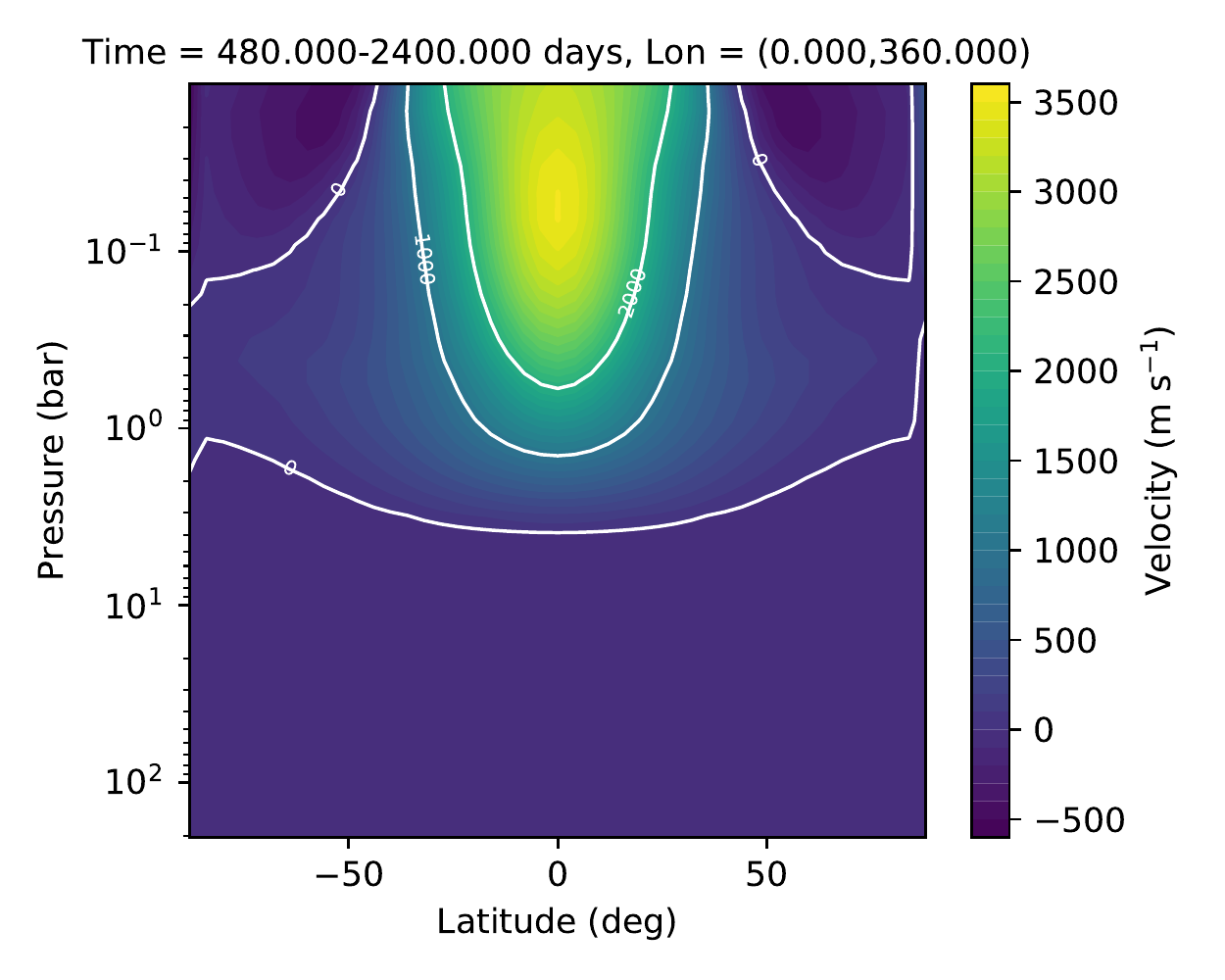}
\caption{Zonally- and temporally-averaged temperature (left) and zonal wind speed (right) in the deep hot Jupiter simulation. The time average was performed over days 480 to 2400 of the simulation. The top altitude is limited to pressures $\sim 10^{-2}$ bar at hottest location on the planet.} \label{fig:deephjTu}
\end{figure*}

\subsection{Acoustic wave experiment}
\label{sec:awexp}
\begin{table*}
\caption{Model parameters for wave simulations}
\centering
\begin{tabular}{lp{0.45\linewidth}ccc}
\hline\hline \\ [-1.5ex]
Symbol & Description & Units & Acoustic waves & Gravity waves\\ [0.5ex]
\hline \\ [-1.5ex]
$r_0$ & Planet radius & m & $6371000$ & 6371000\\ 
$g$ & Gravity & m s$^{-2}$ & $9.8$ & 9.8 \\
$\Omega$ & Rotation rate & rad s$^{-1}$ & 0 & 0 \\
$R_d$ & Gas constant & J K$^{-1}$ kg$^{-1}$ & $287$ & 287 \\
$C_P$ & Atmospheric heat capacity & J K$^{-1}$ kg$^{-1}$  & $1005$ & 1005 \\
$P_{\text{ref}}$ & Reference pressure (bottom boundary) & bar & 1 & 1 \\
$T_{\text{init}}$ & Initial temperature of atmosphere & K & 300 & 300 (at $P_{\text{ref}}$) \\
\hline \\ [-1.5ex]
$\Delta t_M$ & Time step & s & $1800$ & 1800  \\
$z_{\text{top}}$ & Altitude of model top & m & 10000 & 10000 \\
$g_{\text{level}}$ & Grid refinement level & - & 5 & 5 \\
$v_{\text{level}}$ & Number of vertical levels & - & 20 & 20/40 \\
$D_{\text{hyp}}$ & Hyperdiffusion coefficient & - & 0 & 0.01 \\
$D_{\text{div}}$ & Divergence damping coefficient & - & 0/0.02 & 0.01  \\
\hline 
\end{tabular}
\label{tab:modelparaWave}
\end{table*}

Here we demonstrate the \texttt{THOR} GCM's representation of acoustic waves. The purpose of this test is to characterize how well the model is able to represent the propagation of acoustic waves and provide a tool to isolate coding errors that may be hard to diagnose in more complicated scenarios. The setup is almost identical to Section 4.1 of \cite{Tomita2004}, but we describe it here for completeness. The atmosphere is initialized with a background isothermal state, for an Earth-radius planet with no rotation and no forcing (radiation or Newtonian cooling). At the beginning of the first time-step, a pressure perturbation is applied over a spatial distribution centered on longitude $\lambda_0 = 0^{\circ}$ and latitude $\phi_0 = 0^{\circ}$. The distribution in latitude $\phi$ and longitude $\lambda$ is given by
\begin{equation}
    \xi(\lambda,\phi) = \begin{cases}
                            \frac{1}{2}(1+\cos{(\pi x/L)})  & x < L,\\
                            0 & x> L,
                            \end{cases} \label{eqn:xi}
\end{equation}
where $L$ is a representative horizontal length and $x$ is the horizontal distance along a great circle from the point $(\lambda_0,\phi_0)$, and is given by
\begin{equation}
    x = r_0 \cos^{-1}{\left[\sin{\phi_0}\sin{\phi}+\cos{\phi_0}\cos{\phi}\cos{(\lambda-\lambda_0})\right]}
\end{equation}
The vertical distribution is
\begin{equation}
    \zeta(z) = \sin{\left( \frac{n_v \pi z}{z_{top}} \right)}, \label{eqn:zeta}
\end{equation}
where $z$ is the altitude, $z_{top}$ is the height of the model top, and $n_v$ is the vertical wave mode. The total initial perturbation field is
\begin{equation}
    {p}'(t = 0) = \delta p~\xi(\lambda, \phi) \zeta(z),
\end{equation}
where $\delta p$ is the amplitude of the perturbation. Here, as in \cite{Tomita2004}, we set $\delta p = 100$ Pa, $r_0 = 6371$ km, $L = r_0/3$, $n_v = 1$, and $z_{top} = 10$ km. Further input parameters are given in Table \ref{tab:modelparaWave}. 

\begin{figure*}
\includegraphics[width=0.33\textwidth]{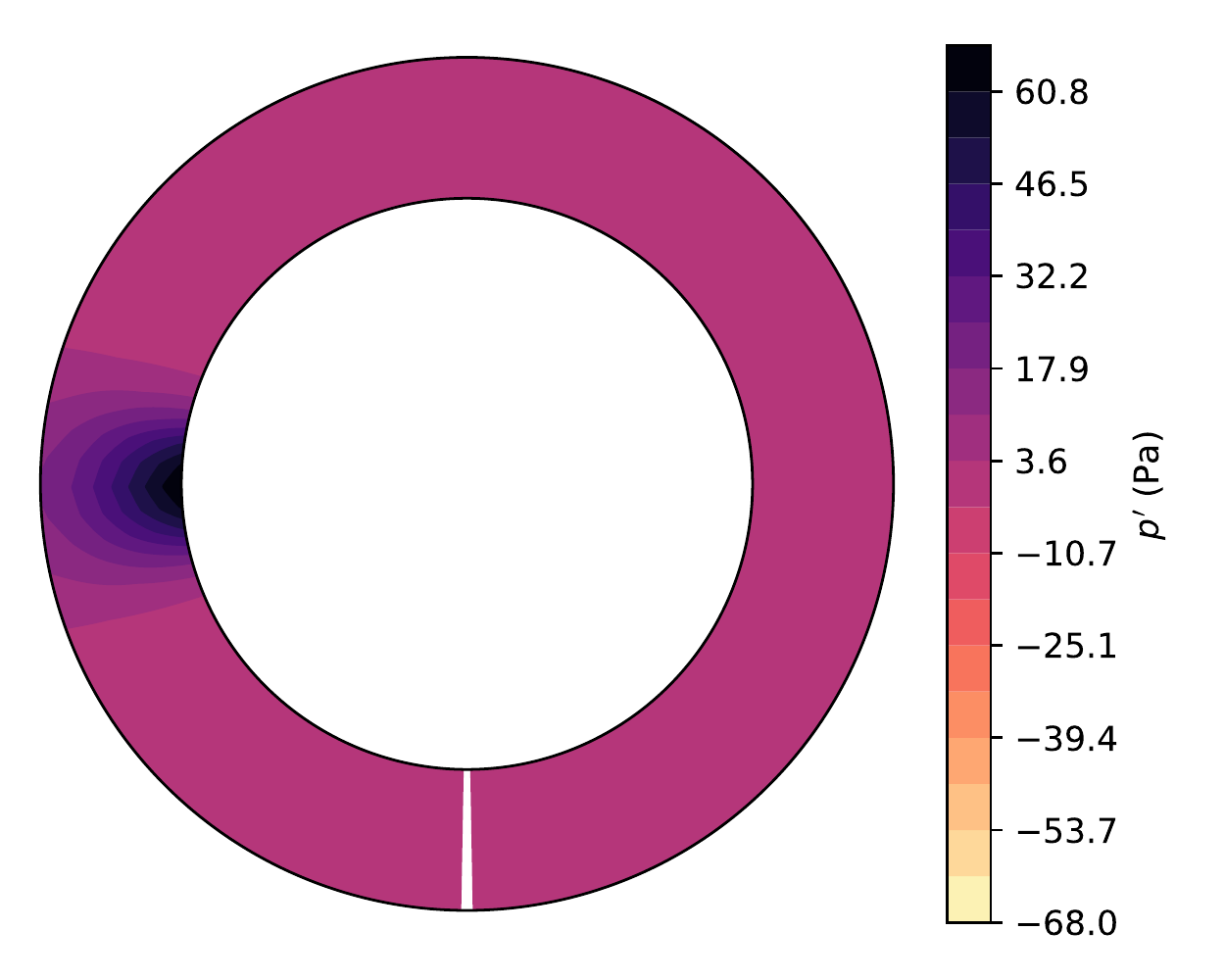}
\includegraphics[width=0.33\textwidth]{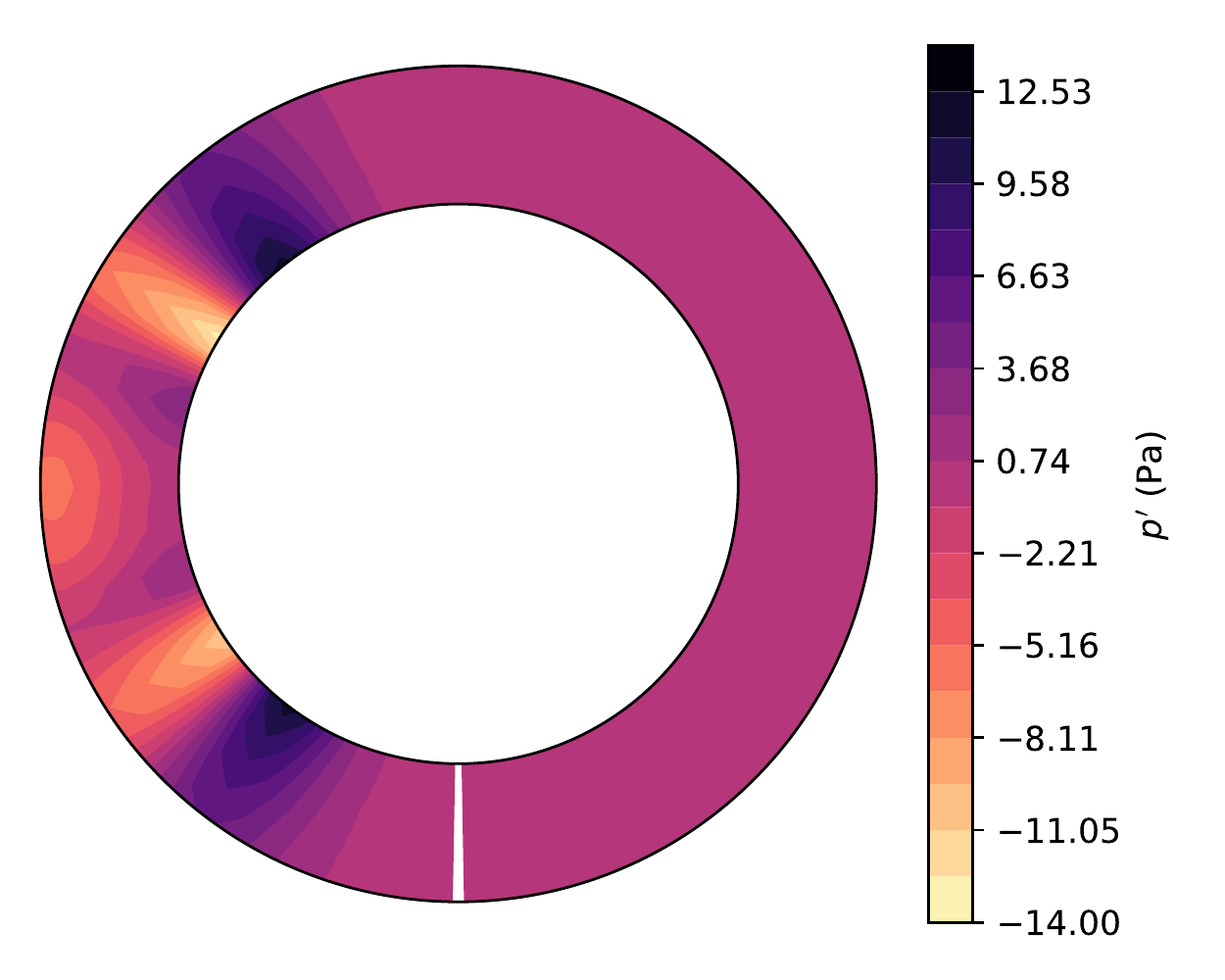} 
\includegraphics[width=0.33\textwidth]{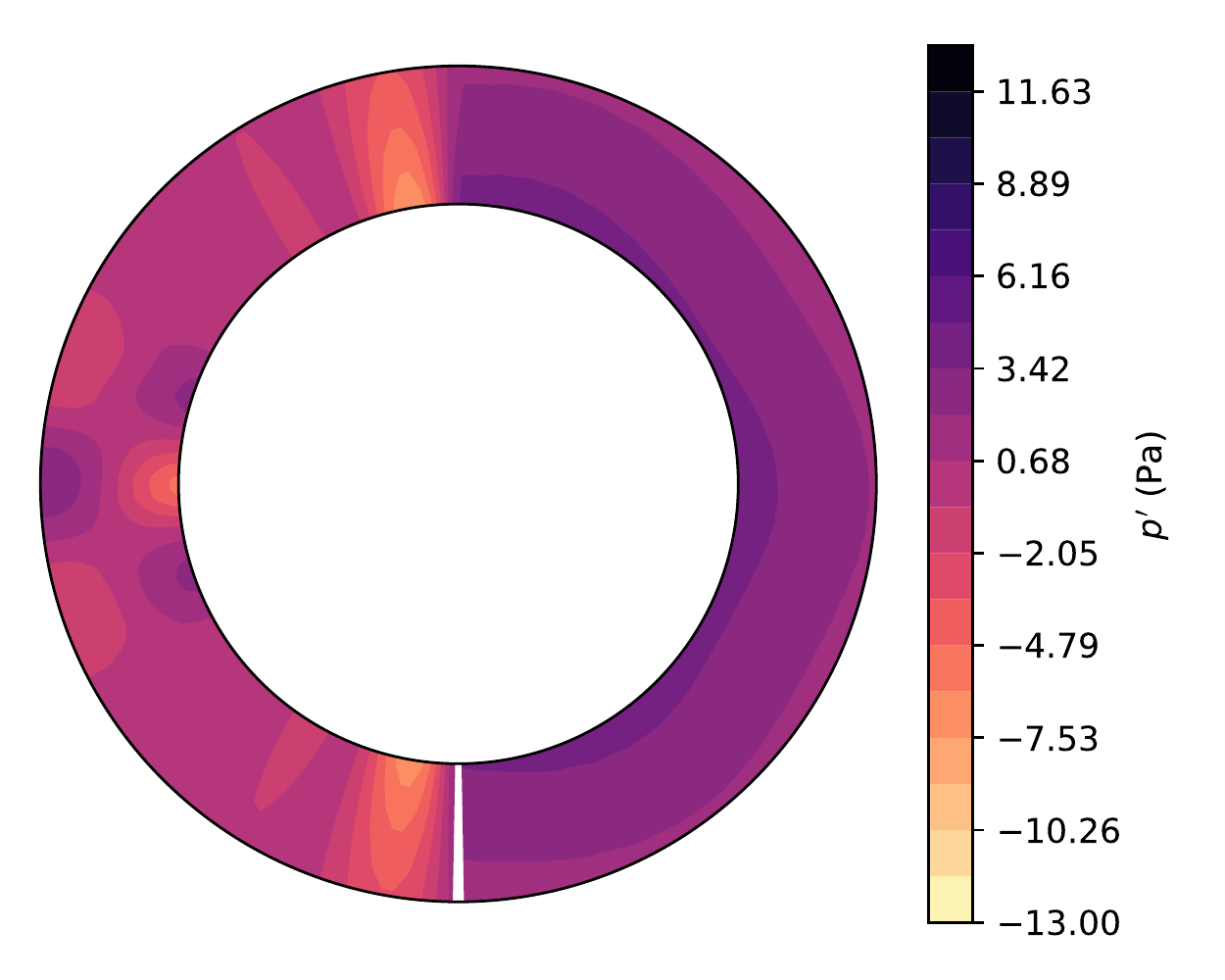} \\
\includegraphics[width=0.33\textwidth]{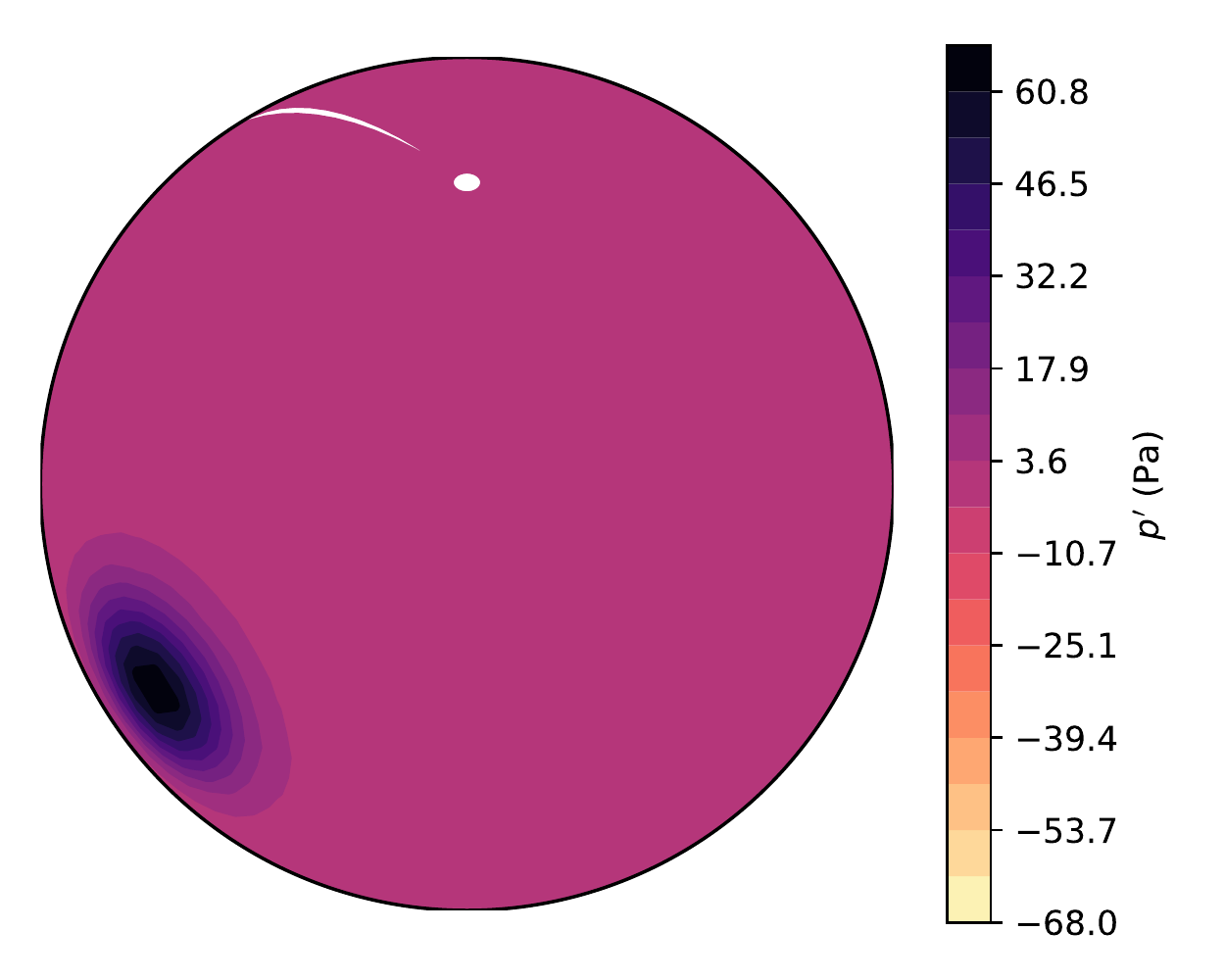}
\includegraphics[width=0.33\textwidth]{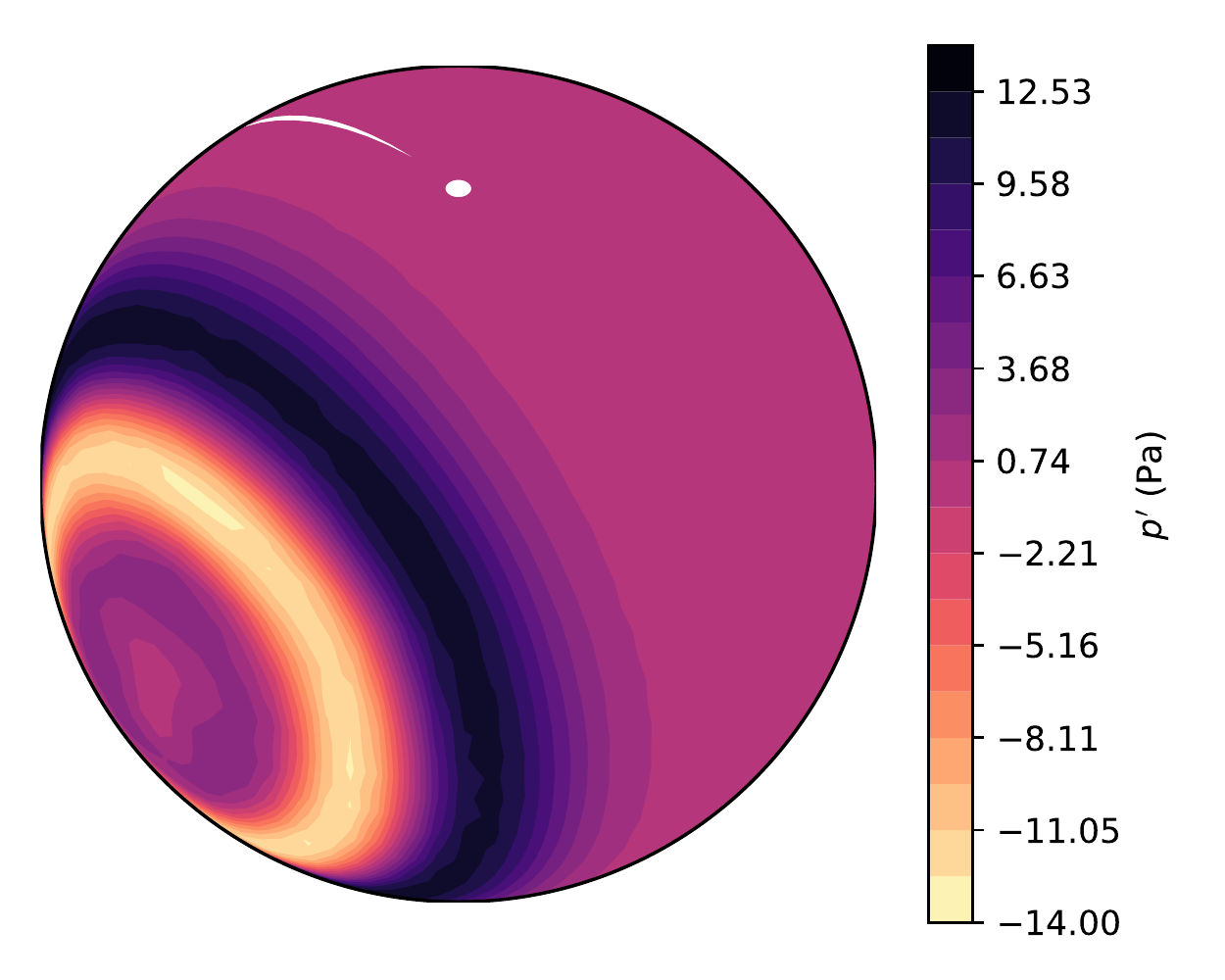}
\includegraphics[width=0.33\textwidth]{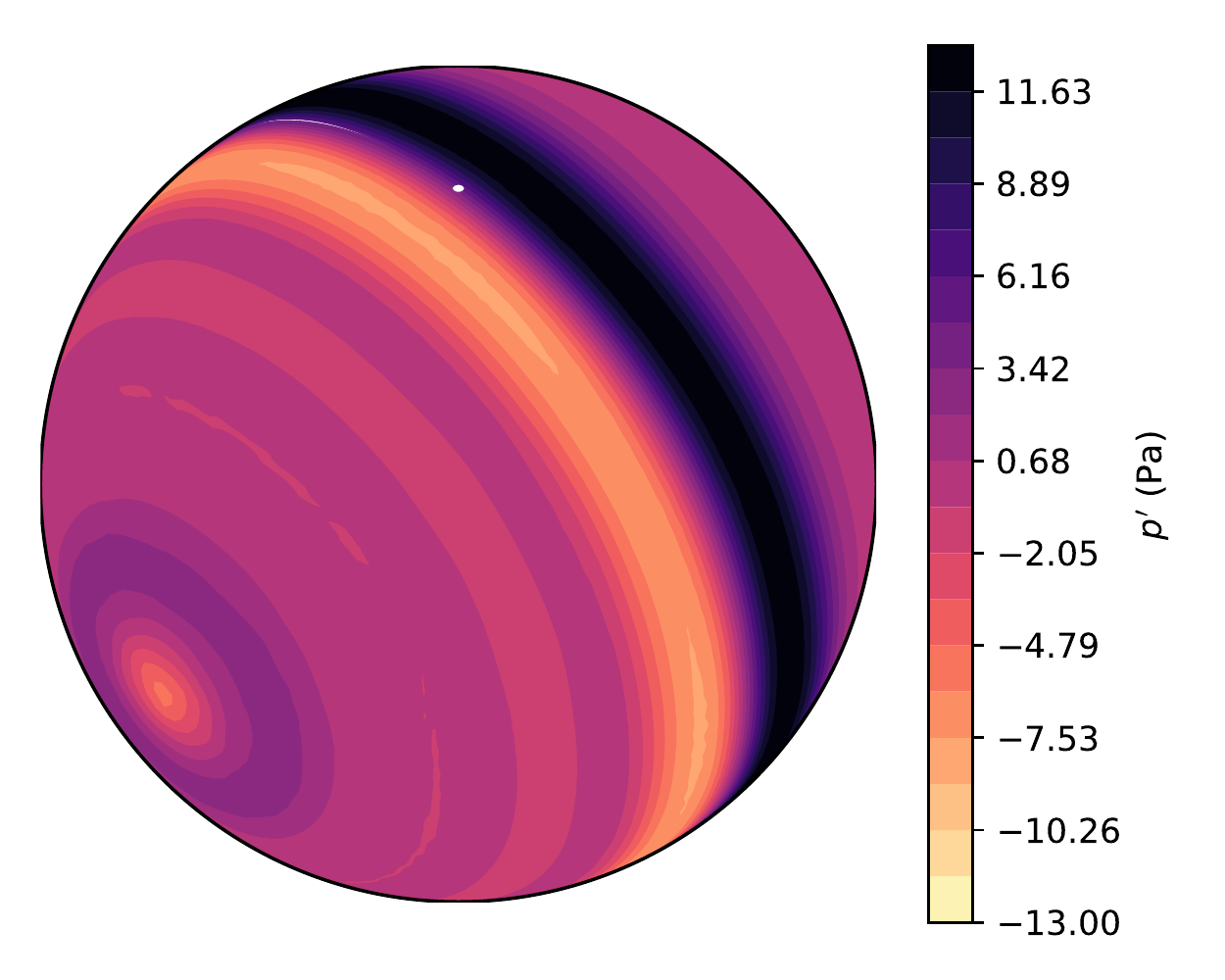}
\caption{The density perturbation in the acoustic wave experiment. The upper panels show the vertical profile around the planet along longitudes $0^{\circ}$ and 180$^{\circ}$. The lower panels show the lowest horizontal level (altitude 250 m). The columns correspond to times $t=(0,4,8)$ hours, from left to right. The plot style and perspective are chosen to facilitate direct comparison with \cite{Tomita2004}. The density perturbation begins at $(\lambda, \phi) = (0^{\circ},0^{\circ})$ and propagates around the planet, reaching the opposite side of the planet in $\sim 17$ hours. \label{fig:aw_pfield}}
\end{figure*}

We compare results from two simulations. In the first, both hyperdiffusion and divergence damping are omitted (by ``divergence damping'' we are referring to 
the terms which depend on the divergence of momentum in Equations 44-46 of \cite{Mendonca2016}, Equation \ref{eqn:divdamp} in this work). In the second, only divergence damping is enabled, with a coefficient $D_{\text{div}} = 0.02$ (hyperdiffusion is still disabled). 
In Figure \ref{fig:aw_pfield}, we show the resulting pressure field in the lowest altitude level at different times for the case with divergence damping. These compare well with Figure 3 of \cite{Tomita2004}. We note, however, that the amplitude of the pressure perturbation in their figure appears much larger than in ours, peaking at $p'\approx 1000$ Pa, which seems inconsistent with an initial perturbation of $\delta p = 100$ Pa as stated in the text. We also see that the amplitude decreases as the wave spreads away from the point of origin (evident in the ranges of the color scales in Figure \ref{fig:aw_pfield}), which is not seen in the \cite{Tomita2004} figure, unless their color scale is mislabeled.

\begin{figure}
\centering
    \includegraphics[width=0.5\textwidth]{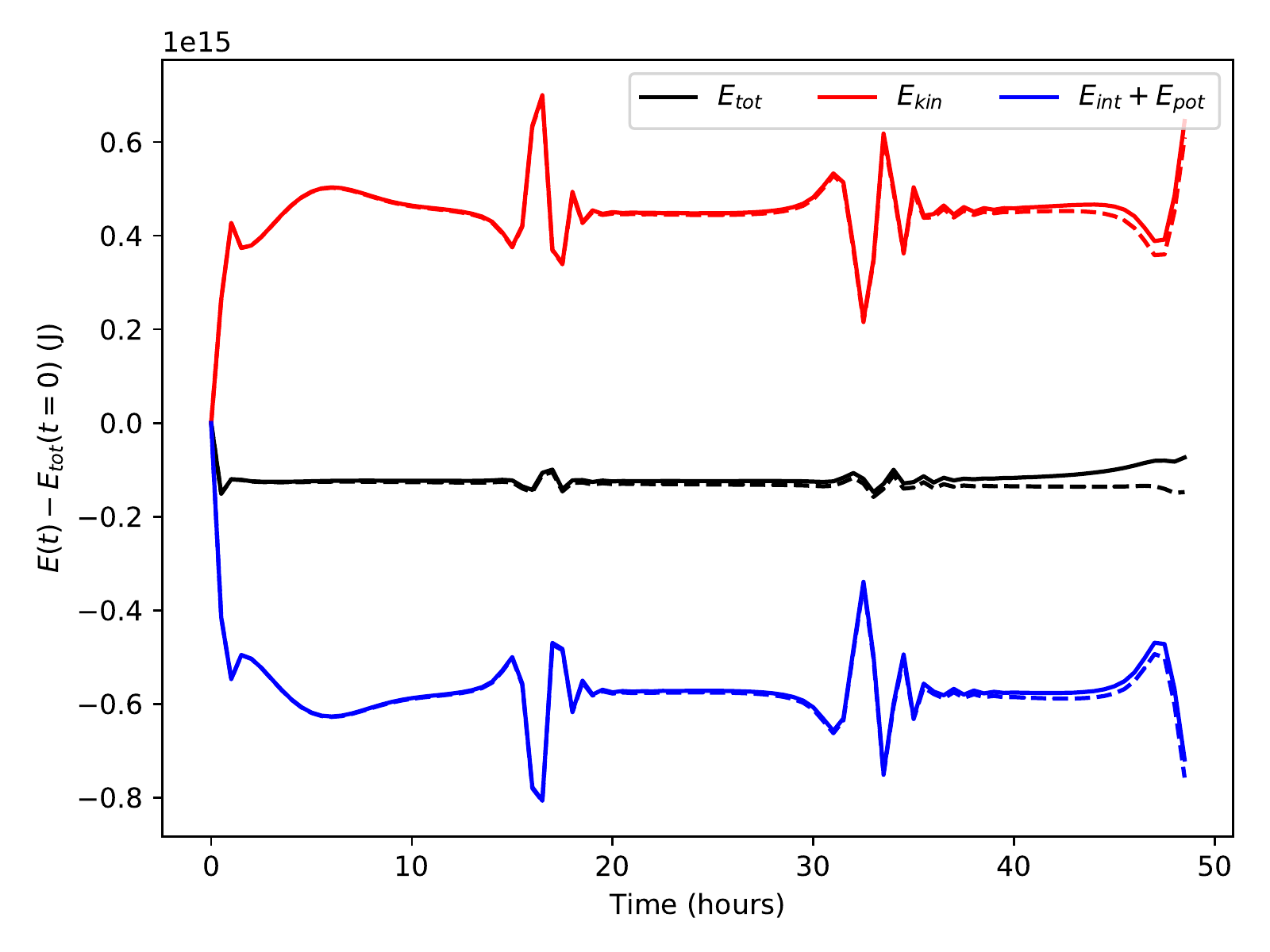}
    \caption{Total, internal plus potential, and kinetic energy in the acoustic wave experiment, as a function of time. Solid curves are for the simulation with no dissipation, dashed are for the simulation with divergence damping. There is some slight adjustment to hydrostatic equilibrium at the beginning of the simulation that results in temperature changes of $<0.1$ K everywhere, and small variations in the energies when the waves meet at $\sim17$ and $\sim 34$ hours. The simulation with no damping has a large energy error that begins to accumulate rapidly after $40$ hours and ultimately crashes the model. With divergence damping enabled, there is a small amount of energy loss. The changes in energy ($\sim 10^{14}$ J) are quite small compared to the total energy of the atmosphere ($10^{24}$ J).}
    \label{fig:aw_energy}
\end{figure}

In Figure \ref{fig:aw_energy}, we plot the globally integrated total energy, internal plus potential energy, and kinetic energy as a function of time. Compare with Figure 4 of \cite{Tomita2004}--though note that their figure has different units. There appears to be a slight mismatch between the exact hydrostatic initial conditions and the \texttt{THOR} algorithm's representation of hydrostatic balance. This leads to a jump in the total energy (on the order of $10^{14}$ J) on the first time step as all columns of the atmosphere adjust very slightly. At present the origin of the discrepancy is unclear, but in any case, the error is quite small compared to the total energy of the atmosphere and should not noticeably affect the results of simulations that include forcing, which produces a much larger change in the overall energy budget compared to the initial conditions. The pressure perturbation is applied at the end of the first time step, reaches the opposite side of the planet in $\approx 17$ h, and returns to the original location in $\approx 33$ h, indicating a sound speed of $\sim 337$ m s$^{-1}$, as found in \cite{Tomita2004}. The theoretical sound speed is $c_s = \sqrt{\gamma R_d T} \approx 347$ m s$^{-1}$, where $\gamma = C_P/C_V$.

\begin{figure*}
\includegraphics[width=0.5\textwidth]{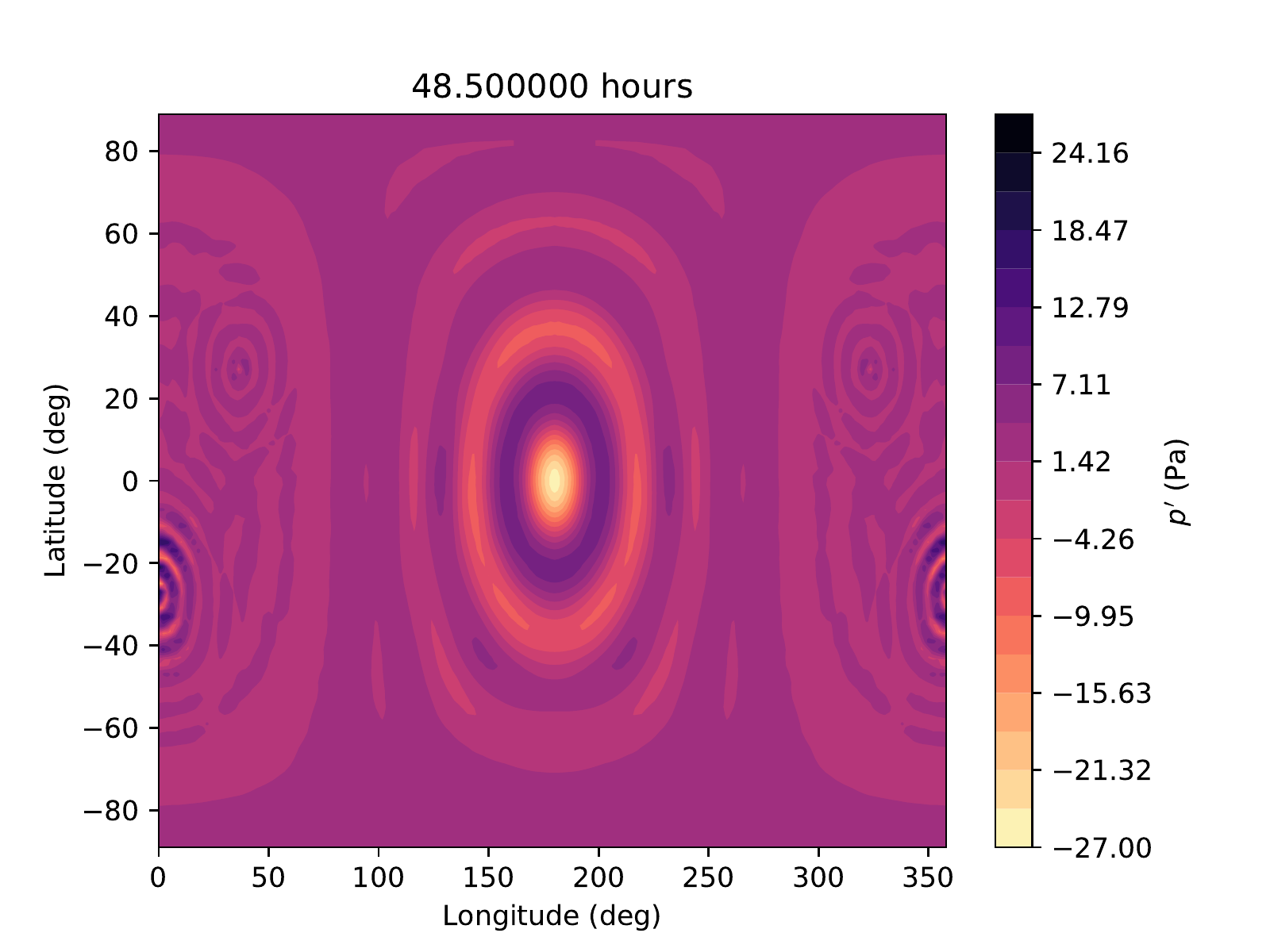}
\includegraphics[width=0.5\textwidth]{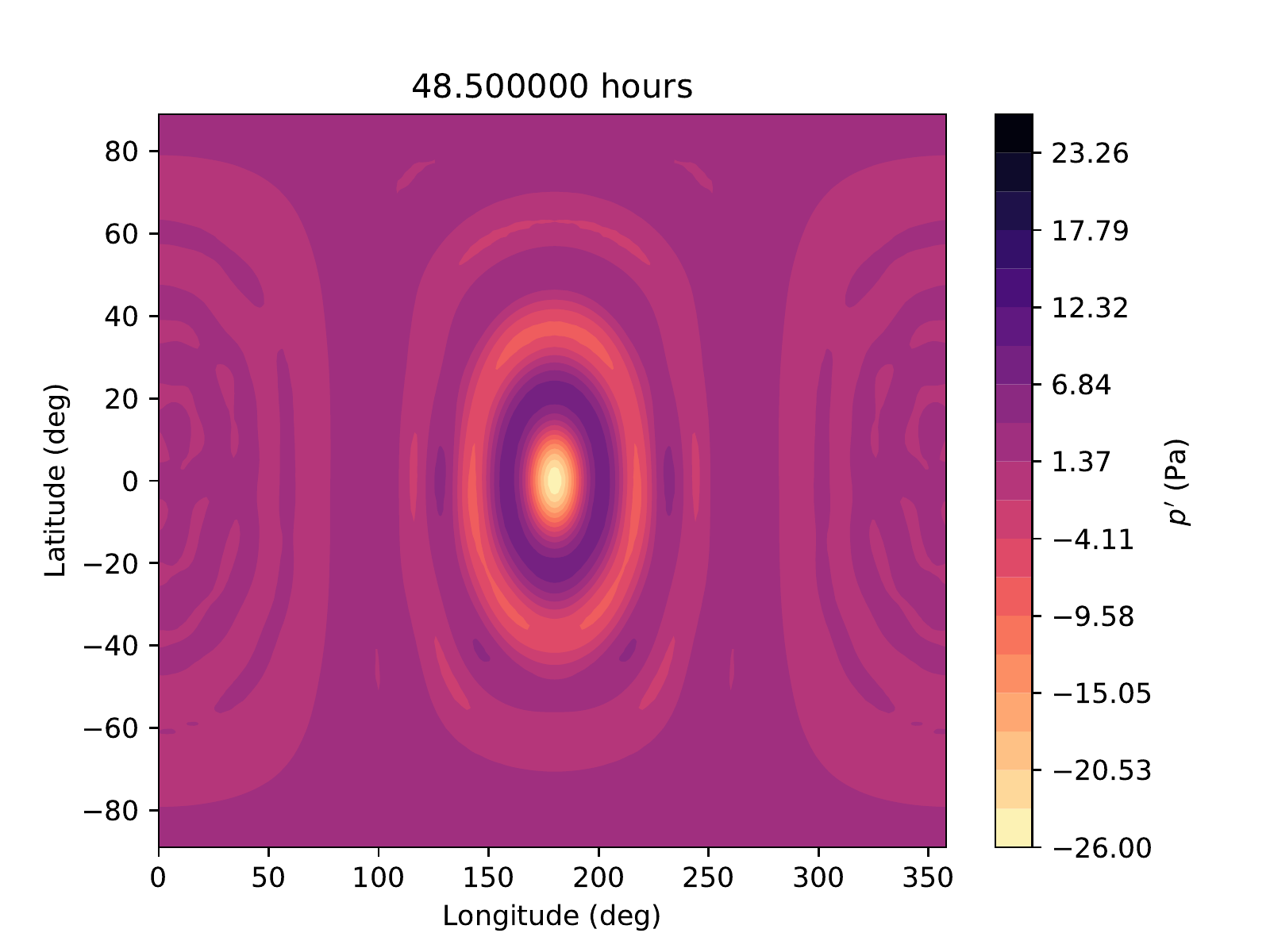}
\caption{Comparison of pressure field for the acoustic wave simulation at $z=250$ m with no damping (left) and with divergence damping (right). }\label{fig:aw_error}
\end{figure*}

From Figure \ref{fig:aw_energy}, we can see that the variation of kinetic energy associated with the sound wave is well compensated by the variation in internal plus potential energy. In the case without divergence damping, the total energy begins to increase erroneously after $\sim 35$ h. The simulation crashes not long after this time. This is the result of grid-scale noise that would be well eliminated by the divergence damping terms. A comparison between the damped and undamped cases of the global pressure field at $48.5$ h is shown in Figure \ref{fig:aw_error}. In the undamped case, we see spurious waves originating from the pentagonal grid points, but these waves are fully eliminated in the damped case. 

The case with divergence damping conserves the energy much better, though there is a slight energy loss over the course of the simulation. Comparing with \cite{Tomita2004}, \texttt{THOR} does not conserve energy as well as their model \texttt{NICAM}. Our choice of entropy as the fundamental quantity in the thermodynamic equation rather than total energy means that the total energy is not conserved as precisely \citep{Satoh2002,Satoh2003}. \cite{Mendonca2016} tested an energy correction scheme in \texttt{THOR} based on \cite{Williamson2009} that, while improving total energy conservation, had little impact of the overall behavior of the model. Thus, until we can develop a more robust general approach, that is physically well described at coarse resolutions, we choose to live with the gradual loss of energy to the dissipation scheme. 

\subsection{Gravity wave experiment} \label{sec:gwexp}
A further benchmark compares the representation of gravity waves in \texttt{THOR} with \texttt{NICAM}. The motivation behind this type of simulation is to characterize the model's representation of gravity waves. This test is originally presented in \cite{Tomita2004}. We set up the atmosphere in the same manner as in that paper and in the acoustic wave experiment of the previous section, using the same planet radius, model top, and shape of the perturbation. See Table \ref{tab:modelparaWave} for the input parameters. The key differences between these simulations and that in Section \ref{sec:awexp} are the atmosphere is given a thermal profile corresponding to a constant Brunt-V\"{a}is\"{a}l\"{a} frequency, instead of an isothermal profile, and the perturbation is applied to the potential temperature, given by
\begin{equation}
\theta'(t=0) = \delta \theta~\xi(\lambda, \phi) \zeta(z),
\end{equation}
where $\theta'$ is the potential temperature perturbation function and $\delta \theta$ is its amplitude. The horizontal and vertical distributions, $\xi$ and $\zeta$, are again given by Equations (\ref{eqn:xi}) and (\ref{eqn:zeta}). In all cases, $\delta \theta = 10$ K. 

The atmosphere for this case is initialized with a constant Brunt-V\"{a}is\"{a}l\"{a} frequency, $N$. This can be defined in terms of the potential temperature as
\begin{equation}
    N^2  = \frac{g}{\theta}\frac{\partial \theta}{\partial z}.
\end{equation}
In order to reduce numerical instabilities and to minimize motion introduced by the initial conditions, we initialize the atmosphere from hydrostatic equilibrium. For an atmosphere with a constant $N$, we must numerically solve the hydrostatic equation for the pressure of each layer as a function of the layer below. In our case, we begin at the reference pressure $P_{\text{ref}}$, which is the pressure at the bottom of the lowest layer at the beginning of the simulation. From there, we progress upward, numerically determining the pressure of each successive layer. More specifically, we use Newton-Raphson iteration to find the pressure in the $i$th layer that satisfies the equation
\begin{equation}
    \frac{P_i - P_{i-1}}{\Delta z} = \frac{1}{2} (\rho_i + \rho_{i-1}) g.
\end{equation}
The relationship between the pressure and density is given by the ideal gas law. An additional constraint must then be made on the temperature, which can be derived from the definition of $N$:
\begin{equation}
    T_i = \frac{1+\beta(N)}{1-\beta(N)}{T_{i-1}},\label{eqn:Tbvfreq}
\end{equation}
where,
\begin{equation}
    \beta(N) = \frac{1}{2}\frac{N^2}{g}\Delta z +\frac{R_d}{C_P}\frac{P_i-P_{i-1}}{P_i+P_{i-1}}.
\end{equation}
The quantities $P_{i-1}$, $\rho_{i-1}$, and $T_{i-1}$ are the pressure, density, and temperature of the layer below the $i$th, except in the case of the lowest layer, where $P_{i-1} = P_{\text{ref}}$, $\rho_{i-1} = P_{\text{ref}}/(R_d T_{\text{init}})$, and $T_{i-1} = T_{\text{init}}$. $T_{\text{init}}$ is a user set initial reference temperature. The value of $\Delta z$ is the distance between the centers of the layers except in the case of the lowest layer, where it is simply the height of the lowest layer. After making the initial guess that $\rho_i = \rho_{i-1}$, we do one Newton-Raphson step to find a new value of $P_i$, update $T_i$ and $\rho_i$ (via Equation \ref{eqn:Tbvfreq}), then repeat the process until the change in $P_i$ is below some threshold (we use $10^{-8}$).

\begin{figure*}
\centering
\includegraphics[width=0.8\textwidth]{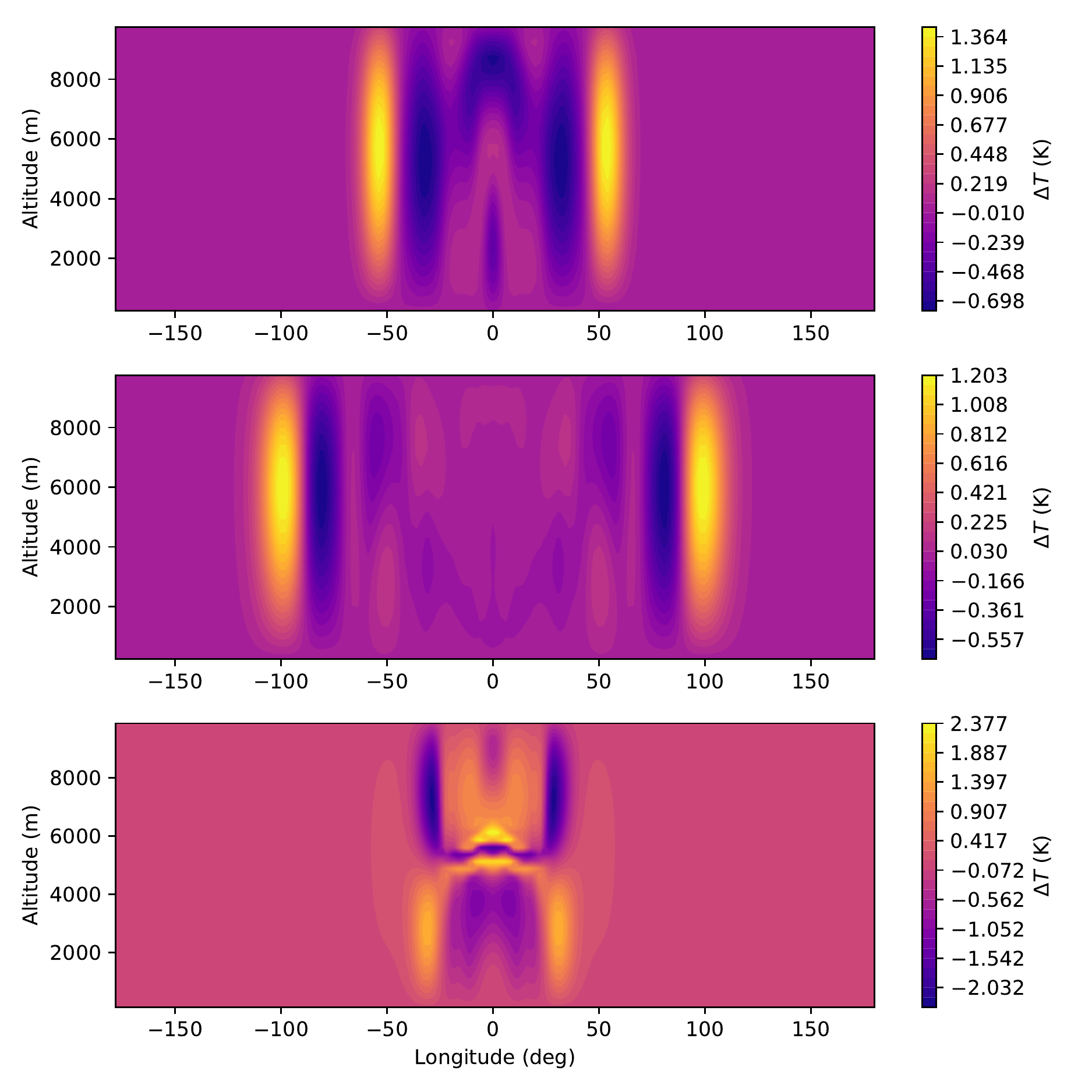}
\caption{Temperature perturbation along the equator in the gravity wave simulations at 48 hours. $\Delta T$ is the difference in temperature from the initial temperature field. The top panel has a Brunt-V\"{a}is\"{a}l\"{a} frequency, $N = 0.01$ s$^{-1}$ and vertical mode $n_v = 1$, the middle has $N = 0.02$ s$^{-1}$ and $n_v = 1$, and the lower panel has $N = 0.01$ s$^{-1}$ and $n_v = 2$. Compare with Figure 5 in \cite{Tomita2004}.}
\label{fig:gwpert}
\end{figure*}

In Figure \ref{fig:gwpert}, we show the results of three simulations. The first has $(N,n_v) = (0.01$ s$^{-1}$,$1)$, the second  $(N,n_v) = (0.02$ s$^{-1}$,$1)$, and the third $(N,n_v) = (0.01$ s$^{-1}$,$2)$. The number of vertical levels is 20 in the $n_v=1$ cases and 40 in the $n_v =2 $ case.  The color contours show the temperature perturbation, $\Delta T$, at the equator after 48 hours of integration. \cite{Tomita2004} give a theoretical estimate for the gravity wave speed, $c_g = N z_{\text{top}}/\pi n_v$, which is 31.8 m s$^{-1}$, 63.7 m s$^{-1}$, and 15.9 m s$^{-1}$ in the three respective cases. The leading peaks in the three cases are located at $\lambda \approx (55^{\circ}, 95^{\circ}, 30^{\circ})$, indicating speeds $c_g \approx (35, 61, 19)$ m s$^{-1}$. \cite{Tomita2004} noted, in particular, the larger relative error in their $n_v=2$ case ($c_g \approx $18 m s$^{-1}$), and theorized that this was caused by insufficient vertical resolution---they used 20 vertical levels in this case. However, we have run the same test with 20 levels and 40 levels (the lower panel of \ref{fig:gwpert} shows the 40 level case), and the locations of the wave peaks are virtually identical in both cases. Most likely, the error in the speed is dominated simply by the difficulty in estimating the locations of the wave fronts.  
 
\begin{figure}
\centering
    \includegraphics[width=0.5\textwidth]{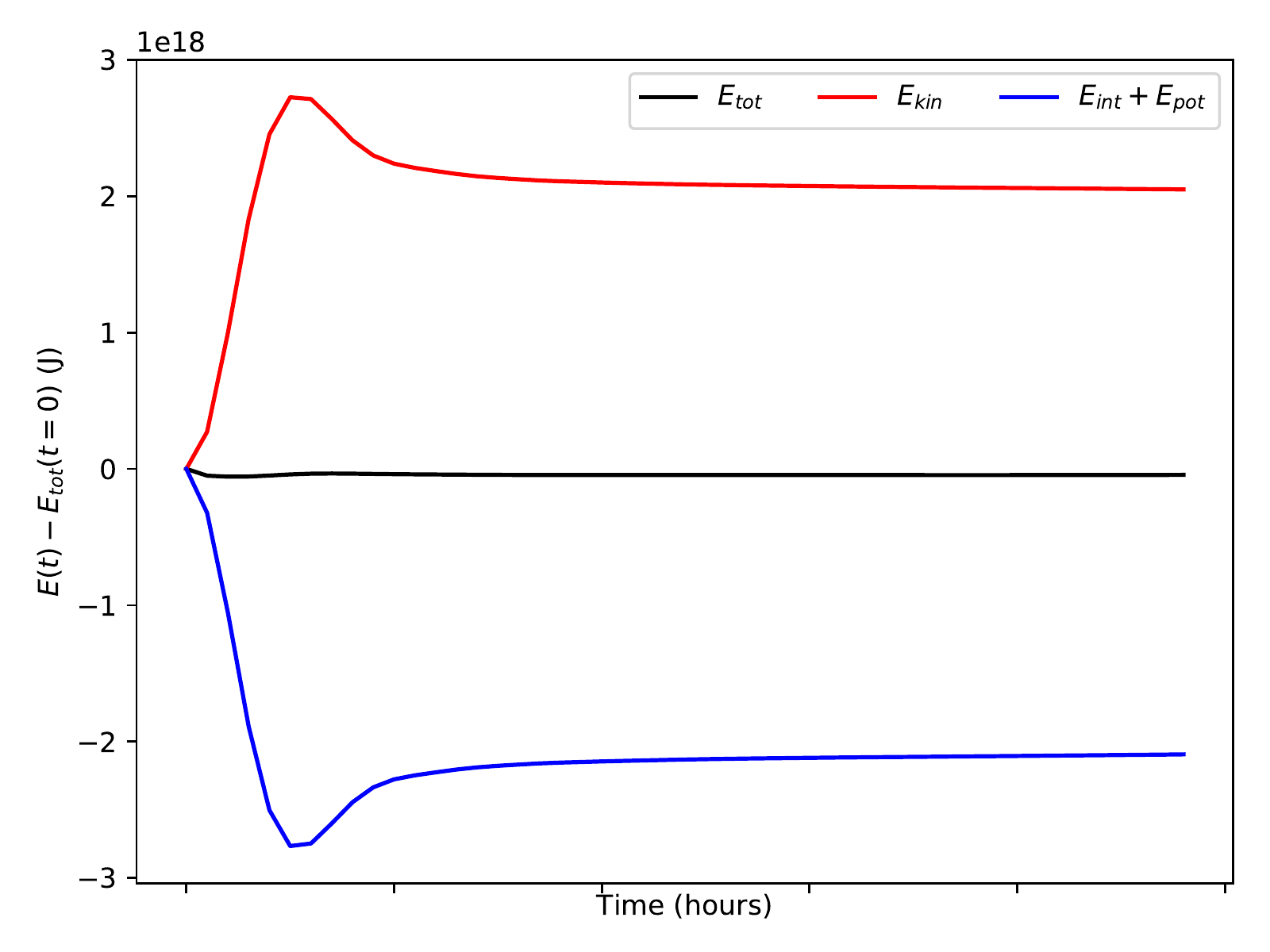}
    \caption{Total, internal + potential, and kinetic energy in the gravity wave experiment, as a function of time, for $N = 0.01$ s$^{-1}$ and $n_v = 1$. Compare with Figure 6 in \cite{Tomita2004}.}
    \label{fig:gw_energy}
\end{figure}

Figure \ref{fig:gw_energy} shows the evolution of the total, internal plus potential, and kinetic energy. As in the acoustic wave case, the kinetic energy mirrors the change in internal plus potential energy. The oscillation in the total energy is orders of magnitude smaller than the others, indicating that the total energy is conserved well. This compares well with Figure 6 in \cite{Tomita2004}, though, as noted in the acoustic wave case, we plot the absolute energy (in J). 

\section{Comparison of THOR with shallow and quasi-hydrostatic approximations}
\label{sec:approx}

\subsection{Earth-like case with double gray opacity} \label{sec:earthrt}

Here, we present a dry Earth-like case to compare to the Held-Suarez test and validate the double gray radiative transfer scheme. Input parameters are given in Table \ref{tab:modelpara}. We utilize optical depth profiles for the short-wave and long-wave radiation in the same style as 
\cite{Frierson2006} and \cite{Heng2011b}, with $n_{\text{sw}} = 2$, $n_{\text{lw}} = 4$, $\tau_{\text{sw}} = 0.2$, $\tau_{\text{lw,eq}} = 6$, and $\tau_{\text{lw,pole}} = 1.5$. In this case, the insolation, or the distribution of incident solar/stellar radiation at the top of the atmosphere, is given by 
\begin{equation}
    Q = S_0 \cos{\phi} \cos{(\lambda-\alpha)},
\end{equation}
where $S_0 = 1367$ W m$^{-2}$, $\phi$ is the latitude, $\lambda$ is the longitude, and the angle $\alpha$ at time $t$ is,
\begin{equation}
    \alpha = (\Omega - n_{\text{orb}}) t.
\end{equation}
Here, $\Omega = 7.292 \times 10^{-5}$ is the rotation rate and $n_{\text{orb}} =  1.98\times 10^{-7}$ is the orbital mean motion. This insolation pattern assumes a zero eccentricity orbit and zero obliquity, but resolves the diurnal cycle. With zero eccentricity, the mean motion $n$ is simply the angular velocity of the planet about the sun. In a future work, we will generalize the insolation to arbitrary orbital and rotation states.

\begin{table*}
\caption{Model parameters for gray RT simulations}
\centering
\begin{tabular}{lp{0.45\linewidth}ccc}
\hline\hline \\ [-1.5ex]
Symbol & Description & Units & Earth & HD 189733-b \\ [0.5ex]
\hline \\ [-1.5ex]
$r_0$ & Planet radius & m & $6371000$ & 79698540 \\ 
$g$ & Gravity & m s$^{-2}$ & $9.8$ & 21.4 \\
$\Omega$ & Rotation rate & rad s$^{-1}$ & $7.292\times10^{-5}$ & $3.279\times10^{-5}$ \\
$R_d$ & Gas constant & J K$^{-1}$ kg$^{-1}$ & $287$ & $3779$ \\
$C_P$ & Atmospheric heat capacity & J K$^{-1}$ kg$^{-1}$  & $1005$ & $13226.8$ \\
$P_{\text{ref}}$ & Reference pressure (bottom boundary) & bar & 1 & 220 \\
$T_{\text{init}}$ & Initial temperature of atmosphere & K & 300 & 1600 \\
\hline \\ [-1.5ex]
$\Delta t_M$ & Time step & s & 500/300 & 150/100  \\
$z_{\text{top}}$ & Altitude of model top & m & 36000 & $3.6\times10^6$ \\
$g_{\text{level}}$ & Grid refinement level & - & 5/6 & 4/5 \\
$v_{\text{level}}$ & Number of vertical levels & - & 32 & 40 \\
$D_{\text{hyp}}$ & Hyperdiffusion coefficient & - & $4.8\times10^{-3}$ & $9.97\times10^{-3}$ \\
$D_{\text{div}}$ & Divergence damping coefficient & - & $4.8\times10^{-3}$  & $9.97\times10^{-3}$  \\
\hline \\ [-1.5ex]
$S_0$ & Incident stellar flux & W m$^{-2}$ & 1367 & 467072 \\
$A_0$ & Albedo & - & 0.3135 & 0.18 \\
$n_{\text{orb}}$ & Orbital mean motion & rad s$^{-1}$ & $1.98\times10^{-7}$ & $3.279\times10^{-5}$ \\
$n_{\text{lw}}$ & Power-law index for long-wave optical depth & - & 4 &  2\\
$f_l$ & Strength of well-mixed absorber (long-wave) & - & 0.1 & 0.5 \\
$\tau_{\text{lw,eq}}$ & Long-wave optical depth at $P_{\text{ref}}$ at equator & - & 6 & 4680 \\
$\tau_{\text{lw,pole}}$ & Long-wave optical depth at $P_{\text{ref}}$ at poles & - & 1.5 & 4680 \\
$n_{\text{sw}}$ & Power-law index for short-wave optical depth & - & 2 & 1 \\
$\tau_{\text{sw}}$ & Short-wave optical depth at $P_{\text{ref}}$ & - & 0.2 & 1170 \\
$\mathcal{D}$ & Diffusivity factor & - & 1 & 2 \\
$C_{\text{surf}}$ & Heat capacity of surface (lower boundary) & J K$^{-1}$ m$^{-2}$ & $10^7$ & - \\
\hline \\ [-1.5ex]
$k_{\text{surf}}$ & Friction coefficient of lower boundary & s$^{-1}$ & $1.1574\times10^{-5}$ & - \\ 
$\sigma_b$ & Boundary layer top (fraction of $P_{\text{ref}}$) & - & 0.7 & - \\
$k_{\text{sp}}$ & Sponge layer strength & s$^{-1}$ & - & $10^{-3}$ \\
$\eta_{\text{sp}}$ & Bottom of sponge layer (fraction of $z_{\text{top}}$) & - & - & 0.8 \\
$n_{\text{lats}}$ & Number of latitude bins used in sponge layer & - & - & 20 \\
\hline 
\end{tabular}
\label{tab:modelpara}
\end{table*}

\begin{figure*}
\includegraphics[width=0.5\textwidth]{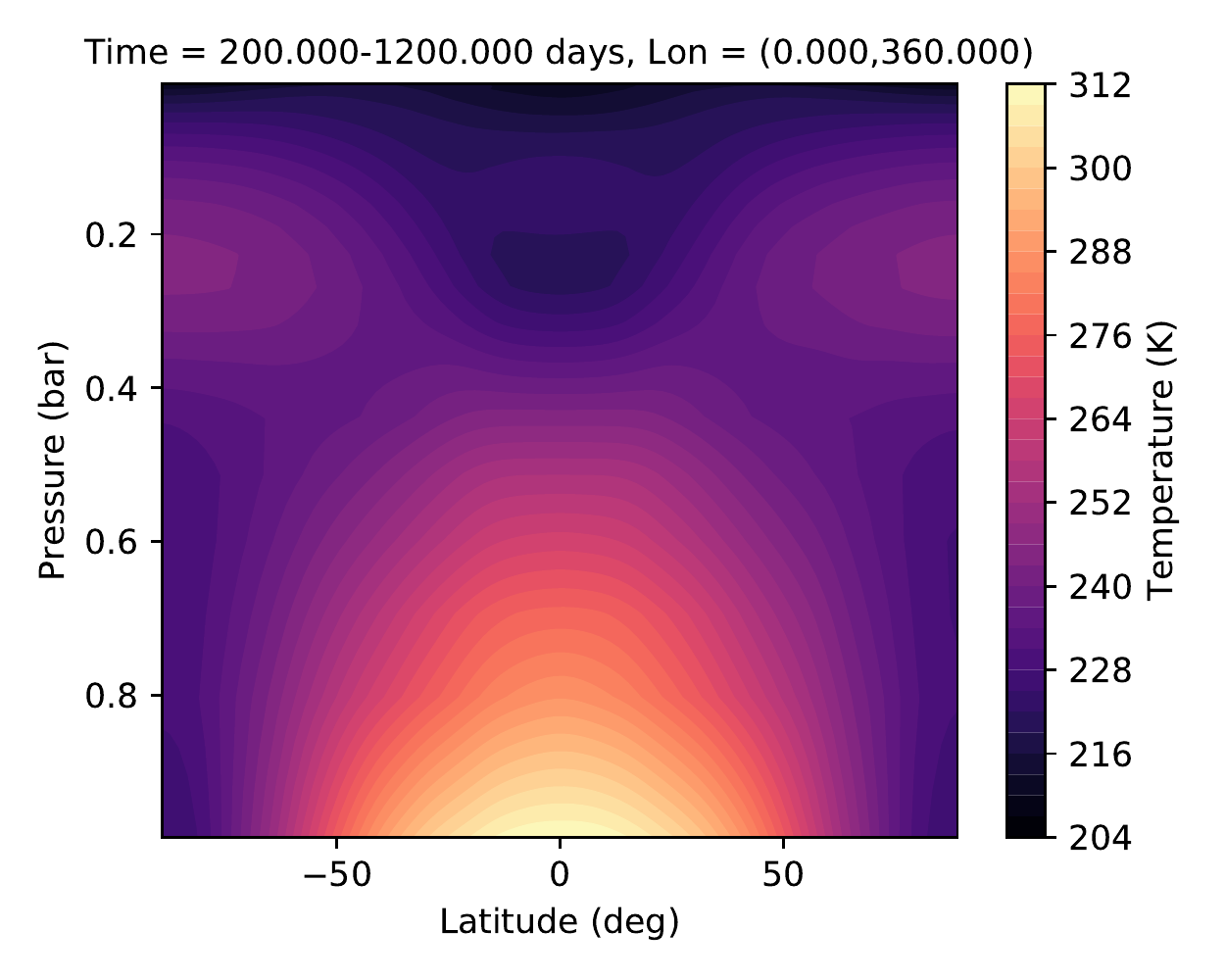}
\includegraphics[width=0.5\textwidth]{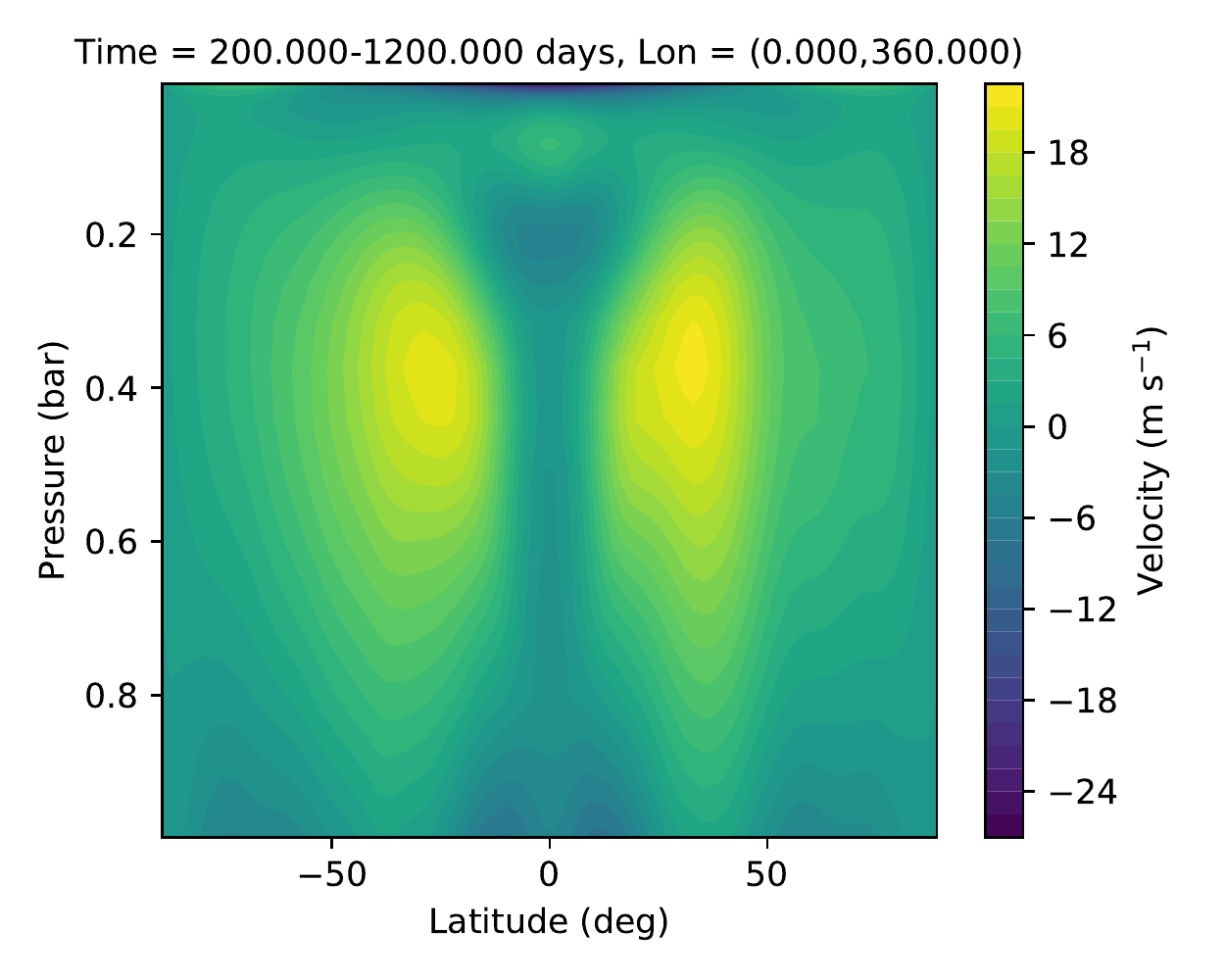} \\
\includegraphics[width=0.5\textwidth]{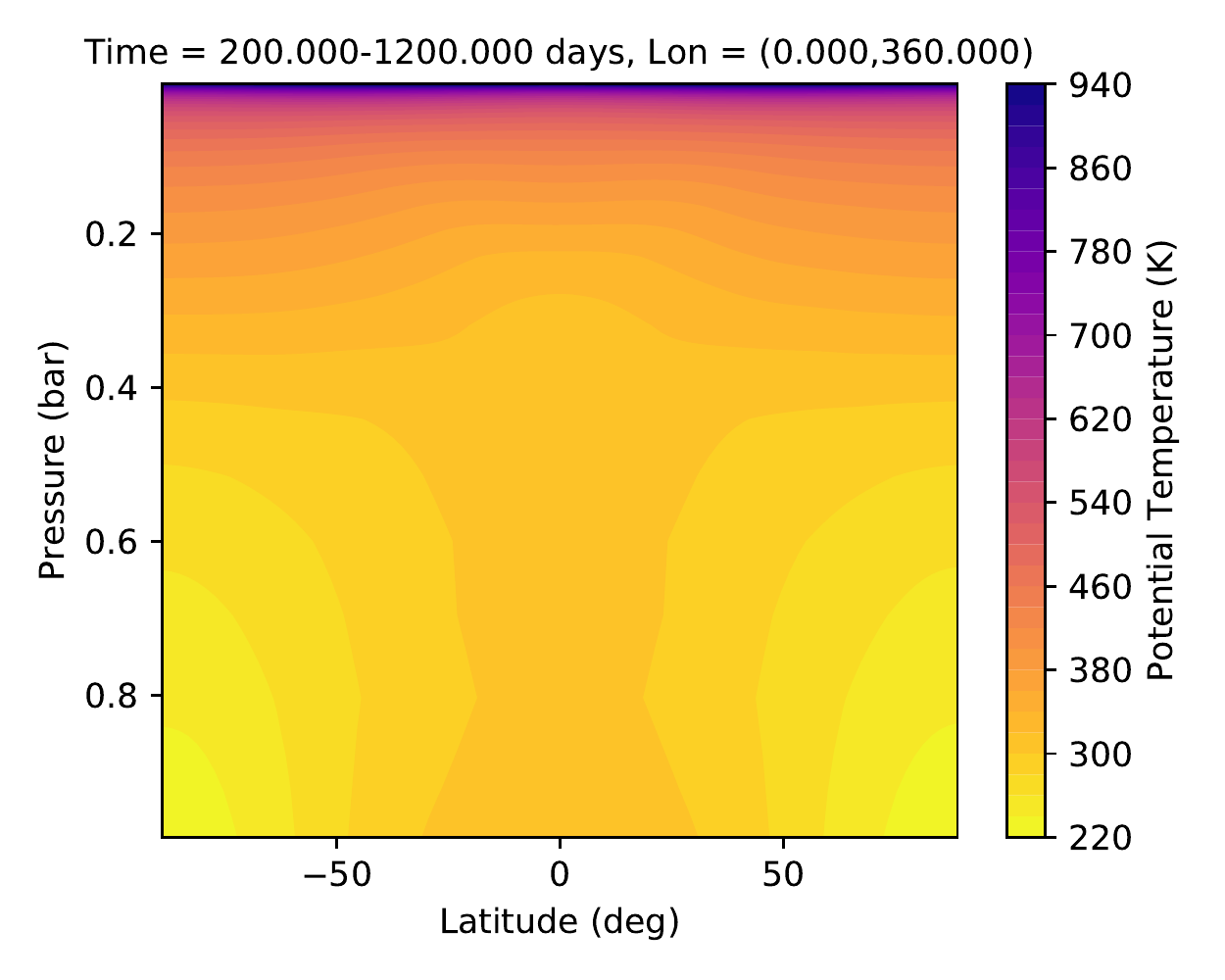}
\includegraphics[width=0.5\textwidth]{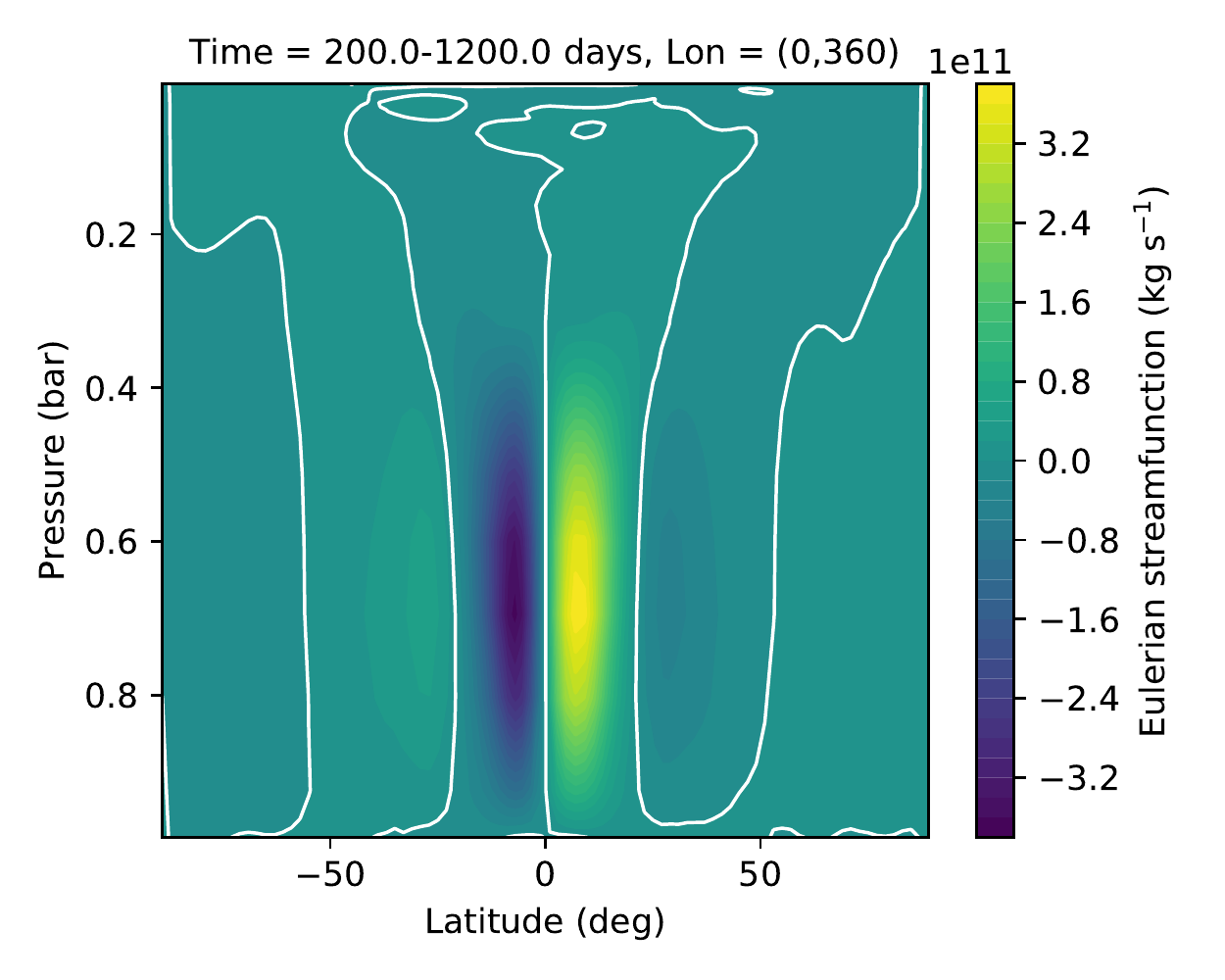}
\caption{Zonal- and time-averaged quantities from the Earth-like, non-hydrostatic deep (NHD) simulation, without convective adjustment enabled. Upper left shows the temperature, upper right the zonal wind speed, lower left the potential temperature, and lower right the mass stream function (positive values indicate clockwise motion). The white line in the lower right panel is the zero pass contour.} \label{fig:earthnodccomp}
\end{figure*}

\begin{figure*}
\includegraphics[width=0.5\textwidth]{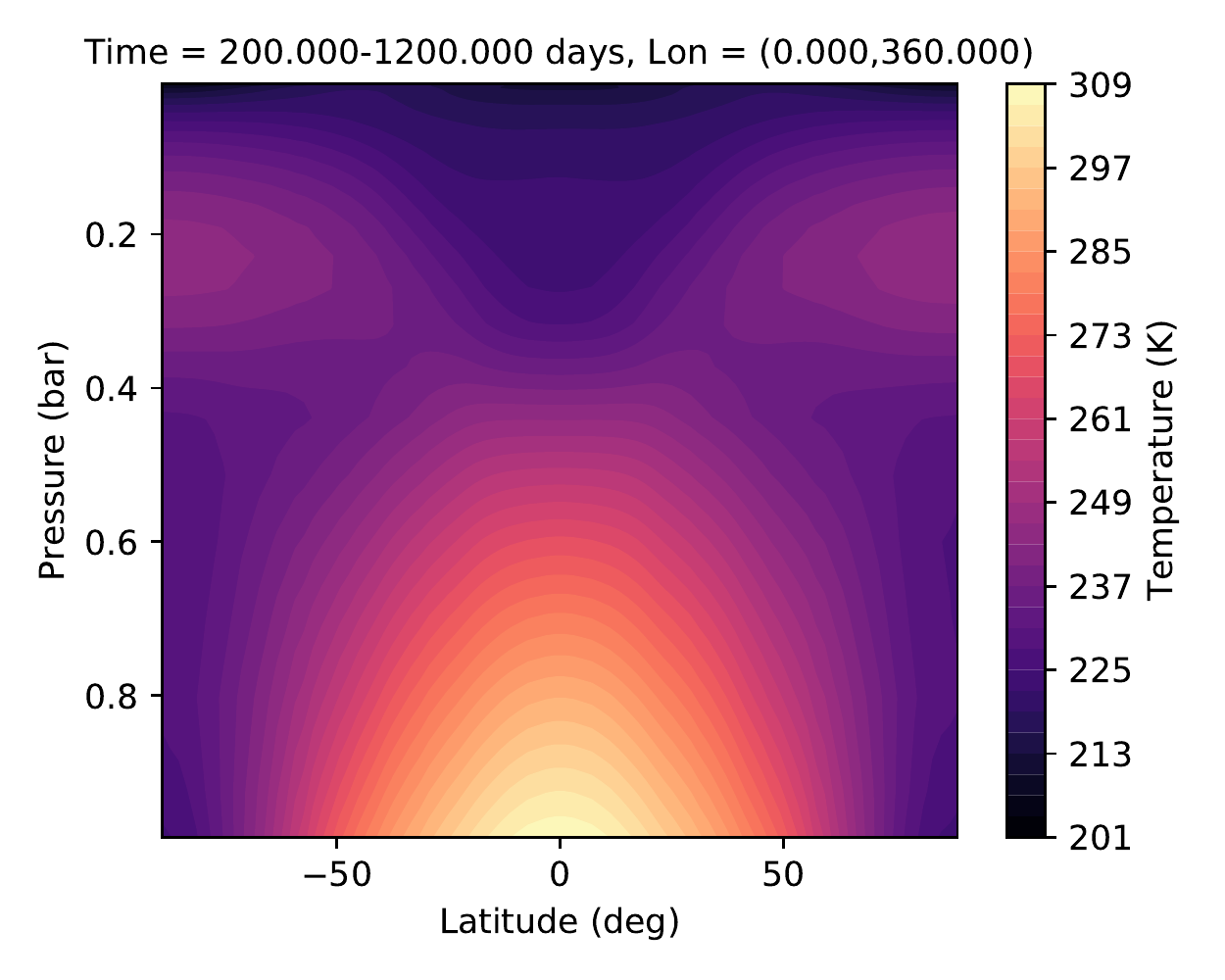}
\includegraphics[width=0.5\textwidth]{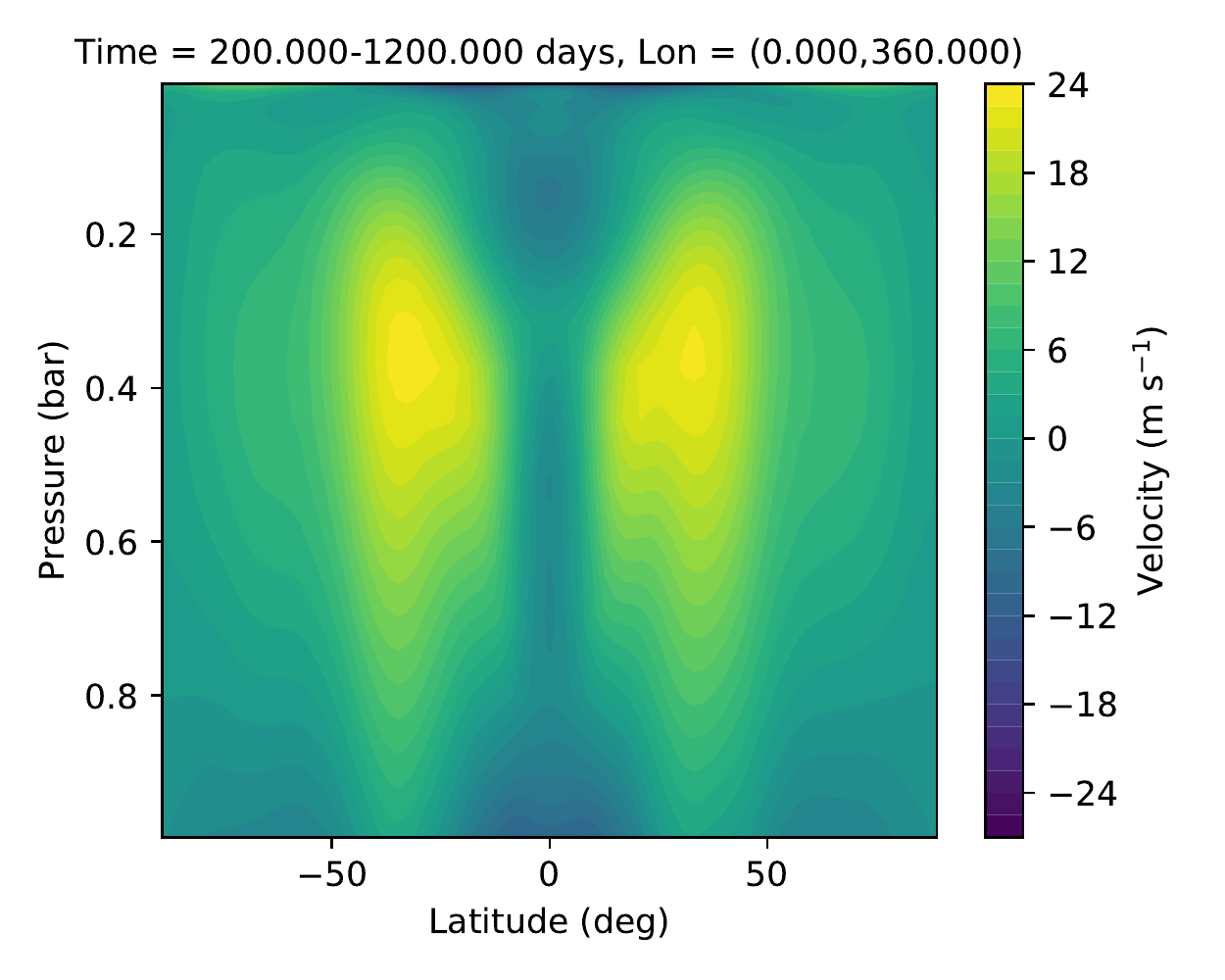} \\
\includegraphics[width=0.5\textwidth]{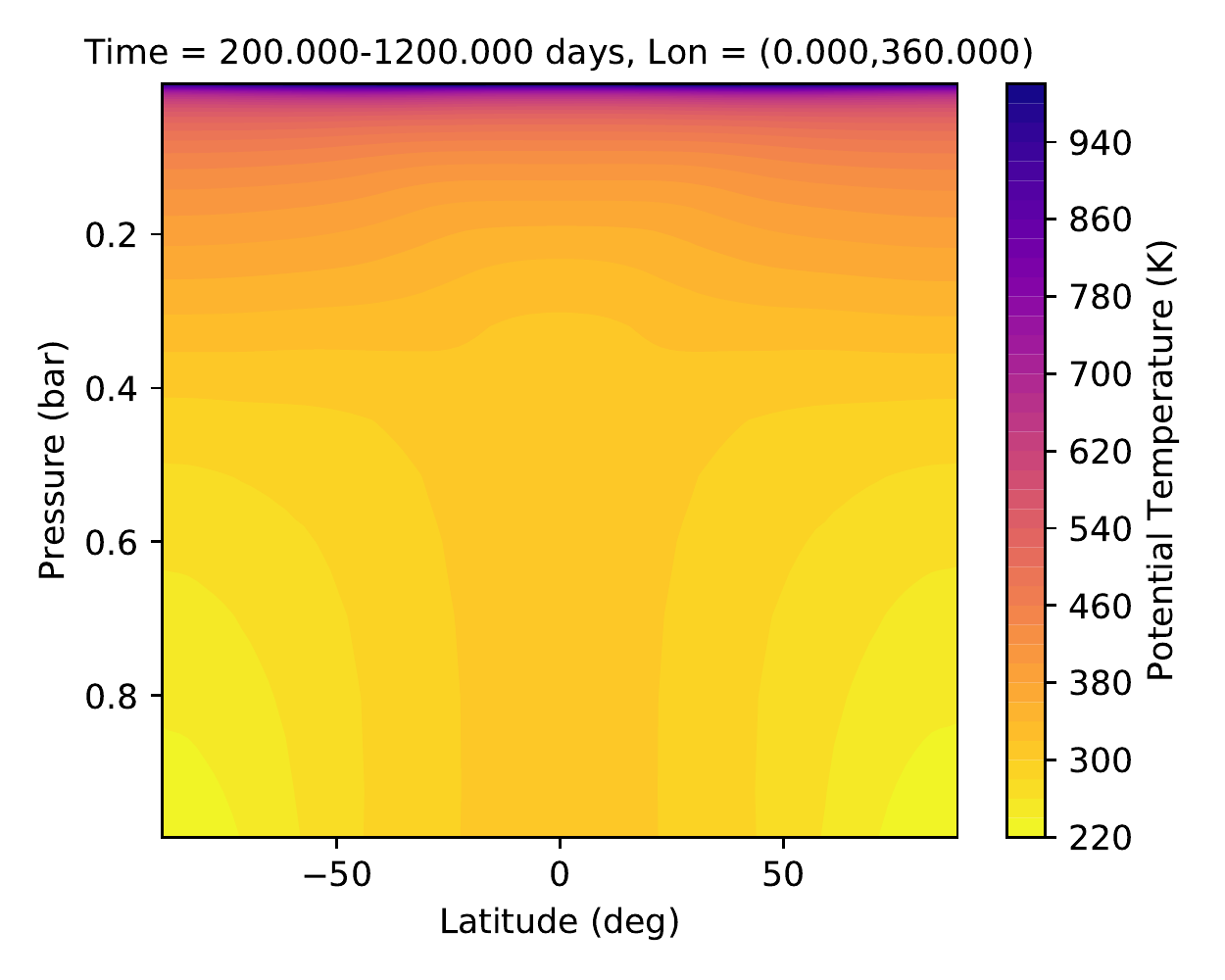}
\includegraphics[width=0.5\textwidth]{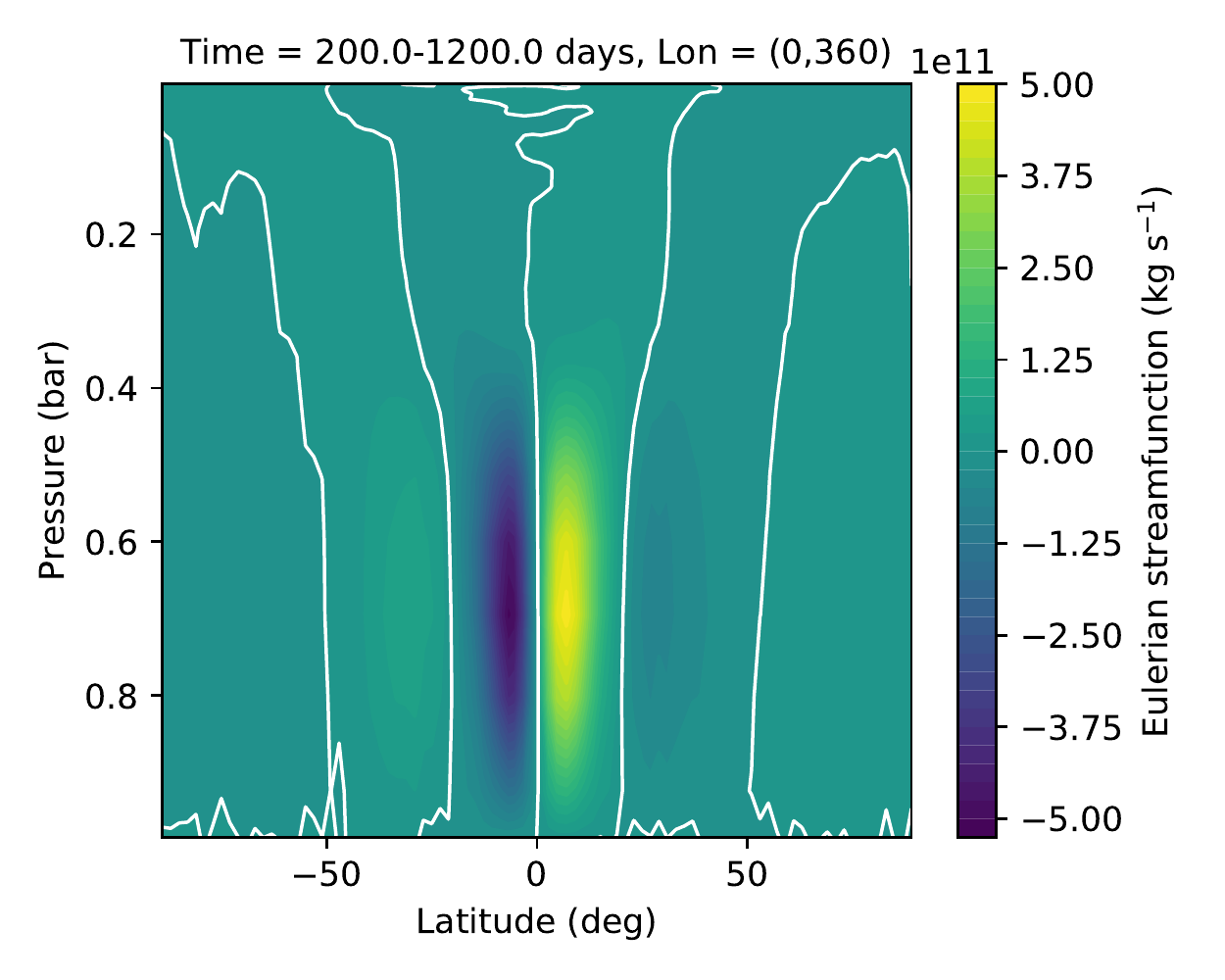}
\caption{Zonal- and time-averaged quantities from the Earth-like, non-hydrostatic deep (NHD) simulation, with convective adjustment enabled. Upper left shows the temperature, upper right the zonal wind speed, lower left the potential temperature, and lower right the mass stream function (positive values indicate clockwise motion). The white line in the lower right panel is the zero pass contour.} \label{fig:earthdccomp}
\end{figure*}

\begin{figure*}
\includegraphics[width=0.5\textwidth]{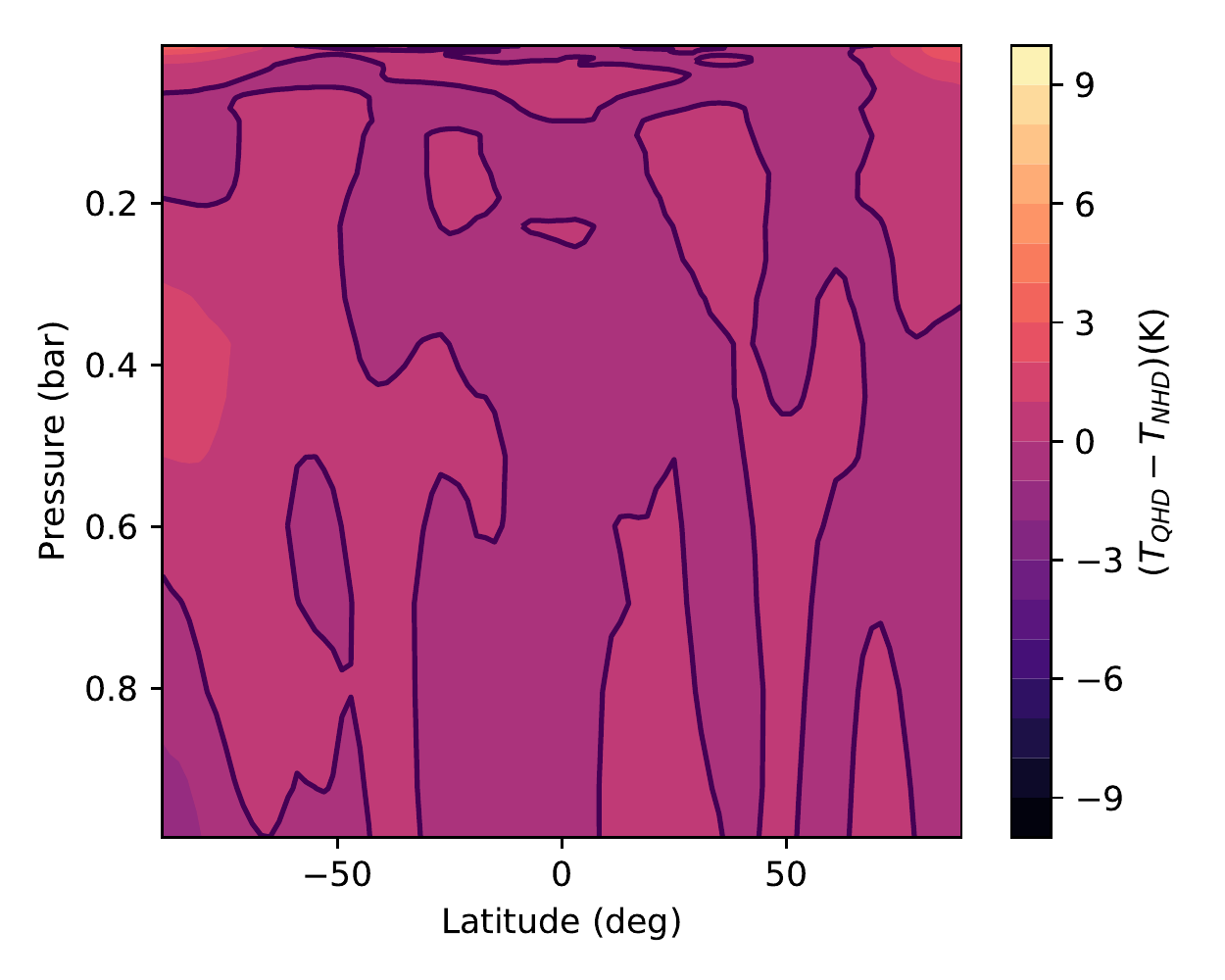}
\includegraphics[width=0.5\textwidth]{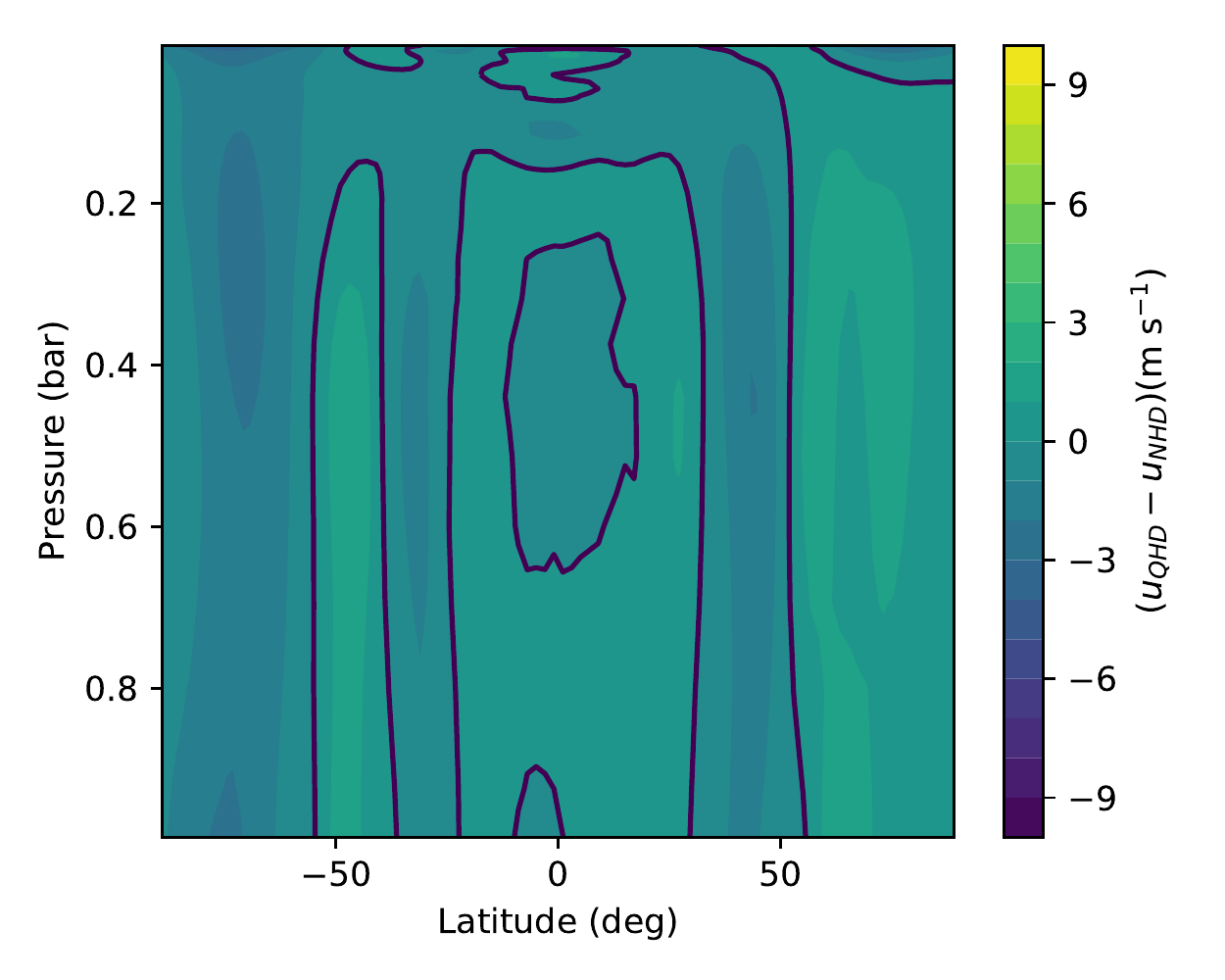}\\
\includegraphics[width=0.5\textwidth]{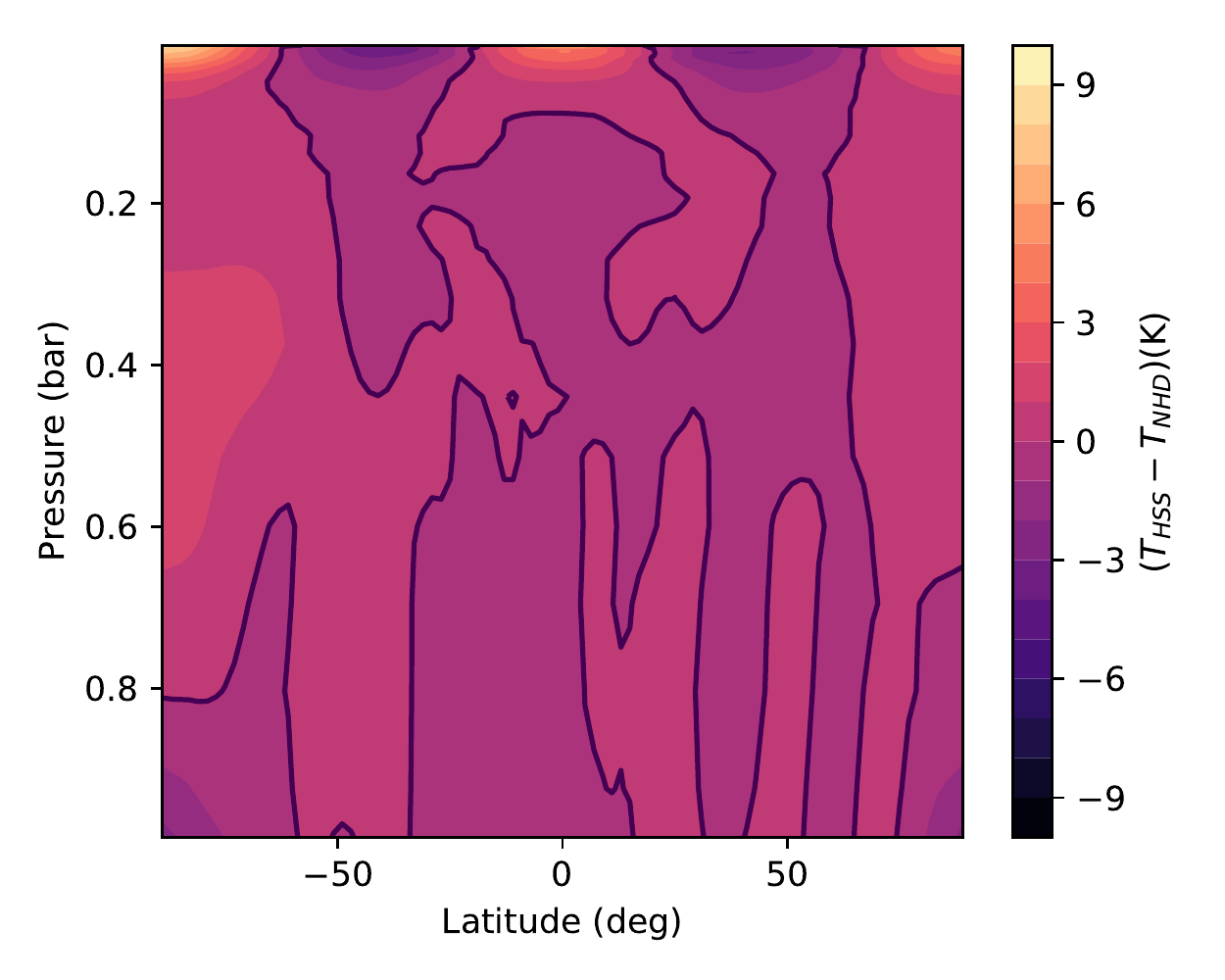}
\includegraphics[width=0.5\textwidth]{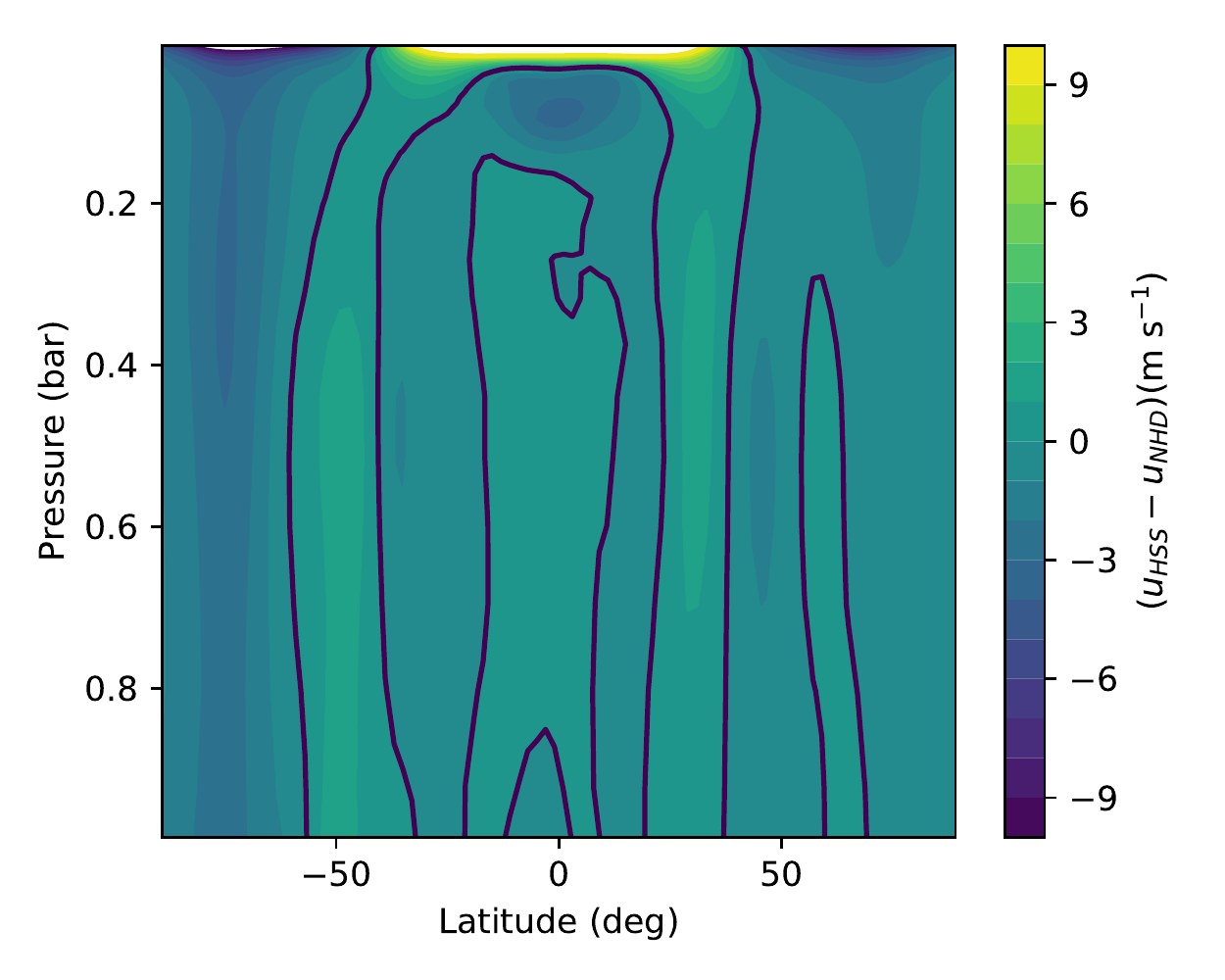}\\
\caption{Residuals between the quasi-hydrostatic deep (QHD) or hydrostatic shallow (HSS) and the non-hydrostatic deep (NHD) Earth-like simulations. The upper panels compare the temperature (left) and zonal wind (right) for the QHD case, the lower panels compare the same for the HSS case. Departures from the NHD simulation are greatest at low pressures; in the lower atmosphere, the average temperatures and wind speeds are very similar. \label{fig:earthhydrostat}}
\end{figure*}

The first case is non-hydrostatic, deep (NHD) without dry convective adjustment (Figure \ref{fig:earthnodccomp}). The second case is NHD with dry convective adjustment (Figure \ref{fig:earthdccomp}). Comparing the two cases, we can see that without dry convection, there is clear spurious heating at $\sim 20^{\circ}-30^{\circ}$ latitude. With convective adjustment enabled, we produce a temperature profile much more similar to the cases in \cite{Heng2011b}. As in that work, we see that the potential temperature profile, in the case without convective adjustment, is unstable near the surface at latitudes $\lesssim 50^{\circ}$. The convective adjustment scheme rectifies the situation and cools the near surface region near $\sim 20^{\circ}-30^{\circ}$ latitude. 

In both cases, we see jet streams form at $\phi \sim 30^{\circ}$ and $P \sim 0.4$ bar (upper right of Figures \ref{fig:earthnodccomp} and \ref{fig:earthdccomp}). There is a hint of a second, weaker stream in each hemisphere at $\sim 60^{\circ}$. The Eulerian mass streamfunction (lower right of  Figures \ref{fig:earthnodccomp} and \ref{fig:earthdccomp}) is similar in the two cases. We see strong thermally-direct cells (Hadley cells) at the equator, and weaker indirect cells adjacent to these. These overturning cells are narrower in latitude than the real Hadley and Ferrel cells seen on Earth. We speculate that this may be a consequence of the lack of a hydrologic cycle in the current version of our model, as narrow Hadley cells were also observed in the dry simulations of \cite{Heng2011a}, compared with, for example, \cite{Merlis2010}. 

We calculate the Eulerian mass stream function using the following relation:
\begin{equation}
    \Psi = \frac{2 \pi r_0 \cos{\phi}}{g} \int_{0}^{P}\overline{[v]} dP,    \label{eqn:streamf}
\end{equation}
where $\Psi$ is the stream function, $r_0$ is planet radius (at the model bottom), $g$ is the gravity, and $\overline{[v]}$ is the time and zonally averaged meridional velocity. Implicit in this definition is hydrostatic balance; in the non-hydrostatic model, this condition is not exactly satisfied. However, the simulation is at all times near enough to hydrostatic equilibrium that there is little difference between in $\Psi$ as calculated above compared to a calculation based on Equation 4.1 of \cite{Peixoto1984}, which does not assume hydrostatic balance.  

We compare the model run using the quasi-hydrostatic, deep (QHD) and hydrostatic, shallow (HSS) approximations to the non-hydrostatic, deep (NHD) case, all with convective adjustment enabled. Figure \ref{fig:earthhydrostat} shows the difference in temperature and zonal wind for the QHD and HSS simulations from the NHD case. Over most of the model domain, the differences are relatively small, on average. The largest differences occur in wind speeds in the upper atmosphere between the HSS and NHD cases. This suggests that non-hydrostatic effects and the geometric corrections for a deep atmosphere are unimportant in this regime, at least in terms of the global average state of the atmosphere, but that the geometric corrections are more important than non-hydrostatic effects. 

We have run a single NHD case at $g_{\text{level}} = 6$, corresponding to a horizontal resolution of $\sim 1^{\circ}$. The output (not shown) is qualitatively very similar to the $g_{\text{level}}=5$ case, though, as we discuss below, the simulation conserves mass to slightly less precision than the lower resolution cases. 

Lastly, we compare several diagnostics for the Earth-like simulations. Figure 16 shows the evolution of the global atmospheric mass, energy, and axial angular momentum. We hope to conserve the total mass; the energy and angular momentum evolve due to input from the stellar radiation, and are thus more diagnostic of convergence. The simulations at resolution $\sim 2^{\circ}$ all conserve mass to a similar precision, a few parts in $10^{12}$ over 1200 days. The simulation at resolution of $\sim 1^{\circ}$ does slightly worse in this respect. In all cases, the model reaches a steady state in $\sim 200-300$ days, after which the total energy and axial angular momentum stay roughly constant.

\begin{figure*}
\includegraphics[width=\textwidth]{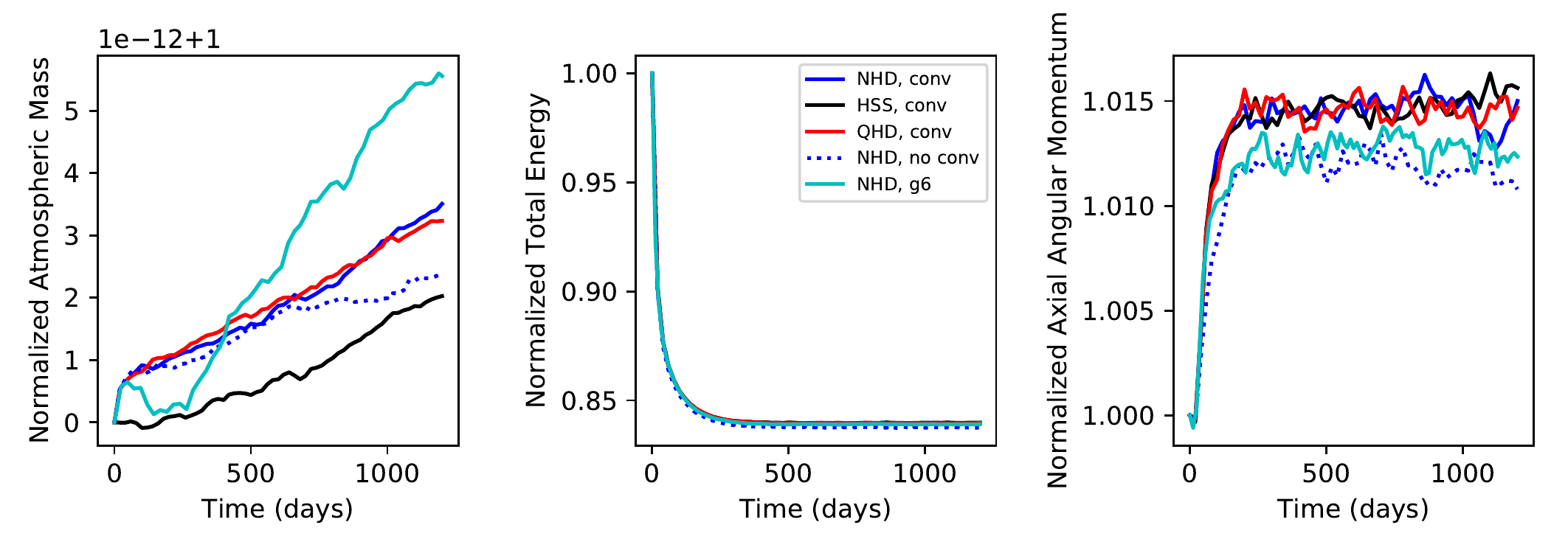}
\caption{Evolution of the total mass, energy, and axial angular momentum in the Earth-like cases. Solid blue is the NHD case with convective adjustment, dotted blue is the NHD case without convective adjustment, red is the QHD case with convective adjustment, and black is HSS case with convective adjustment. Cyan shows the NHD case at $g_{\text{level}}=6$, or a horizontal resolution of $\sim 1^{\circ}$; all other cases had $g_{\text{level}}=5$ ($\sim 2^{\circ}$ resolution).}
\end{figure*}

\subsection{HD 189733 b} \label{sec:hd189rt}

Here we show a comparison of QHD and NHD simulations of the hot Jupiter HD 189733 b. The input parameters are summarized in Table \ref{tab:modelpara}. Simulations including the shallow approximation (HSS) are very similar to the QHD simulations, thus we stick here to a comparison of QHD and NHD. The largest differences in flow for this planet occur between the QHD and NHD simulations, indicating that the primary source of the differences described below is the term $Dv_r/Dt$, the Lagrangian derivative of the vertical velocity. 

\begin{figure*}
\includegraphics[width=0.5\textwidth]{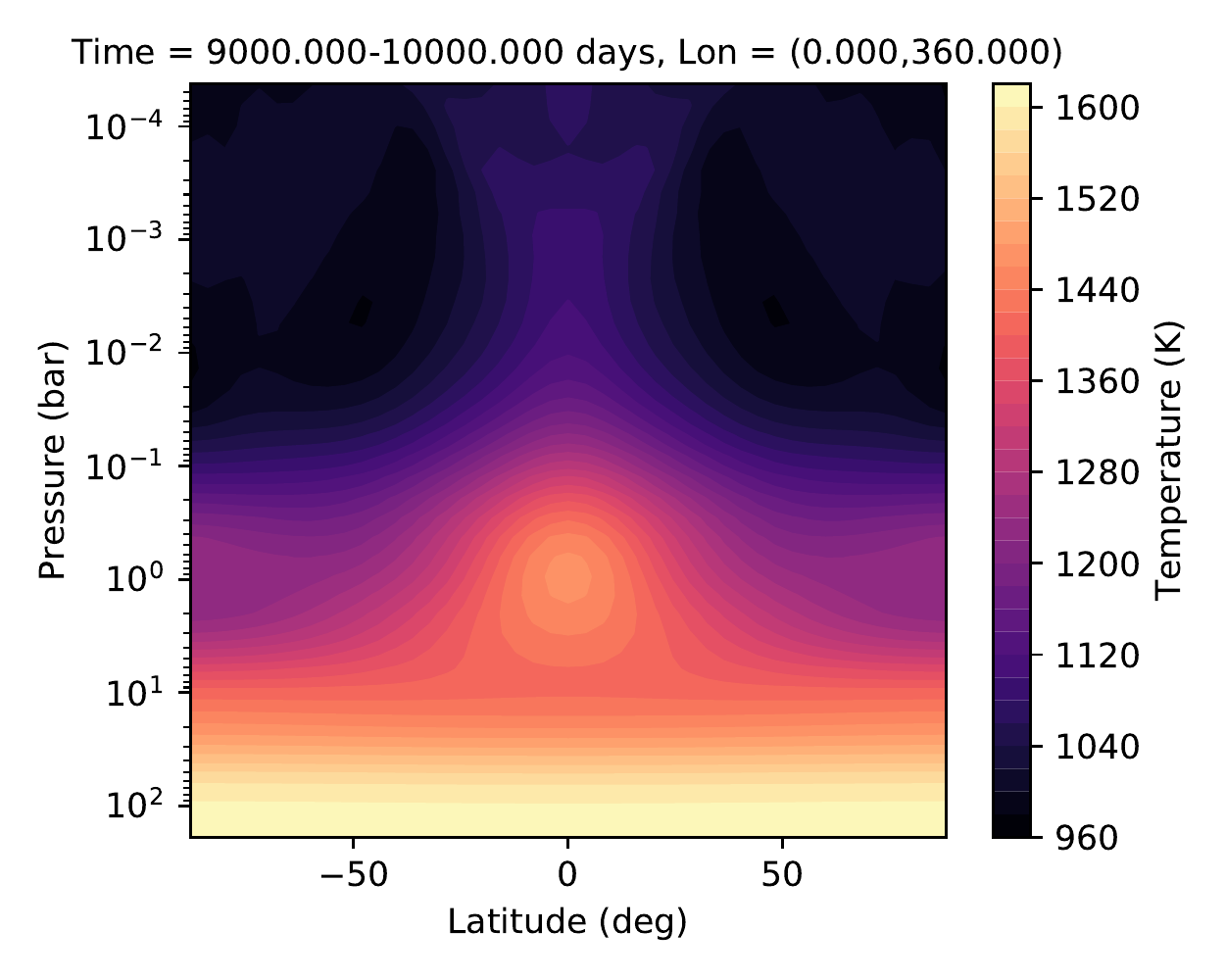}
\includegraphics[width=0.5\textwidth]{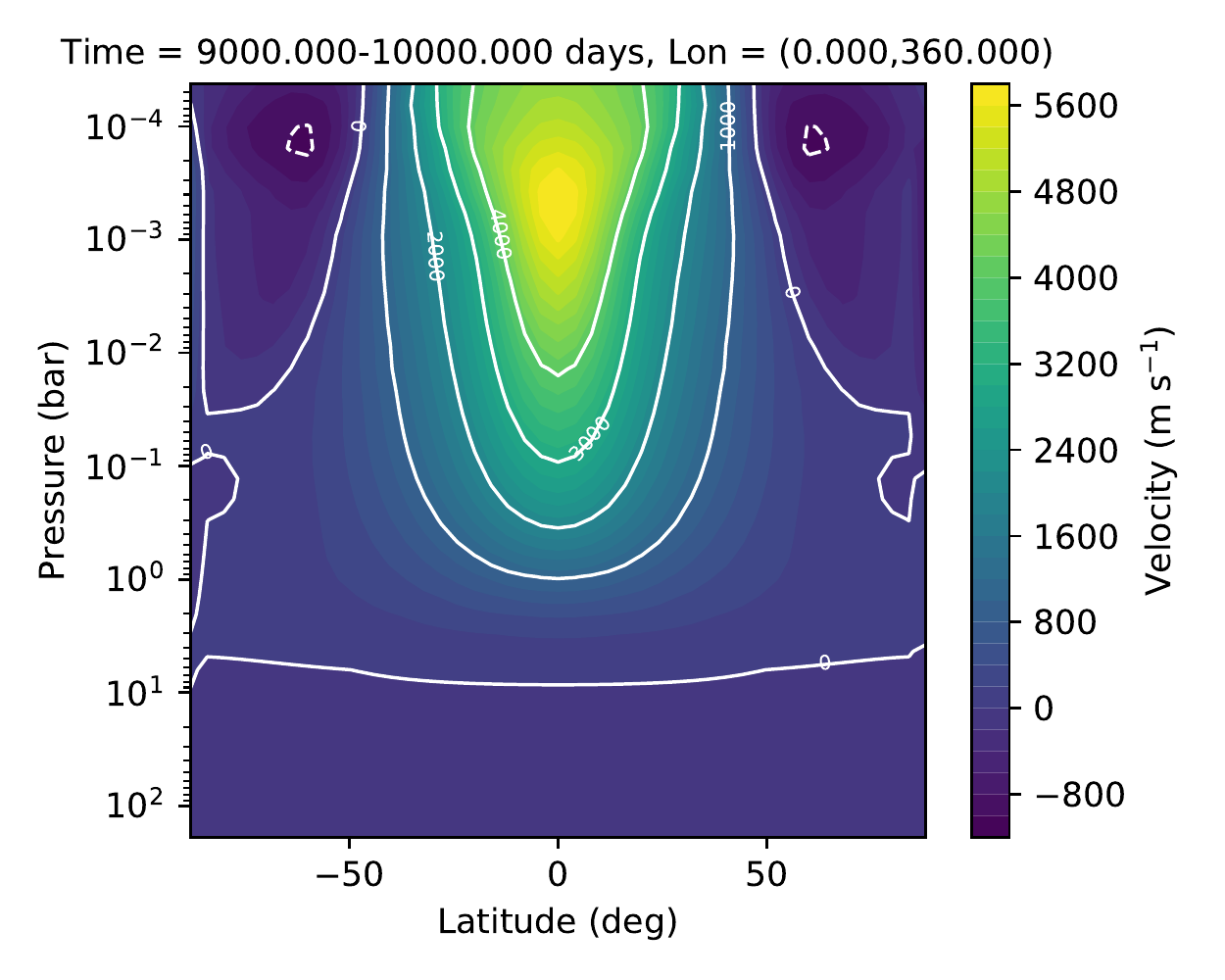}\\
\includegraphics[width=0.5\textwidth]{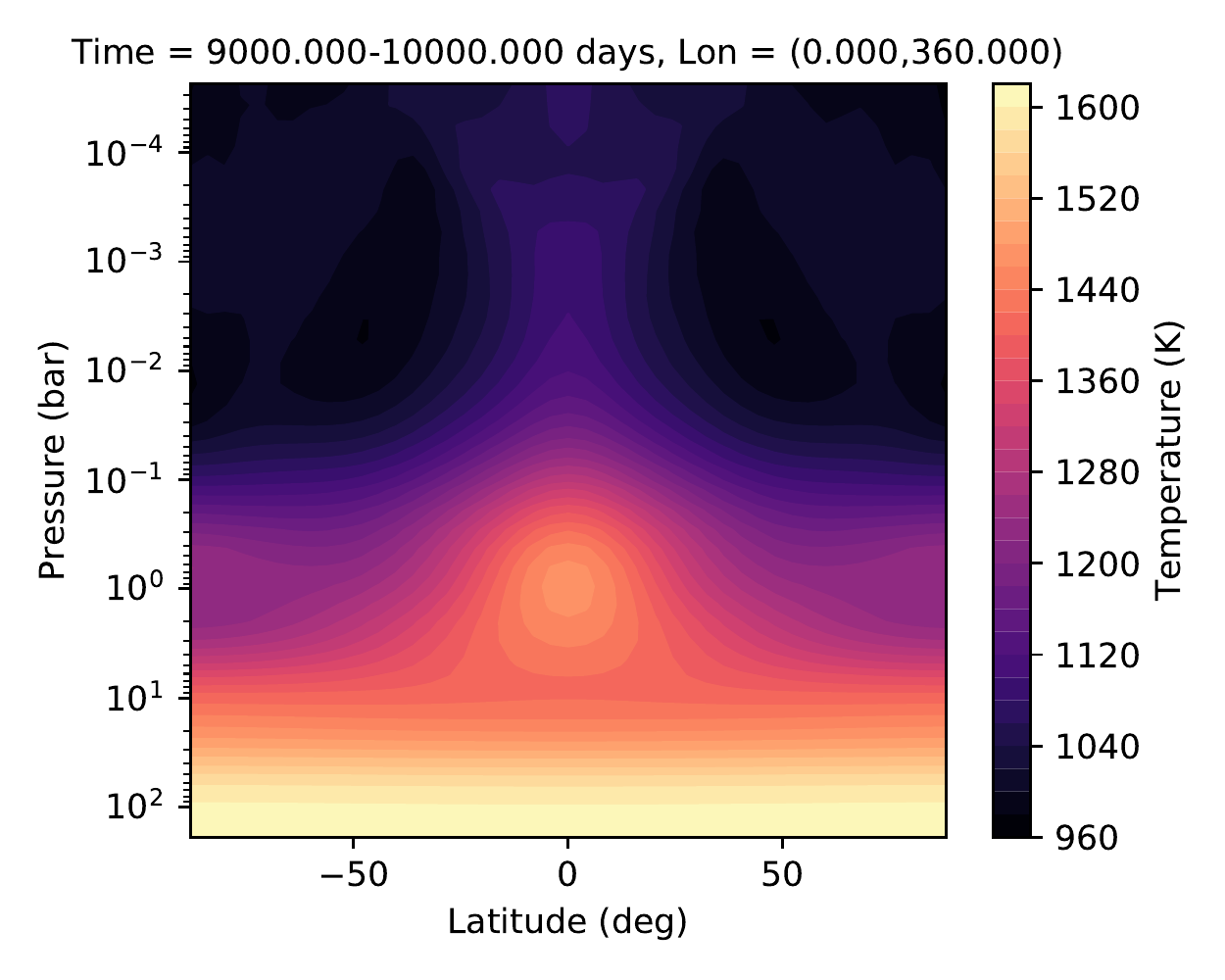}
\includegraphics[width=0.5\textwidth]{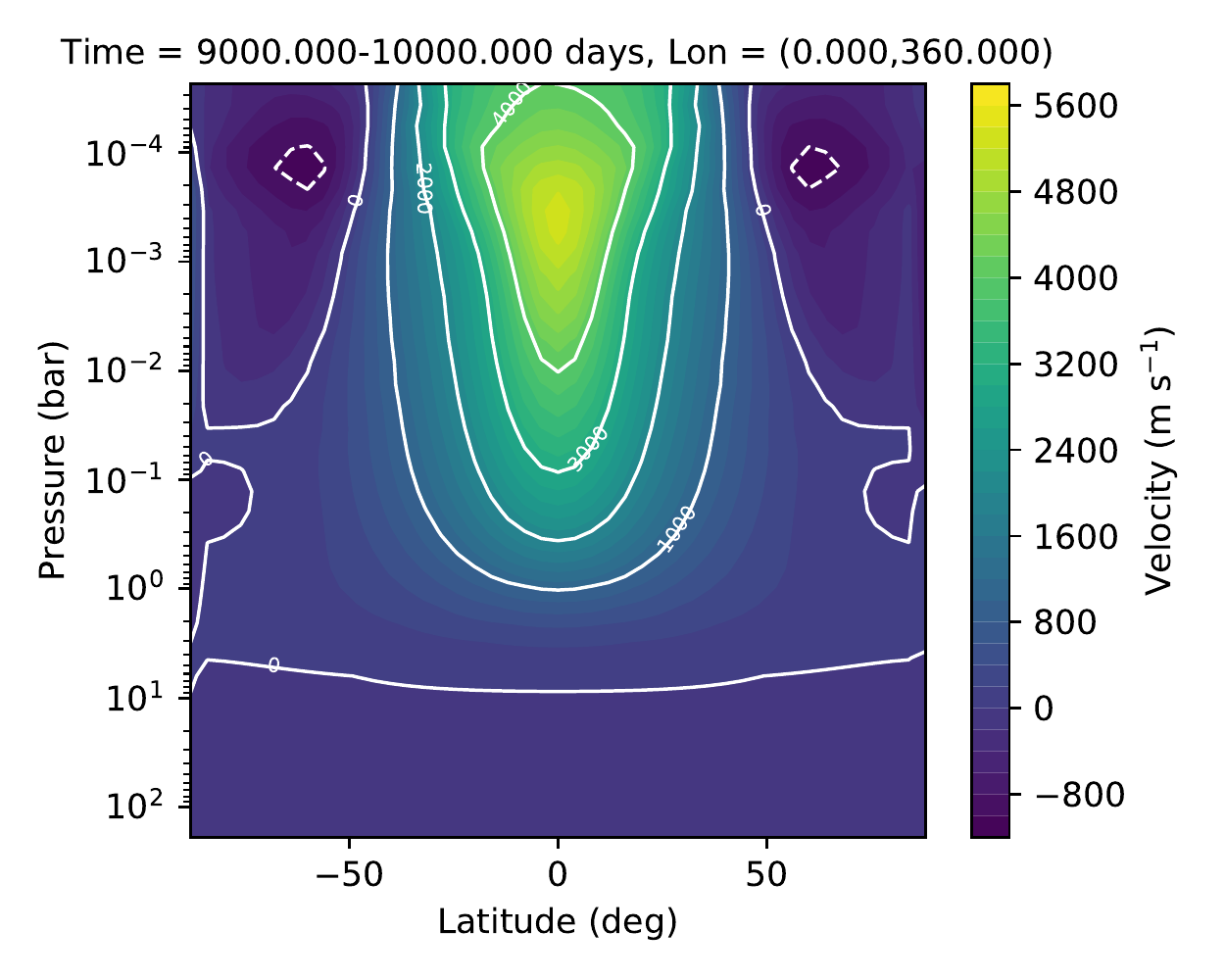}
\caption{Zonally averaged temperature and zonal wind speed for simulations of HD 189733 b at $\sim 4^{\circ}$ resolution. The upper panels show the non-hydrostatic, deep (NHD) case, the lower panels the quasi-hydrostatic, deep (QHD) case. All quantities are averaged over the last 1000 (Earth) days of the 10000  day simulation. \label{fig:189g4zonal1}}
\end{figure*}

Figure \ref{fig:189g4zonal1} shows the zonally and temporally averaged temperature and zonal wind speed during the last 1000 days of the NHD and QHD simulations. 
The overall temperature structure is very similar between the two cases. The zonal wind speed plots indicate the presence of superrotation, as expected, and the overall structure is similar. However, the maximum velocities of the jets differ by $\sim 5\%$, broadly consistent with the comparison between the ``Prim'' and ``Full'' simulations of \cite{Mayne2017}. While that study compared simulations using the primitive equations with the full non-hydrostatic equations, the sole difference between our simulations here is the neglect of the material derivative of the vertical velocity, $Dv_r/Dt$, and the hyper-diffusive term, $\mathcal{F}_{v_r}$, in the QHD case. This results in a difference of the maximum jet speed, likely because the 3-D representation of waves has been modified. Though the difference is small at this resolution, it becomes more pronounced in the higher resolution simulations, as we describe shortly.

\begin{figure*}
\includegraphics[width=0.5\textwidth]{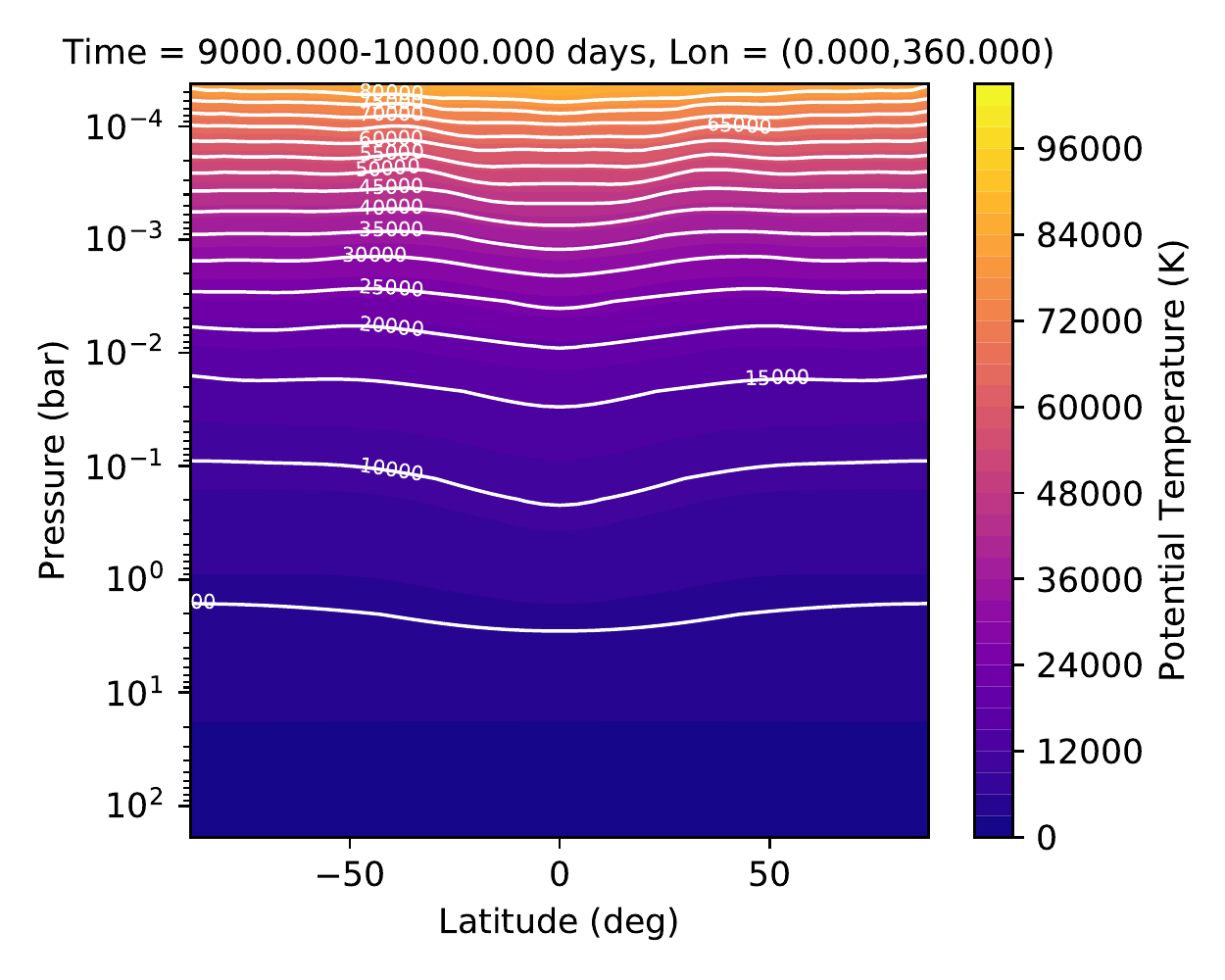}
\includegraphics[width=0.5\textwidth]{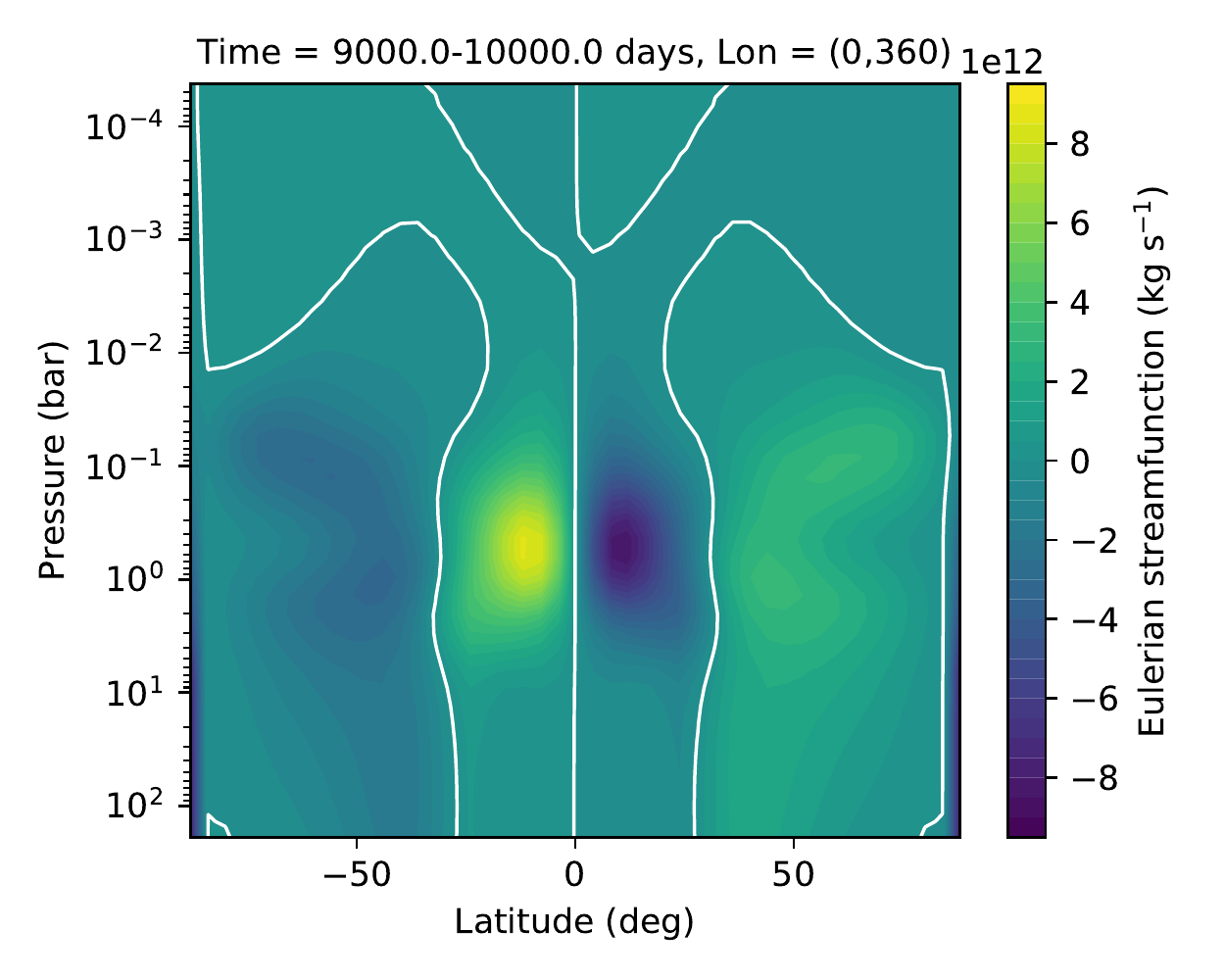}\\
\includegraphics[width=0.5\textwidth]{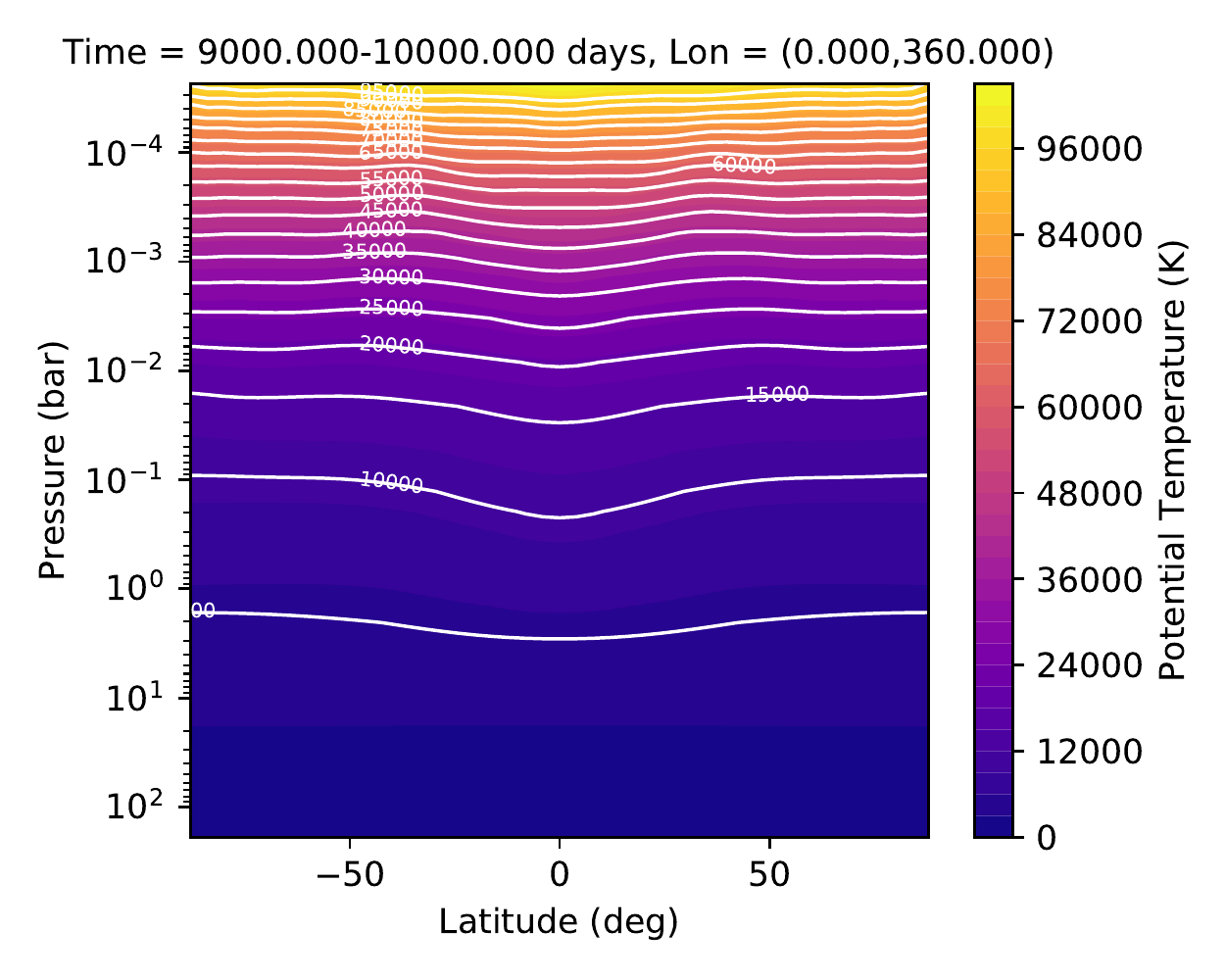}
\includegraphics[width=0.5\textwidth]{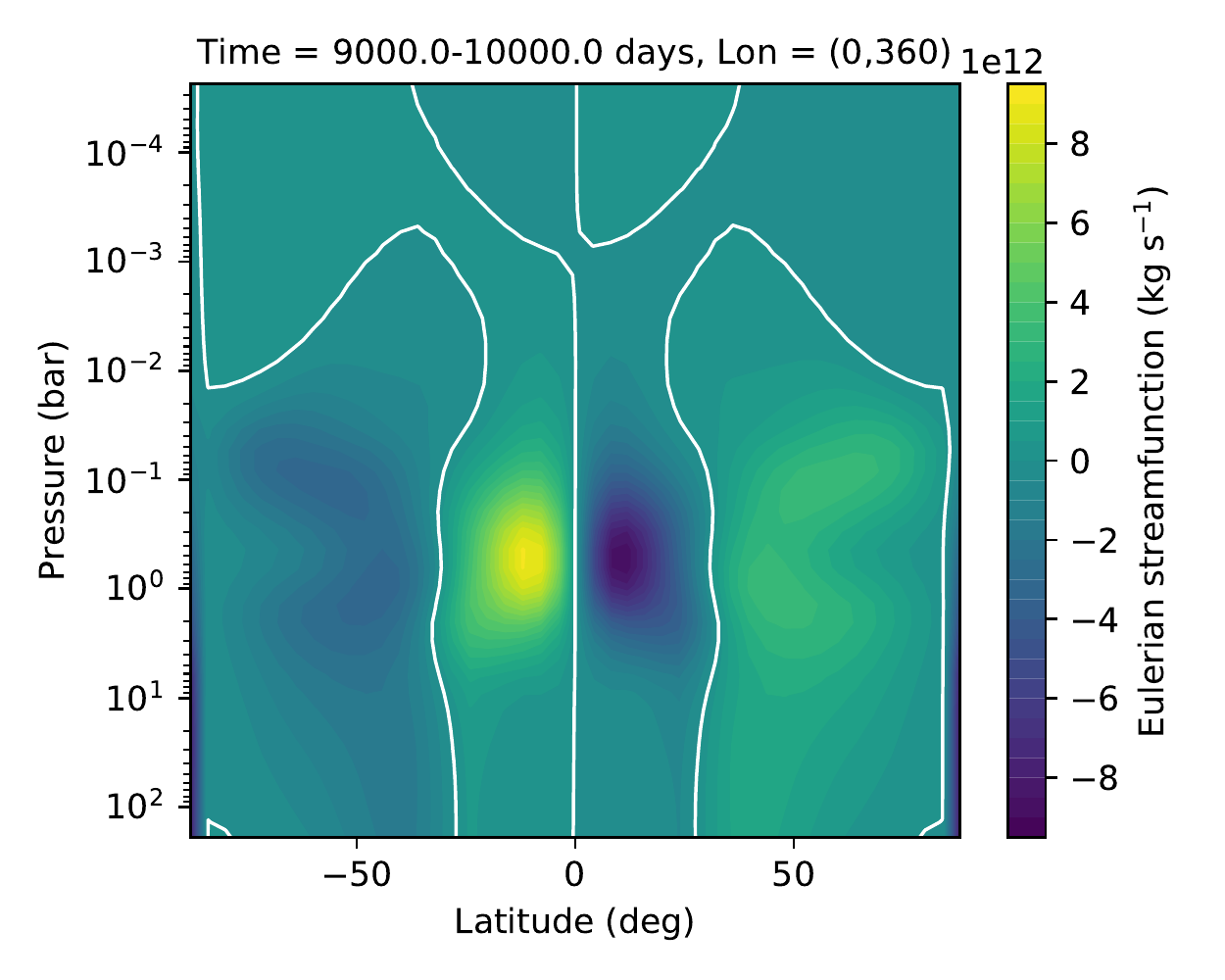}
\caption{Zonally averaged potential temperature and stream function for simulations of HD 189733 b at $\sim 4^{\circ}$ resolution. In the stream function plots, positive values indicate clockwise motion. The upper panels show the non-hydrostatic, deep (NHD) case, the lower panels the quasi-hydrostatic, deep (QHD) case. All quantities are averaged over the last 1000 (Earth) days of the 10000 day simulation. \label{fig:189g4zonal2}}
\end{figure*}

Figure \ref{fig:189g4zonal2} shows the zonally and temporally averaged potential temperature and Eulerian mass streamfunction of the NHD and QHD cases. Though the dry convection scheme is enabled in these simulations, it likely has very little effect as everywhere the atmosphere is quite stable, as indicated by the potential temperature structure. Though this quantity is averaged over longitude, local profiles of the potential temperature at different longitudes near the equator also show stability. In the plots for the mass streamfunction, we see thermally-indirect overturning cells at the base of the equatorial jet, as also seen in \cite{Charnay2015} and \cite{Mendonca2018a}. Thermally-direct cells appear adjacent to the indirect cells at higher latitudes. Overturning in the upper atmosphere is weak in comparison and indiscernible on this scale. 

\begin{figure*}
\includegraphics[width=0.5\textwidth]{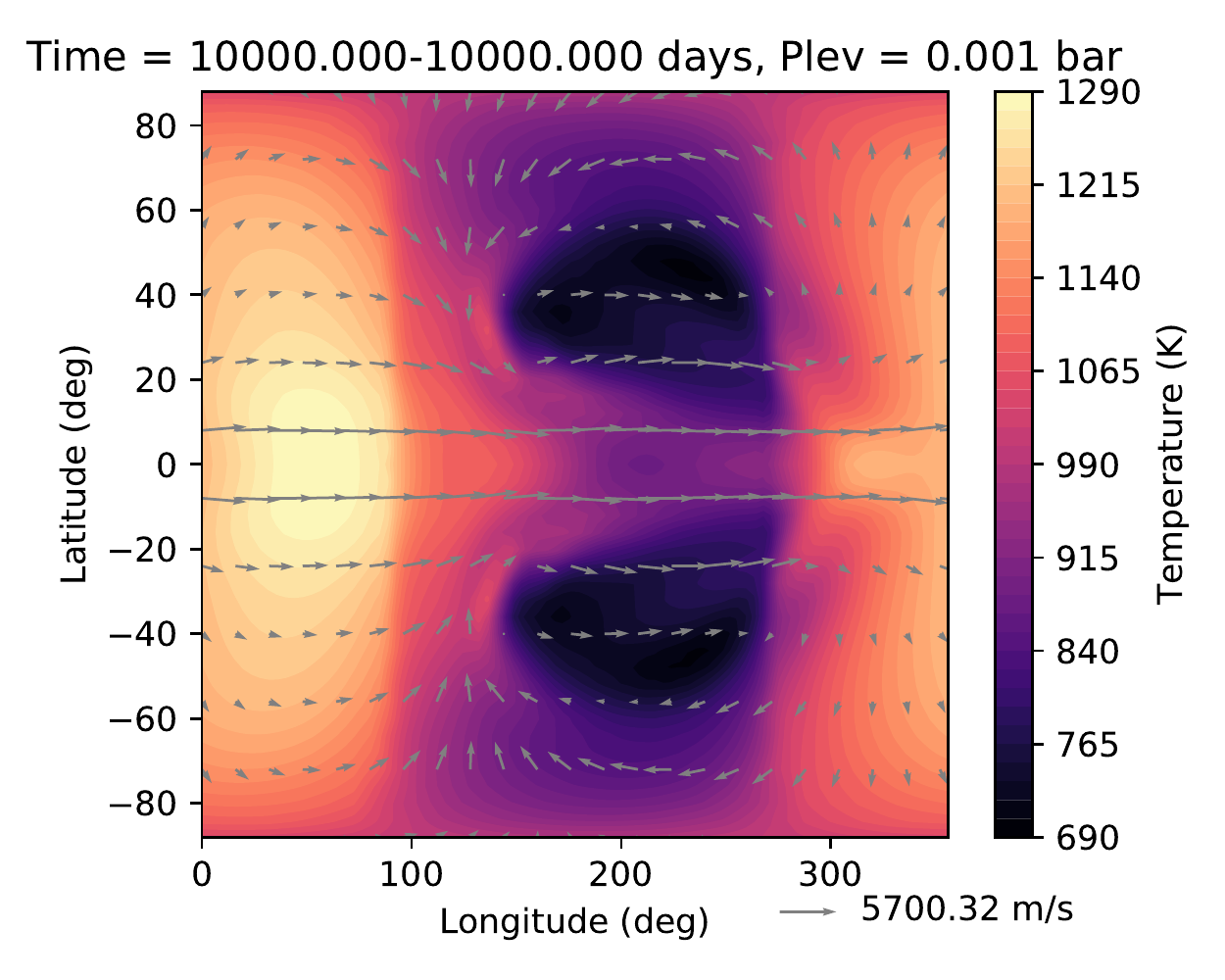}
\includegraphics[width=0.5\textwidth]{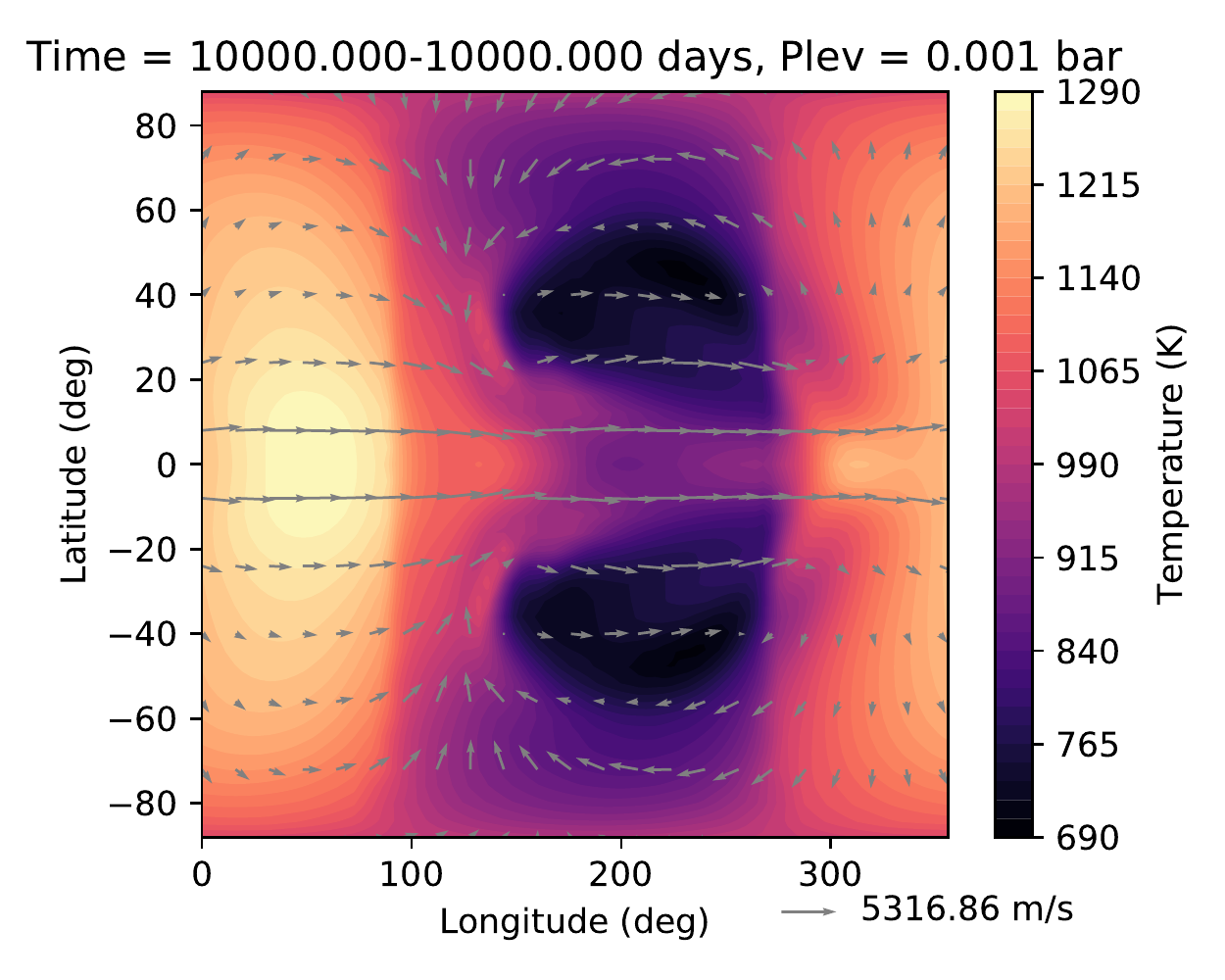}\\
\includegraphics[width=0.5\textwidth]{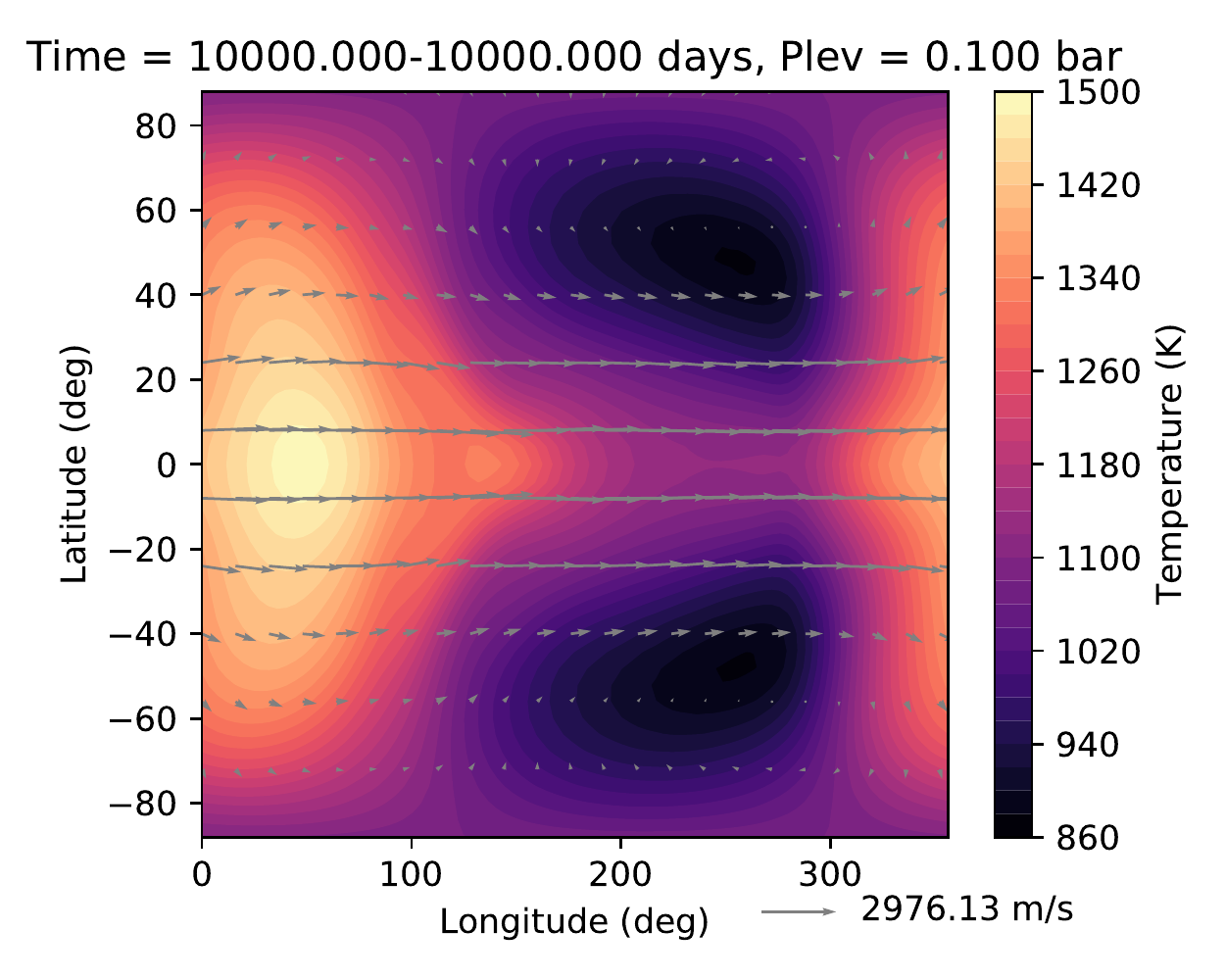}
\includegraphics[width=0.5\textwidth]{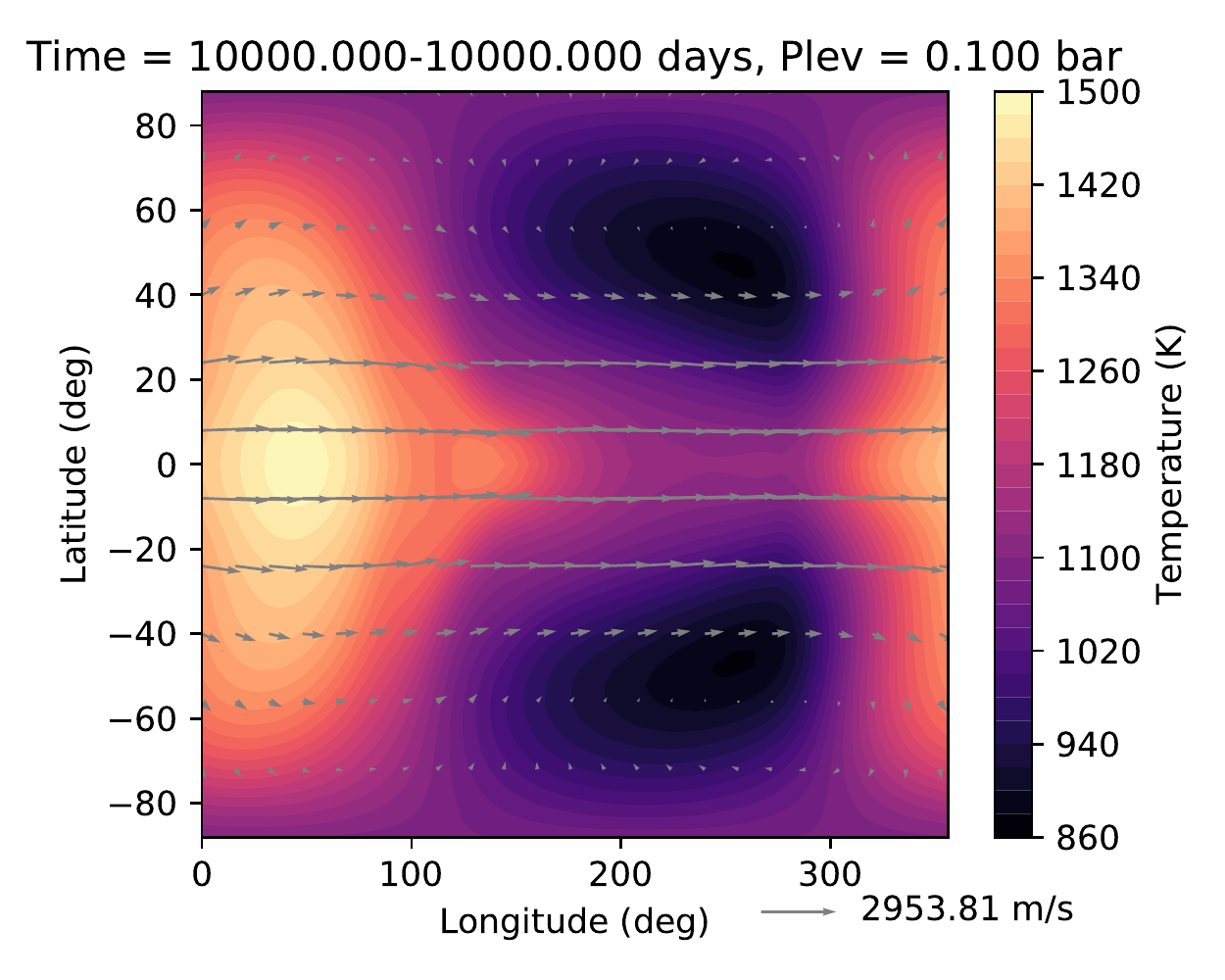}\\
\includegraphics[width=0.5\textwidth]{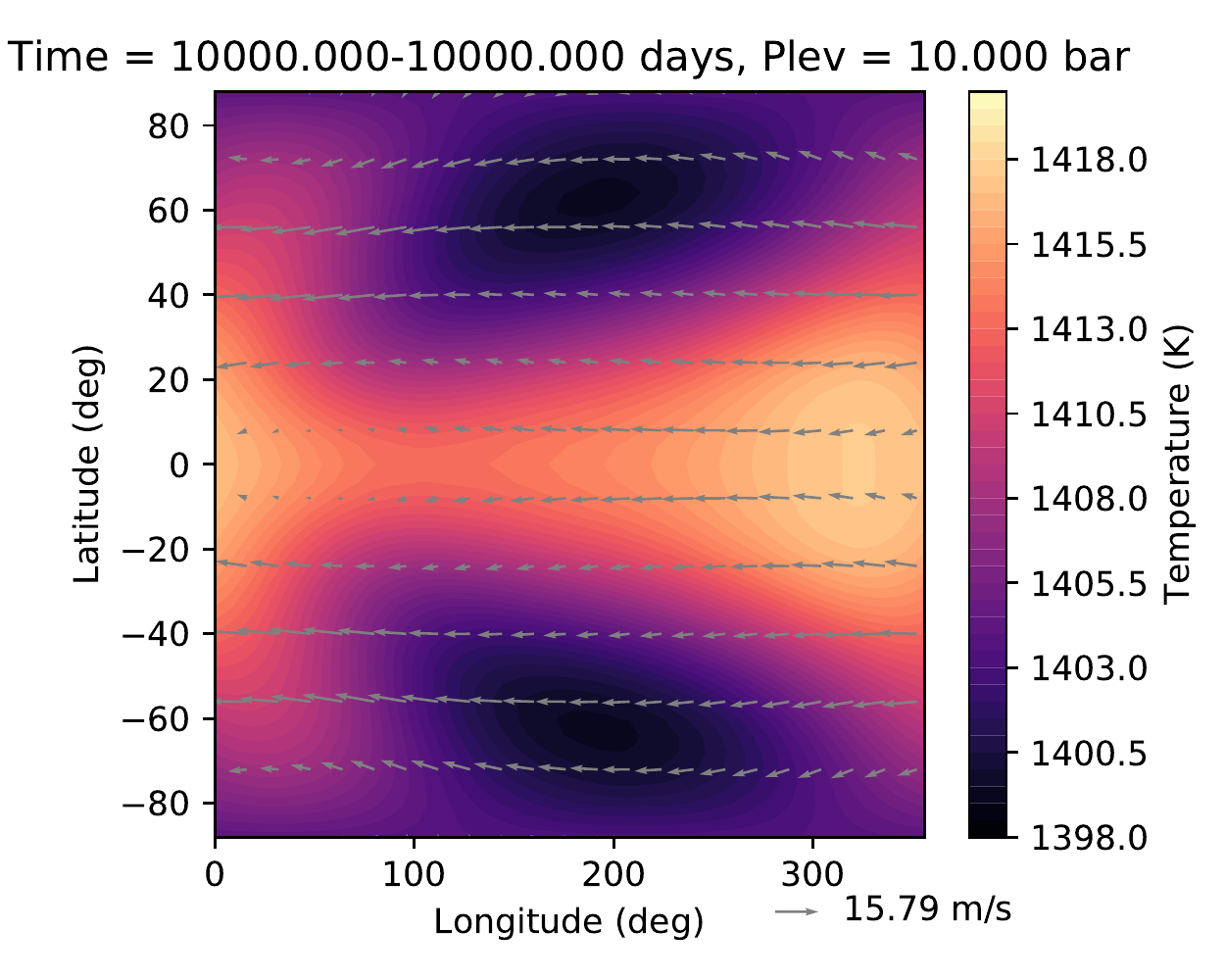}
\includegraphics[width=0.5\textwidth]{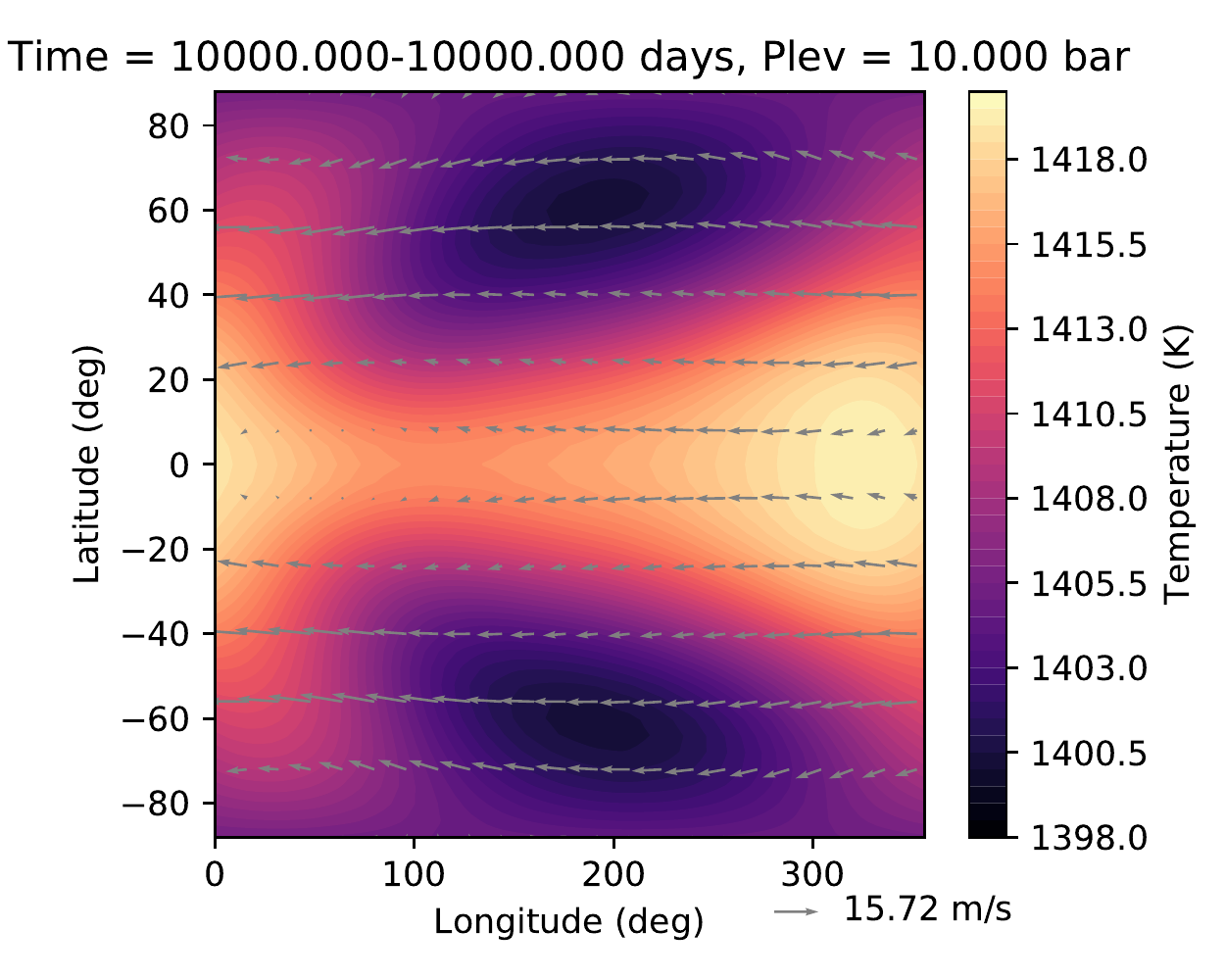}
\caption{Snapshots at 10000 days of the temperature (color) and horizontal winds (arrows) on isobaric surfaces for simulations of HD 189733 b at $\sim 4^{\circ}$ resolution. The left panels are the NHD simulation, the right are the QHD simulation. The substellar point is at $0^{\circ}$ longitude, $0^{\circ}$ latitude. \label{fig:189g4horiz}} 
\end{figure*}

Figure \ref{fig:189g4horiz} shows snapshots (at 10000 days) of the temperature and horizontal wind speeds along isobars for the NHD and QHD simulations. At $0.1$ bar and above (in altitude), in the region of the superrotational jet, we see the characteristic chevron shape and east ward hot spot offset associated with hot Jupiters. The wind vectors indicate eastward motion and that the gas is pushed away from the sub-stellar point (at $0^{\circ}$ longitude). At 1 mbar, we see return flow on the night side at high latitudes, which results in convergence on the equator at the eastern terminator and the recognizable chevron shape. The NHD and QHD simulations show only minor differences in temperature, but the velocities are slightly higher in the NHD case. Standing Rossby waves are easily discernible at these pressure levels. At the $10$ bar level, flow at the equator is retrograde, though comparatively sluggish ($\sim$ 10 m s$^{-1}$). 

\begin{figure*}
\includegraphics[width=0.5\textwidth]{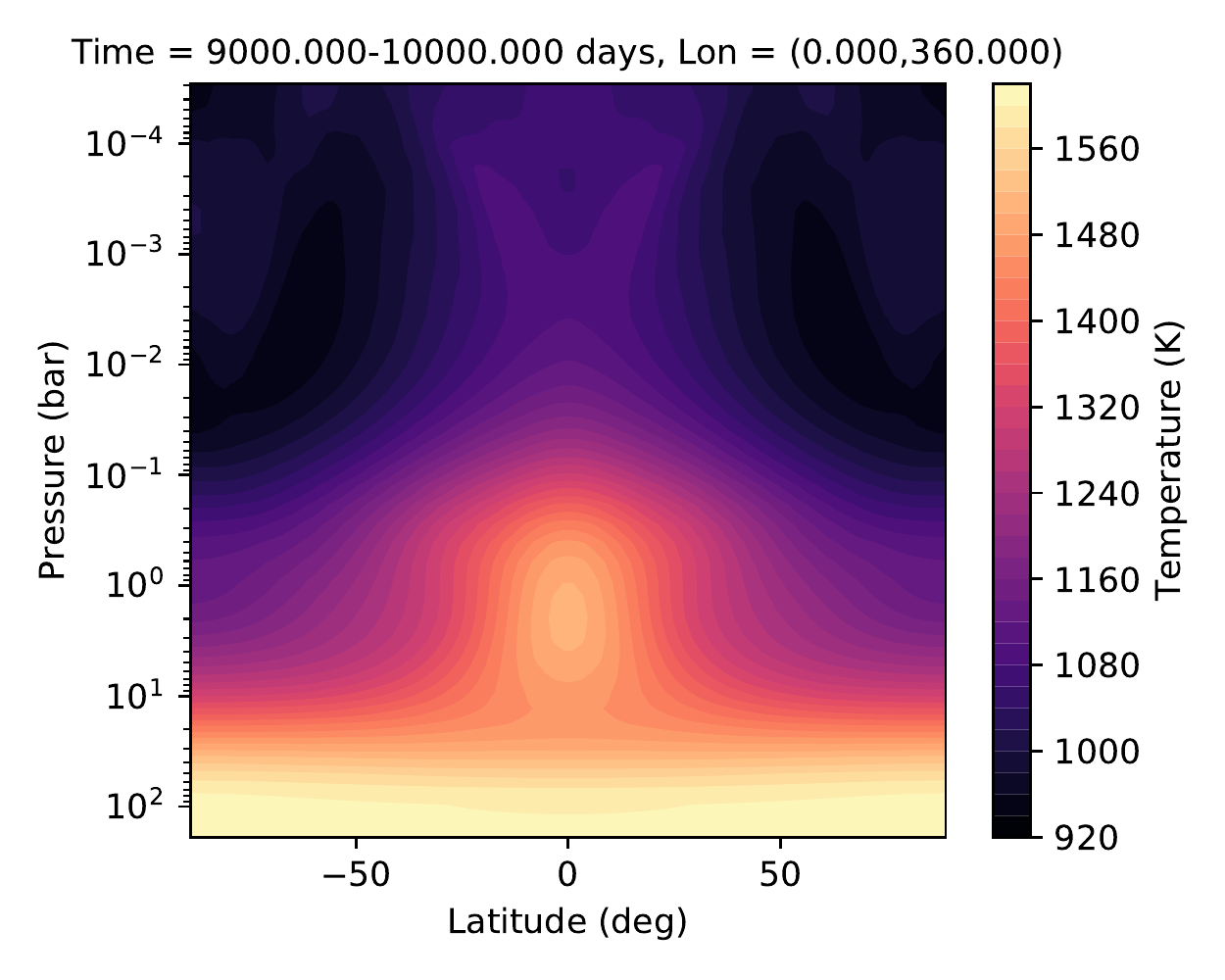}
\includegraphics[width=0.5\textwidth]{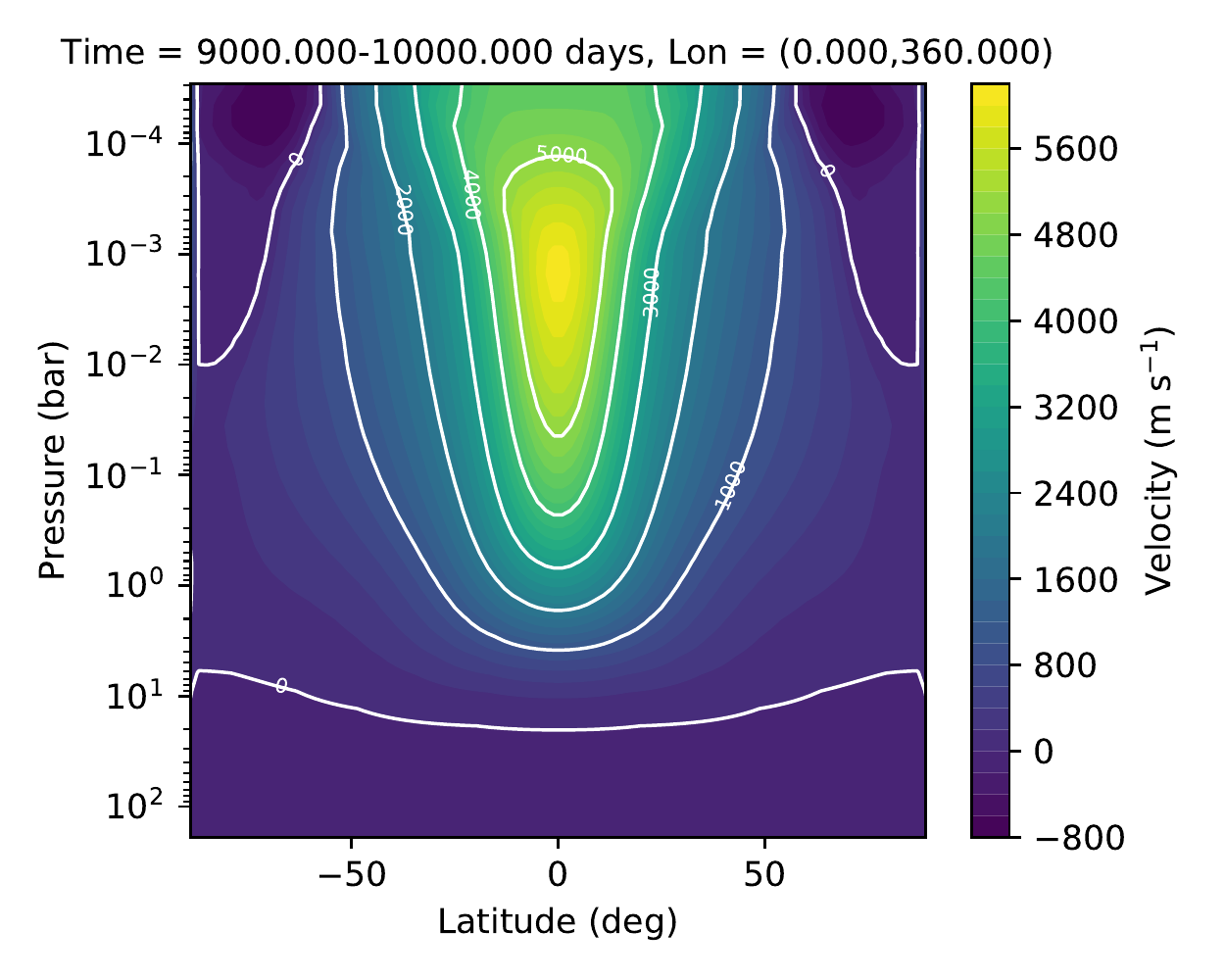}\\
\includegraphics[width=0.5\textwidth]{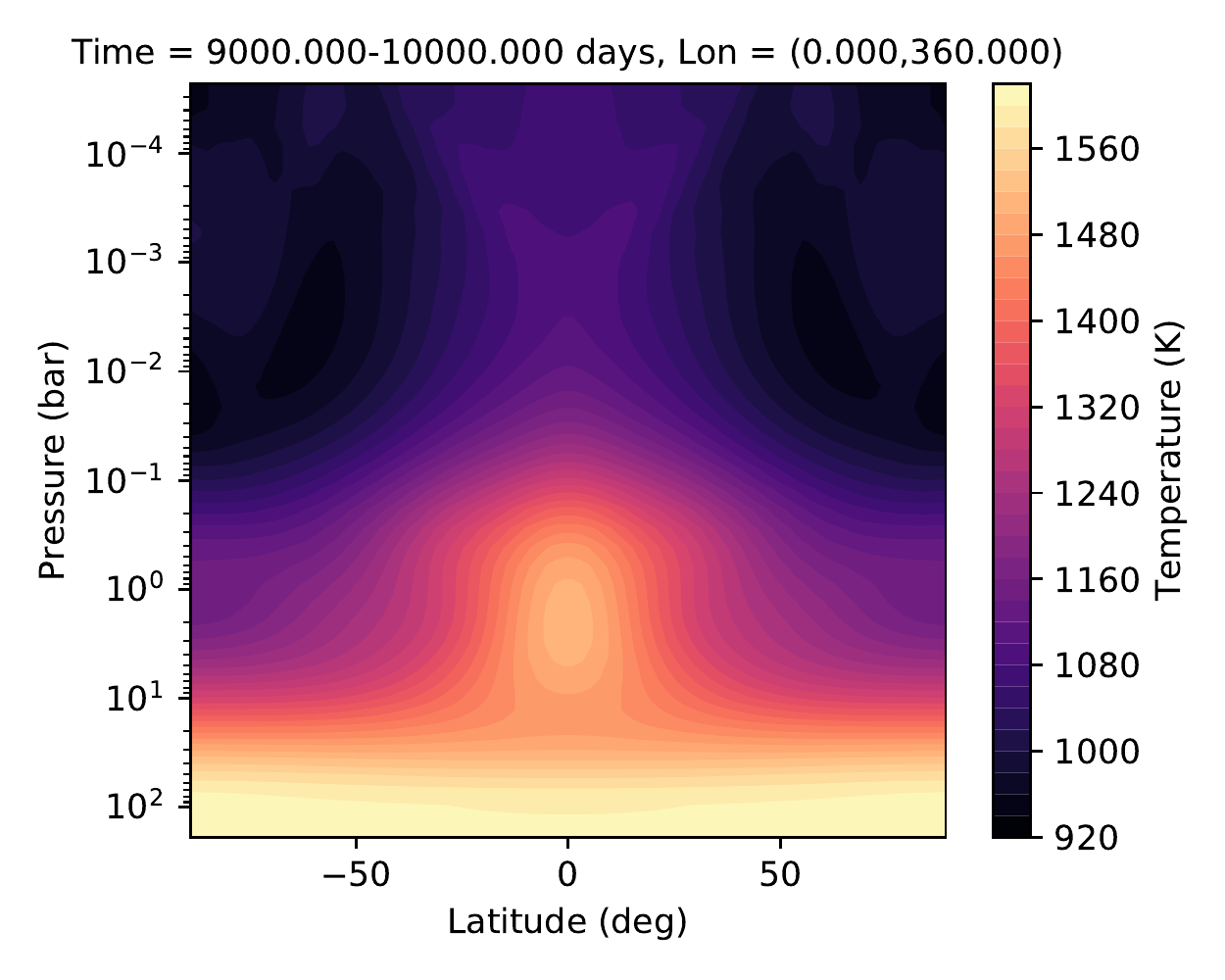}
\includegraphics[width=0.5\textwidth]{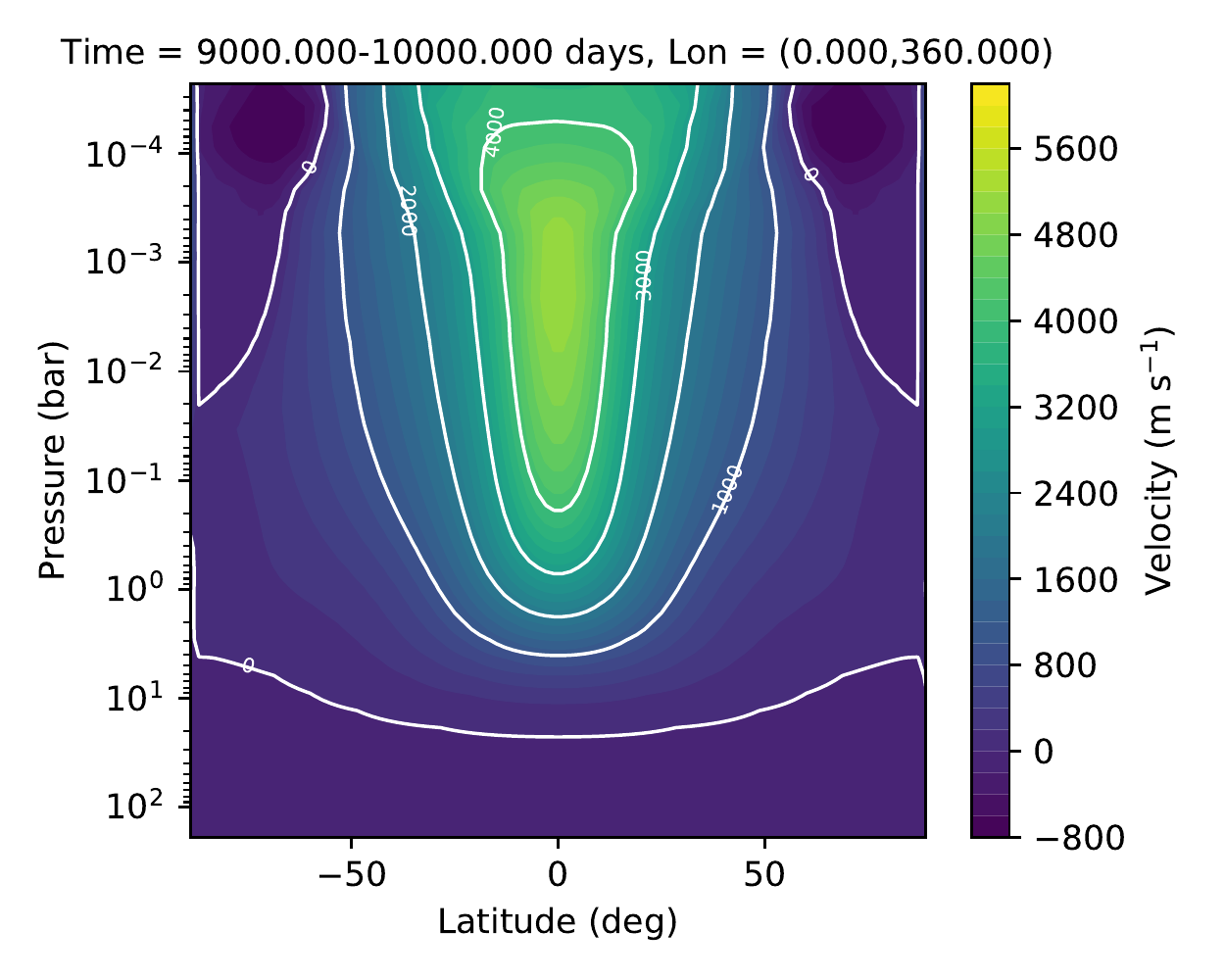}
\caption{Zonally averaged temperature and zonal wind speed for simulations of HD 189733 b at $\sim 2^{\circ}$ resolution. The upper panels show the non-hydrostatic, deep (NHD) case, the lower panels the quasi-hydrostatic, deep (QHD) case. All quantities are averaged over the last 1000 (Earth) days of the 10000  day simulation. \label{fig:189g5zonal1}}
\end{figure*}

\begin{figure*}
\includegraphics[width=0.5\textwidth]{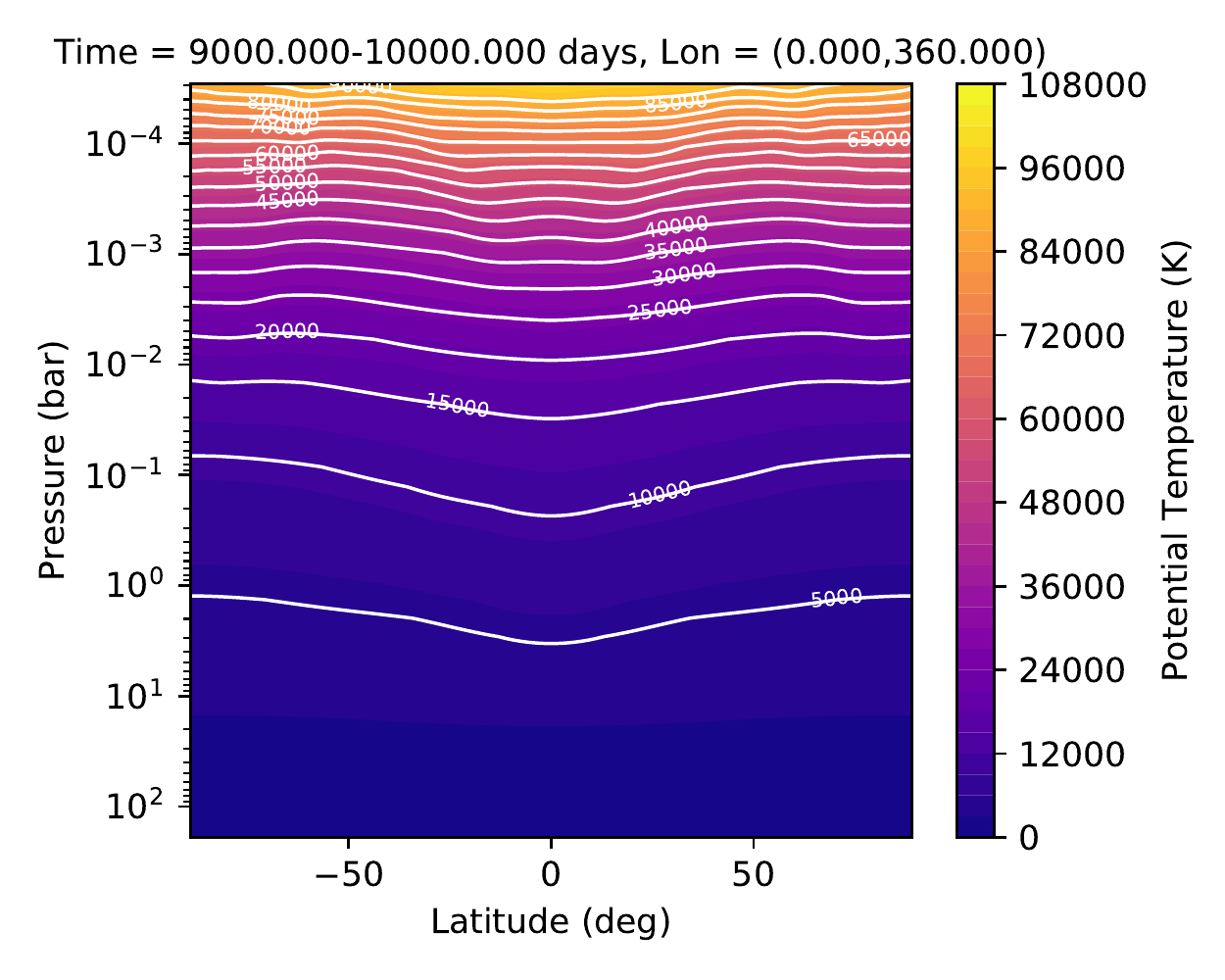}
\includegraphics[width=0.5\textwidth]{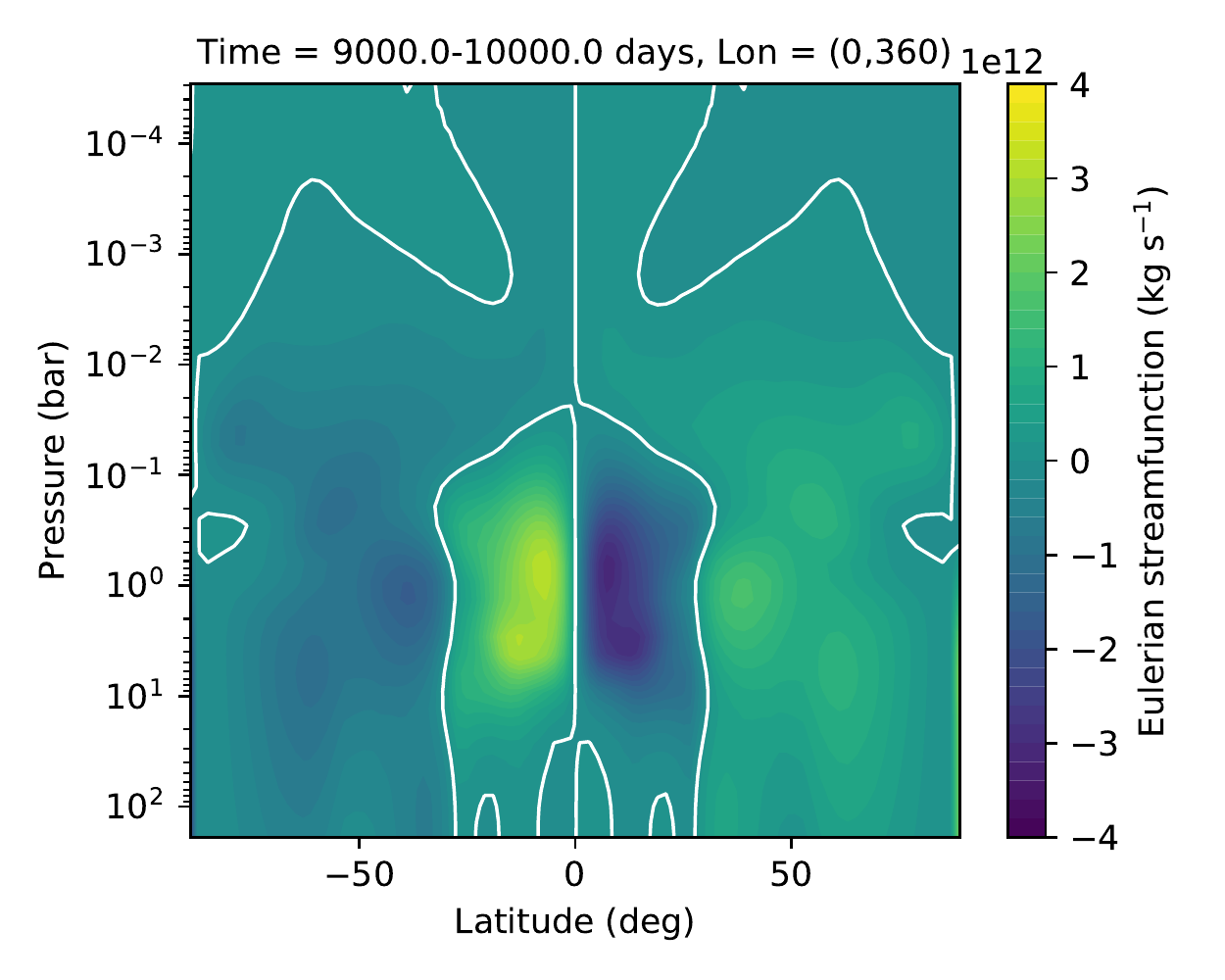}\\
\includegraphics[width=0.5\textwidth]{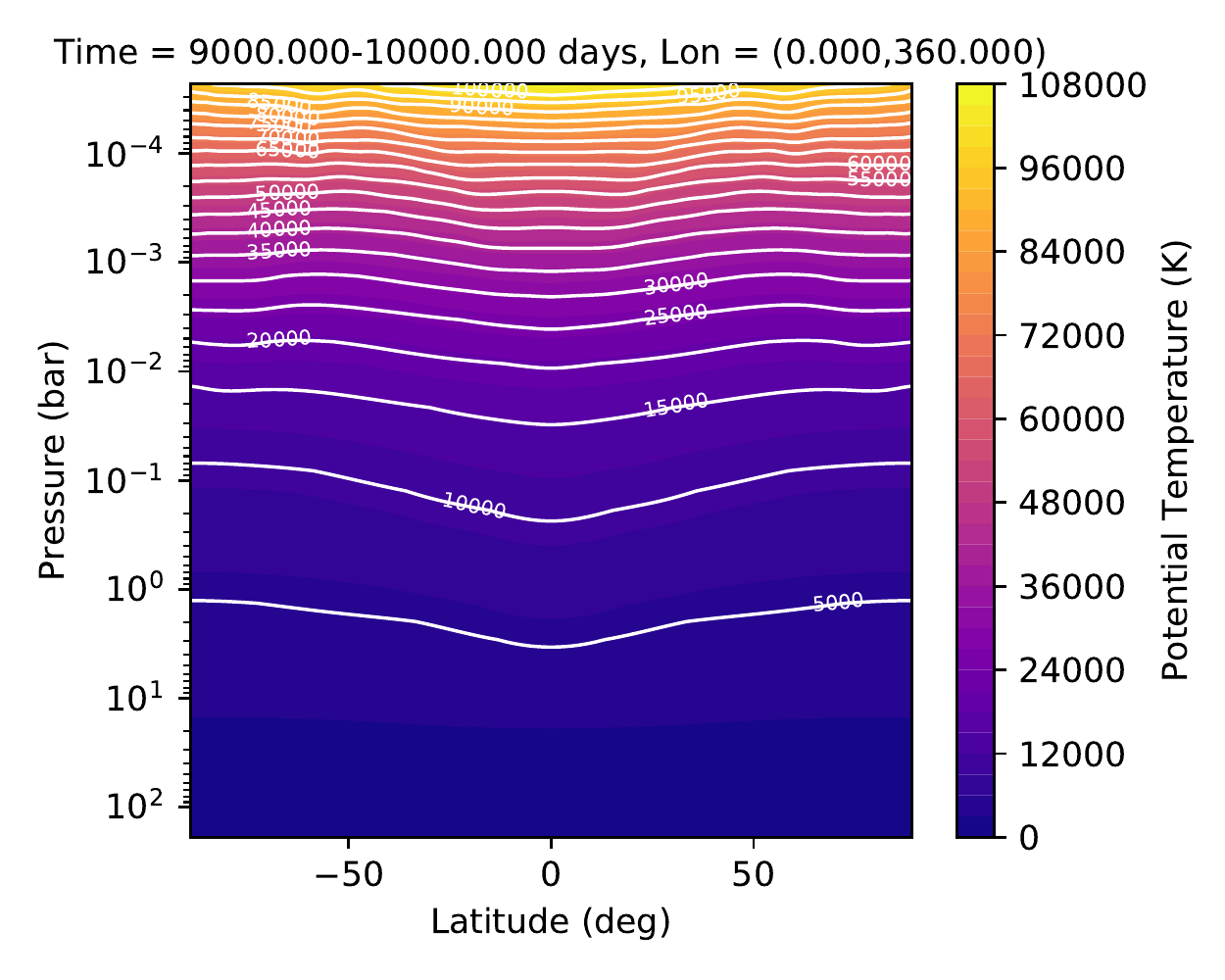}
\includegraphics[width=0.5\textwidth]{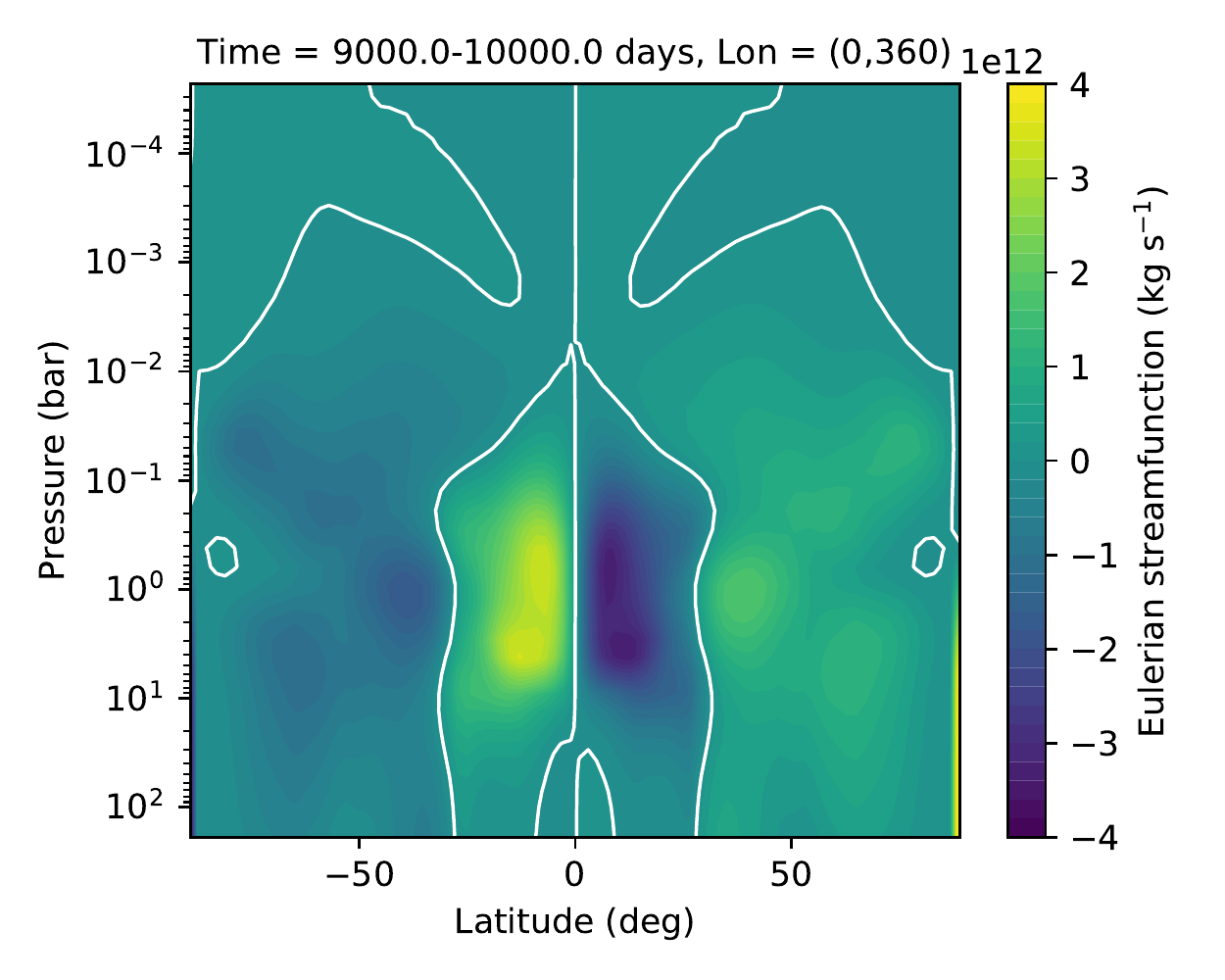}
\caption{Zonally averaged potential temperature and stream function for simulations of HD 189733 b at $\sim 2^{\circ}$ resolution. In the stream function plots, positive values indicate clockwise motion. The upper panels show the non-hydrostatic, deep (NHD) case, the lower panels the quasi-hydrostatic, deep (QHD) case. All quantities are averaged over the last 1000 (Earth) days of the 10000 day simulation. \label{fig:189g5zonal2}}
\end{figure*}

\begin{figure*}
\includegraphics[width=0.5\textwidth]{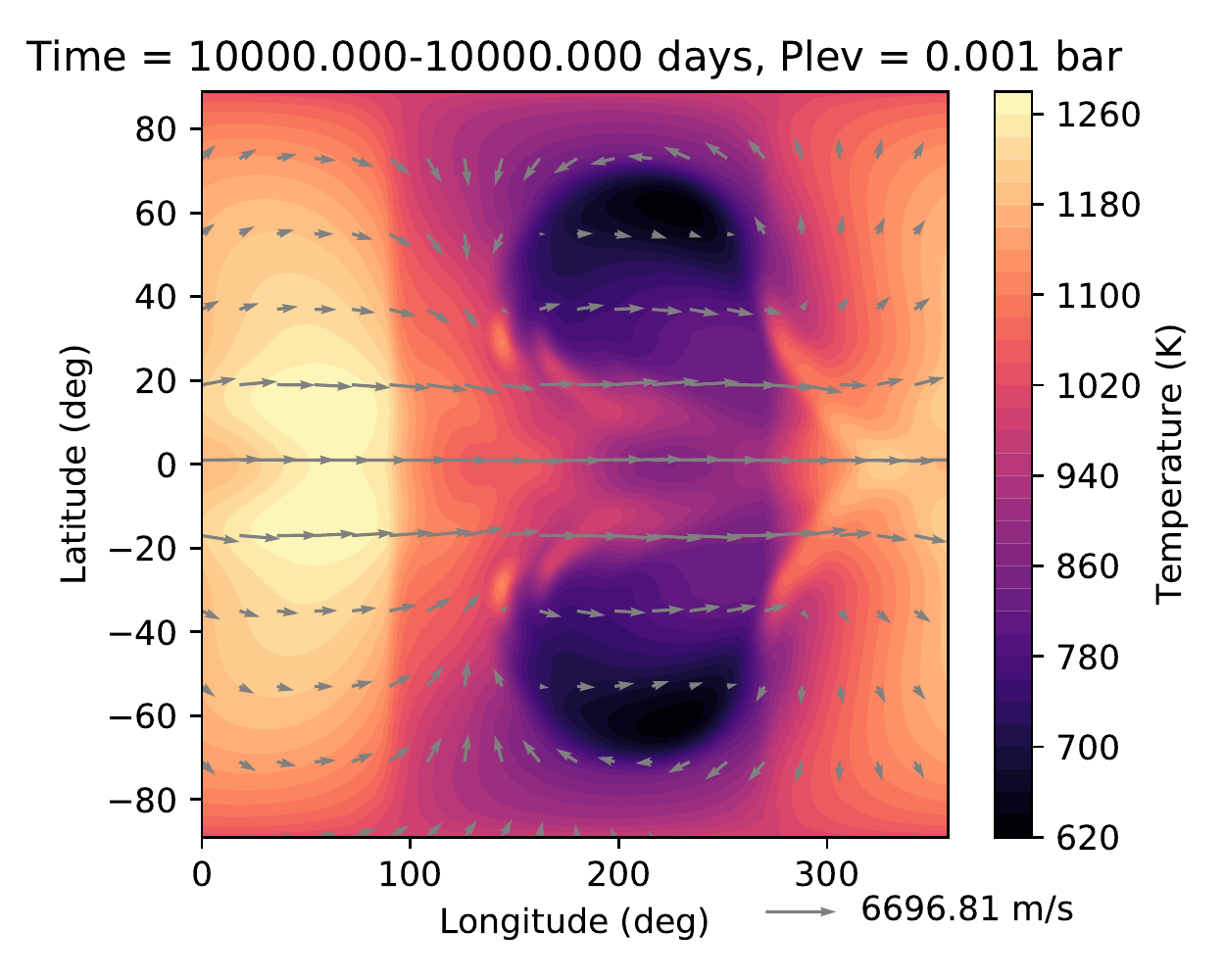}
\includegraphics[width=0.5\textwidth]{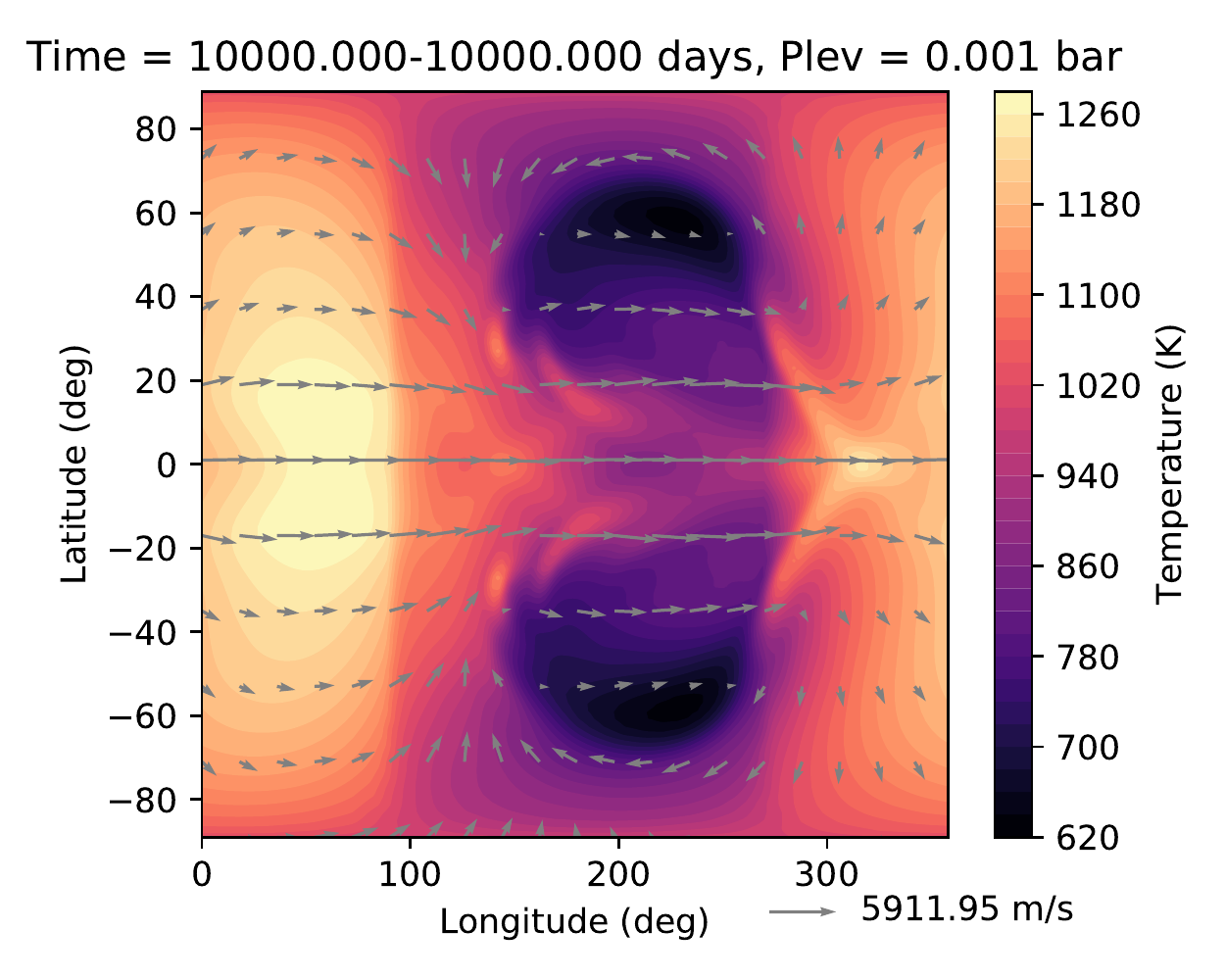}\\
\includegraphics[width=0.5\textwidth]{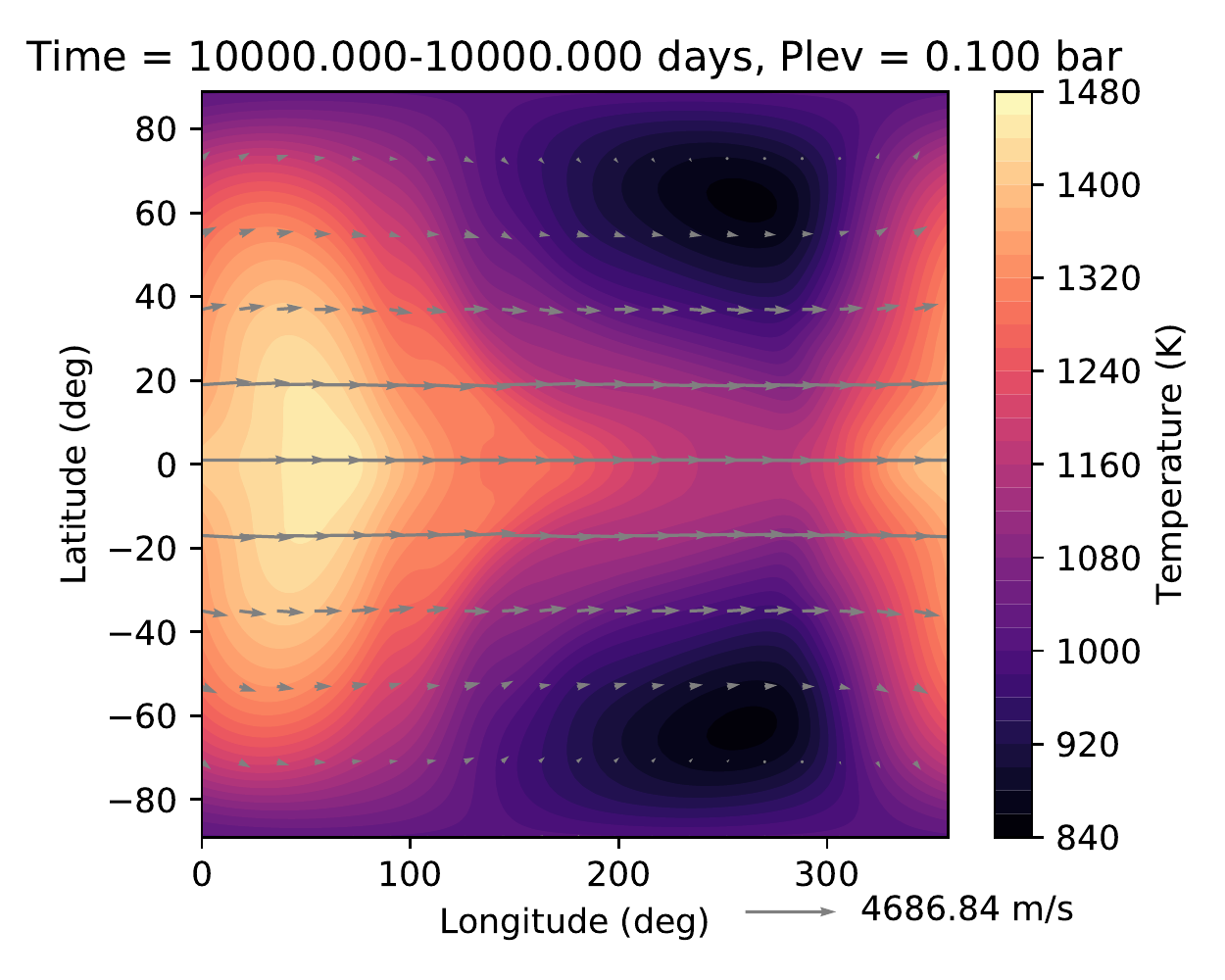}
\includegraphics[width=0.5\textwidth]{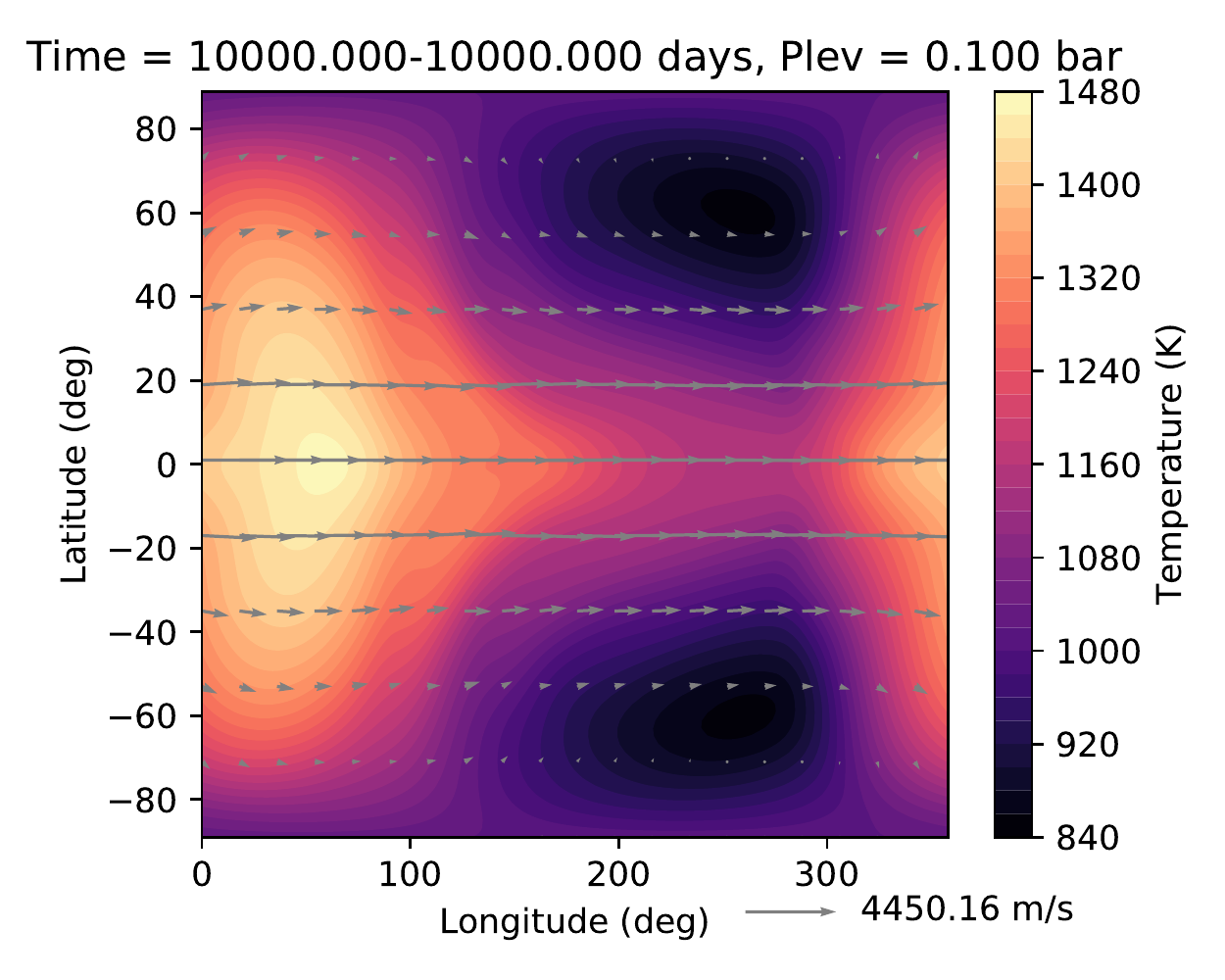}\\
\includegraphics[width=0.5\textwidth]{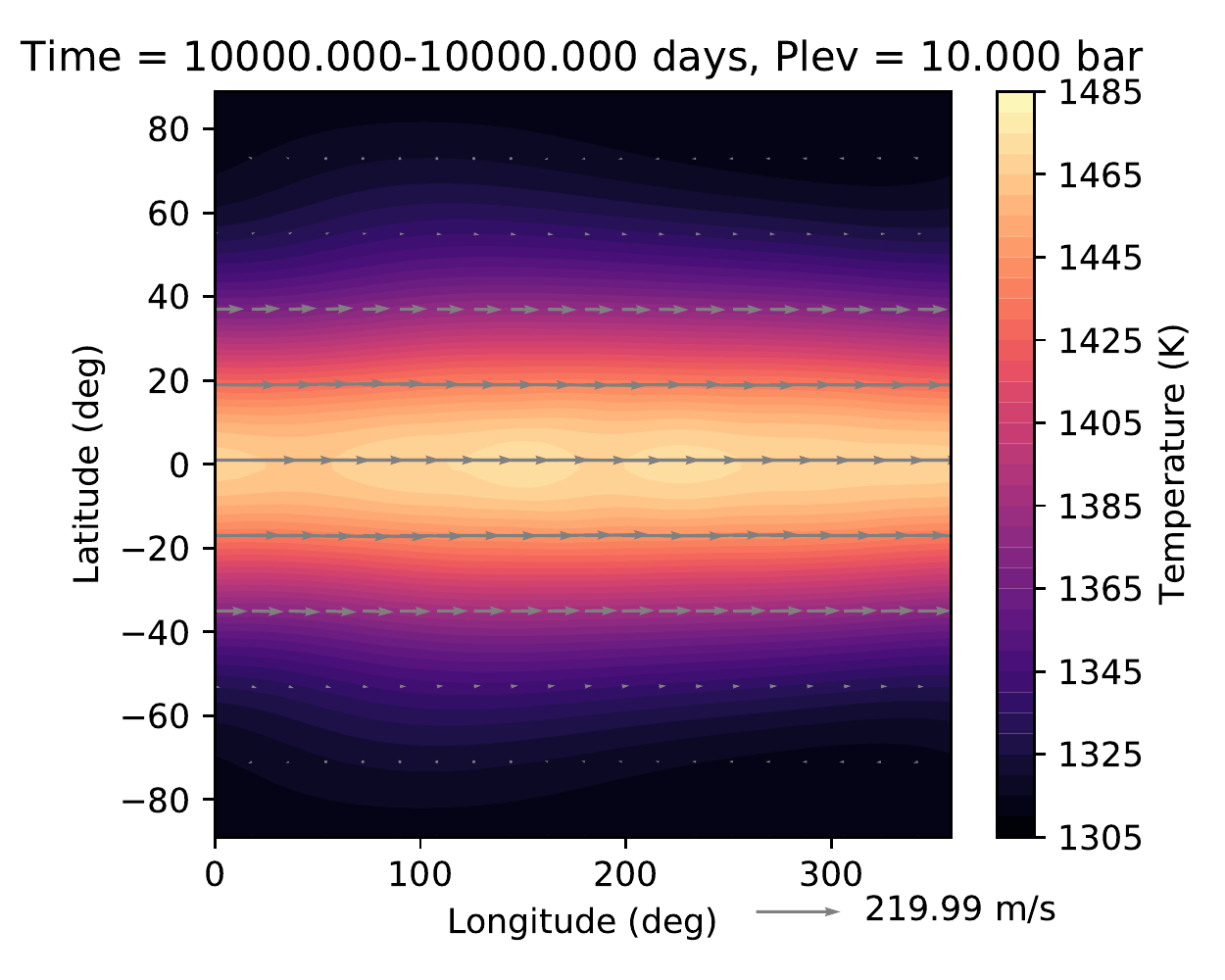}
\includegraphics[width=0.5\textwidth]{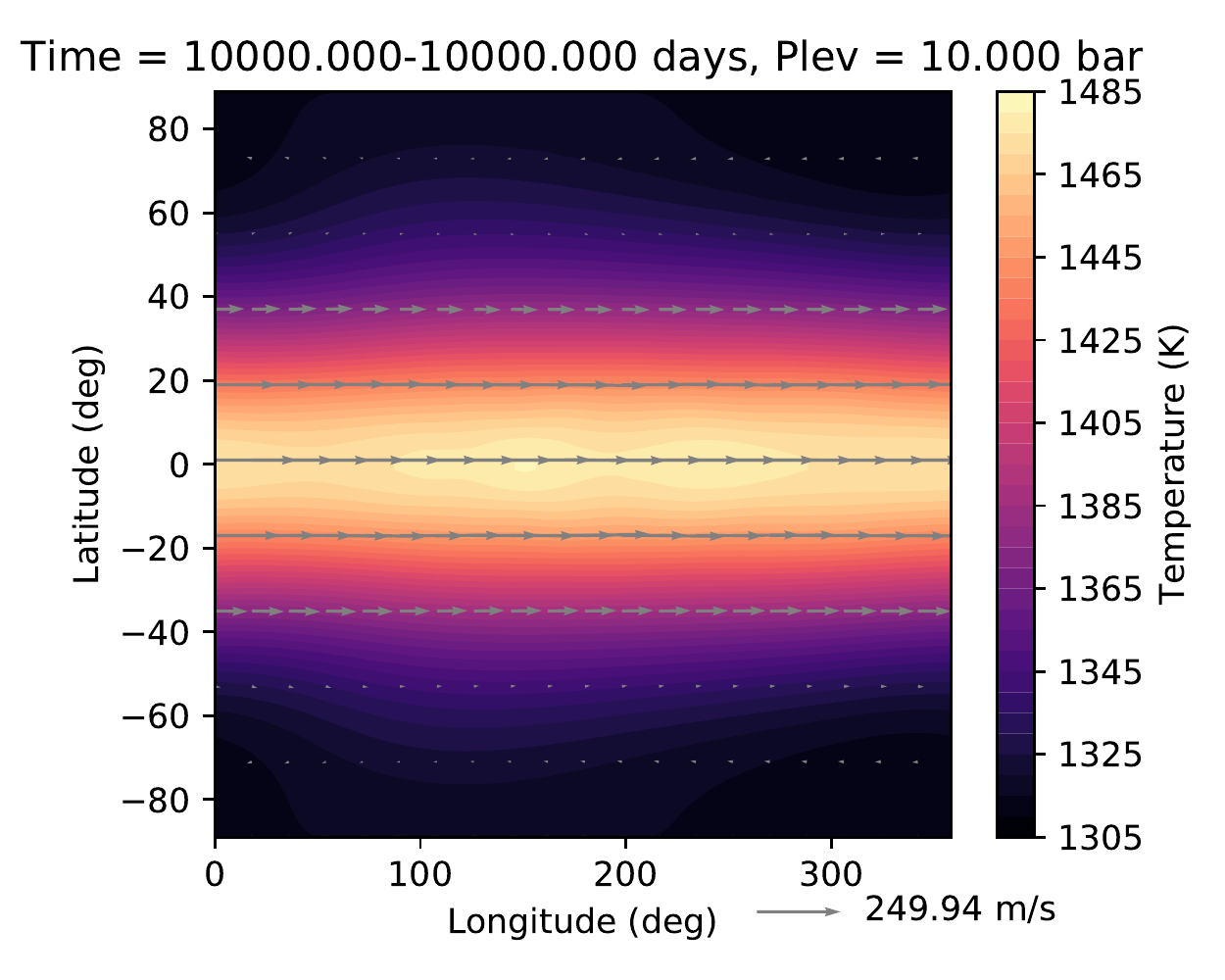}
\caption{Snapshots at 10000 days of the temperature (color) and horizontal winds (arrows) on isobaric surfaces for simulations of HD 189733 b at $\sim 2^{\circ}$ resolution. The left panels are the NHD simulation, the right are the QHD simulation. The substellar point is at $0^{\circ}$ longitude, $0^{\circ}$ latitude.\label{fig:189g5Tlevels}} 
\end{figure*}

Figure \ref{fig:189g5zonal1} shows the temperature and zonal wind speed, as in Figure \ref{fig:189g4zonal1}, but for simulations with a resolution of $\sim 2^{\circ}$. Temperatures shows a similar pattern to the $\sim 4^{\circ}$ resolution simulations, but the superrotational jet has increased in speed and has broadened in latitude, pushing the return flow toward higher latitudes. The increase in velocity is likely due to better resolution of waves that carry angular momentum toward the equator. The peak velocity of the NHD simulation is $\sim 20\%$ greater than in the QHD simulation, similar to the effect seen at $\sim 4^{\circ}$ resolution. The potential temperature and stream function for the $\sim 2^{\circ}$ simulations are shown in Figure \ref{fig:189g5zonal2}. The temperature and wind speeds along isobars are shown in Figure \ref{fig:189g5Tlevels}.

\begin{figure*}
\includegraphics[width=0.5\textwidth]{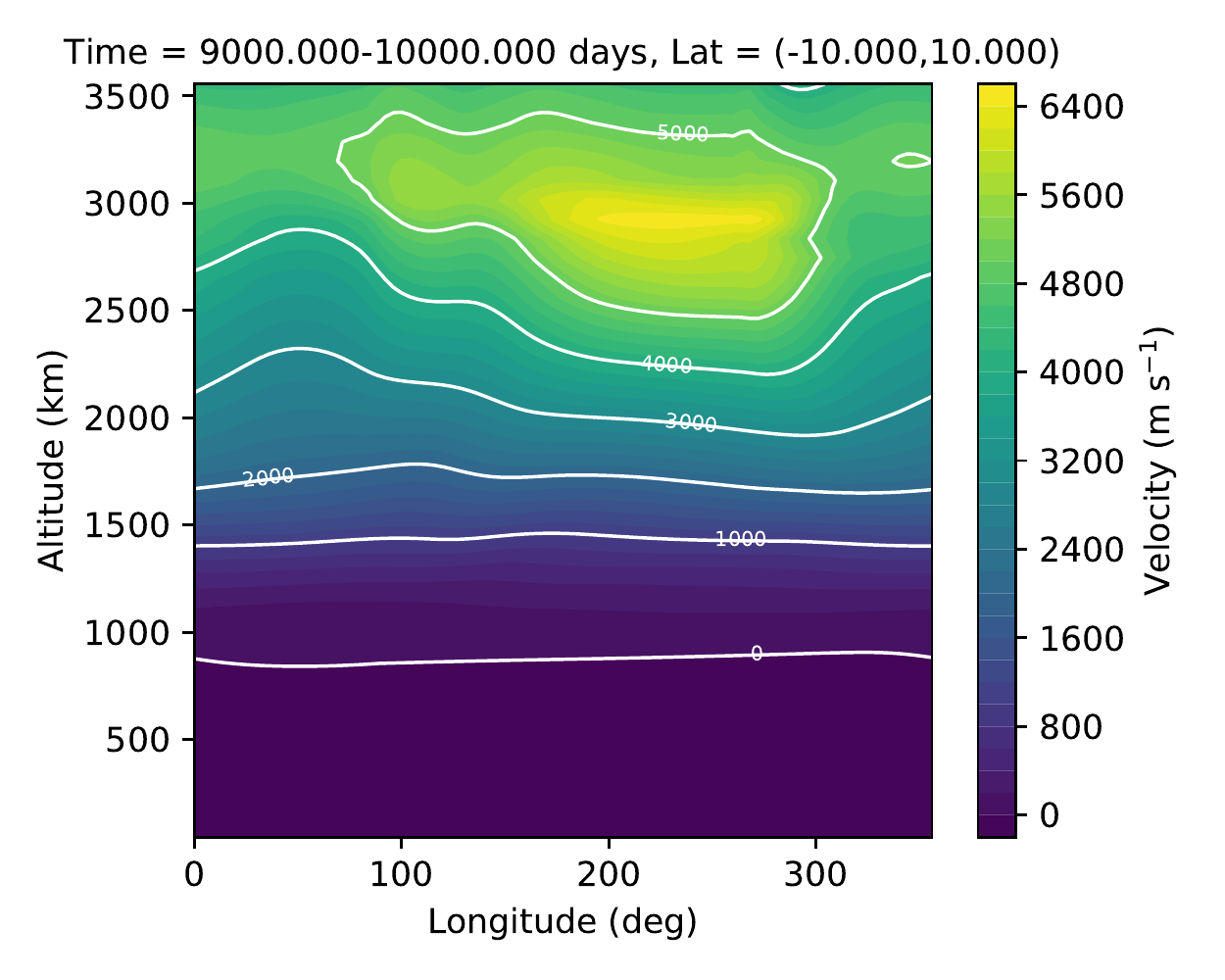}
\includegraphics[width=0.5\textwidth]{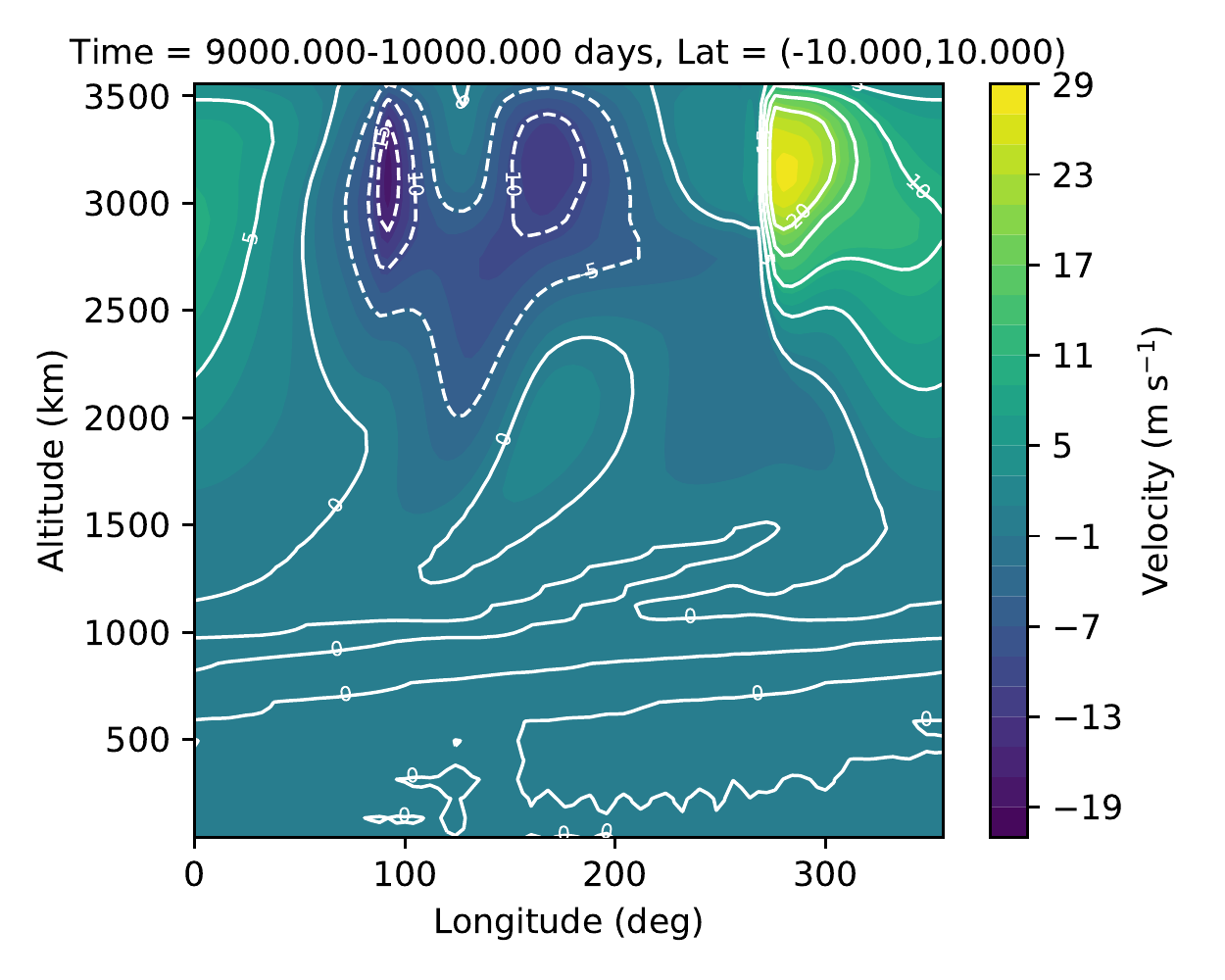}\\
\includegraphics[width=0.5\textwidth]{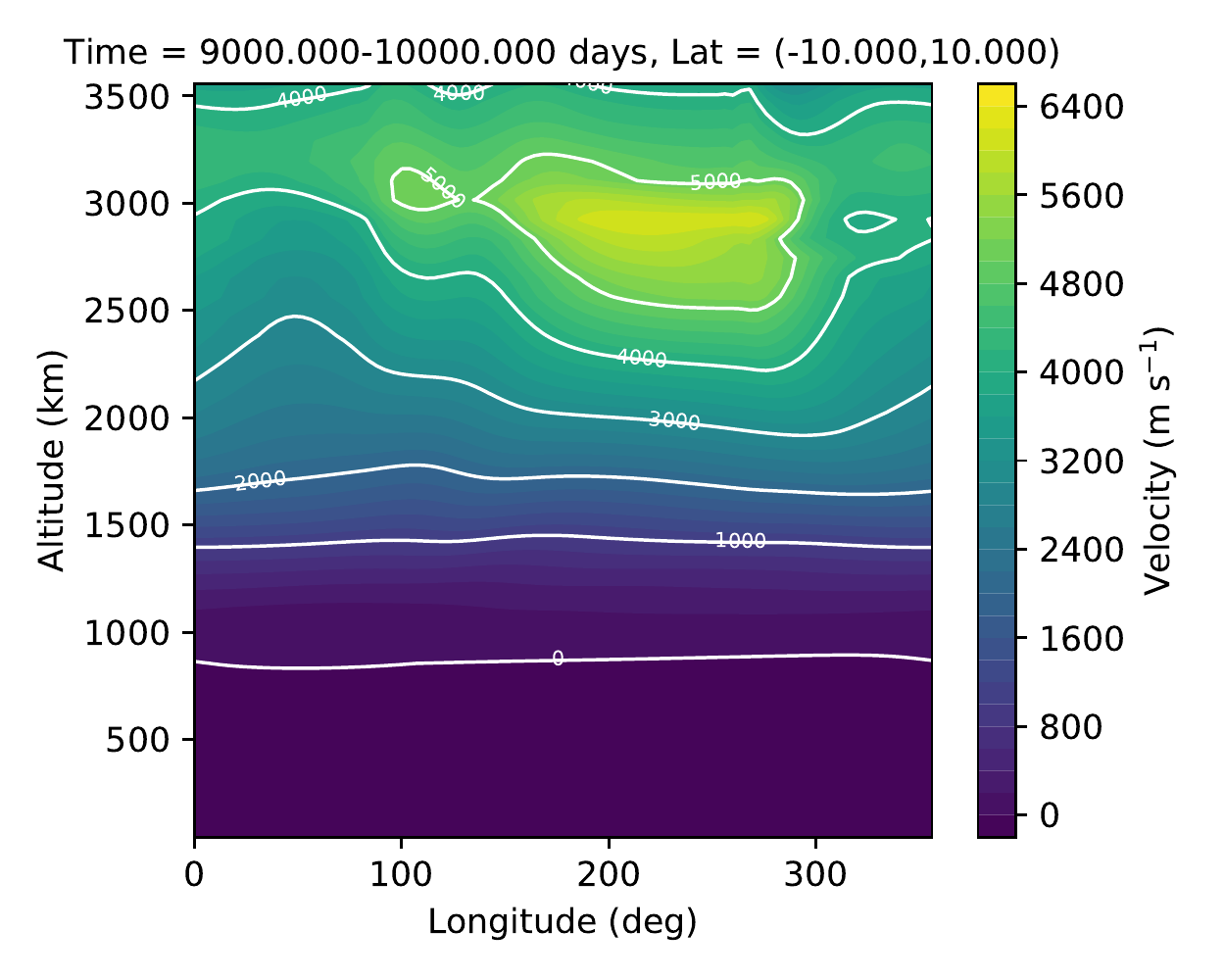}
\includegraphics[width=0.5\textwidth]{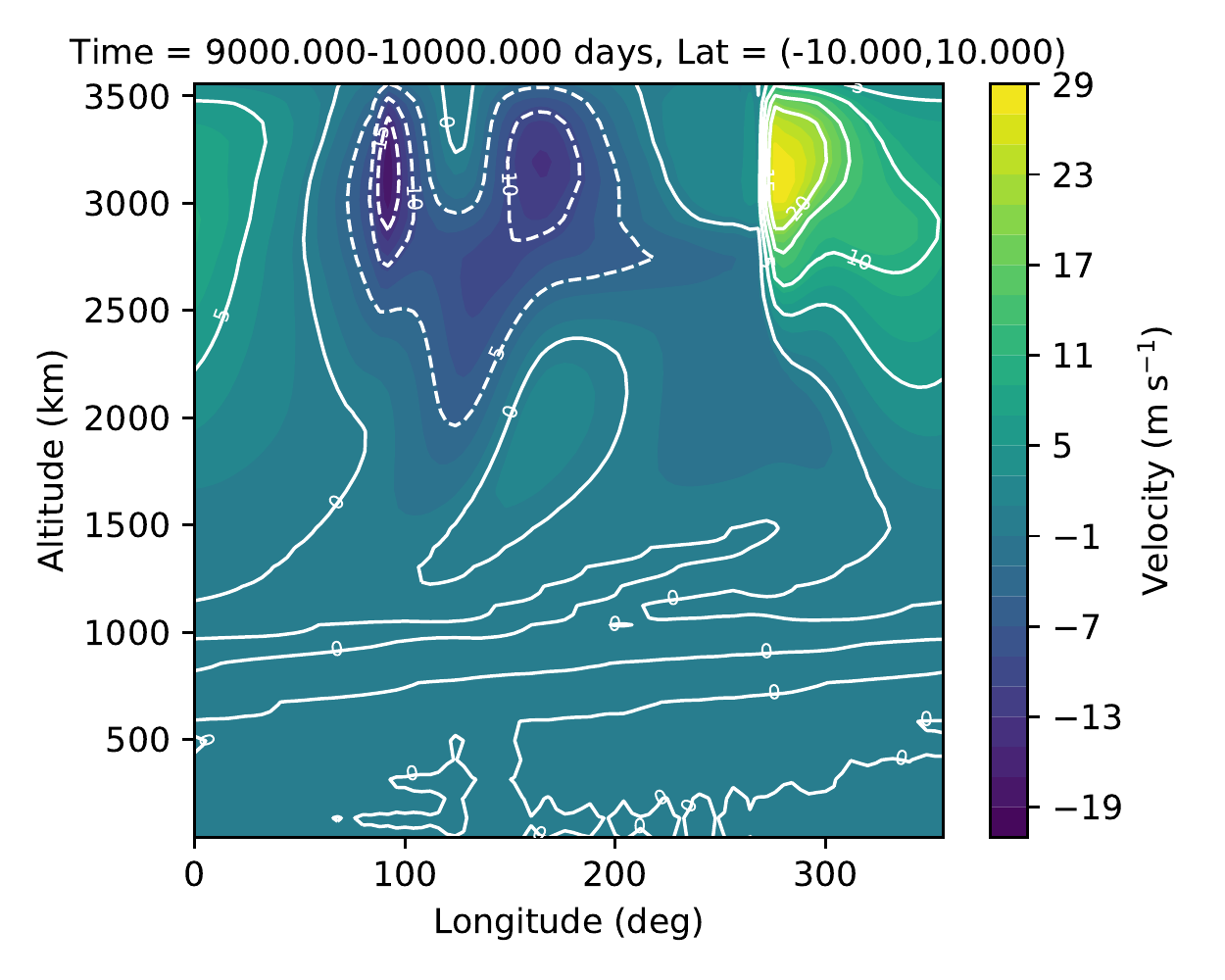}
\caption{Zonal (left) and vertical (right) wind speeds averaged over a $20^{\circ}$ degree latitude band centered about the equator for simulations of HD 189733 b at $\sim 4^{\circ}$ resolution. The upper panels are the NHD case, the lower panels the QHD case. All quantities are averaged over the last 1000 (Earth) days of the 10000 day simulation. \label{fig:189g4eqtwind}}
\end{figure*}

\begin{figure*}
\includegraphics[width=0.5\textwidth]{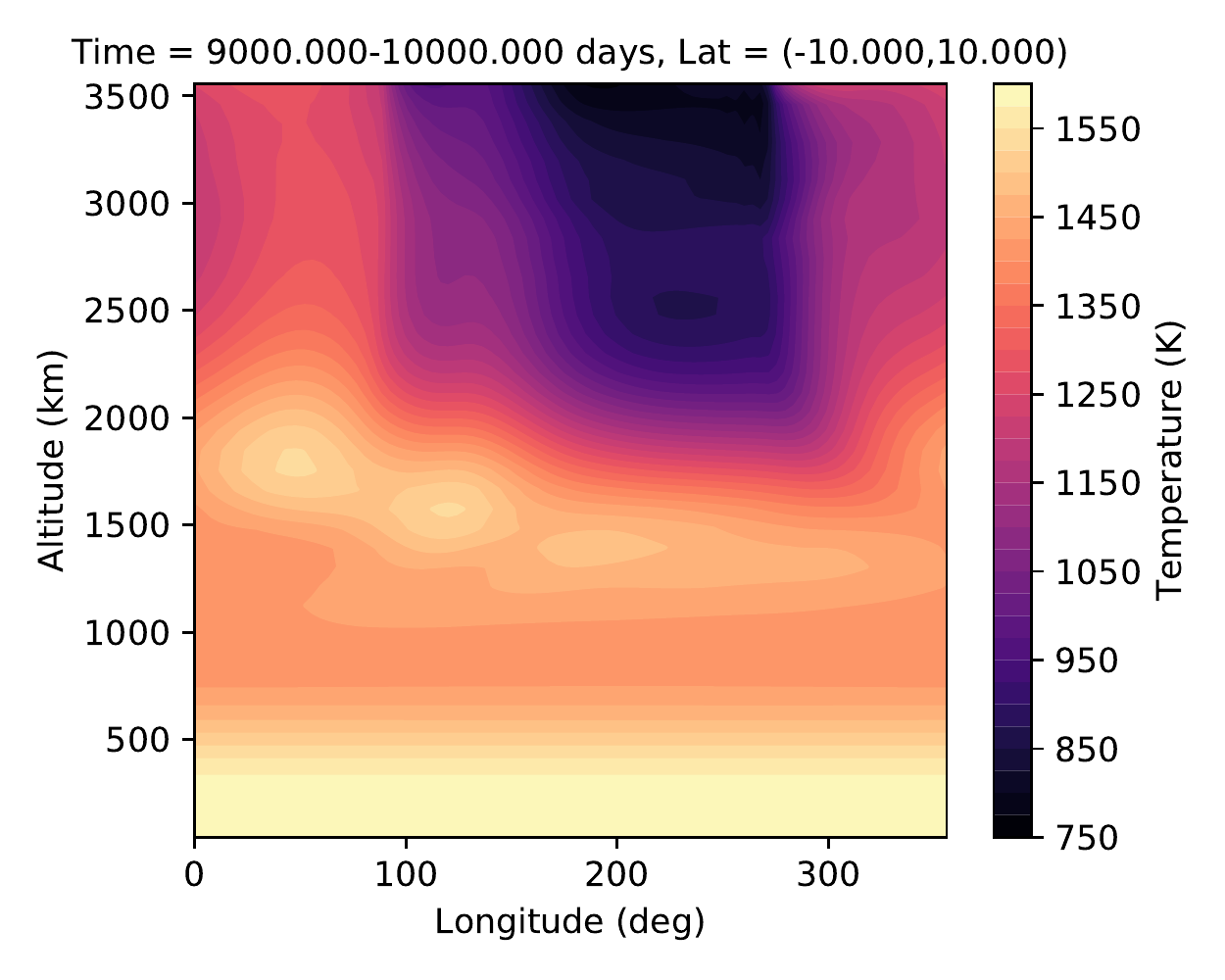}
\includegraphics[width=0.5\textwidth]{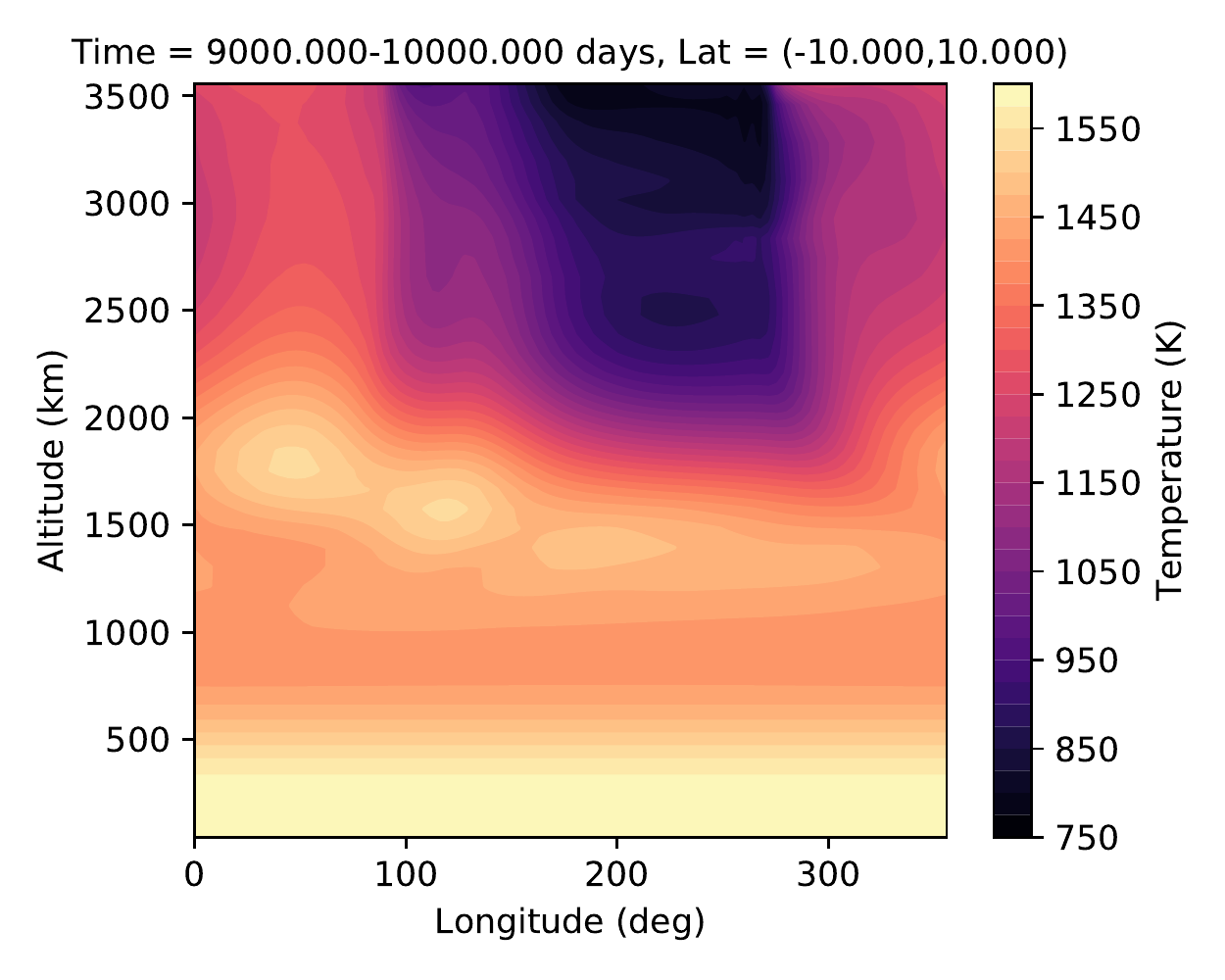}
\caption{Temperatures averaged over a $20^{\circ}$ degree latitude band centered about the equator for simulations of HD 189733 b at $\sim 4^{\circ}$ resolution. The left panel is the NHD case, the right panel is the QHD case. Temperatures are averaged over the last 1000 (Earth) days of the 10000 day simulation. \label{fig:189g4eqttemp}}
\end{figure*}

Figure \ref{fig:189g4eqtwind} shows the zonal and vertical velocities in a latitudinal band along the equator, averaged over the last 1000 days, for the NHD and QHD simulations at $\sim 4^{\circ}$ resolution. Here we have used altitude as our vertical coordinate to avoid extrapolation at the top of the model. The values are averaged over a $20^{\circ}$ latitudinal band weighted by $\cos{\phi}$, where $\phi$ is the latitude. Zonal winds are highest on the night side (longitudes $90^{\circ}-270^{\circ}$) toward the western terminator. As the flow approaches the day side, it slows because of the increase in pressure. Vertical velocities are slow everywhere in comparison to the horizontal winds, though the figures show a lot of structure. Upwelling and downwelling occur, for example, side by side along the western terminator. Upwelling dominates on the day side of the planet, with downwelling on the nightside, most prominantly east of the eastern terminator. Figure \ref{fig:189g4eqttemp} shows equatorial temperatures in the same style, comparing the NHD and QHD cases.

\begin{figure*}
\includegraphics[width=0.5\textwidth]{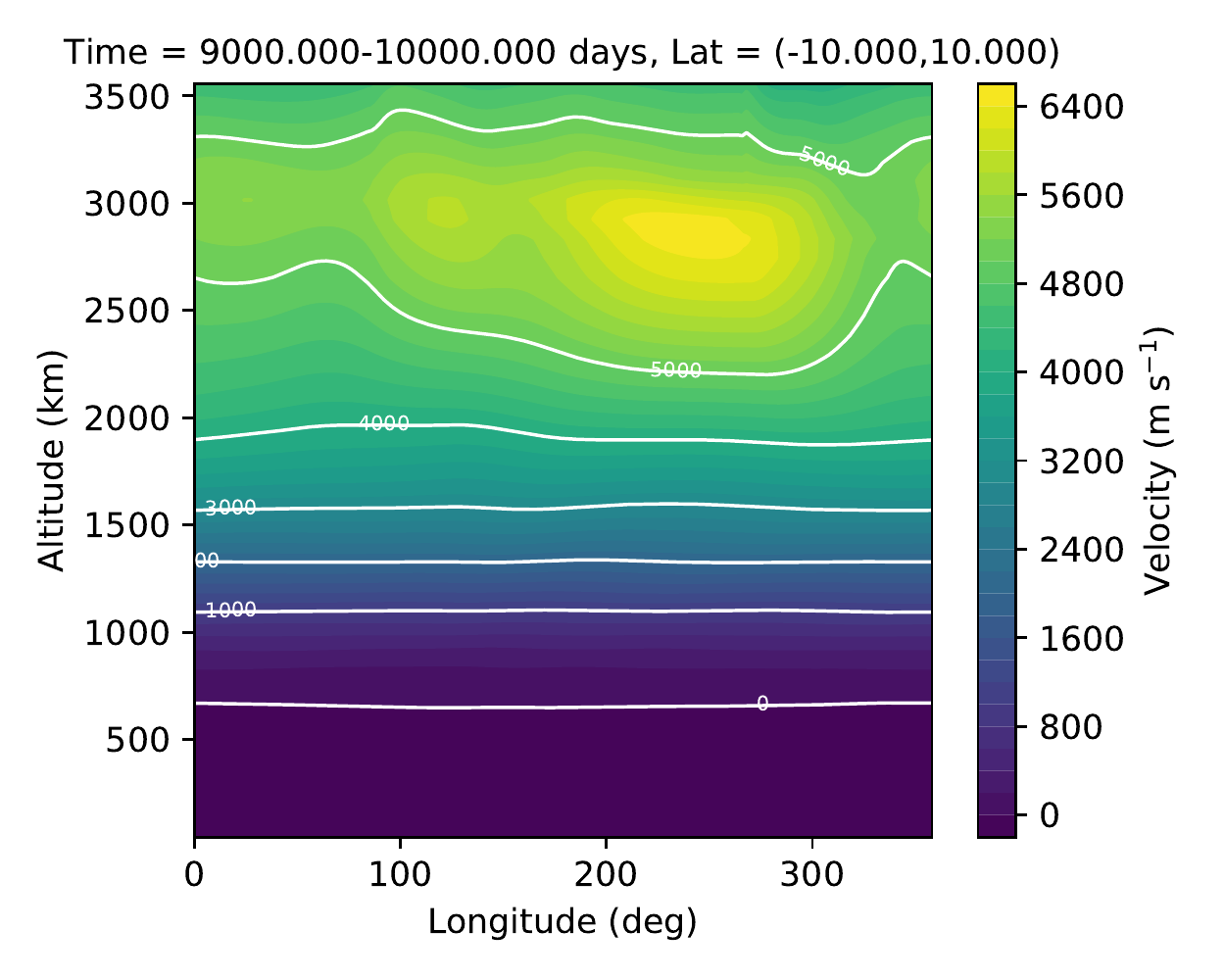}
\includegraphics[width=0.5\textwidth]{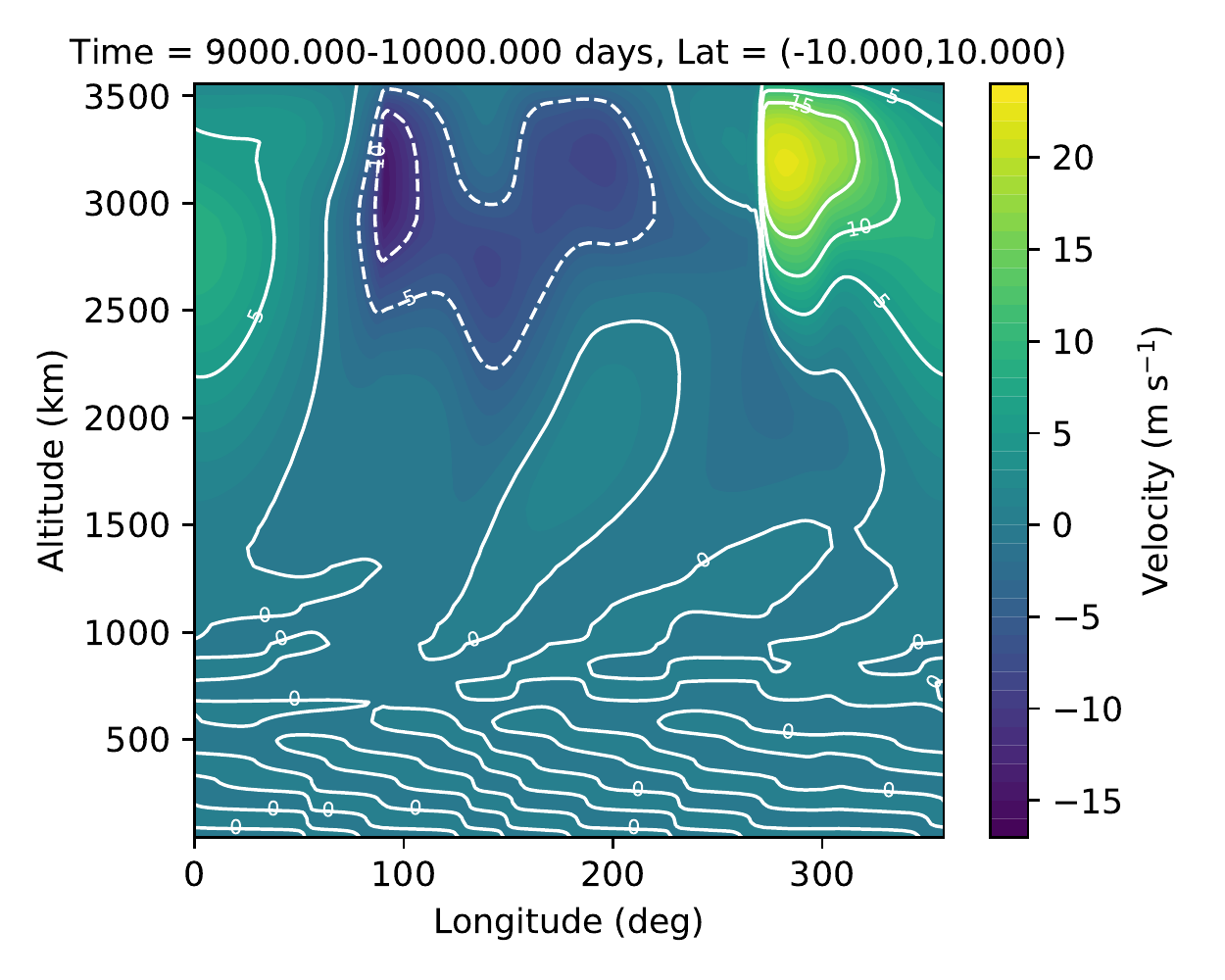}\\
\includegraphics[width=0.5\textwidth]{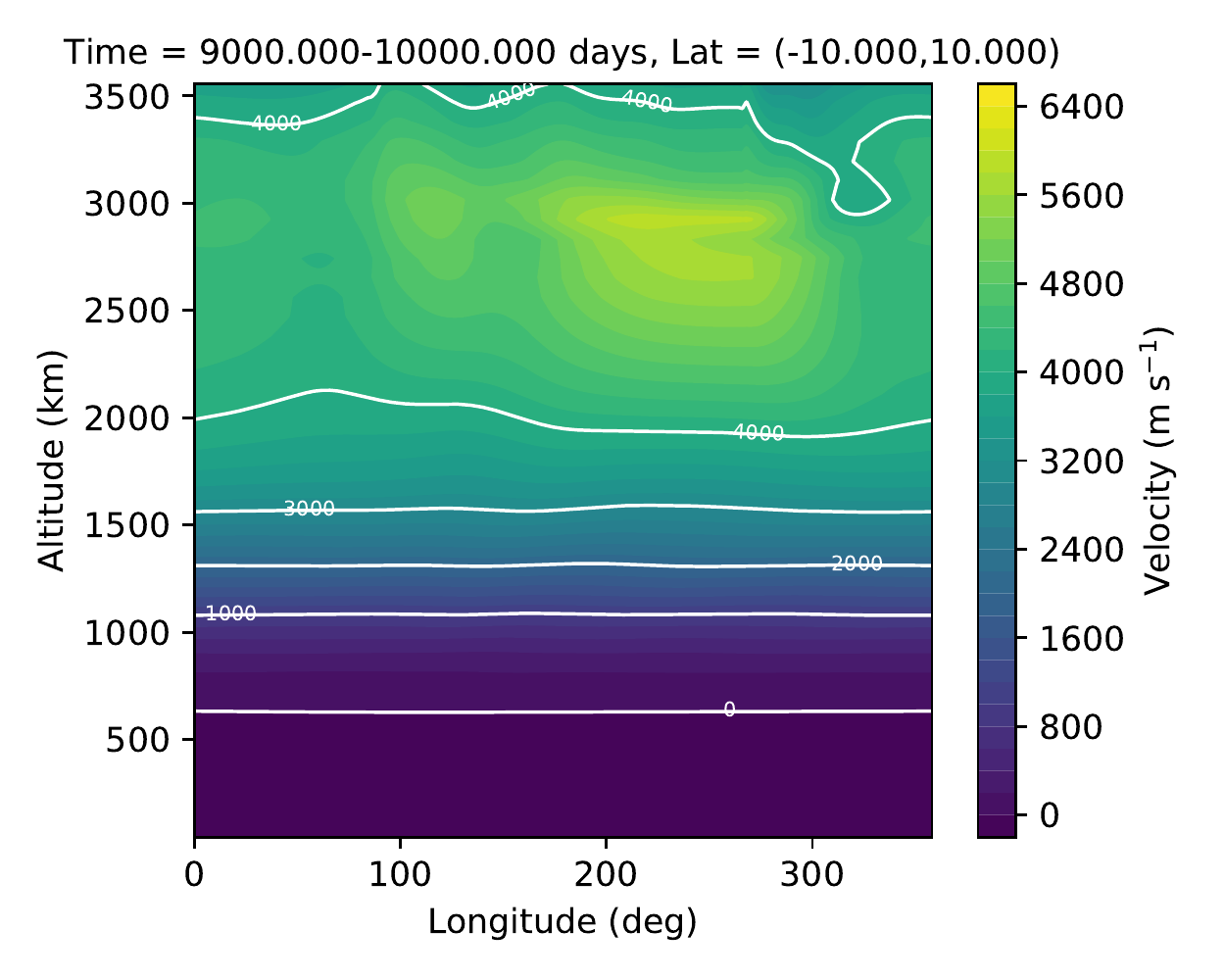}
\includegraphics[width=0.5\textwidth]{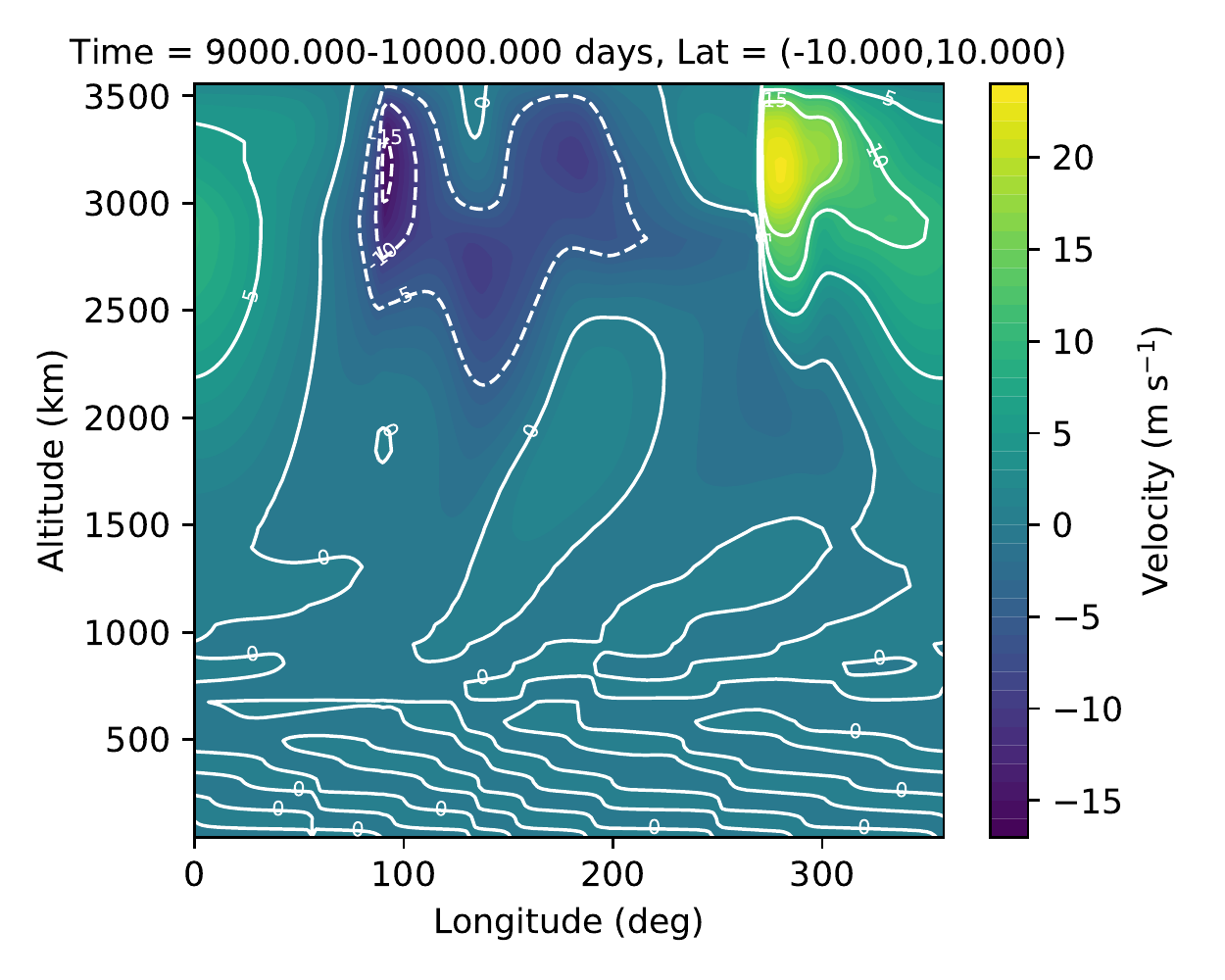}
\caption{Zonal (left) and vertical (right) wind speeds averaged over a $20^{\circ}$ degree latitude band centered about the equator for simulations of HD 189733 b at $\sim 2^{\circ}$ resolution. The upper panels are the NHD case, the lower panels the QHD case. All quantities are averaged over the last 1000 (Earth) days of the 10000 day simulation. \label{fig:189g5eqtwind}}
\end{figure*}

\begin{figure*}
\includegraphics[width=0.5\textwidth]{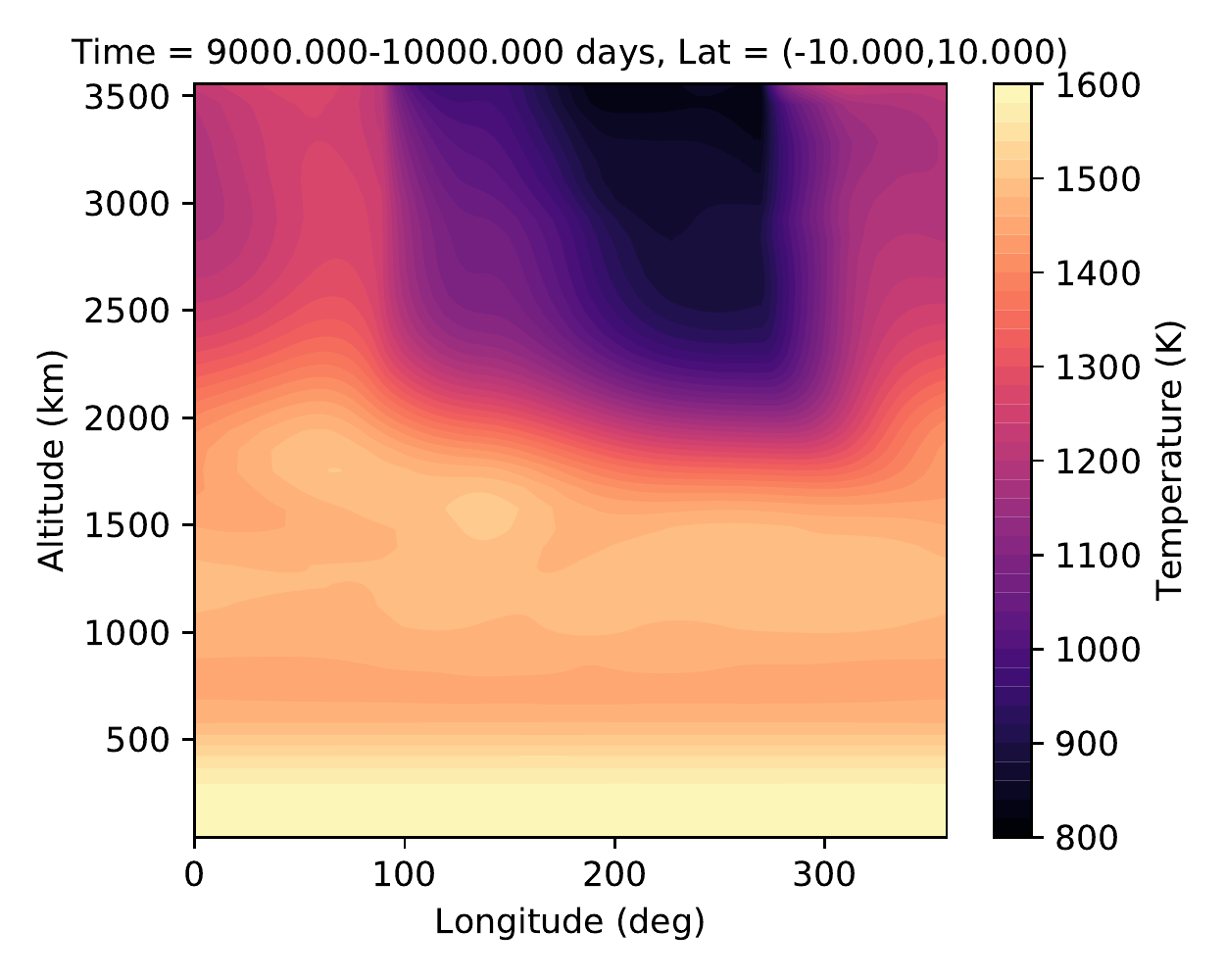}
\includegraphics[width=0.5\textwidth]{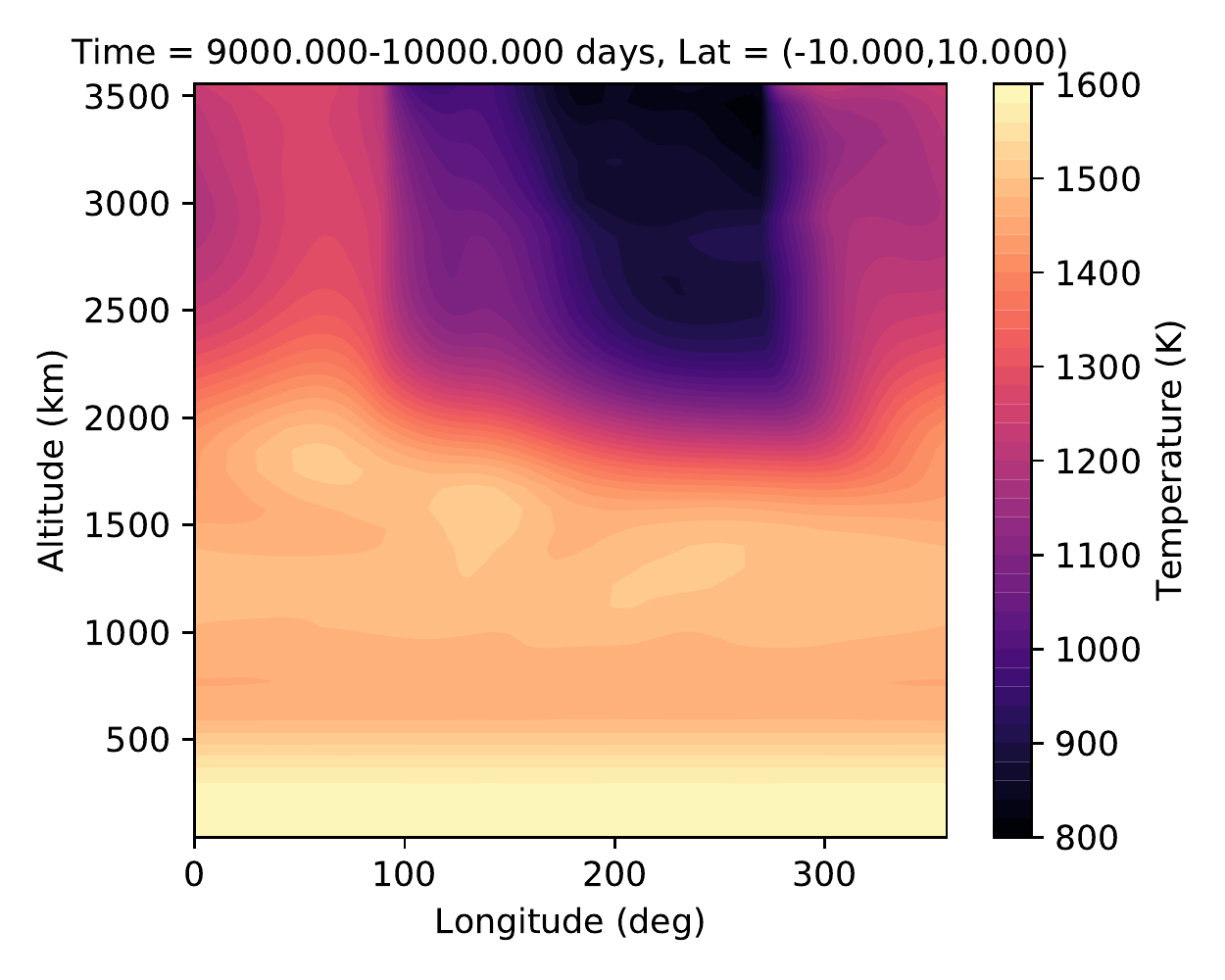}
\caption{Temperatures averaged over a $20^{\circ}$ degree latitude band centered about the equator for simulations of HD 189733 b at $\sim 2^{\circ}$ resolution. The left panel is the NHD case, the right panel is the QHD case. Temperatures are averaged over the last 1000 (Earth) days of the 10000 day simulation. \label{fig:189g5eqttemp}}
\end{figure*}

The same quantities are shown for the $\sim 2^{\circ}$ simulations in Figures \ref{fig:189g5eqtwind} and \ref{fig:189g5eqttemp}. While the temperature structure is largely unchanged from the $\sim 4^{\circ}$ cases, the velocities have changed more significantly. In particular, there is now a strong difference between the peak zonal winds in the NHD and QHD cases. The zonal wind speeds have increased, the vertical wind speeds have slightly decreased, and the flow now extends deeper into the atmosphere, as also seen in Figure \ref{fig:189g5zonal1}.

\section{Effects of Numerical Dissipation} \label{sec:diffcomp}

\begin{figure*}
\includegraphics[width=0.5\textwidth]{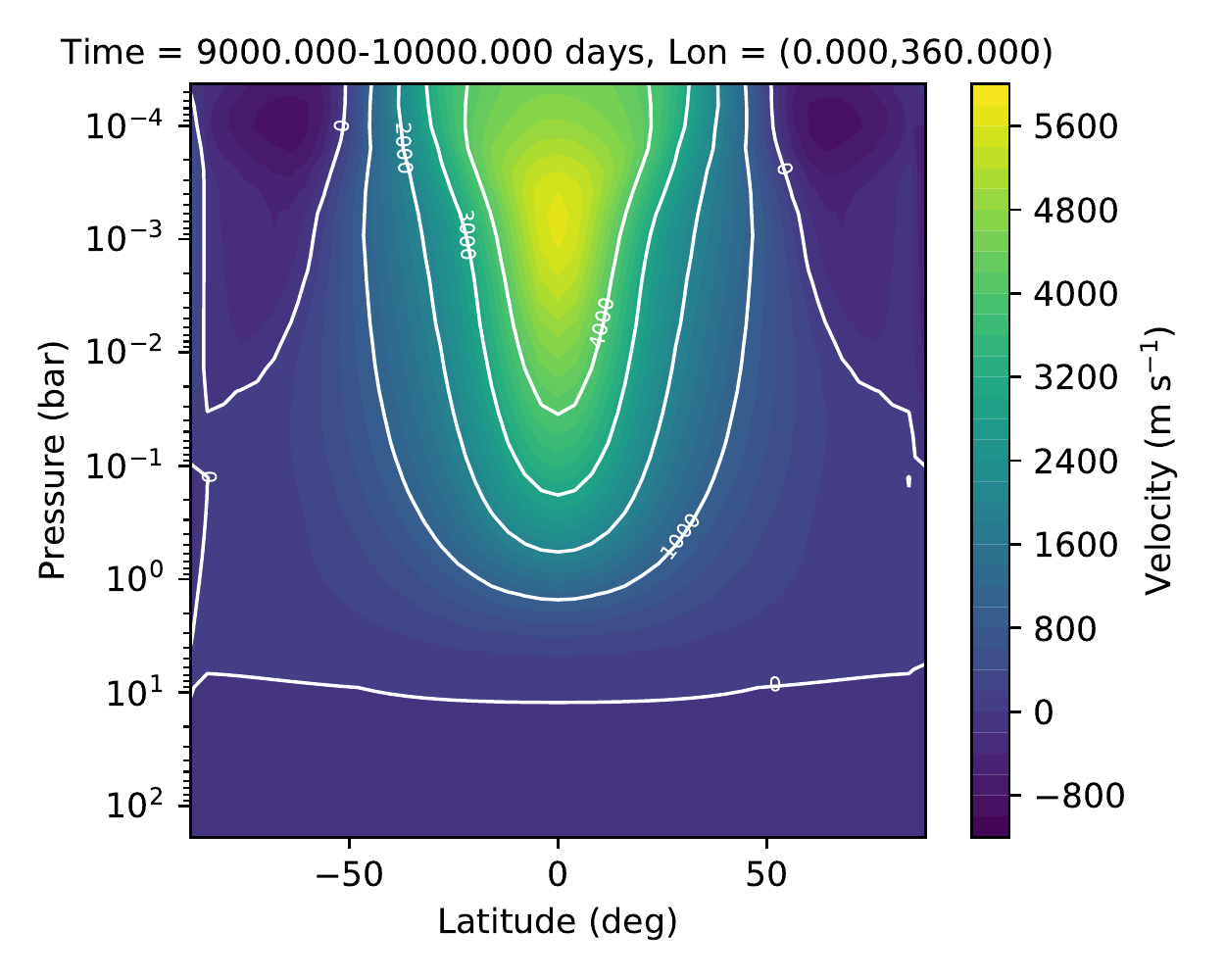}
\includegraphics[width=0.5\textwidth]{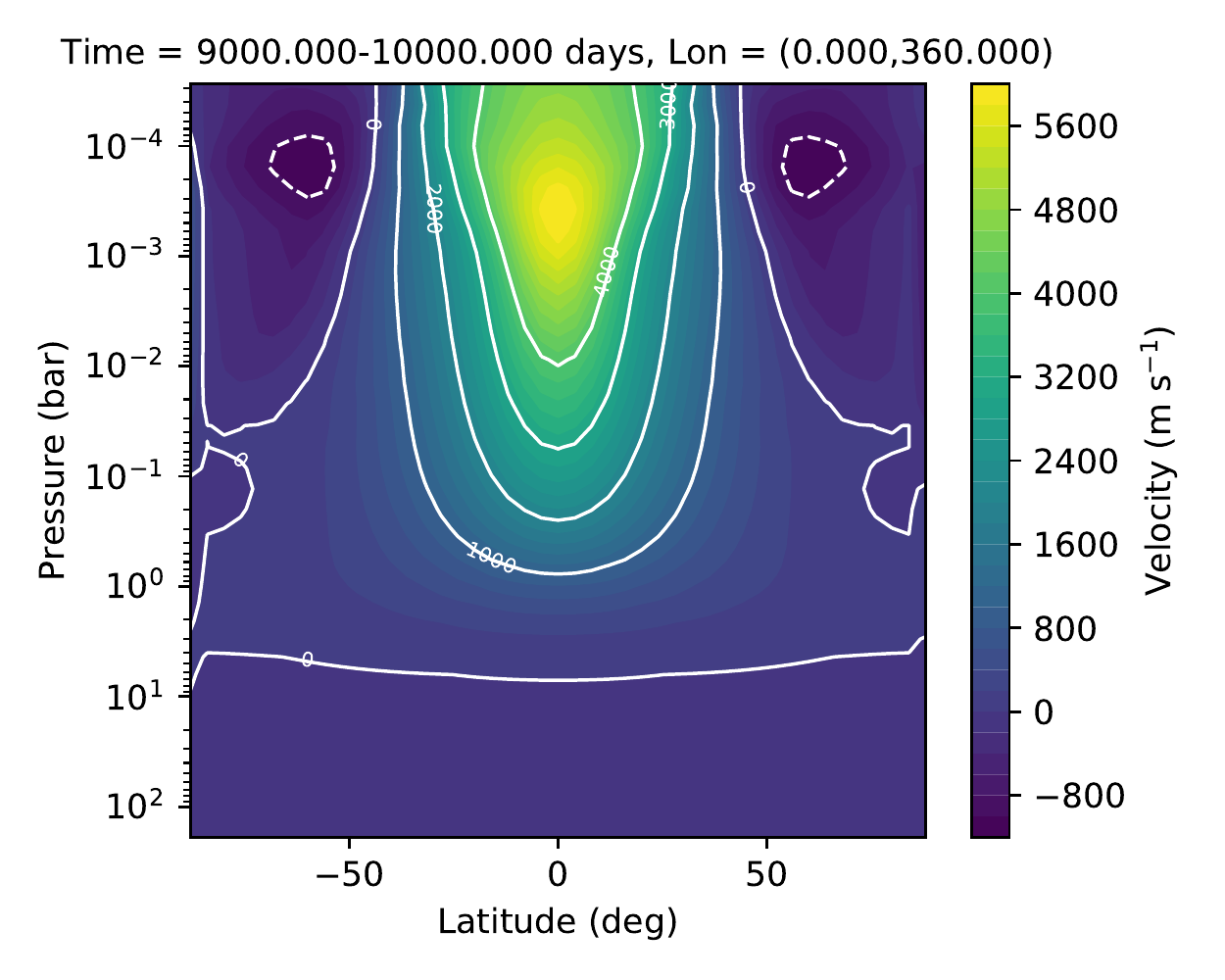}\\
\includegraphics[width=0.5\textwidth]{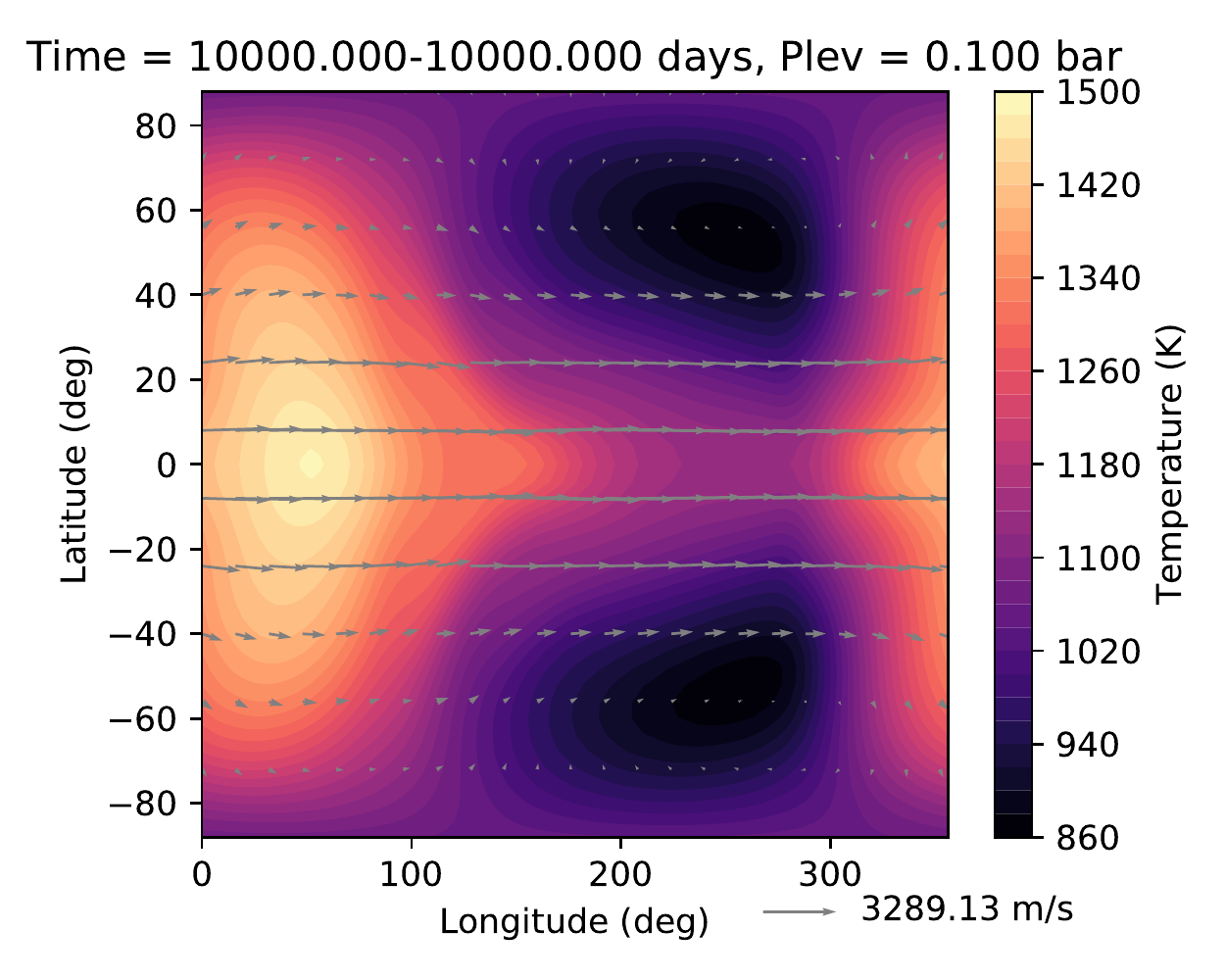}
\includegraphics[width=0.5\textwidth]{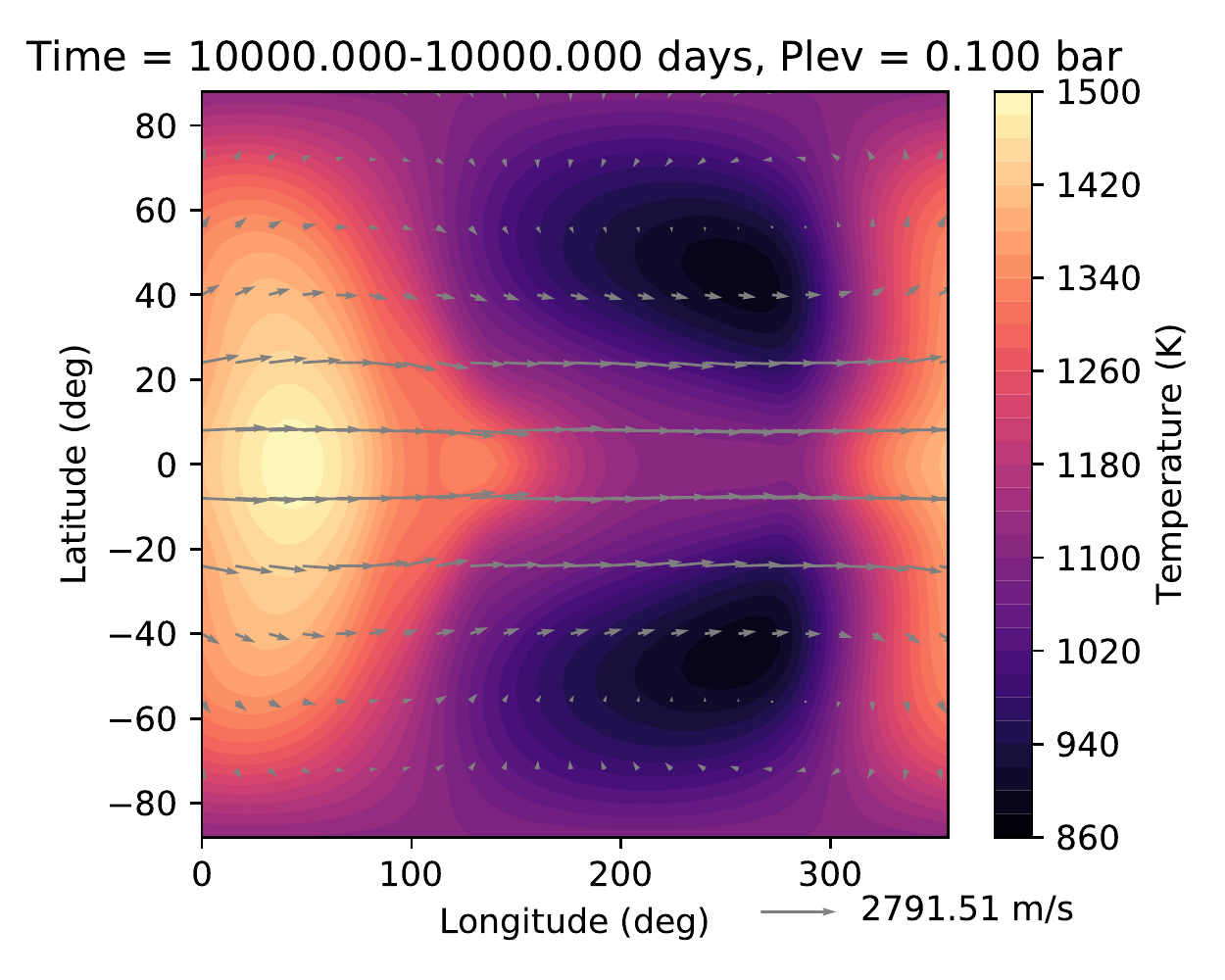}
\caption{Zonally averaged wind speed (top) and temperature at the photosphere (bottom) for simulations of HD 189733 b using different numerical dissipation strengths. The left panels correspond to $D_{\text{hyp}} = D_{\text{div}} = 4.99 \times 10^{-3}$, the right to $D_{\text{hyp}}=D_{\text{div}} = 1.499 \times 10^{-2}$. Compare to the NHD simulations in Figures \ref{fig:189g4zonal1} and \ref{fig:189g4horiz}.  \label{fig:diffcomp}}
\end{figure*}

Here, we examine the effects of numerical dissipation on the simulations of HD 189733 b. We perform two additional simulations, at resolution $\sim 4^{\circ}$, with $D_{\text{hyp}}=D_{\text{div}} = 4.99 \times 10^{-3}$ and $D_{\text{hyp}}=D_{\text{div}} = 1.499 \times 10^{-2}$ (or 0.5 and 1.5 $\times$ the original value of $9.97 \times 10^{-3}$). Zonal winds and temperatures at 0.1 bar are shown in Figure \ref{fig:diffcomp}. The peak averaged zonal wind speed is similar in these cases to the simulation presented in Section \ref{sec:approx}, however some differences are apparent. The equatorial jet appears wider and penetrates deeper into the atmosphere in the case with weaker dissipation. 
Calculation of the phase curve indicates that the change in diffusion produces a small shift ($\approx 3^{\circ}$) in the hot spot offset. Thus this feature appears to be largely insensitive to the numerical dissipation, as also found in \cite{Heng2011b}.

\begin{figure*}
\includegraphics[width=0.5\textwidth]{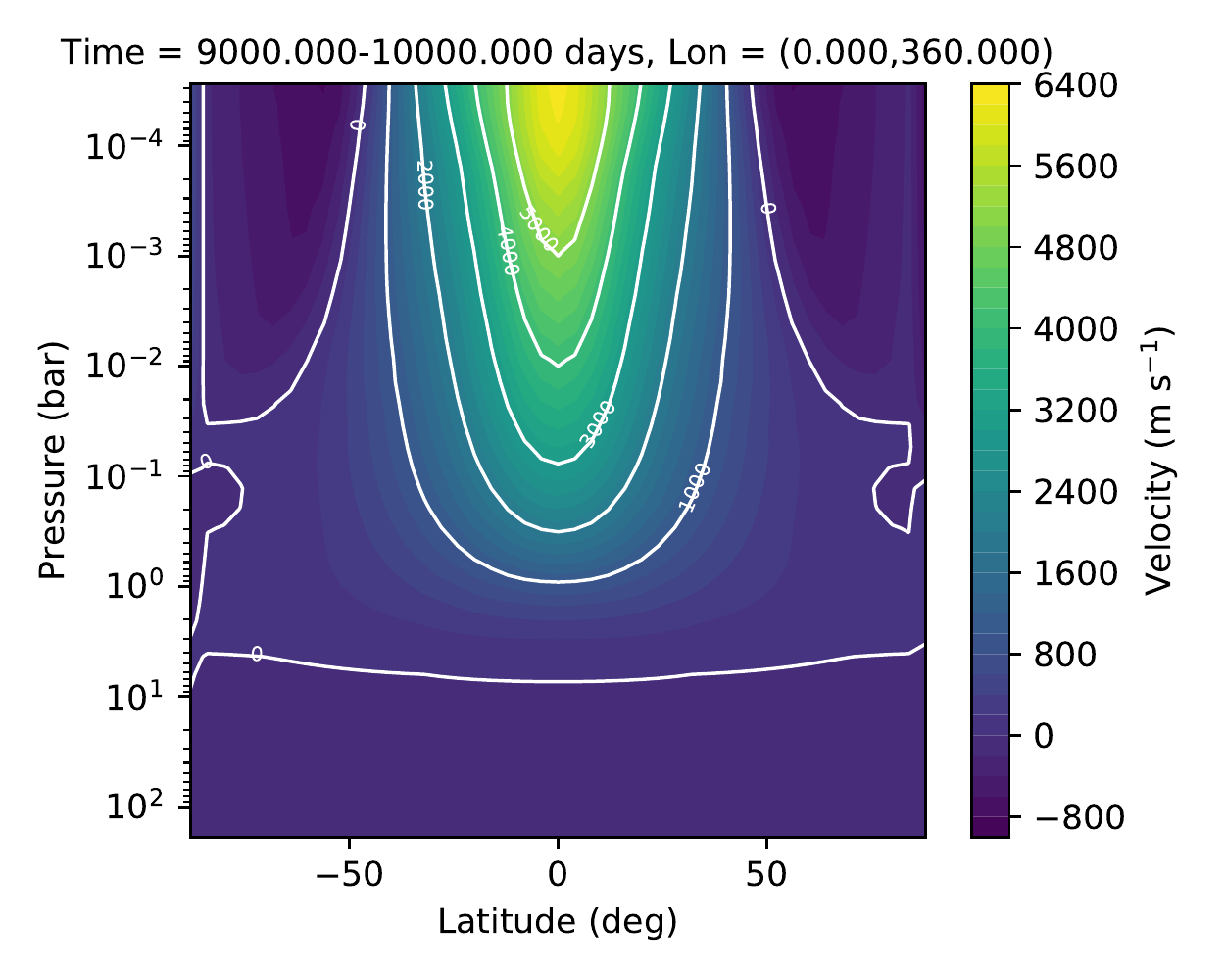}
\includegraphics[width=0.5\textwidth]{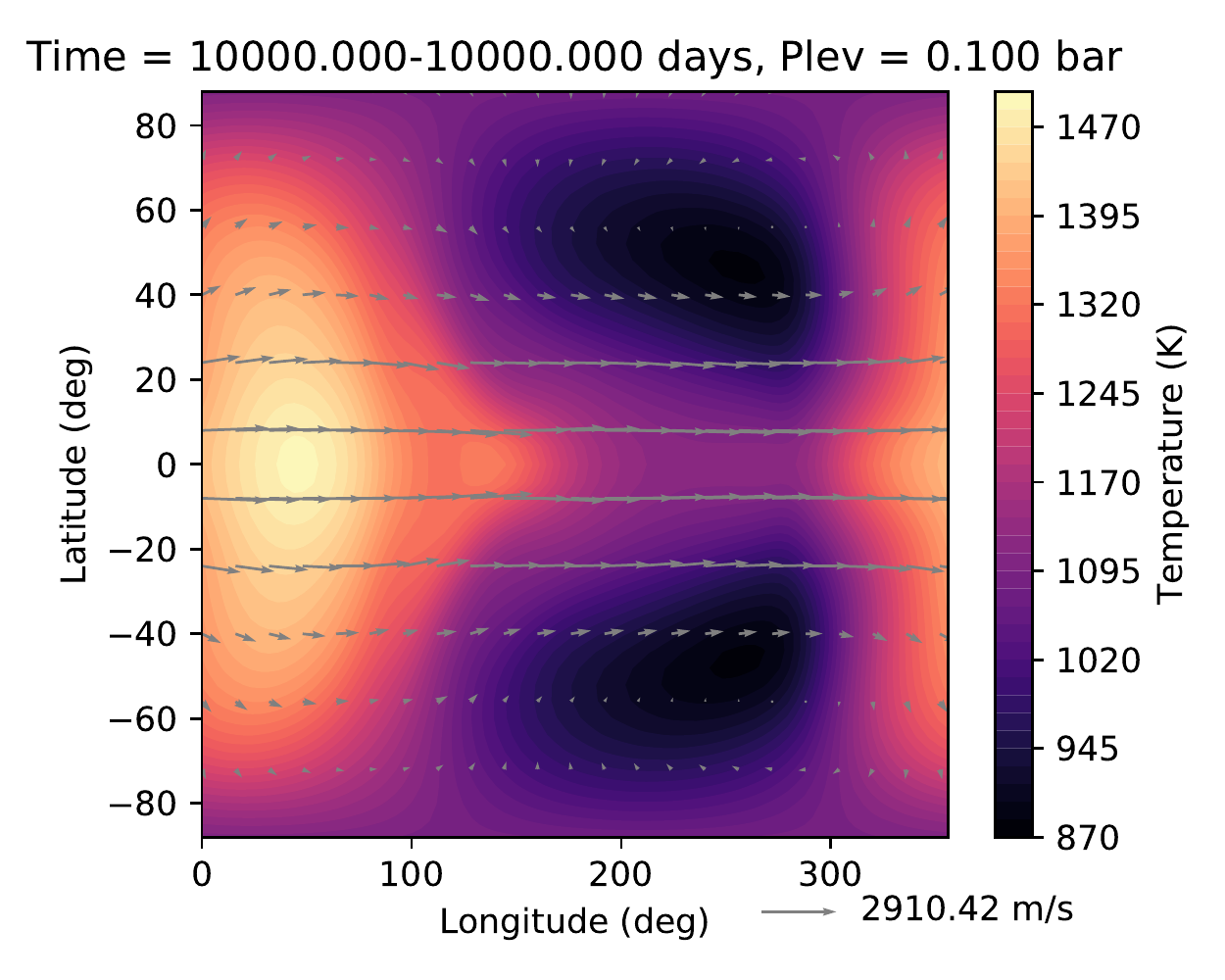}\\
\includegraphics[width=0.5\textwidth]{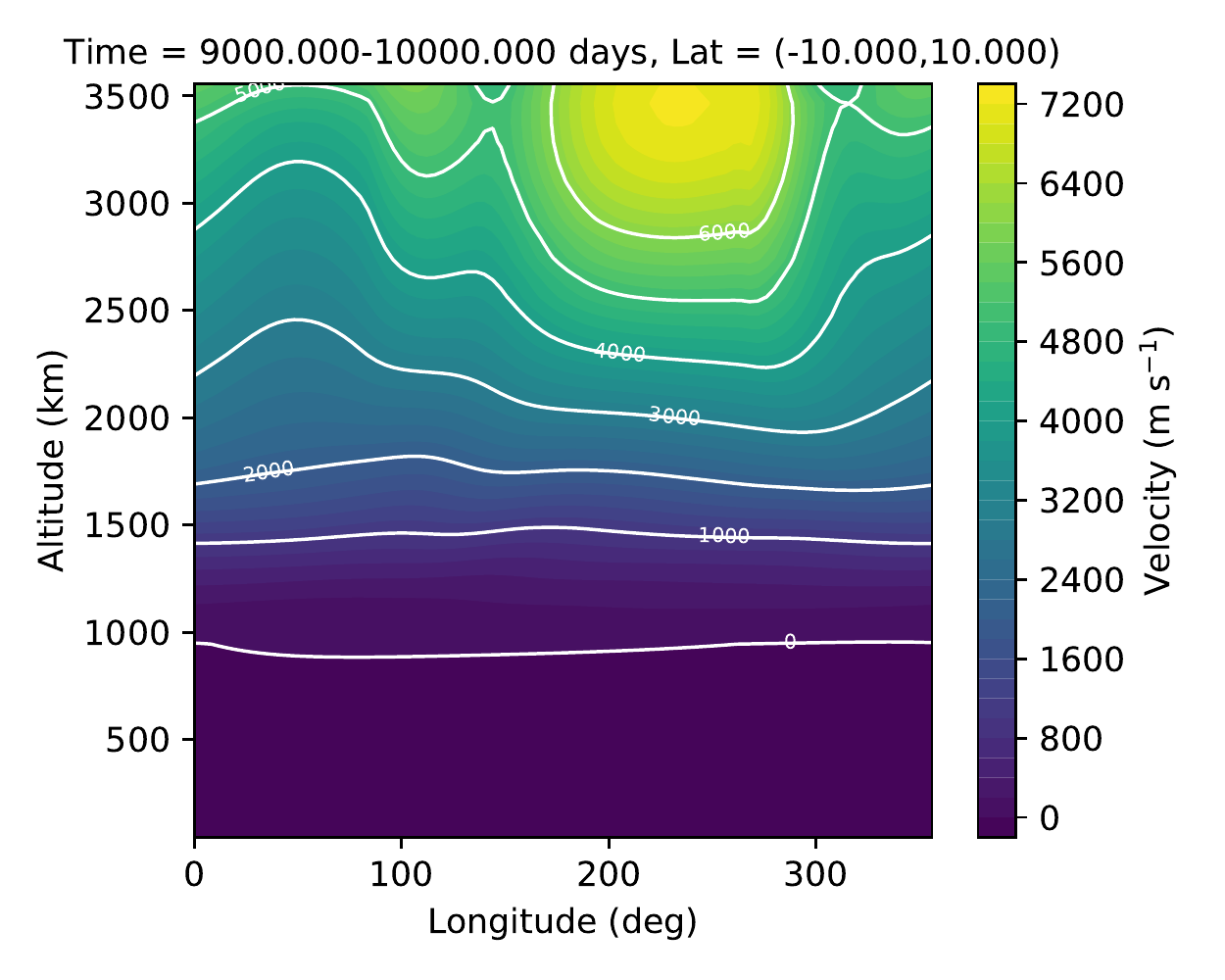}
\includegraphics[width=0.5\textwidth]{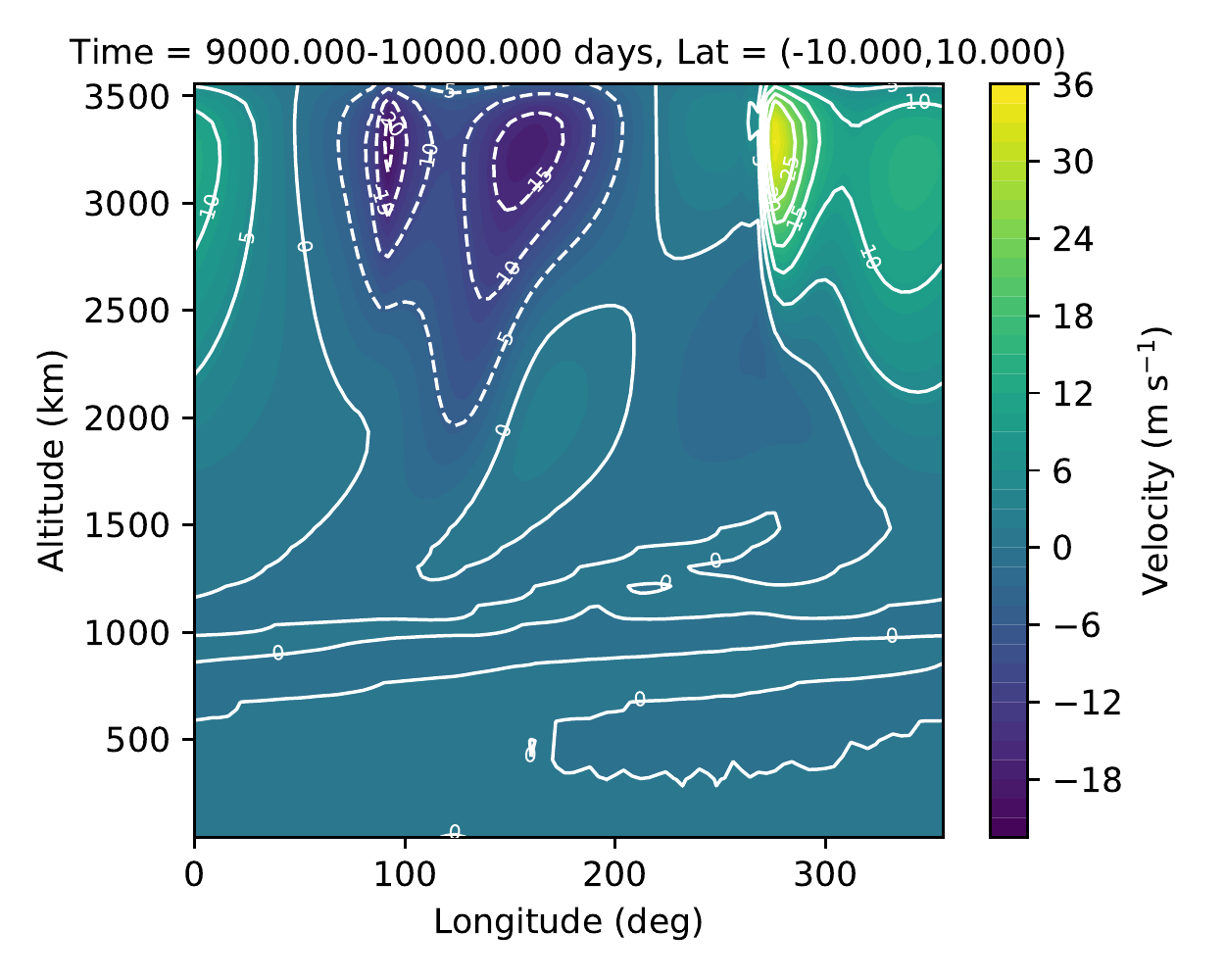}
\caption{Zonal-mean zonal velocity (upper left), temperature and horizontal winds at $P=0.1$ bar (upper right), zonal winds along the equator (lower left), and vertical winds along the equator (lower right), for HD 189733 b, when the sponge layer drag is removed after 5000 days. } \label{fig:rmsp189}
\end{figure*}

We have run one final simulation to test effect of sponge layer drag at the top boundary. In this simulation, the sponge layer was removed after 5000 days; everything is otherwise identical to the NHD case at $\sim 4^{\circ}$ resolution. The purpose here is to establish whether the sponge layer can be removed once the model has spun up and flow is established or if the model simply crashes. Then, if the model remains stable, we would like to see how the flow changes in the absence of this numerical damping. In this case, the model does not crash, and the results are shown in Figure \ref{fig:rmsp189}. 

The zonal mean zonal wind speed in this case is increased compared to the case in Figure \ref{fig:189g4eqtwind}. The primary contribution occurs on the night side of the planet. Further, the wind speed increases monotonically with height and the maximum appear right against the upper boundary. Otherwise, the basic structures of the flow remain similar to the case with the sponge layer continuously enabled. There are minor changes in the pattern of the vertical winds, because of the increased night-side zonal wind speed, though the main trends remain the same: downwelling occurs on the night side of the planet, upwelling occurs on the day side and is strongest at the western terminator.

\begin{figure*}
\includegraphics[width=\textwidth]{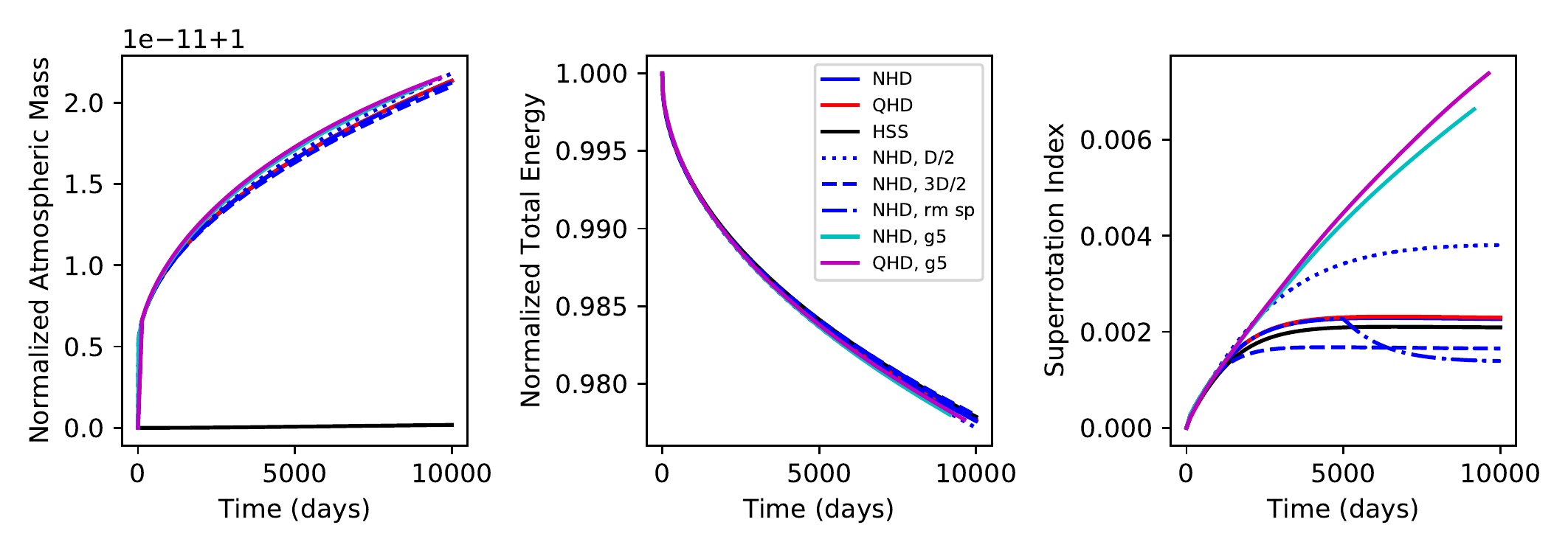}
\caption{Evolution of the total mass, energy, and superrotation index for the HD 189733 b simulations. Solid curves correspond to simulations with $D_{\text{hyp}}=D_{\text{div}} = 9.97\times 10^{-3}$. Dark blue lines are the NHD simulations at $\sim 4^{\circ}$ resolution, cyan is the NHD simulation at $\sim 2^{\circ}$, red and black are, respectively, the QHD and HSS simulations at $\sim 4^{\circ}$, and magenta is the QHD simulation at $\sim 2^{\circ}$. Three additional simulations are shown: NHD with $D_{\text{hyp}} = D_{\text{div}} = 4.99 \times 10^{-3}$ (dark blue, dotted), NHD with $D_{\text{hyp}}=D_{\text{div}} = 1.499 \times 10^{-2}$ (dark blue, dashed), and NHD with $D_{\text{hyp}}=D_{\text{div}} = 9.97 \times 10^{-3}$ and the sponge layer removed after 5000 days (dark blue, dash-dot). \label{fig:hd189glob}}
\end{figure*}

In Figure \ref{fig:hd189glob} we compare the evolution of global quantities for all of our simulations of HD 189733 b: the total atmospheric mass, the total energy (internal, kinetic and potential), and the superrotation index \citep{Read1986}. The superrotation index, a measure of the axial angular momentum in excess of the solid body rotation, is calculated at each point in time as 
\begin{equation}
    S_I = \frac{\sum_{i,j} \mathbf{l}_{ij} \cdot \mathbf{\hat{e}_3}}{\sum_{i,j} \mathbf{l}^{\text{sb}}_{ij}\cdot \mathbf{\hat{e}_3}} -1,
\end{equation}
where the angular momentum at the $i$th grid point and $j$th level, $\mathbf{l}_{ij}$, is given in Equation \ref{eqn:angmom}, and the solid body rotation angular momentum (\emph{i.e.}, the angular momentum at each location if the wind speed was zero) is 
\begin{equation}
\mathbf{l}^{\text{sb}}_{ij} = \rho_{ij} \mathbf{r}_{ij} \times (\mathbf{\Omega} \times \mathbf{r}_{ij}) V_{ij}.
\end{equation}

Ideally, the atmospheric mass should be conserved. In practice, there is a slow drift due to numerical errors. In all of our simulations of HD 189733 b, mass is conserved to a few parts in 10$^{11}$, as shown in Figure \ref{fig:hd189glob}. This is close to machine precision for hundreds of thousands of time steps. Conservation is more than an order of magnitude better in the HSS simulation---it would seem that there is some increased numerical error associated with the curvature components of the operators. In any case, the errors are reasonably small.

The total energy is not expected to be conserved over the entire simulation, but ideally would approach steady state values as the simulation advances. The main reason for this is that the initial conditions are not in radiative equilibrium and it takes many thousands of days of simulation time to bring the entire atmosphere to this equilibrium. In practice, energy is also continually lost because of the numerical dissipation \text{(see Section \ref{sec:awexp})}---as explained earlier, we do not artificially inject dissipated energy back into heat \citep[see e.g.,][]{Rauscher2012}. However, this error is likely to be small compared to the radiative imbalance. We have several key developments in progress to address energy conservation in \texttt{THOR}. First, we are developing better initial conditions, which will start the model closer to radiative equilibrium. Second, we are exploring the use of alternate forms of the thermodynamic equation, for example, the equation for total energy. We will present these developments in a future work.

Axial angular momentum would also be conserved in an ideal simulation. As with the mass and energy, numerical dissipation and integration errors lead to a gradual drift of the total axial angular momentum. This can be seen in the superotation index, which is a measure of the change in axial angular momentum in time. The drift in angular momentum is positive because the numerical dissipation is largest in the deep atmosphere, where the flow is retrograde \citep{Mendonca2019}. In all our cases, the total change is less than 10\%, which is acceptable in comparison with other GCMs \citep[see, e.g.,][]{Polichtchouk2014}. 

The superrotation index can be interpreted as a measure of convergence \citep{Mendonca2019}. This value plateaus (in log-space) as the model approaches steady state. In the low resolution simulations, this is reached at $\sim 2500-3000$ days. The superrotation index of the high resolution simulations continues to increase, indicating that after 10000 days, these simulations are still not fully converged, consistent with the fact that the equatorial jet continues to deepen in time. 

Interestingly, the NHD simulation in which the sponge layer is removed at 5000 days begins to lose angular momentum much faster than the others. It seems that allowing for greater reflection of waves of the top boundary causes an increase in the total dissipation occurring in the atmosphere. It may thus be more desirable to retain the sponge layer damping in such simulations, despite the fact that it is an additional artificial component to the model. 

\begin{figure*}
\centering
\includegraphics[width=0.5\textwidth]{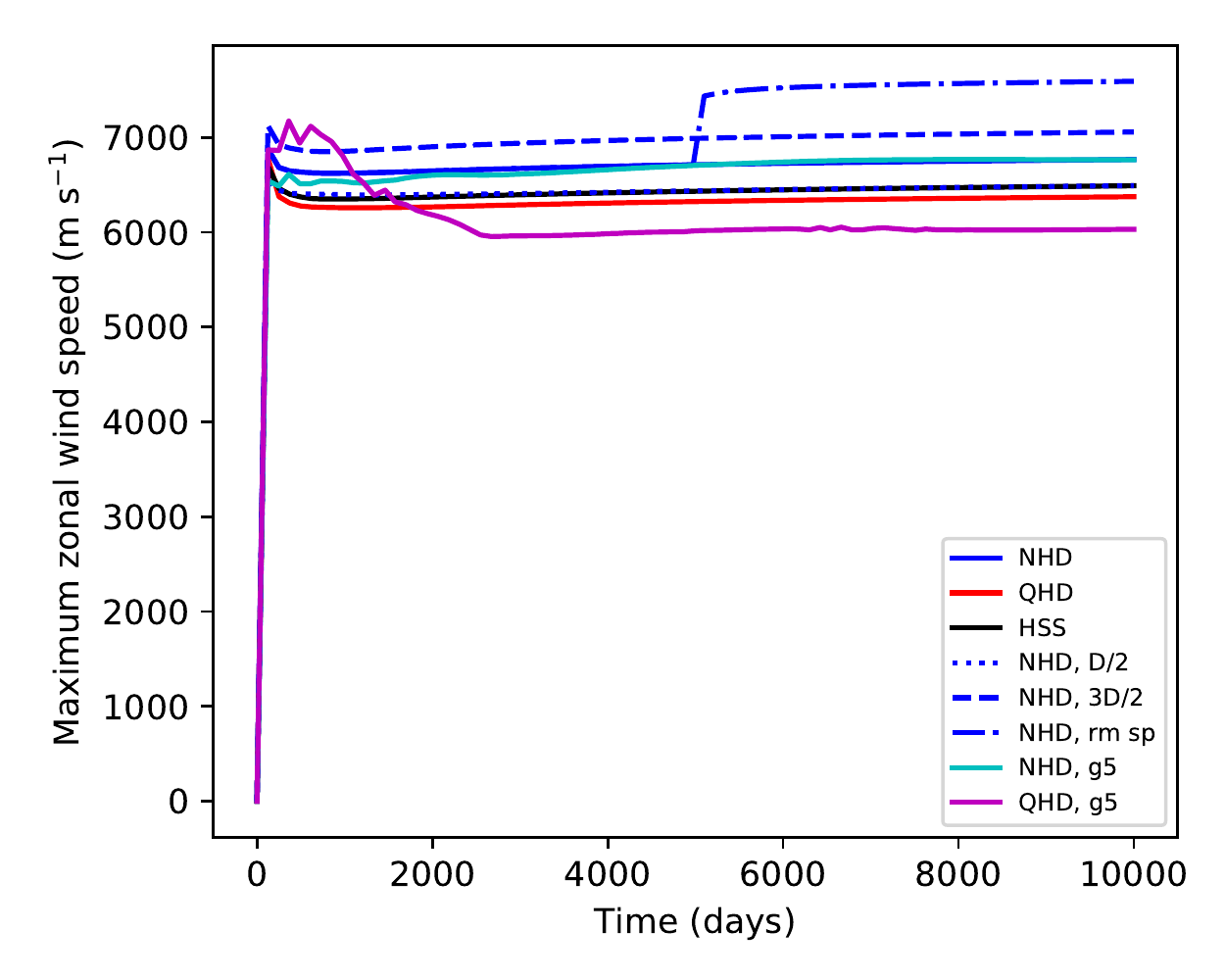}
\caption{Maximum zonal wind speed as a function of time for the HD 189733 b simulations. \label{fig:hd189umax}}
\end{figure*}

Figure \ref{fig:hd189umax} compares the maximum zonal wind speed across the suite of HD 189733 b simulations. The zonal flow develops very quickly in the upper atmosphere and changes very little after $\sim2000$ days in all simulations. In the $\sim2^{\circ}$ resolution ($g_{\text{level}} = 5$) simulations, the peak zonal winds are unaffected by the continued spin up of the deep regions, indicating that flow in the upper atmosphere is converged, even though the lower atmosphere is not.  

\section{Summary} \label{sec:summary}
Here, we have presented a suite of simulations using \texttt{THOR}. We have demonstrated how the model performs under a range of conditions. Despite the relative simplicity of the model (for a GCM), we reproduce important features of Earth's atmosphere, such as the temperature structure, zonal flow, and Hadley circulation. We have also reproduced the general features of several benchmarks for dynamical cores: a synchronously rotating Earth, a deep hot Jupiter, and wave tests previously presented in \cite{Tomita2004}. We can also reproduce the dominant features of hot Jupiter atmospheres, such as the day/night temperature contrast and the equatorial superrotation. 

The flexibility of \texttt{THOR} has allowed us to test the impact of commonly used approximations, like hydrostatic balance and shallow geometry, for an Earth-like case and a hot Jupiter. For the Earth-like case, the shallow approximation makes the largest difference in the results, though neither the QHD nor the HSS solution departs strongly from the full NHD simulation. While the approximations have minor consequences for the Earth-like case, we see a 15\%-20\% change in the peak zonal winds when the QHD approximation is made in our hot Jupiter case. Scale analysis indicates that the $Dv_r/Dt$ term (neglected in the QHD approximation) is four order of magnitude smaller than the pressure gradient and gravitational acceleration, suggesting that the QHD approximation is valid. Comparing the Earth-like and hot Jupiter cases, $Dv_r/Dt \sim 10^{-7}$ for Earth and $\sim 10^{-3}$ for HD 189733 b, while the dominant terms ($1/\rho~dP/dz$ and $g$) are $\sim 10$ m s$^{-2}$ for both planets. Thus while the QHD approximation is good in both cases, the error is larger for the hot Jupiter. We suspect that the neglect of this term changes the behavior of waves that transport angular momentum, resulting in a difference in zonal wind speed. The vertical velocities are relatively unchanged. 

We have also explored the consequences of numerical dissipation in our hot Jupiter case. Numerical dissipation makes a small difference in the overall zonal flow and the location of the peak in the thermal emission. Though the effects of modeling choices appear relatively minor, as observational data improves, it may be possible to constrain physical processes in the models, such as whether non-hydrostatic effects are significant enough to influence the bulk atmospheric circulation of hot Jupiters.

This is the first major upgrade to the open-source \texttt{THOR} GCM. The version consolidates physics modules that were developed in \cite{Mendonca2018a} and \cite{Mendonca2018b} and makes them available to the public, along with major improvements in code design, user friendliness, and plotting tools. The simulations presented in this work are not intended as a step forward in scientific knowledge, but as benchmarks or signposts for the \texttt{THOR} model. Our simulations of a dry Earth-like planet compare well with previously published works \citep{Merlis2010,Heng2011a,Heng2011b}. More realistic simulations of terrestrial planets will require a more sophisticated scheme for turbulence in the lower atmosphere \citep{Obukhov1971,Mellor1982,Galperin1988,Frierson2006}, a more realistic representation of convection \citep{Betts1986a,Betts1986b,Ding2016}, and the effects of condensation \citep{Frierson2007,Ogorman2008}. Our hot Jupiter simulations also compare well with prior works \citep{Showman2002,Cooper2005,Rauscher2010,Heng2011b,Mayne2017}, in that we observe equatorial super-rotation with wind speeds $\sim5$ km s$^{-1}$. We have not included algorithms capable of resolving shock formation in the atmosphere or sub-grid shear-driven turbulence \citep{Goodman2009,Li2010,Heng2012shock,Fromang2016}, clouds \citep{Heng2012cloud,Parmentier2013}, or magnetically-induced drag \citep{Perna2010,Rauscher2012,Menou2012,Batygin2013} that may be important in these planets. 


\appendix

\section{Deep hot Jupiter temperature profiles}
\label{sec:deephjapp}

The equilibrium profiles utilized in Section \ref{sec:deephj} and shown in Figure \ref{fig:deephjprof}, for the deep hot Jupiter benchmark test are given by:

\begin{equation}
    T_{\text{night}} = \begin{cases}
                    \max(T_{\text{night}}^{\star} e^{0.1(\log{P}-\log{P_l})},250 \text{ K}) & P<P_l,\\
                    T_{\text{night}}^{\star} & P > P_l,
                    \end{cases}
\end{equation}
and
\begin{equation}
    T_{\text{day}} = \begin{cases}
                    \max(T_{\text{day}}^{\star} e^{0.015(\log{P}-\log{P_l})},1000 \text{ K}) & P<P_l,\\
                    T_{\text{day}}^{\star} & P > P_l,
                    \end{cases}
\end{equation}

where $P_l = 1$ mbar. Polynomials for $T_{\text{night}}^{\star}$ and $T_{\text{day}}^{\star}$ in the equations above are

\begin{align}
    \begin{aligned}
    T_{\text{night}}^{\star} = & ~1388.77348  + 279.575848 \tilde{P} - 213.835822 \tilde{P}^2 + 21.0010475 \tilde{P}^3 + 100.938036 \tilde{P}^4 \\
    &+ 12.7972336\tilde{P}^5- 13.9266925 \tilde{P}^6  - 3.70783272 \tilde{P}^7 + 0.522370269 \tilde{P}^8 \\
    & + 0.320837882 \tilde{P}^9  + 0.0451831612 \tilde{P}^{10} + 2.18195583\times10^{-3} \tilde{P}^{11}\\
    &+ 3.98938097\times10^{-6} \tilde{P}^{12},
    \end{aligned}
\end{align}
and
\begin{align}
    \begin{aligned}
    T_{\text{day}}^{\star} = &~ 2152.06036 + 29.3485512 \tilde{P} - 183.318696\tilde{P}^2 + 46.3893130 \tilde{P}^3 + 19.8116485 \tilde{P}^4\\
    &- 28.5473177\tilde{P}^5 - 2.52726545 \tilde{P}^6+ 8.43627538\tilde{P}^7 + 2.62945375\tilde{P}^8\\
    &- 0.297098168\tilde{P}^9- 0.286871487 \tilde{P}^{10} - 0.0590629443 \tilde{P}^{11}\\
    &- 5.38679474\times10^{-3} \tilde{P}^{12} - 1.89972415\times10^{-4}\tilde{P}^{13},
     \end{aligned}
\end{align}
where $\tilde{P} = \log{(P/1\text{ bar})}$.

\section{Code improvements} \label{sec:codeimp}

We have implemented a number of coding improvements since the original release of \texttt{THOR}. Chief amongst these is the insurance of binary reproducibility, \emph{i.e.}, separate runs using identical initial conditions will produce identical results down to machine precision. Briefly, we describe coding procedures that ensure this property. 

The primary change is the elimination of atomic addition. Atomic addition can be used in CUDA code to ensure that parallel threads operating on the same data (and thus the same memory location) do not interfere with each other. The problem with standard addition in parallel is that reading and writing are done as separate operations, so that if multiple threads read and write to the same location, the threads may read the data at the same time (thus beginning operations with the same values) but only the last thread to write will update the sum---this is known as a ``race condition''. Atomic addition ensures that reading and writing are treated as a single operation, thus threads do not interfere with each other and the computation can be done correctly.

The trouble with atomic addition is not that it is inaccurate, it is simply that there is no way for the machine to guarantee the order that the threads operate. Thus, because of machine round-off error, one may get a slightly different end result each time the code is run, depending on the pseudo-random order in which the threads perform the operations. Further, atomic addition can result in code slow-down as threads are forced to wait for other threads to finish their operations. 

\begin{figure*}
\includegraphics[width=0.51\textwidth]{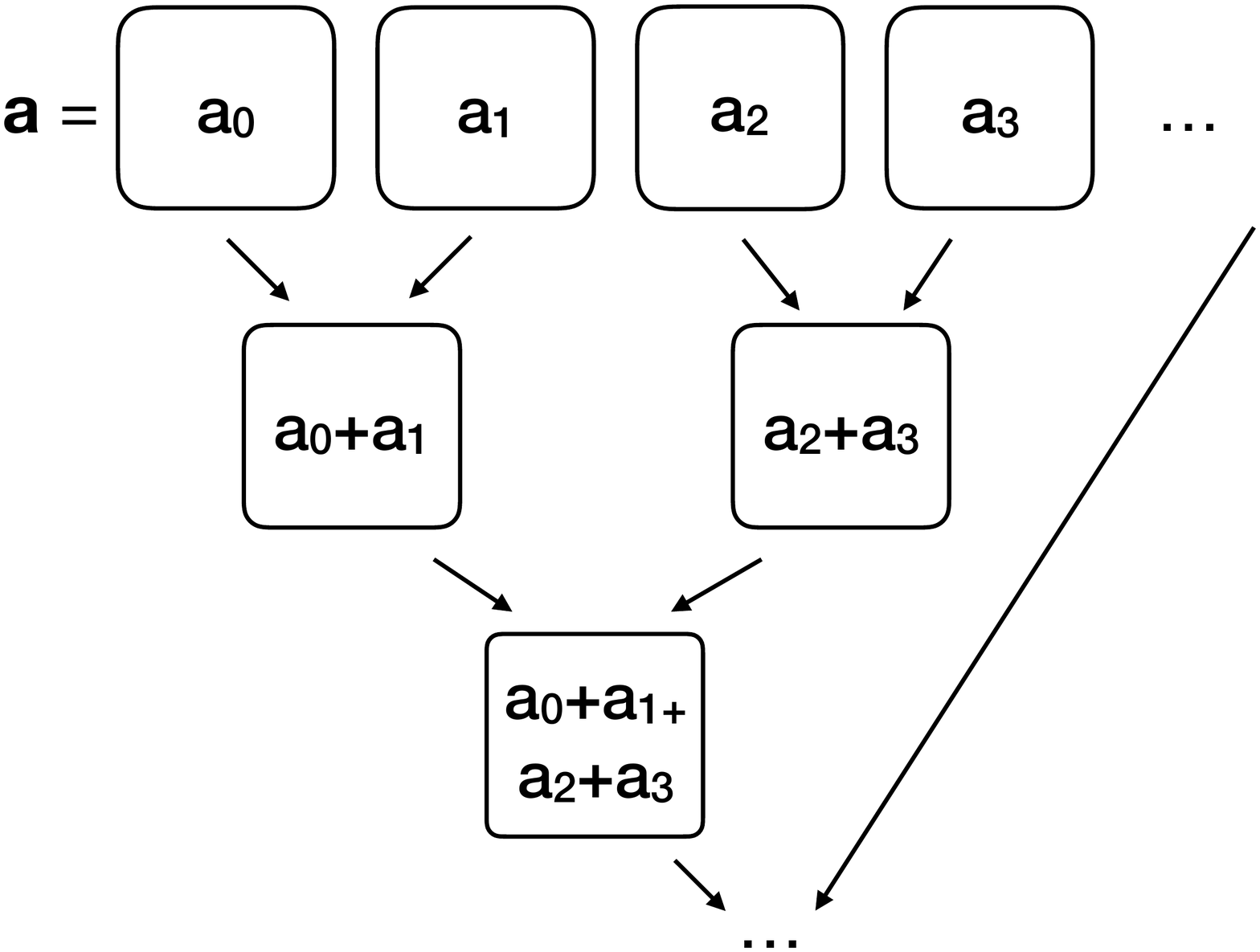}
\includegraphics[width=0.49\textwidth]{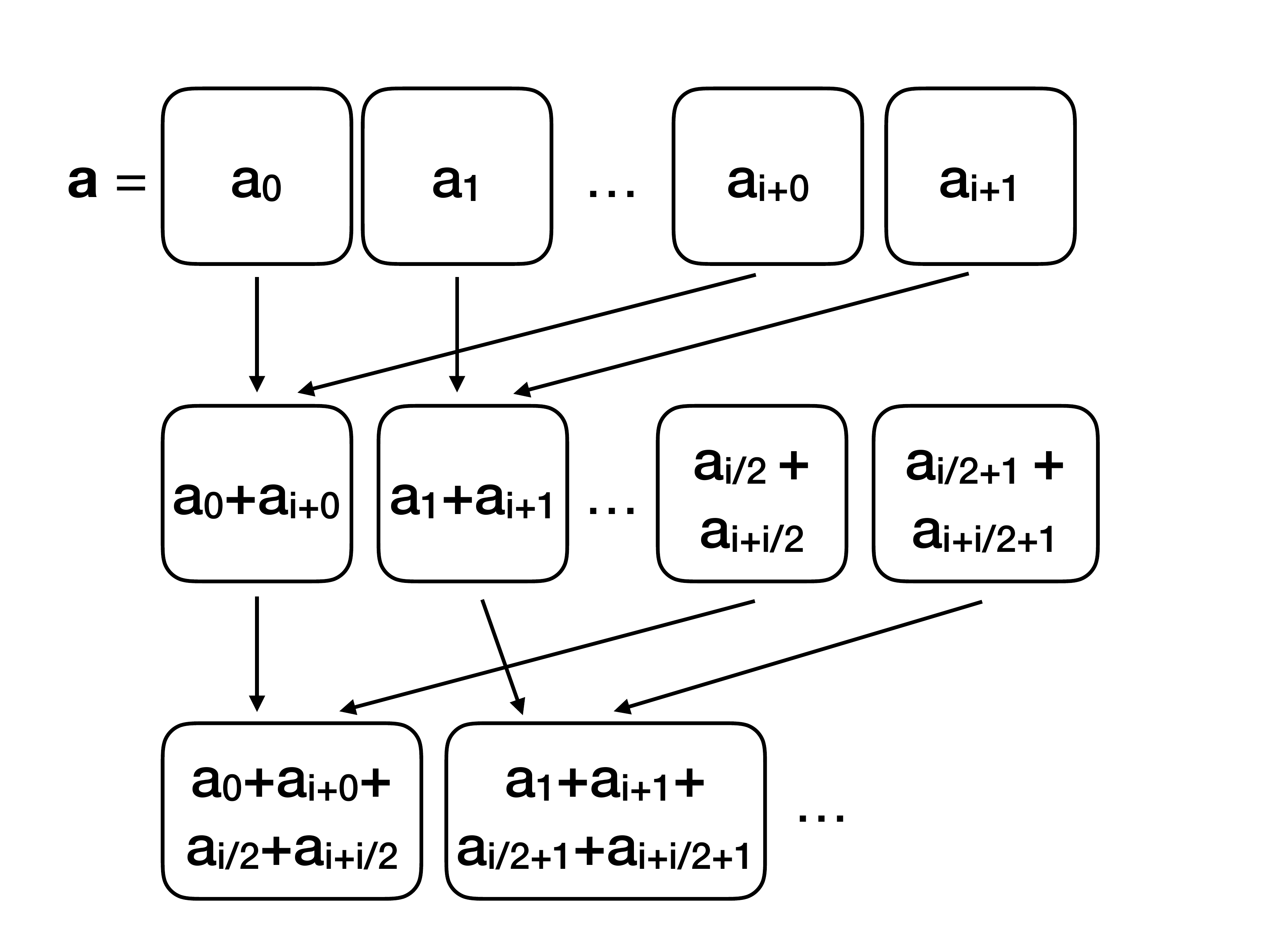}
\caption{The concept of reduction addition used for summing over an array. Left: pairs of array elements are added iteratively, reducing the size of the array at each iteration, until one element remains (which is the total sum). Right: reduction addition on one block of the GPU adds the array elements separated by $i$, where $i$ is a large power of 2, until the only total in each block remains. The totals from the blocks are then summed on the CPU to produce the total sum of the array.} \label{fig:redadd}
\end{figure*}

The alternative, which ensures that the summation is done correctly and that the order of operations is always the same, is ``reduction addition'' \citep[see, for example, Appendix A.1 of][]{SandersBook}. The concept is illustrated in Figure \ref{fig:redadd} (left). For an array $\mathbf{a}$ consisting of elements $a_0, a_1, a_2, ..., a_n$, we first add pairs of elements (not necessarily adjacent as shown in Figure \ref{fig:redadd}), resulting in an array with half the length of the original. Then, we again sum pairs of elements. The process can be repeated until the one element remains, which is final sum of the array. 

 Parallelizing the process on a GPU is somewhat more complicated. First, we subdivide the array of length $n$ into subarrays of length $2i$, where $i$ must be a power of two. The block-size on the GPU is then $i$. The number of blocks must be long enough to cover the entire array, that is, $2*N_{blocks}*i \geq n$. For $2*N_{blocks}*i > n$, the blocks extend past the length of the array, so the end of the array is padded with zeros. Each block will execute $i$ threads which each add two numbers (not adjacent pairs, in this case), $a_j + a_{i+j}$, where $j$ represents the thread number. On each block, we then have an array of length $i/2$. We then execute $i/2$ threads in each block which add the pairs of numbers ${a}'_j+{a'}_{i/2+j}$, where ${a}'$ is the new array, and repeat until only one element is left in each block, dividing the array length by two each iteration. Finally, the sum from each block is added sequentially on the CPU to produce the final summation of the array. The first two iterations of this process are illustrated in Figure 1 (right). 

In \texttt{THOR}, reduction addition is used, for example, in the computation of global quantities (Section \ref{sec:newphys}) and of the zonal averages used in sponge layer damping (Section \ref{sec:diff}). Parallel reduction is performed in $\log_2(n)$ steps.

Additional code improvements include:
\begin{enumerate}
    \item Physical processes independent of the dynamical core (fluid equations) have been modularized. Currently, the physics module consists only of gray radiative transfer, but future releases will also contain chemical tracers and boundary layer drag (in development). The purpose of the module structure is to facilitate coupling with external models or to allow the development of alternatives for the same physical processes. In the near future, this will be used to couple \texttt{THOR} with the radiative transfer model \texttt{HELIOS} \citep{Malik2017,Malik2018}. Each module is given its own set of configuration options and outputs, and is responsible for reading and writing these. At designated points throughout primary code of \texttt{THOR}, the code checks to see whether any physics modules need to be called, and the necessary values (state variables) are passed to these modules. The dynamical core then receives back a set of fluxes that are incorporated into the fluid equations. This structure allows for modules to be modified, omitted, or replaced without the need to modify the primary code of \texttt{THOR}.
    
    \item Initial conditions are now set entirely in external configuration files. The previous version of the model required separate compilations for every change to the initial conditions, no matter how minor. Basic parameters, such as the physical characteristics of a planet or modelling choices, are set in a text file, while more involved properties, such as a non-isothermal T-P profile or a 3-D wind field, can be specified by modifying binary input files using Python code in our repository. 
    
    \item Python code for plotting and regridding the icosahedral grid is now included in the repository. This code utilizes the \texttt{NumPy} \citep{Oliphant2006guide}, \texttt{SciPy} \citep{Jones2001}, \texttt{Matplotlib} \citep{Hunter2007matplotlib}, \texttt{h5py}\footnote{https://www.h5py.org}, \texttt{PyCUDA} \citep{Kloeckner2012}, and \texttt{PySHTools} \citep{Wieczorek2018shtools} libraries. In order to produce data that plotting algorithms can utilize, it is necessary to interpolate from the icosahedral grid onto a latitude-longitude grid. This is done utilizing our own \texttt{PyCUDA} implementation of the M\"oller-Trumbore algorithm \citep{Moller1997} and properties of the icosahedral grid indexing. If desired, the vertical coordinate can be changed from altitude to pressure via interpolation, prior to the horizontal regridding, such that the horizontal interpolation is done along isobars. We often use this feature in this work.
    
    \item Compilation of the model is now more flexible and user friendly. The compiling process can now auto-detect the necessary GPU specifications and handles dependencies in a more robust fashion. The location of the HDF5\footnote{https://www.hdfgroup.org} libraries, for example, can be auto-detected. 
    
    \item A number of debugging and performance testing tools are now built into the model and are selectable at compile time. These can be used to test for binary reproducibility or to check the magnitude of differences produced by code changes. 
\end{enumerate}

\acknowledgements

We acknowledge financial support
from the Swiss National Science Foundation, the European
Research Council (via a Consolidator Grant to KH; grant
number 771620), the PlanetS National Center of Competence
in Research (NCCR), the Center for Space and Habitability
(CSH) and the Swiss-based MERAC Foundation.

We would also like to thank Daniel Kitzmann, Yohai Kaspi, Jonathan Mitchell, Graham Lee, Mark Hammond, 
and Nathan Mayne for helpful discussions. Enormous thanks to the anonymous referee 
for the extremely constructive review. Thanks also to Brett Morris for help with and management of the Github repository.

\bibliographystyle{aasjournal}
\bibliography{thorv2}
\end{document}